\documentclass[rmp,aps,twocolumn,showpacs,nofootinbib]{revtex4}
\usepackage{amsmath}  
\usepackage{graphicx} 
\usepackage{epsfig}
\usepackage{dcolumn}  
\usepackage{bm}       

\def\ii{{\rm i}} \def\ee{{\rm e}} \def\vb{{\bf v}} \def\jb{{\bf j}}
\def\Eb{{\bf E}} \def\Rb{{\bf R}} \def\rb{{\bf r}}  \def\Bb{{\bf B}}
\def\zt{{\hat{\bf z}}} \def\Rt{{\hat{\bf R}}} \def\nt{{\hat{\bf n}}}
\def\qb{{\bf q}} \def\pb{{\bf p}} \def\gb{{\bf g}} \def\Hb{{\bf H}}
\def\Fb{{\bf F}} \def\Ab{{\bf A}} \def\sb{{\bf s}} \def\hb{{\bf h}}
\def\Db{{\bf D}} \def\ft{{\hat{\bf \varphi}}}  \def\fb{{\bf f}}

\def\QQ{{q_\parallel}}
\def\QQv{{{\bf q}_\parallel}}
\def\ksp{{q_\parallel^{\rm SP}}}
\def\th{{\hat{\bf \varphi}_{\bf q}}}
\def\fh{{\hat{\bf \varphi}}}  \def\Rh{{\hat{\bf R}}}  \def\fb{{\bf f}}
  \def\rh{{\hat{\bf r}}}

\def\kb{{\bf k}}
\def\xx{\hat{\bf x}}
\def\yy{\hat{\bf y}}
\def\zz{\hat{\bf z}}

\begin{document}
\title{Optical excitations in electron microscopy}

\author{F. J. Garc\'{\i}a de Abajo}

\affiliation{Instituto de \'Optica - CSIC, Serrano 121, 28006 Madrid, Spain}

\email{jga@cfmac.csic.es}

\begin{abstract}
This review discusses how low-energy, valence excitations created by swift electrons can render information on the optical response of structured materials with unmatched spatial resolution. Electron microscopes are capable of focusing electron beams on sub-nanometer spots and probing the target response either by analyzing electron energy losses or by detecting emitted radiation. Theoretical frameworks suited to calculate the probability of energy loss and light emission (cathodoluminescence) are revisited and compared with experimental results. More precisely, a quantum-mechanical description of the interaction between the electrons and the sample is discussed, followed by a powerful classical dielectric approach that can be in practice applied to more complex systems. We assess the conditions under which classical and quantum-mechanical formulations are equivalent. The excitation of collective modes such as plasmons is studied in bulk materials, planar surfaces, and nanoparticles. Light emission induced by the electrons is shown to constitute an excellent probe of plasmons, combining sub-nanometer resolution in the position of the electron beam with nanometer resolution in the emitted wavelength. Both electron energy-loss and cathodoluminescence spectroscopies performed in a scanning mode of operation yield snap shots of plasmon modes in nanostructures with fine spatial detail as compared to other existing imaging techniques, thus providing an ideal tool for nanophotonics studies.
\end{abstract}
\pacs{68.37.Lp,79.20.Uv,78.60.Hk,73.20.Mf}
\date{\today}
\maketitle \tableofcontents



\section{Introduction}
\label{introduction}

Investigating minute physical processes has been essential for
advancing knowledge and generating the right questions all the way
from the beginning of modern science up to recent developments in
nanotechnology. Far-field optical microscopes have contributed to
this end, but they are limited by diffraction to a
spatial resolution of about half the light wavelength in practice.
Actually, the best resolution ($<1$\,{\AA}) is currently achieved using
electrons rather than light \cite{R1987_2,NCD04}, for which the de
Broglie wavelength \cite{L1925} is well below $0.1$\,{\AA} at
energies above 20 keV.

\begin{figure}
\includegraphics[width=80mm,angle=0,clip]{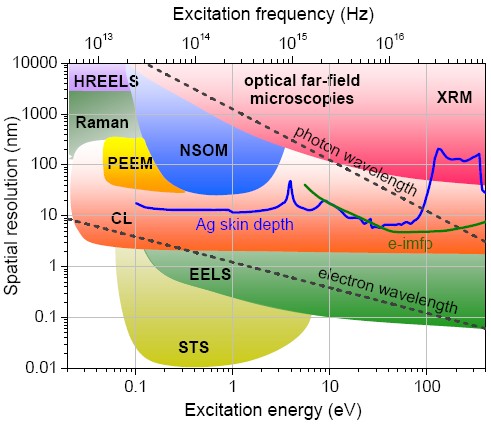}
\caption{\label{techniques} (Color online) Atlas of spatially-resolved spectroscopy techniques, organized according to their space and energy resolution (see Appendix\ \ref{listofsymbols} for a list of acronyms). The relation between wavelength and energy is represented by dashed curves for photons and electrons. The universal electron inelastic-mean-free-path (e-imfp) is given as a function of electron energy. The skin depth of Ag is calculated from optical data \cite{P1985}.}
\end{figure}

Besides acquiring static images of the nanoworld, we are interested
in finding out about the excitations that small objects can sustain,
which inform us on their dynamical evolution and are relevant for
encoding and manipulating information and for exploring a myriad of
applications in fields such as molecular biology. Thus, our ultimate goal is to
perform spectroscopy at the shortest possible length scale with the
highest energy resolution. With this focus in mind, we can classify
the available experimental techniques as shown in
Fig.~\ref{techniques}, clearly indicating that electron-based
spectroscopies offer the best choice for optimizing spatial resolution.

Two routes have been devised so far for dealing with the diffraction
limit:\footnote{The existence of a diffraction limit has been questioned by recent investigations
of super-oscillating functions \cite{Z08}.}
(1) to reduce the wavelength of the probe and (2) to employ
near-field detection techniques. (1) Moving to shorter photon
wavelengths is difficult due to the lack of versatile lenses and
mirrors beyond the ultraviolet. Nonetheless, recent advances in
x-ray microscopy (XRM) have allowed 15\,nm imaging resolution
\cite{CHL05} at energies above 250\,eV. Alternatively, shifting from
photons to electrons of the same energy encompasses a significant
decrease in wavelength and improved spatial resolution.\footnote{Heavier particles such as protons have been argued to provide good spatial resolution compared to electrons \cite{DM04}. Actually, a helium ion microscope has been recently released \cite{WNE06} and achieved 2.4\,{\AA} resolution (Carl Zeiss 2008 announcement).}
(2) Exploiting the near field is another option, particularly when relying on evanescent
components, the fast decay of which provides an extra handle for enhancing
resolution using localized probes. For instance, near-field scanning
optical microscopy (NSOM, based on a subwavelength tip at the end
of a fiber that brings, collects, or scatters light) and tip-enhanced NSOM can push
spatial resolution down to tens of nanometers \cite{BFW92,H08}. Even
better detail in the sub-{\AA} domain is routinely achieved
using the formerly developed scanning tunneling microscope (STM), on
which scanning tunneling spectroscopy (STS) yields 0.1 eV energy
resolution \cite{H94}. However, STS only works with metals, and its
extreme spatial accuracy ($\sim 0.01$\,{\AA} in the vertical
direction) requires close proximity between the tip and the sample surface, so that single-electron
excitations dominate the spectra and mask collective modes relevant
to optics, such as plasmons.

Electron microscopes are thus the best option for resolving both
localized and extended excitations with sub-nanometer spatial detail and $<0.1\,$eV
energy resolution in any type of material \cite{LBZ06}. These instruments
are sufficiently versatile to be surface-sensitive and to simultaneously procure
information on bulk properties. Their performance has
considerably improved in recent years due to extraordinary advances in energetic
electron sources and optics. A number of studies in different fields
benefit from the unprecedented spatial resolution of electron microscopy, which no other
technique can currently match.

\begin{figure}
\includegraphics[width=80mm,angle=0,clip]{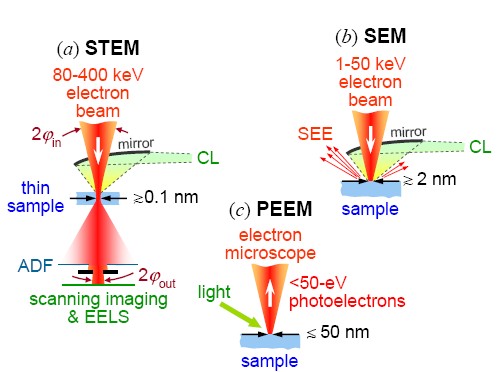}
\caption{\label{EM} (Color online) Schematic
representation of various types of electron microscopes equipped to perform
spectroscopy. {\bf (a)} Scanning transmission electron microscopes (STEMs)
allows analyzing the energy distribution of electrons transmitted
through a thin specimen ($\lesssim 100$\,nm thick), on which the incident
beam is focused down to a $<0.1$\,nm spot. Large-momentum transfers are
collected through annular dark-field detectors (ADFs).
Cathodoluminescence (CL) light detection is also possible on a STEM. {\bf (b)}
Scanning electron microscopes (SEMs) collect secondary electrons (SEs) to
form images or CL emission to perform spectroscopy, with spatial
resolution down to $\sim 2$\,nm. {\bf (c)} Photoemission electron microscopes
(PEEMs) can render spectroscopic information on the photo-excitation
process and electron-hole (e-h) dynamics with $\lesssim 50$\,nm
resolution.}
\end{figure}

Electron microscopy has different flavors depending on the type of
signal that we measure [inelastic losses, cathodoluminescence (CL) light emission,
secondary electrons (SEs), etc.]. Several widely available types of setups are
schematically represented in Fig.~\ref{EM}. The transmission
electron microscope (TEM) provides by far the most powerful
combination of spectral and energy resolution, mainly via analysis
of loss events. This type of microscope can operate like
a conventional optical microscope, projecting a bright image of an
object on a phosphor screen or a CCD camera to picture the magnified sample.
Energy-filtered TEM (EFTEM) allows constructing images out of those
electrons that have donated a determined amount of energy to the specimen,
which is useful to visualize selected losses [for
instance, core excitations identifying the chemical environment of
atomic species \cite{B93}]. Excellent spatial resolution is obtained
using TEMs operated in scanning mode (STEM), in which the beam is
focused and scanned over the sampled area, either to form
images out of large-momentum transfers recorded by annular dark-field
electron detectors or to collect loss spectra at specific locations.

Most TEMs use energetic electrons in the range of 50-400 keV and require
very thin specimens ($\lesssim 100$ nm) that are relatively transparent to
these electrons. Sample preparation is thus an important issue for
achieving optimum spatial resolution.

In contrast, scanning electron microscopes (SEMs) can work with any
sample that is covered by a thin metal layer ($\sim 1-2$ nm),
forming images by collecting SEs upon bombardment
with a primary beam of $\gtrsim 2$ nm in diameter. Furthermore,
one can perform spectroscopy on SEMs through wavelength-resolved
CL emission. Although CL spectroscopy can be also
collected on a TEM \cite{YST96}, it is most commonly available in
SEMs equipped with a light collector system (e.g., an ellipsoidal
mirror) and a photon spectrometer.

Low-energy electron microscopy (LEEM) \cite{R95} relies on the use of 1-100\,eV electrons and can achieve $<50\,$nm spatial resolution accompanied by sub-electronvolt energy resolution. As an example of application, LEEM has been recently used to observe acoustic plasmons \cite{DPV07}. In contrast, photoemission electron microscopy (PEEM) \cite{B94} is particularly useful to map the light intensity in illuminated nanostructures, as recently demonstrated for lithographically patterned Ag nanoparticle arrays \cite{CGN05,paper120}. The femtosecond dynamics of plasmon excitations has been also resolved using multiphoton PEEM with pump-probe illumination \cite{KOP05,KPP07}.

Finally, there are other configurations that make use of electrons to perform spectroscopy. For instance, reflection electron energy-loss spectroscopy (REELS) has been recently used to determine optical properties of noble metals \cite{W06,WWV07,WVW08} and graphite \cite{CFF08} after careful data analysis.

\subsection{Spectroscopy using electron microscopes}
\label{spectroscopyusing}

A swift electron impinging on a microscope specimen
gives rise to secondary electron emission (SEE) and CL that can be
energy-analyzed, thus yielding information on the excitation
spectrum of the sample. Electron energy-loss spectroscopy (EELS)
performed in STEMs adds up to this suite of
probes. The main advantage of electron microscopes is that these
types of spectroscopies can be performed with truly nanometer
spatial resolution.

Unlike freely propagating light, the bare field of an electron is evanescent, as we show in Sec.\ \ref{anevanescentsource}. This is advantageous to investigate localized excitations, involving wave-vector components of the electromagnetic field that lie outside the light cone. For example, electron beams were instrumental in discovering and characterizing collective excitations of conduction electrons in metals \cite{R1948,W1956,R1957,PS1959,CS1975_2,R1980}, known as plasmons because they are quasi-particles associated to oscillations of the electron gas {\it plasma}. Specifically, bulk plasmons in a homogenous metal are conspicuous in EELS, since they are electrostatic and longitudinal in nature,\footnote{Bulk plasmons in a source-free metal are characterized by zero magnetic field and longitudinal electric field ($\nabla\times\Eb=0$), so that they trivially satisfy Maxwell's equations under the condition of vanishing permittivity. In contrast, surface plasmons are confined to metal-dielectric interfaces, they involve nonzero magnetic fields, and they have transversal character ($\nabla\cdot\Eb=0$ in each homogeneous region of space separated by the interface on which plasmons are defined).} and for this reason they couple very efficiently to moving charges.

Similarly, plasmons can be confined at the surface of a metal, and they are actually the source of interesting phenomena and applications that configure the field of plasmonics (see Sec.\ \ref{prospectsforplasmonics}). Surface plasmons (SPs) are versatile entities: they can be trapped in metal particles that are much smaller than the wavelength \cite{paper155}, and they can hybridize with light extending over larger metallic structures \cite{CNB01}. More precisely, surface plasmon polaritons (SPPs), which are a subset of SPs capable of propagating on planar surfaces \cite{BDE03,ZSC06,O06} or along 1D waveguides \cite{BVD06}, hold great promise to become the natural link between current nanoelectronics and future integrated nanophotonics, operating at frequencies that are $\sim 10^5$ times higher than microchip clocks.

\begin{figure}
\includegraphics[width=80mm,angle=0,clip]{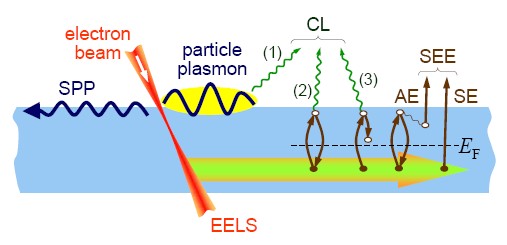}
\caption{\label{EM-2} (Color online) Schematic representation of some of the excitation processes triggered in a solid by a swift passing electron, and their connection to measured signals in electron microscopes: CL, secondary electron
emission (SEE), and electron energy-loss spectroscopy (EELS). The electron can launch propagating surface plasmon polaritons (SPPs; see Sec.\ \ref{plasmonemission}). It can also produce surface plasmons localized in nanoparticles or in surface features (Sec.\ \ref{prospectsforplasmonics}). Localized surface plasmons typically decay by coupling to radiation, thus giving rise to contribution (1) to CL (Sec.\ \ref{eire}). Electronic excitations in the sample (e.g., e-h pairs) can decay radiatively, back to the initial state [coherent CL emission (2), automatically included in the random-phase approximation (RPA) dielectric function as a bubble diagram \cite{L1954}; see Sec.\ \ref{eire}], or to a different excited state (incoherent CL emission; Sec.\ \ref{incoherentluminescence}). In metals, secondary electronic excitations constitute the dominant decay channel, producing Auger electrons (AEs), sometimes above the vacuum level, so that they contribute to the detected SEE. Direct excitation of electrons from the target is also possible to produce {\it true} SEs.}
\end{figure}

As illustrated in Fig.~\ref{EM-2}, the passage of a fast electron can excite localized plasmons in metallic nanoparticles, but it can also launch SPPs in planar metallic surfaces \cite{H1977,BJK06,VVP06} or in metal nanowires \cite{VDK07}. Quite different from SPPs, localized plasmons can decay radiatively, thus contributing to CL, although all sorts of plasmons can partially decay via inelastic channels that involve electronic excitations, including electron-hole (e-h) pair creation and SEE if the electron is near the surface and its energy above the vacuum level. We discuss this matter more thoroughly in Sec.\ \ref{mechanismsoflight}.

The electron can directly excite e-h pairs too, the decay of which gives rise to the emission of CL and SEs. This produces in general a complex cascade of SEs, which a transport equation approach is suited to model reliably \cite{RB91}. The cascade includes energetic electrons that generate further SEs and CL, adding complexity to this scenario. However an element of simplicity comes from the separation between coherent and incoherent processes, from the point of view of the emitted light, as we explain in more detail in Sec.\ \ref{cathodoluminescence}. In particular, the coherent CL signal is dominant in metals.

\begin{figure}
\includegraphics[width=80mm,angle=0,clip]{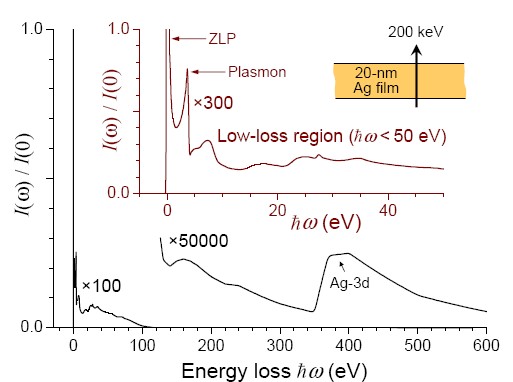}
\caption{\label{overview} (Color online) Typical electron
energy-loss spectrum showing a narrow zero-loss peak
(ZLP; FWHM$\approx$0.2\,eV), collective modes in the valence loss region
(e.g., the 3.7\,eV plasmon of Ag in the inset), and much weaker core
excitations at higher lost energies. The spectrum has been
calculated for 200\,keV electrons traversing a 20\,nm Ag film using optical data for Ag \cite{P1985} and assuming a
collection angle of 5\,mrad.}
\end{figure}

We are interested in low-energy excitations, which involve
electrons and holes in the valence and conduction bands, as well as
collective modes (e.g., localized plasmons and SPPs). This is by far
the most intense part of the loss spectrum (see Fig.\
\ref{overview}). Its analysis yields information on the
materials optical response with the kind of spatial resolution that
is currently dreamed of in the context of nanophotonics. It is our purpose to
review historical developments, to summarize recent advances in this
area, to present an overview of theoretical methods, and to point
out some opportunities opened by electron microscopy in order to
expand and complement nanophotonics studies in a way that can be
particularly beneficial for emerging areas such as plasmonics.

\section{Interaction of Swift Electrons with Matter}
\label{interactionofelectron}

We first consider some fundamental aspects of the interaction between fast electrons a matter, as well as theoretical approaches suitable to simulate EELS.

\subsection{An evanescent source of light in matter}
\label{anevanescentsource}


The electromagnetic field that accompanies a point charge moving in
vacuum can be regarded as an evanescent source of radiation which permits
exploring regions of momentum-energy space that lie outside the
light cone. This has interesting consequences: fast electrons
generate SPPs when passing near a metal
surface, as we show in Sec.\ \ref{planarsurfaces}, but they can
also excite non-dipolar modes in small particles \cite{paper164}, which would be
difficult to resolve using external light instead.

It is helpful to examine first the electric field produced by a swift charged particle
moving inside a homogeneous medium. The field can be conveniently
decomposed into different frequency contributions using the Fourier transform
\begin{eqnarray}
\Eb(\rb,t)=\int\frac{d\omega}{2\pi}\,\Eb(\rb,\omega)\,\ee^{-\ii\omega
t}, \label{Fourierwt}
\end{eqnarray}
and also into momentum components with $\exp(\ii\qb\cdot\rb)$
spatial dependence. We assume that the electron describes a
straight-line trajectory with constant velocity vector $\vb$,
crossing the origin at time $t=0$. This is a reasonable assumption
for the swift electrons and relatively thin specimens typically examined with TEMs. Then, the electron charge density becomes
$-2\pi e\delta(\omega-\qb\cdot\vb)$ in $\qb-\omega$ space. Direct
solution of Maxwell's equations yields
\begin{eqnarray}
\Eb(\rb,\omega)=\frac{\ii e}{\pi}\int
d^3\qb\,\,\frac{\qb/\epsilon-k\vb/c}{q^2-k^2\epsilon}
\,\,\ee^{\ii\qb\cdot\rb}\,\, \delta(\omega-\qb\cdot\vb), \label{Eqw}
\end{eqnarray}
where $k=\omega/c$ is the light wavenumber in free space and
$\epsilon$ is the dielectric function of the homogeneous medium.
It should be noted that $\epsilon$ can depend on both $\qb$ and $\omega$. The
consequences of the wavevector dependence are discussed later in
Sec.\ \ref{bulklosses}, but they are ignored in the remainder of
this section.

Equation\ (\ref{Eqw}) contains some interesting elements. Retardation effects show up both through $k^2\epsilon$ in the denominator, which modifies the strength of the field (producing, for example, weaker interaction in metals and stronger interaction in low-index dielectrics), and through a term proportional to the velocity vector in the numerator, which adds transversal components to the field.

\begin{figure}
\includegraphics[width=75mm,angle=0,clip]{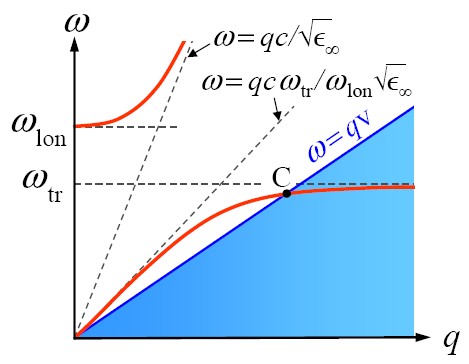}
\caption{\label{homogeneous} (Color online)
Wavevector-frequency diagram showing the light dispersion relation $q^2=k^2\epsilon$ in a
polaritonic material described by Eq.\ (\ref{epspol}) (thick solid curves)
and its intersection with allowed transfers coming from an electron moving
with velocity $\vb$ (shaded region). Point C signals the frequency threshold of Cherenkov radiation (CR) emission.}
\end{figure}

The delta function inside the integral of Eq.\ (\ref{Eqw}) expresses energy conservation for
transfers of frequency $\omega$ and wavevector $\qb$ from the electron to
the material. Neglecting relativistic corrections, the electron
energy is reduced from $m_ev^2/2$ to $|m_e\vb-\hbar\qb|^2/2m_e$ during the
transfer, so that the energy difference $\hbar\omega$ leads to
\begin{eqnarray}
\omega=\qb\cdot\vb-\hbar q^2/2m_e\approx\qb\cdot\vb. \label{wqv}
\end{eqnarray}
The right-most expression is the non-recoil approximation, which
works extremely well under the usual condition $\hbar q\ll
m_ev$.\footnote{Electrons with typical TEM energies above 80 keV have
velocities $v>0.5c$. When they undergo valence losses $\hbar\omega<50$\,eV, the momentum transfer is $\hbar q_z\approx\hbar\omega/v<0.0004\,m_ev$ along the direction of the trajectory.} This
approximation remains valid if the energy transfer is computed using
relativistic expressions. The kinematically allowed transfers span a
solid area for all possible relative orientations between $\vb$ and
$\qb$, as shown by the shaded region in Fig.\ \ref{homogeneous} with upper boundary $\omega=qv$.

\begin{figure}
\includegraphics[width=80mm,angle=0,clip]{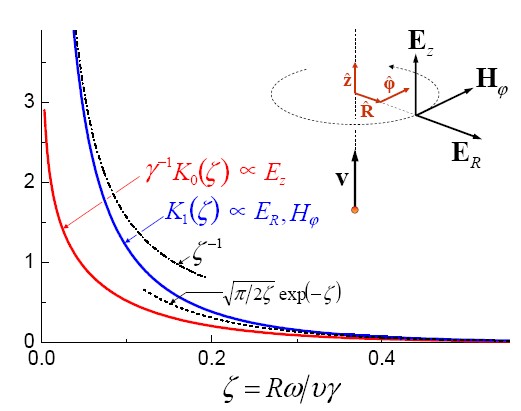}
\caption{\label{EHbare} (Color online) Transverse spatial-direction dependence of the $\exp(-\ii\omega t)$ contribution to the electromagnetic field set up by an electron moving in vacuum with velocity $v=0.7c$ ($\gamma\approx 1.4$ and kinetic energy $\approx 200\,$keV) along the positive $z$ axis. The only non-vanishing components ($E_R$, $E_z$, and $H_\varphi$) decay exponentially at large distance $R$ from the trajectory. The inset shows the orientation of these components relative to the electron velocity vector. The small $R$ limit is dominated by the $1/R$ divergence of $E_R$ and $H_\varphi$.}
\end{figure}

The zeros of the denominator in the integrand of Eq.\ (\ref{Eqw})
signal the dispersion relation of light in the medium,
$q=k\sqrt{\epsilon}$. This has been represented in Fig.\
\ref{homogeneous} for a polaritonic material described by the
dielectric function of Eq.\ (\ref{epspol}) (thick solid curves). The figure shows two
different frequency domains separated by the condition $v^2\epsilon=c^2$ (point C in Fig.\ \ref{homogeneous} is a graphical solution of this equation). At lower frequencies with respect to C, the electron does not couple to excitations in the medium and the spectral components of the electric field decay exponentially away from the trajectory, as explicitly shown by solving the integral in Eq.\ (\ref{Eqw}). We find \cite{J99}
\begin{eqnarray}
\Eb(\rb,\omega)=\frac{2e\omega}{v^2\gamma_\epsilon\epsilon}\,\gb(\rb),
\label{Ehomo1}
\end{eqnarray}
where
\begin{eqnarray} \gb(\rb)=\ee^{\ii\omega z/v}
\left[\frac{\ii}{\gamma_\epsilon}K_0\left(\frac{\omega
R}{v\gamma_\epsilon}\right)\zt-K_1\left(\frac{\omega R}{v\gamma_\epsilon}\right) \Rt\right],
\label{Ehomo2}
\end{eqnarray}
$\gamma_\epsilon=1/\sqrt{1-\epsilon v^2/c^2}$ is the Lorentz
contraction factor,\footnote{Interestingly, the Lorentz factor $\gamma_\epsilon$ involves the velocity of light in the material, $c/\sqrt{\epsilon}$, which is in turn frequency-dependent.}$^,$\footnote{The square roots are chosen to yield positive real parts in this work. Notice that ${\rm Im}\{\epsilon\}$ is always positive in the retarded response formalism followed here, and it becomes a positive infinitesimal in non-lossy dielectrics.} and the notation $\rb=(\Rb,z)$ with $\Rb=(x,y)$ has been employed (see inset of Fig.\ \ref{EHbare}), with the velocity taken along $\zz$. Similarly, the magnetic field reduces to
\begin{eqnarray}
\Hb(\rb,\omega)=\frac{-2e\omega}{vc\gamma_\epsilon} K_1\left(\frac{\omega R}{v\gamma_\epsilon}\right) \ee^{\ii\omega z/v} \ft,
\label{Ehomo3}
\end{eqnarray}
where $\ft$ is the azimuthal unit vector (see Fig.\ \ref{EHbare}). The announced exponential decay with $R$ of both $\Eb(\rb,\omega)$ and $\Hb(\rb,\omega)$ arises from the asymptotic behavior of the modified Bessel functions $K_m$ for large arguments \cite{AS1972}. The non-vanishing components of these fields are represented in Fig.\ \ref{EHbare}. The electromagnetic field extends up to distances of the order of $\sim v\gamma_\epsilon/\omega$, the Bohr cutoff. Notice however that one cannot assign this value to a characteristic Coulomb delocalization distance, since the field diverges at the origin as $\sim 1/R$, so that large interaction contrast is expected across small distances in the region close to the trajectory.

When the electron moves faster than light in the medium, under the
condition
\begin{eqnarray}
v>c/\sqrt{\epsilon}, \label{Cherenkovcond}
\end{eqnarray}
there is overlap between the photon dispersion relation and the
shaded region in Fig.\ \ref{homogeneous}, so that the electron can
emit Cherenkov radiation (CR) (see Sec.\
\ref{cherenkovradiation}). The field displays oscillatory behavior
and decays as $1/\sqrt{R}$ away from the trajectory [this stems from the modified Bessel functions $K_0$ and $K_1$ for imaginary argument in Eq.\ (\ref{Ehomo2})].

It should be stressed that Eqs.\ (\ref{Ehomo1})-(\ref{Ehomo3}) and Fig.\ \ref{EHbare} refer to each monochromatic component of the electromagnetic field, evolving with time as $\exp(-\ii\omega t)$ [see Eq.\ (\ref{Fourierwt})]. An electron moving in vacuum can be actually regarded as an evanescent source of super-continuum light with the spectral dependence shown in Fig.\ \ref{EHbare}.

\subsection{Classical dielectric formalism}
\label{classicaldielectric}

The pioneering work of \textcite{F1940} on the stopping of fast
charged particles in dielectric materials opened up the application
of classical electrodynamics to describe the interaction of swift
electrons with matter. Following this useful tradition, we now
extend the dielectric formalism of the previous section to cope with
inhomogeneous samples and discuss in particular the spectral loss
probability, which is relevant to EELS experiments.

The energy loss suffered by a fast electron moving with constant velocity $\vb$ along a straight line trajectory $\rb=\rb_e(t)$ can be related to the force exerted by the induced electric field $\Eb^{\rm ind}$ acting back on the electron as
\cite{R1957}
\begin{eqnarray}
\Delta E &=& e \int dt \; \vb\cdot\Eb^{\rm ind}\left[\rb_e(t),t\right] \nonumber \\
         &=& \int_0^\infty \hbar\omega\, d\omega \;
               \Gamma_{\rm EELS}(\omega),
\nonumber
\end{eqnarray}
where the $-e$ electron charge has been included (i.e., $\Delta
E>0$) and
\begin{eqnarray}
\Gamma_{\rm EELS}(\omega) = \frac{e}{\pi\hbar\omega} \int dt \,
                       {\rm Re} \left\{ \ee^{-\ii\omega t}
              \vb\cdot\Eb^{\rm ind}\left[\rb_e(t),\omega\right] \right\}
\label{Ggen}
\end{eqnarray}
is the so-called loss probability, which is given per unit of transferred frequency $\omega$. The problem of calculating the
loss probability reduces then to solving the electric field set up
by the electron. A large deal of work has been devoted to obtaining the electric field
for many geometries, including planar surfaces, isolated spheres,
neighboring spheres, circular cylinders, wedges, and more complex
shapes, using both analytical and fully-numerical methods, either
within the non-retarded approximation, based upon solutions of
Poisson's equation, or with full inclusion of retardation effects
by solving Maxwell's equations (see Sec.\ \ref{electronenergy} and
references therein). Next, we outline the general features of this formalism.

\subsubsection{Non-retarded approximation}
\label{nonretarded}

In the non-retarded approximation, we neglect the delay experienced by the electromagnetic signal that mediates the electron-sample interaction. Then, the electric field admits the form $\Eb(\rb,\omega)=-\nabla\phi(\rb,\omega)$, and we can disregard $\Hb$ in the absence of magnetic response.

It is useful to express the electric potential $\phi$ in terms of the screened
interaction $W(\rb,\rb',\omega)$, defined as the potential created
at $\rb$ by a unit point charge located at $\rb'$ [an implicit $\exp(-\ii\omega t)$ time dependence is understood].
This quantity has to be combined with the charge density
corresponding to the moving electron. Considering a non-recoiled
straight line trajectory, and assuming without loss of generality
that the velocity vector is directed along the positive $z$
axis [i.e., the trajectory is $\rb_e(t)=\rb_0+\vb t$ and $\vb=v\zz$], the electron charge density in frequency space $\omega$ reduces to
\begin{eqnarray}
      \rho(\rb,\omega)&=&-e\int dt \,\ee^{\ii\omega t}\,
      \delta(\rb-\rb_0-\vb t)\nonumber\\&=&\frac{-e}{v}\delta(\Rb-\Rb_0)\,\ee^{\ii\omega(z-z_0)/v},
\label{eq12}
\end{eqnarray}
where the notation shown in the inset of Fig.\ \ref{EHbare} has been used. From here, the potential reads
\begin{eqnarray}
\phi(\rb,\omega) = \frac{-e}{v}\int dz'\;W(\rb,\Rb_0,z',\omega)\,\,\ee^{\ii\omega (z'-z_0)/v}. \nonumber
\end{eqnarray}
Finally, plugging these expressions into Eq.\ (\ref{Ggen}), the non-retarded (NR) loss probability is found to be
\begin{eqnarray}
&&\Gamma^{\rm NR}_{\rm EELS}(\Rb_0,\omega)=\frac{e^2}{\pi\hbar v^2}\int dz\,dz'\cos\left[\frac{\omega(z-z')}{v}\right] \nonumber\\
&&\,\,\,\,\,\,\,\,\,\times\,\,  {\rm Im}\left\{-W(\Rb_0,z,\Rb_0,z',\omega)\right\}, \label{Gfos}
\end{eqnarray}
where the dependence of $\Gamma^{\rm NR}_{\rm EELS}$ on the electron impact parameter $\Rb_0$ is explicitly shown.\footnote{It should be noted that the reciprocity theorem [$W(\rb,\rb',\omega)=W(\rb',\rb,\omega)$] and the fact that the bare Coulomb interaction is a real function have been utilized in the derivation of Eq.\ (\ref{Gfos}).}$^,$\footnote{The induced field in Eq.\ (\ref{Ggen}) can be safely replaced by the total field, because the bare field of the moving charge does not produce stopping. We have accordingly dropped the superscript ${\rm ind}$ in Eq.\ (\ref{Gfos}).}

The loss probability can be thus derived from the knowledge of the
frequency-dependent screened interaction $W$. A local description of
the sampled materials in terms of a frequency- and space-dependent
dielectric function $\epsilon(\rb,\omega)$ provides often reliable
results. Detailed expressions of $W$ are offered in Appendix\
\ref{screenedinteraction} for planar, spherical, and cylindrical
geometries. Nevertheless, Eq.\ (\ref{Gfos}) is valid beyond the
local response approximation, and subtle effects such as local-field
corrections can be incorporated via more realistic quantum-mechanical
representations of the screened interaction.

\subsubsection{Retardation effects}
\label{retardationeffects}

In high-voltage TEMs, retardation may become important because the speed of the charged projectiles is a sizeable fraction of the speed of light. For instance, the Lorentz contraction factor at 200 keV takes already a value significantly different from one, $\gamma=1.4$. Retardation has two noticeable consequences for spectroscopy: (1) it increases the range of interaction of the electron probe in directions normal to the trajectory, as clearly illustrated by Eq.\ (\ref{Ehomo2}) and Fig.\ \ref{EHbare}; and (2) it produces redshifts in excitation mode energies \cite{paper155}. The latter are discussed below in further detail, but we can anticipate than this effect turns out to be important when the excitations extend over specimen distances that cannot be neglected in front of the corresponding light wavelength. A proper description of these effects requires calculating the electric field of Eq.\ (\ref{Ggen}) from Maxwell's equations in the presence of the moving electron and the sample under consideration.

The electromagnetic response of a structured material is fully captured
in its electric Green tensor. In particular, the electric field
produced by an external current density $\jb(\rb,\omega)$ in an
inhomogeneous medium of permittivity $\epsilon(\rb,\omega)$ can be
written in frequency space $\omega$ as
\begin{eqnarray}
\Eb(\rb,\omega)=-4\pi\ii\omega \int d\rb'\,G(\rb,\rb',\omega)\cdot\jb(\rb',\omega) \label{eq6}
\end{eqnarray}
in terms of $G$, the electric Green tensor of Maxwell's equations in
Gaussian units, satisfying
\begin{eqnarray}
&& \nabla\times\nabla\times G(\rb,\rb',\omega) - k^2\,
\epsilon(\rb,\omega)\, G(\rb,\rb',\omega) \nonumber \\ && =
\frac{-1}{c^2}\,\delta(\rb-\rb') \label{defgreen}
\end{eqnarray}
and vanishing far away from the sources ($|\rb-\rb'|\rightarrow\infty$ limit).

For the electron charge density of Eq.\ (\ref{eq12}),
the external current density reduces to $\jb=\vb\rho$,
which upon insertion into Eq.\ (\ref{eq6}), and this in
turn into Eq.\ (\ref{Ggen}), allows us to write the
loss probability as
\begin{eqnarray}
      && \Gamma_{\rm EELS}(\Rb_0,\omega)=\frac{4e^2}{\hbar}\int dz dz'\,\cos\left[\frac{\omega(z-z')}{v}\right]\nonumber\\
      &&\,\,\,\,\,\,\times\,\,{\rm Im}\left\{-G_{zz}[\Rb_0,z,\Rb_0,z',\omega]\right\},
\label{eq14}
\end{eqnarray}
where $G_{zz}=\zt\cdot G\cdot\zt$.\footnote{In the retarded case, the reciprocity theorem states that
$G(\rb,\rb',\omega)=G^T(\rb',\rb,\omega)$, a fact that we have used
to recast exponential factors involving $z$ and $z'$ into a cosine function
in the derivation of Eq.\ (\ref{eq14}). Moreover, the
free-space Green function is entirely made of plane-wave
components lying inside the light cone, so that it cannot
contain wavevectors $\omega/v>k$. This guarantees that the integral in
Eq.\ (\ref{eq14}) yields zero for an electron moving in vacuum, and that we are allowed to utilize the total rather than the
induced field to obtain this equation.} Interestingly, this expression works for any sign of $v$, and therefore the loss probability is independent on whether the electron moves towards positive or negative $z$'s.

\subsection{Quantum approach}
\label{quantumapproach}

The quantum nature of both the electron probe and the excitations
sustained by the targeted materials permeate many aspects of the electron-sample
interaction. However, \textcite{RH1988} showed that a
quantum mechanical description of EELS yields the same results as a
semiclassical formalism if all the inelastic signal is collected,
and therefore readers that are more interested in applications and
in a general overview of the subject can overlook this section,
unless they feel motivated by Sec.\
\ref{selfinteractionofasingleelectron}.

We work for simplicity in the non-retarded limit, in which the
validity of Eq.\ (\ref{Gfos}) is demonstrated next under very common experimental
conditions \cite{RH1988}.


\subsubsection{Quantum description of the target}
\label{quantumdescription}

The last century has witnessed important developments in the field of
interaction of fast charges with solids, particularly in the community
of electronic and atomic collisions (see, for instance, \onlinecite{PR92}, \onlinecite{Z99},
\onlinecite{W02}, and references therein), from which our
theoretical understanding of electron microscopy has benefited
considerably. In that context, ion and electron stopping was of
great importance to understand dynamical screening in solids, with
interesting developments such as the description of plasmon excitations
using second quantization schemes in planar \cite{LS1971} and
spherical surfaces \cite{AF1976}. Moreover, a non-retarded, fully
quantum-mechanical expression for the loss probability was derived
using a self-energy formalism \cite{EBR1987,EFR1990}.
A quantum treatment of the target has been shown to be necessary for a correct assessment of delocalization in the excitation of core levels \cite{OA98,AFL03}.
We here obtain a general expression for the loss probability, starting from the more widely used Fermi golden rule, and assess the conditions under which it agrees with the semi-classical formalism presented above, as previously discussed by \textcite{RH1988}.

\begin{figure}
\includegraphics[width=70mm,angle=0,clip]{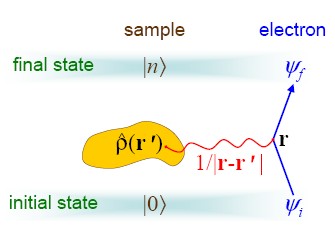}
\caption{\label{quantum} (Color online) Schematic representation of the Coulomb interaction between a swift electron and a specimen, showing the elements involved in Eq.\ (\ref{FGR}), including the target electron-density operator $\hat\rho$ at point $\rb'$.}
\end{figure}

During the interaction of a fast electron with a target, the latter
can undergo transitions from its ground state $|0\rangle$ of energy
$\hbar\omega_0$ to excited states $|n\rangle$ of energy $\hbar\omega_n$, while
the incoming electron of energy $\hbar\varepsilon_i$ and wave function
$\psi_i(\rb)$ acquires components $\psi_f(\rb)$ of lower energy
$\hbar\varepsilon_f$. Since the interaction with very energetic electrons
is generally small, the transition rate is well described within first-order perturbation theory (Fermi's golden rule):
\begin{eqnarray}
\frac{d\Gamma^{\rm NR}}{dt}=\frac{2\pi e^2}{\hbar} \sum_{f,n}&& \left|\int d^3\rb d^3\rb'
\frac{\psi_f^*(\rb)\psi_i(\rb) \, \langle
n|\hat\rho(\rb')|0\rangle}{|\rb-\rb'|} \right|^2 \nonumber
\\ &\times& \,\, \delta\left(\varepsilon_f-\varepsilon_i+\omega_n-\omega_0\right),
\label{FGR}
\end{eqnarray}
where we have used the target-probe Coulomb interaction and
$\hat\rho$ is the target electron-density operator (see Fig.\ \ref{quantum}). We can now
recast this expression into a more convenient form by relating the
target matrix elements to the linear-response susceptibility
\cite{PN1966}
\begin{eqnarray}
&& {\rm Im}\left\{\chi(\rb,\rb',\omega)\right\} \nonumber \\ &&=\frac{-\pi}{\hbar} \sum_n
\langle 0|\hat\rho(\rb)|n\rangle\,\langle n|\hat\rho(\rb')|0\rangle\,\delta(\omega_n-\omega_0-\omega),\nonumber
\end{eqnarray}
valid for $\omega>0$, and this in turn to the screened interaction
\begin{eqnarray}
W^{\rm ind}(\rb,\rb',\omega)=\int d^3\rb_1 d^3\rb_2\,\frac{\chi(\rb_1,\rb_2,\omega)}{|\rb-\rb_1| |\rb'-\rb_2|}. \nonumber
\end{eqnarray}
Then, the sum of Eq.\ (\ref{FGR}) can be separated into specific values of the
frequency transfer $\omega=\varepsilon_i-\varepsilon_f$  as
\begin{eqnarray}
\frac{d\Gamma^{\rm NR}}{dt}=\int_0^\infty d\omega \; \frac{d\Gamma^{\rm NR}(\omega)}{dt}, \nonumber
\end{eqnarray}
where
\begin{eqnarray}
\frac{d\Gamma^{\rm NR}(\omega)}{dt}&=& \frac{2e^2}{\hbar} \sum_f \int d^3\rb d^3\rb' \;
\psi_f(\rb)\psi_i^*(\rb)\psi_f^*(\rb')\psi_i(\rb') \nonumber \\
&\times& {\rm Im}\left\{-W(\rb,\rb',\omega)\right\}
\delta(\varepsilon_f-\varepsilon_i+\omega). \label{PQM}
\end{eqnarray}
This expression is general for incident electrons of well-defined
energy, and we remind the reader that $W$ contains all
quantum-mechanical details of the sample response, although
expressions obtained from dielectric theory such as those offered in
Appendix\ \ref{screenedinteraction} yield reliable results in most situations encountered in practice.

\subsubsection{Quantum effects in the fast electrons}
\label{quantumeffects}

Several authors have analyzed the formation of images in electron
microscopes from a quantum-mechanical viewpoint, considering the
influence of instrumental parameters, as for example the beam aperture and the
collection angle \cite{K1983,B1985,RH1988,MS95}. Particular emphasis
has been made on taming delocalization, which is relevant
to devise ways of improving the spatial resolution of the inelastic
signal \cite{OA98,AFL03}. We are not going to enter into details
of the instruments here, but it is instructive to consider effects related to
partial detection of the transmitted electrons.

We can regard each electron in a TEM as consisting of a coherent superposition of
plane waves, yielding for instance a narrow focus close to the sample
in STEM mode. A very accurate approximation can be adopted
if we keep in mind that the angular aperture of the beam has remained in the range
of a few milliradians since the early days of electron microscopy
\cite{R1987_2}. The components of the incident electron wavevector perpendicular to
the beam direction $z$ are thus $p_{i\perp}\lesssim 10^{-2}p_i$, where
$p_i$ is the total wavevector. The parallel components become $p_{i\parallel}=(p_i^2-p_{i\perp}^2)^{1/2}\approx p_i$, where we are neglecting $\approx p_{i\perp}^2/2p_i\lesssim 10^{-4}p_i$ in
front of $p_i$. We conclude that the incident charge can be reliably described near the sample by the
wave function
\begin{eqnarray}
\psi_i(\rb)=\frac{1}{L^{1/2}}\,\ee^{\ii p_iz}\,\psi_{i\perp}(\Rb), \label{psii}
\end{eqnarray}
where $L$ is the quantization length of the trajectory and the $\Rb$
dependence reflects the variation perpendicular to $z$. This should
be valid for typical sample thickness $\Delta z\sim 50$ nm, along
which the lateral divergence of the beam can be quantified as
$\Delta R\lesssim 0.5$ nm, a small value compared to typical dimensions of
common photonic and plasmonic structures.

The transmitted electrons can be again described by plane waves of
wavevector $\pb_f$. Then, the frequency transfer reduces to
$\omega\approx \qb\cdot\vb$ under the approximation of Eq.\
(\ref{wqv}), where $\qb=\pb_i-\pb_f$ is the wavevector transfer and
$\vb=(\hbar/m_e)p_i\zz$ is the incident electron velocity.

We are now prepared to recast Eq.\ (\ref{PQM}) into a more practical
formula. Using delta-function normalization for the final states, multiplying the transition rate by the interaction time
$L/v$, and making use of the above considerations, we find
\begin{eqnarray}
\Gamma^{\rm NR}(\omega)=\int d^2\pb_{f\perp}\;\frac{d\Gamma^{\rm NR}(\omega)}{d\pb_{f\perp}},
\nonumber
\end{eqnarray}
where
\begin{eqnarray}
\frac{d\Gamma^{\rm NR}(\omega)}{d\pb_{f\perp}}&=& \frac{e^2}{4\pi^3\hbar
v^2} \int d^3\rb\,d^3\rb' \;
\psi_{i\perp}^*(\Rb)\psi_{i\perp}(\Rb') \label{PQM2} \\
&\times&\,\, \ee^{\ii\pb_{f\perp}\cdot(\Rb-\Rb')}
\,\,\ee^{\ii\omega(z'-z)/v}\,\,
{\rm Im}\left\{-W(\rb,\rb',\omega)\right\} \nonumber
\end{eqnarray}
is the loss probability resolved in $\pb_{f\perp}$ (the lateral
wavevector of the transmitted electron $\perp\vb$), or
equivalently, the outgoing direction. This formula
indicates that STEMs can be used to retrieve the full nonlocal
dependence of ${\rm Im}\left\{-W(\rb,\rb',\omega)\right\}$, but we
defer a detailed discussion of this point to Sec.\ \ref{selfinteractionofasingleelectron}, in which we explore vicinage
effects.

In practice, electron beams are polychromatic. However, a trivial extension of the above formalism leads to the intuitive result that the loss probability is then the average of Eq.\ (\ref{PQM2}) over the incident-beam spectrum, thus precluding interference between components of different primary energy.

Equation\ (\ref{PQM2}) leads to a powerful result, established
by \textcite{RH1988}, regarding the validity of the classical
dielectric formalism employed in Sec.\ \ref{nonretarded}. The
unrestricted integral over all possible values of $\pb_{f\perp}$
yields $\delta(\Rb-\Rb')$, so that the loss probability reduces to
\begin{eqnarray}
\Gamma^{\rm NR}_{\rm EELS}(\omega)=\int d^2\Rb \; \left|\psi_{i\perp}(\Rb)\right|^2 \; \Gamma^{\rm NR}_{\rm EELS}(\Rb,\omega), \nonumber
\end{eqnarray}
where $\Gamma^{\rm NR}_{\rm EELS}(\Rb,\omega)$ is given by Eq.\ (\ref{Gfos}). In other words,
the EELS probability is well described by Poisson's equation if all
the inelastic signal is collected (i.e., using a wide acceptance
angle in the spectrometer), but it needs to be averaged over electron impact
parameters weighed by the spot intensity $|\psi_{i\perp}(\Rb)|^2$.

\section{Electron Energy-Loss Spectroscopy}
\label{electronenergy}

\textcite{HB1944} were the first to propose and demonstrate EELS in TEMs, although earlier pioneering experiments reported  energy losses of transmitted electrons in thin films \cite{L1904}. This technique has become
standard in the electron microscopy community and is capable of
providing information on electronic band structures and plasmons in the low-energy loss
region, as well as atomically-resolved chemical-identity encoded in core losses \cite{BCP93}. During its prolific
existence, valence EELS has contributed to fields as varied as
biochemistry (for example, in the study of excitations sustained by nucleic acid
bases, reported by \onlinecite{CIJ1971}), interplanetary science
(for instance, in the explanation of a 5\,eV strong absorption feature in
cosmic dust found by \onlinecite{BDE05}), and microelectronics (in particular, in
the investigation of the resistivity of CMOS elements performed by
\onlinecite{PPC06}).

Since the early days of EELS, transmission electron
microscopes have undergone a tremendous series of improvements that
permit currently achieving $\sim 0.1$ eV energy resolution for a
sub-nanometer-size electron beam. This opens up new vistas in the
low-energy loss region, such as addressing the optical properties of photonic
structures with unprecedented spatial detail.

The excitation spectrum of a typical material is illustrated in
Fig.~\ref{overview} for bulk Ag. The EELS intensity for a given
energy loss $\hbar\omega$ directly reflects the strength of specimen modes
corresponding to that energy \cite{E96}. The loss spectrum has been traditionally separated into valence- and core-loss
regions, with the division between them arbitrarily established at
$\sim 50$ eV. With the customary use of thin samples, the main feature in an
EELS spectrum is the zero-loss peak (ZLP) of unscattered electrons,
in which unresolved, very-low-energy excitations (e.g., phonons)
are buried. The intensity of valence features is over two
orders of magnitude lower than the ZLP in the 20\,nm Ag film
considered in Fig.~\ref{overview}, while higher-energy core
excitations are even less probable. This presents a typical scenario
in which the inelastic signal is superimposed into the tail of the ZLP.
Removal of the ZLP is thus important, and various sophisticated
procedures have been devised that produce very reliable results
\cite{LBZ06,paper125}.

Valence losses are generally more intense than core losses and allow collecting any
given amount of inelastic signal with less electron
irradiation, therefore producing minimum damage to the sample.
However, the interpretation of low-energy-loss images is less
direct because it involves the excitation of delocalized modes \cite{H03,ELB08}.
In this respect, it is useful to rely on theoretical simulations, which are
now advanced to the point of dealing with virtually any sample
geometry in a predictive way.

The books of \textcite{R1980} and \textcite{E96} are classic
references of EELS that extensively cover these subjects and provide
detailed considerations on the operation and principles of TEMs.
Several review papers present shorter summaries \cite{B06} and
collections of experimental spectra \cite{A04}. Comprehensive
reports on valence EELS \cite{W96_2} and its theoretical analysis
\cite{RZA00} are also available. The present article intends to supplement
this field by providing a more extended overview of low-energy
losses, their characteristics and theoretical understanding, and
several examples of novel application to nanophotonics, for which they might
become particularly useful.

\subsection{Space, momentum, and energy resolution}
\label{spatialresolution}

Nowadays, some commercial TEMs incorporate the latest achievements in space and energy resolution. The latter has been
tremendously improved over the last decade by the development of aberration correctors in the electron optics, particularly the
spherical aberration \cite{UH98}. This has reduced the size of the beam spots from the 1-2 nanometers commonly achieved without such correctors down to sub-{\AA} dimensions \cite{KDL99}, which allows for instance imaging closely spaced atoms in crystal samples using core losses \cite{BDK02,NCD04,VFL04,VOR07}, and even identifying details of chemical bonding \cite{MKM08}. However, the interpretation of atomically-resolved images is sometimes difficult and requires to rely on simulations \cite{WDF08,OP08}. The search for improved spatial resolution through aberration-correction is actually a work in progress, for example through theoretical analysis for improved incoherent imaging \cite{IXM08} and through new sub-Angstrom beam characterization techniques \cite{DEE08}.

The factors that limit the spatial resolution of a microscope are typically
known as delocalization effects \cite{E03,EB05,E07}. There are
several of them, of very different nature, that can be grouped into
three distinct categories: (1) instrumental or lens delocalization,
which as noted above can be reduced below 1\,{\AA}; (2) Coulomb
delocalization associated to the finite range of the field that
accompanies a fast electron (see Fig.\ \ref{EHbare} and discussion in Sec.\ \ref{anevanescentsource}); and (3) the extended nature of the excitations that are probed, varying from macroscopic
distances in the case of low-energy SPPs and CR losses, to
nanometers in particle plasmons, and less than 1\,{\AA} for core
losses. For example, sub-atomic resolution is currently limited by the
size of the probe rather than by the extension of the ionizing
interaction that is employed to resolve core levels \cite{AFL03}.

Additionally, spectral resolution has dramatically improved with the arrival
of new spectrometers yielding $\sim 50$\,meV accuracy \cite{BBB03}, and
electron monochromators \cite{SZT03}, which roughly consist in filtering out
incident electrons outside a narrow energy window at the expense of
reducing the beam current. An energy resolution of 0.1 eV has been achieved. Measuring excitations down
to less than $\hbar\omega=0.5$ eV is now possible thanks to the limited
extension of the monochromatized ZLP \cite{TTT99,LBZ06}. This spectral
resolution should be sufficient for studying most collective excitations supported by metallic systems, in which the intrinsic width produced by absorption is generally larger than a tenth of an electronvolt.

Deconvolution techniques, utilized to eliminate the ZLP and enhance
spectral resolution, have progressed tremendously in recent years \cite{BEF01,LBZ06}.
Furthermore, dynamical instabilities in the electron beam energy, which directly
damage spectral details, are now corrected by resorting to fast spectra
acquisition, well below the $50-60\,$Hz of the omnipresent electrical
network, and by subsequent addition of the collected spectra after
repositioning the ZLP maximum \cite{paper125}.

A reasonable degree of momentum-transfer resolution is possible in
TEMs by varying the divergence half-angle of the incident beam
$\varphi_{\rm in}$ and the collection half-angle of the spectrometer
$\varphi_{\rm out}$ [see Fig.\ \ref{EM}(a)]. These parameters are
typically in the range of a few milliradians. Of course, the radius
of the electron spot $\Delta R$, which controls spatial resolution
in STEMs, is related to $\varphi_{\rm in}$ through
the uncertainty principle, $(m_e/\hbar)\Delta R\,\varphi_{\rm in}\ge 1/2v$.
Actual operation conditions in TEMs are very close to this limit.

These impressive achievements are the result of a historical quest, still in progress, for ever better spatial and spectral resolution \cite{E03,EB05,E07}. However, many factors are remaining that limit the performance of electron microscopes, as recently reviewed by \textcite{E07}.


\subsection{Bulk losses and determination of bulk dielectric functions}
\label{bulklosses}

Although the volume of homogeneous materials offers little chances for performing microscopy, bulk losses are a splendid source of information on optical properties \cite{R1980,P1985} and they configure an area of research in which the electron microscopy community has made extremely important contributions. In a homogeneous medium, we can combine Eqs.\
(\ref{Eqw}) and (\ref{Ggen}) to write the EELS probability as
\begin{eqnarray}
\Gamma_{\rm bulk}(\omega)&=&\frac{e^2L}{\pi\hbar v^2}\label{Pbulkinteg}\\&\times&{\rm Im}\left\{
\left(\frac{v^2}{c^2}-\frac{1}{\epsilon}\right) \ln
\left[\frac{q_c^2-k^2\epsilon}{(\omega/v)^2-k^2\epsilon}\right]
\right\}, \nonumber
\end{eqnarray}
where $L$ is the length of the trajectory and $\epsilon$ has been
assumed to be independent of wavevector $q$. This is the local response
approximation, which applies when low-enough momentum transfers are
collected below a certain cutoff
\[\hbar q_c\approx\sqrt{(m_ev\varphi_{\rm out})^2+(\hbar\omega/v)^2}\]
that is determined by the half-aperture collection angle of the
microscope spectrometer $\varphi_{\rm out}$ (assuming $\varphi_{\rm in}\ll\varphi_{\rm out}$).

The non-retarded version ($c\rightarrow\infty$ limit) of Eq.\
(\ref{Pbulkinteg}),
\begin{eqnarray}
\Gamma_{\rm bulk}^{\rm NR}(\omega)=\frac{2e^2L}{\pi\hbar v^2}\,
{\rm Im}\left\{\frac{-1}{\epsilon}\right\}\,\ln(q_cv/\omega),
\label{Pbulkintegnr}
\end{eqnarray}
is commonly employed to perform Kramers-Kronig (KK) analysis and to retrieve bulk optical constants from measured EELS data after absolute calibration of the loss function ${\rm Im}\{-1/\epsilon\}$ \cite{PF91,SSI1980,E96,FMJ98,ZEV08}. However, retardation corrections like CR losses (see Sec.\ \ref{cherenkovradiation}) can be very damaging in the determination
of electronic band gaps and they must be carefully removed from the measured spectra in some cases \cite{JSH04,SFS06,SS07,EB08}. In addition, multiple inelastic scattering events and factors depending on beam divergence and spectrometer acceptance angles, as well as the precision in the alignment of the sample, have to be properly addressed before using KK to retrieve $\epsilon$ \cite{S08,BV08}. Separating volume and surface losses adds another complication, which can be solved by comparing spectra acquired from specimens of different thicknesses \cite{MBM07}.

\begin{figure}
\includegraphics[width=80mm,angle=0,clip]{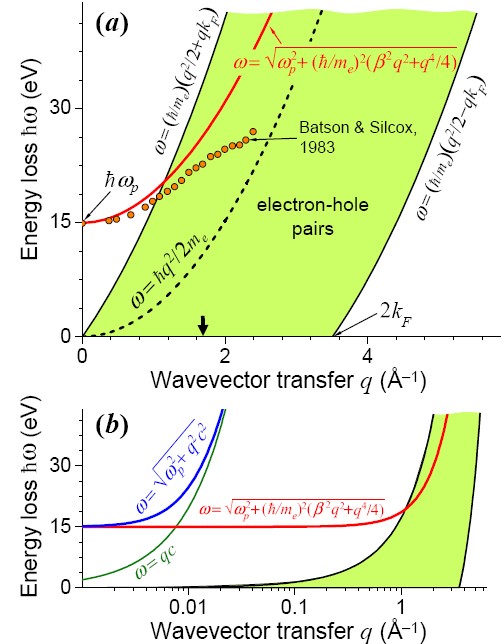}
\caption{\label{bulk} (Color online) Bulk excitations in Al. {\bf (a)} The conduction band of Al is well described by a degenerate free-electron gas of Fermi wavevector $k_{\rm F}=1.75$\,\AA$^{-1}$. The e-h pair excitations supported by this system are contained in the shaded region within the plotted wavevector-energy ($q-\hbar\omega$) diagram. The volume plasmon dispersion relation measured by \textcite{BS1983} (symbols) is compared to the longitudinal plasmon-pole approximation $\omega=[\omega_p^2+(\hbar/m_e)^2(\beta^2q^2+q^4/4)]^{1/2}$ (thick solid curve) derived from the condition $\epsilon_{\rm lon}(q,\omega)=0$, as obtained from Eq.\ (\ref{epsPP}) with $\beta=(3/5)^{1/2}k_F$ and $\epsilon_b=1$. {\bf (b)} Logarithmic-scale representation of some the curves in (a), compared to the light line ($\omega=qc$) and the transversal bulk plasmon dispersion relation ($\omega=[\omega_p^2+q^2c^2]^{1/2}$). The latter, which is obtained by inserting Eq.\ (\ref{epst}) into $q^2=k^2\epsilon_{\rm tr}$, gives rise to light emission in thin films irradiated by electrons \cite{F1958,VS1973}.}
\end{figure}

One is often interested in opening the collection angle to increase
the inelastic signal and to reduce sample damage. Nonlocal effects
are then apparent, for example through e-h pair
excitations, as represented in Fig.\ \ref{bulk}(a) for a free-electron
gas of parameters corresponding to Al. For instance, this region is accessible when using 200 keV electrons and $\varphi_{\rm out}=10\,$mrad, a combination that results in the $q_c$ value indicated by the heavy arrow of Fig.\ \ref{bulk}(a). Under such circumstances, we need to include spatial dispersion in the dielectric function, which has different
forms for longitudinal and transversal fields,
$\epsilon_{\rm lon}(q,\omega)$ and $\epsilon_{\rm tr}(q,\omega)$, respectively, as
pointed out in Appendix\ \ref{homogeneousmedia}. The
momentum-resolved loss probability is then found to be
\begin{eqnarray}
\frac{d\Gamma_{\rm bulk}(\omega)}{d\qb_\perp}&=&\frac{e^2L}{\pi^2\hbar v^2} \label{Pbulk}\\ &\times&{\rm Im}\bigg\{\frac{1}{q^2}\left[\frac{1}{\epsilon_{\rm tr}(q,\omega)}-\frac{1}{\epsilon_{\rm lon}(q,\omega)}\right]\nonumber\\
&& \,\,\,+\left[\frac{v^2}{c^2}-\frac{1}{\epsilon_{\rm tr}(q,\omega)}\right]\frac{1}{q^2-k^2\epsilon_{\rm tr}(q,\omega)}\bigg\}, \nonumber
\end{eqnarray}
where $\qb_\perp$ is the 2D wavevector transfer perpendicular to the electron trajectory, and the total wavevector transfer satisfies $q^2=q_\perp^2+\omega^2/v^2$ in virtue of Eq.\ (\ref{wqv}).

The momentum dependence of the transversal permittivity can be
safely neglected in this discussion, since it constitutes a correction
of order $v_F/c$ to the response, where $v_F$ is the Fermi velocity (e.g., $v_F=1.39\times 10^8\,$cm\,s$^{-1}$ in gold).
Besides, the Lindhard formula \cite{L1954} for the response of a
free-electron gas (see explicit expression in Sec.\ \ref{LindhardMermin}) provides a fair description of the longitudinal
dielectric function in good metals such as Al, although local-field
corrections can be also relevant \cite{VRO02}.

In this context, the response of gold has increasing technological importance, but detailed experimental analysis of the momentum-dependent optical constants is still missing. The local dielectric function determined by \textcite{JC1972} from ellipsometry measurements is routinely employed in plasmonic studies, and it can be well described by the Drude-like formula of Eq.\ (\ref{epst}) in the visible and NIR part of the spectrum (see Table\ \ref{table2}). In a recent report \cite{paper162}, the author has proposed a nonlocal $\epsilon_{\rm lon}$ for gold constructed as
\begin{eqnarray}
  \epsilon_{\rm lon}(q,\omega)=\epsilon^{\rm exp}(\omega)-\epsilon^{\rm D}(\omega)+\epsilon^{\rm M}(q,\omega),
  \label{eqwjpcc}
\end{eqnarray}
where $\epsilon^{\rm D}$ and $\epsilon^{\rm M}$ are the Drude and Mermin dielectric functions given by Eqs.\ (\ref{drude}) and (\ref{epsMermin}), respectively, and corresponding to the values of $\omega_p$ and $\eta$ listed in Table\ \ref{table2}, whereas $\epsilon^{\rm exp}$ is the measured local permittivity. Further experimental effort is still required to contrast this model.

The spectrum in bulk metals is dominated by plasmon losses, first
observed by \textcite{R1948} and \textcite{L1948}, and later
identified as collective oscillations of the conduction-electron gas by \textcite{PB1952}. These modes are signaled by the
vanishing of $\epsilon_{\rm lon}$, as deduced from the non-retarded (NR)
limit of Eq.\ (\ref{Pbulk}),
\begin{eqnarray}
\frac{d\Gamma_{\rm bulk}^{\rm NR}(\omega)}{d\qb_\perp}=\frac{e^2L}{\pi^2\hbar v^2q^2}\,{\rm Im}\left\{\frac{-1}{\epsilon_{\rm lon}(q,\omega)}\right\}.
\label{PNRbulk}
\end{eqnarray}
The dispersion relation of bulk plasmons has been extensively studied in the past using EELS
\cite{W1956,CSG1976,BCS1976,CMK1980,R1980} and is shown in Fig.\ \ref{bulk}(a) for Al, as measured by \textcite{BS1983} (symbols).
When broad collection angles are considered, one can reliably
approximate $\epsilon_{\rm lon}(q,\omega)$ by a
plasmon-pole analytical expression \cite{R1957}, reproduced in Eq.\
(\ref{epsPP}). The plasmon dispersion relation derived from this formula
provides a reasonable interpolation between the measured plasmon
and the Bethe ridge $\omega=\hbar q^2/2m_e$, shown in Fig.\ \ref{bulk}(a) as a solid curve.


Spatially-resolved bulk-plasmon spectroscopy has been used to identify the presence of metals in a specimen [e.g., Na in either Na$_2$O or silicate glasses \cite{JSS08}] and to demonstrate the effect of quantum confinement in carbon nanostructures by tracing local variations in the plasmon energy \cite{SMG06}, which are related to changes in the conduction-electron density through Eq.\ (\ref{wpfor}).

Finally, magnetic circular dichroism has been measured in the response of iron at x-ray energies \cite{SRH06}, suggesting that further work could eventually explore optical activity at visible and NIR frequencies via EELS, for instance in chiral \cite{RFS06} and left-handed \cite{BS08} metamaterials.

\subsection{Planar surfaces}
\label{planarsurfaces}

Surfaces and interfaces can host trapped modes that modify the local band structure and the optical response with respect to bulk materials. For instance, we find electronic surface states confined by directional gaps in noble-metal interfaces \cite{S1985}, but also specific collective modes of conduction electrons (e.g., surface plasmons). As we show later on, these features can be characterized by EELS performed in STEMs with high degree of spatial resolution.

\begin{figure}
\includegraphics[width=80mm,angle=0,clip]{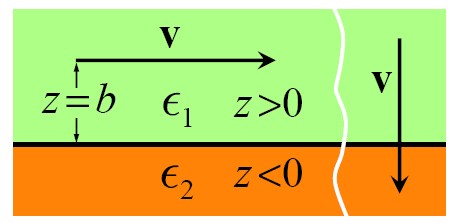}
\caption{\label{planar-surface} (Color online) Planar interface separating two media of dielectric functions $\epsilon_1$ and $\epsilon_2$, respectively. An electron is shown moving parallel to the boundary inside medium 1 (left), with impact parameter $b$ relative to the interface. Another electron is represented crossing the interface at normal incidence (right).}
\end{figure}

A particularly instructive situation is presented when the beam is directed parallel to a planar interface separating two different media (Fig.\ \ref{planar-surface}, left). This configuration has been extensively studied both experimentally
\cite{P1968,LBN1977,WH1989,MBW97} and theoretically \cite{LS1971,EP1975,GGH1985}. The analysis of this geometry in the non-retarded limit can be readily made from the knowledge of the screened interaction $W(\rb,\rb',\omega)$, defined as the potential created at point $\rb$ by a charge oscillating with frequency $\omega$ and placed at $\rb'$. Expressions of $W$ for simple geometries, including that of Fig.\ \ref{planar-surface}, are given in Appendix\ \ref{screenedinteraction}. A closed-form expression for the loss probability of the electron following the parallel trajectory of Fig.\ \ref{planar-surface} (left) is then obtained from Eqs.\ (\ref{Gfos}) and Eqs.\ (\ref{Wplane}). One finds
\begin{eqnarray}
\Gamma^{\rm NR}_{\rm planar}(\omega)&=&\frac{2e^2L}{\pi\hbar v^2}\,\,\bigg\{
\ln\left(\frac{q_cv}{\omega}\right) {\rm Im}\left\{\frac{-1}{\epsilon_1}\right\}
\label{PsurfNR}\\
&+& K_0\left(\frac{2\omega b}{v}\right)\,
\left[{\rm Im}\left\{\frac{-2}{\epsilon_1+\epsilon_2}\right\}
-{\rm Im}\left\{\frac{-1}{\epsilon_1}\right\}\right]\bigg\},
\nonumber
\end{eqnarray}
where $L$ is the length of the trajectory. The impact parameter $b$ and the dielectric functions $\epsilon_1$ and $\epsilon_2$ are defined in Fig.\ \ref{planar-surface}. The first term inside the curly brackets is the bulk loss probability of Eq.\ (\ref{Pbulkintegnr}). The square brackets in Eq.\ (\ref{PsurfNR}) contain the effect of the interface, which dies off exponentially with $b$ as $K_0(\theta)\approx\exp(-\theta)\sqrt{\pi/2\theta}$ for large $\theta=2\omega b/v$ \cite{AS1972}. The first interface term describes the excitation of intrinsic boundary modes, signalled by the condition $\epsilon_1+\epsilon_2=0$. It was first derived by \textcite{EP1975}, who obtained the expression
\begin{eqnarray}
\Gamma^{\rm NR}_{\rm planar}(\omega)=\frac{4e^2L}{\pi\hbar v^2}\;K_0\left(\frac{2\omega b}{v}\right)\;{\rm Im}\left\{\frac{-1}{1+\epsilon}\right\}
\label{EcheniquePendry}
\end{eqnarray}
for an electron moving in vacuum near a medium of dielectric function $\epsilon$. The second interface term in Eq.\ (\ref{PsurfNR}) accounts for a reduction of bulk losses (i.e., the transfer of oscillator strength from volume to surface modes). This is the so-called begrenzungseffekt.

Equation\ (\ref{PsurfNR}) has been successfully applied to explain
surface losses suffered by electrons passing near MgO cubes
\cite{M1982,C1982,C1982_2,WC1988,WH1989} in the so-called aloof
configuration (see below), and also to EELS in a Si/SiO$_2$
interface and in GaAs surfaces \cite{HM1985}.

Inclusion of retardation effects is critical to account for CR losses and to correctly assess the weight of interface effects, as demonstrated by \textcite{MBW97} for the Si/SiO$_2$ interface, and later by \textcite{CKS07} for multilayer structures containing HfO. These studies have been recently extended by \textcite{YCM08} and \textcite{CYM08}, who have measured EELS in slabs and layered stacks formed by Si and SiO$_2$ using electron beams parallel to the interfaces and have demonstrated the important role of retardation corrections for a quantitative comparison with theory.

The retarded counterpart of Eq.\ (\ref{PsurfNR}) has been reported by
several authors \cite{O1967,GGH1985}. A particularly simple
derivation consists in expanding the integrand of Eq.\ (\ref{Eqw})
into $p$ (TM) and $s$ (TE) electromagnetic plane-wave components, which are later reflected at
the interface to act back on the electron \cite{FGF91}. One finds
\begin{eqnarray}
\Gamma_{\rm EELS}(\omega)=\Gamma_{\rm bulk}(\omega)+\Gamma_{\rm ref}(\omega),
\nonumber
\end{eqnarray}
where $\Gamma_{\rm bulk}$ is the bulk loss inside the medium 1 in which the electron is moving, given by Eq.\ (\ref{Pbulkinteg}), and
\begin{eqnarray}
\Gamma_{\rm ref}(\omega) &=& \frac{2e^2L}{\pi\hbar v^2}
            \int_0^\infty \frac{dq_y}{q_\parallel^2} \,\,
   \label{planarret} \\ &\times&
          {\rm Re} \left\{q_{z1} \ee^{2\ii q_{z1}b}
          \left[\left(\frac{q_yv}{q_{z1} c}\right)^2 r_s -
          \frac{1}{\epsilon_1} r_p\right]
          \right\}
\nonumber
\end{eqnarray}
is the loss due to reflection of the electron field at the interface. Here, $q_\parallel=\sqrt{\omega^2/v^2+q_y^2}$, $q_{zj}=\sqrt{k^2\epsilon_j-q_\parallel^2}$, and
\begin{eqnarray}
r_p=\frac{\epsilon_2\;q_{z1}-\epsilon_1\;q_{z2}}{\epsilon_2\;q_{z1}+\epsilon_1\;q_{z2}}
\label{rp}
\end{eqnarray}
and
\[r_s=\frac{q_{z1}-q_{z2}}{q_{z1}+q_{z2}}\]
are Fresnel reflection coefficients for $p$ and $s$ waves, respectively. In the
non-retarded limit, one has $r_s=0$ and
$r_p=(\epsilon_2-\epsilon_1)/(\epsilon_2+\epsilon_1)$, from which
Eq.\ (\ref{PsurfNR}) is easily recovered. The advantage of this
approach, based upon optical reflection coefficients, is that it can
handle more complicated surfaces, like the periodically corrugated
boundary of a confined photonic crystal \cite{PM94,paper075}. For large impact parameters, Eq.\ (\ref{planarret}) provides a retardation correction to Eq.\ (\ref{EcheniquePendry}), consisting in substituting $2\omega b/v\gamma$ for $2\omega b/v$ in the argument of the $K_0$ Bessel function; the apparent impact parameter at very large velocity is contracted to $b/\gamma$.


\subsubsection{Excitation of surface plasmons and surface-plasmon polaritons}
\label{excitationofsurface}

\begin{figure}
\includegraphics[width=80mm,angle=0,clip]{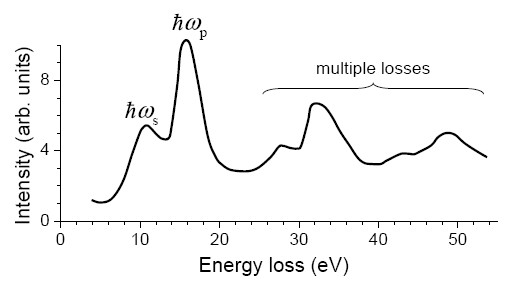}
\caption{\label{Swan} First observation of surface plasmons in the
energy loss distribution of 2020\,eV electrons specularly reflected
on an Al surface under 45$^\circ$ incidence (adapted from
\onlinecite{PS1959}).}
\end{figure}

Surface modes, signalled by the condition $\epsilon_1+\epsilon_2=0$,
are characteristic of the interface between a dielectric having
${\rm Re}\{\epsilon_1\}>0$ and a metal satisfying ${\rm Re}\{\epsilon_2\}<0$.
These excitations were first identified and understood by \textcite{R1957} to
explain anomalous valence losses that were previously observed at
positions differing from bulk plasmons in the spectra of fast
electrons transmitted through thin foils (see, for instance, feature D
in Fig. 2 of \onlinecite{W1956}). Ritchie's description in terms of
a classical hydrodynamic plasma to represent the conduction electron
band led \textcite{SF1960} to baptize these excitations as surface
plasmons, which were later confirmed in electron energy-loss
experiments \cite{PS1959} (see Fig.\ \ref{Swan}). Bulk and surface modes were observed as $\hbar\omega_p\approx 15$\,eV and
$\hbar\omega_s\approx 10.6$\,eV energy losses in electrons reflected from an Al
surface. [Incidently, the spectrum of Fig.\ \ref{Swan} contains multiple plasmon losses (i.e., two or more plasmons being excited by the same electron). The coherent aspects of double plasmon excitation have deserved some attention in the past \cite{SFS87}.]

Aluminum is a prototypical example of nearly-free electron gas for which
the dielectric function is well described by the Drude formula of Eq.\ (\ref{drude}).
The SP of the bare surface, satisfying the condition $\epsilon+1=0$,
has then a frequency $\omega_s=\omega_p/\sqrt{2}$, in excellent
agreement with experimental observations.

Surface plasmons were also observed for other materials in
subsequent reflection experiments performed on melted metals \cite{P1968}, and
even the evolution of these modes through a solid-liquid phase
transition was reported \cite{P1965}. Other pioneering electron transmission
experiments managed to demonstrate the observation of SPs as a characterization
technique \cite{R1967_2} and to determine the dependence of SP
energies on the thickness of thin films \cite{BGI1966}. Evidence of surface and bulk plasmon
excitation has been also observed in the spectra of photoelectrons
and Auger electrons emitted from metals \cite{VT1978,OGH1990,SYT97}.

Similar to the bulk plasmons of Fig.\ \ref{bulk}(a), SPs are affected
by nonlocal effects and exhibit significant dispersion, which was
already reported in early EELS studies \cite{KR1976}. This has stimulated a
rich literature intended to achieve a reasonable first-principles
description of the dynamical response of crystal surfaces (see
\onlinecite{PSC07}, and references therein). Nonlocal
effects involve comparatively short-range interactions extending a
distance $\sim 1/q$ for typical wavevector transfers $q$ in the
{\AA}$^{-1}$ domain [see Fig.\ \ref{bulk}(a)]. However, we are mainly
concerned here with longer-range optical excitations (e.g., SPPs), in which
space dispersion can be generally overlooked, except when sharp metallic structures or narrow gaps between metals are involved \cite{paper162}, as shown in Sec.\ \ref{nonlocaleffects} for electrons passing near metal-sphere dimers. Nonlocal effects can be also significant at small separations $\lesssim 1\,$nm between the passing electron and metal surfaces \cite{ZE1990}.

An interesting scenario is presented by the aloof configuration, in
which the electron trajectory does not even intersect the sample.
This leads to distant interaction, capable of providing surface-specific information, free of close-encounter events
between the fast probe and the target atoms. The exponential decay
of surface losses with $2\omega b/v\gamma$ indicates that low-energy transfers are favored in the aloof configuration. This has the additional benefit of minimizing sample damage.

The aloof configuration was pioneered by \textcite{LBN1977}, who
forced the electrons to describe parabolic trajectories, deflected
from the metal surface by a bias potential. Subsequent studies
managed to aim a STEM aloof beam parallel to the planar
surfaces of MgO cubes \cite{M1982,C1982,C1982_2}. Loss spectra were later recorded after transmission through
perforated metallic channels \cite{WBA1984}. The aloof configuration
has been recently revisited and compared with near-field optical
microscopy \cite{CMT98,EHR99,CMT99,CLI03,ICM08}.

\begin{figure*}
\includegraphics[width=180mm,angle=0,clip]{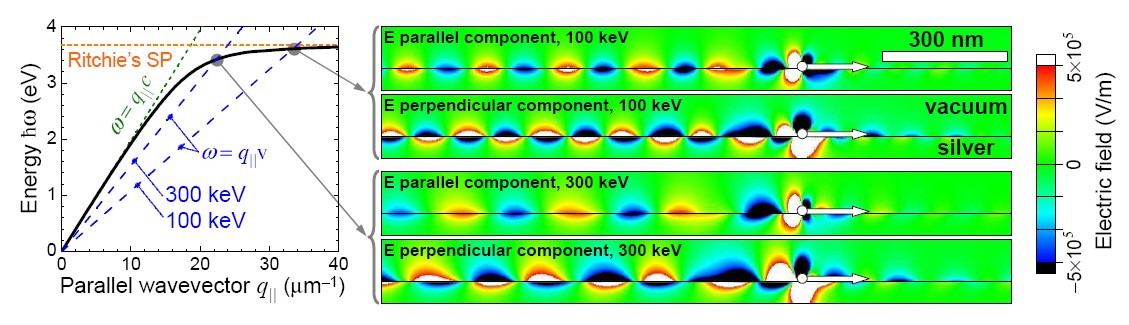}
\caption{\label{SP} (Color online) Surface plasmon dispersion relation in Ag (left) and excitation of SPPs by swift electrons (right). The SPPs saturate at Ritchie's non-retarded frequency for large parallel wavevector and follow the light line $\omega=\QQ c$ in the low-energy limit. An electron moving with velocity $\vb$ parallel to the surface preferentially excites plasmons of frequency and wavevector related by $\omega\lesssim\QQ v$. This condition is indicated by solid dots in the dispersion relation for the two electron energies considered on the right. Faster electrons generate plasmons of lower $\QQ$ and longer wavelength. The impact parameter is 10\,nm in the field plots. The dielectric function of Ag is taken from \textcite{P1985}.}
\end{figure*}

Retardation adds another source of plasmon dispersion. Since SP modes are resonances in the surface response, they must involve a divergence in the reflectivity for incident evanescent waves with the right values of $(\QQ,\omega)$. More precisely, we obtain the plasmon dispersion relation from the vanishing of the denominator of $r_p$ [Eq.\ (\ref{rp})], leading to
\begin{eqnarray}
\ksp=k\,\sqrt{\frac{\epsilon_1\epsilon_2}{\epsilon_1+\epsilon_2}} \label{SPdisp}
\end{eqnarray}
under the condition that ${\rm Re}\{\epsilon_1\}$ and ${\rm Re}\{\epsilon_2\}$ have opposite signs.
The dispersion relation of Eq.\ (\ref{SPdisp}) is represented in Fig.\ \ref{SP} (left) for
a silver-vacuum interface. When the retarded nature of SPs becomes important, they are usually referred to as SPPs, because they acquire a polaritonic character reflecting a mixture of collective electron motion and propagating electromagnetic fields. This tends to happen at low wavevectors (i.e., at long wavelengths according to Fig.\ \ref{SP}), for which SPPs increasingly approach the light line ($\omega=\QQ c$). It is important to stress that this polaritonic regime occurs for momentum transfers that are well below the region exhibiting nonlocal effects [cf. horizontal scales in Figs.\ \ref{bulk}(a) and \ref{SP}]. This is shown in Fig.\ \ref{bulk}(b), where the hybridization of transversal bulk plasmons with light takes place at values of $q$ that are two orders of magnitude smaller than the Fermi wavevector $k_F$. More details on SPPs, including a comprehensive literature survey, can be found in the book of \textcite{R1988}.

The electric and magnetic fields associated to a SPP plane wave moving along $x$ have similar strength in Gaussian units. For instance, in the vacuum region outside a semi-infinite metal of permittivity $\epsilon$, we have
\[\Eb^{\rm SP}(\rb,\omega)=[(\xx q_z^{\rm SP}-\zz\ksp)/k]\;\exp[\ii(\ksp x+q_z^{\rm SP} z)],\]
\[\Hb^{\rm SP}(\rb,\omega)=\yy\;\exp[\ii(\ksp x+q_z^{\rm SP} z)],\]
where $q_z^{\rm SP}=\sqrt{k^2-(\ksp)^2}=-k/\sqrt{\epsilon+1}$. In particular, the electric field satisfies $\nabla\cdot\Eb^{\rm SP}=0$, and therefore it is transversal.\footnote{Although the main $\Eb^{\rm SP}$-component of long-wavelength SPPs is along the surface normal, there is also a finite electric field along the propagation direction, but the normal components of both the wavevector and $\Eb^{\rm SP}$ have a $\pi/2$ phase difference relative to the parallel components, leading to $\nabla\cdot\Eb^{\rm SP}=0$.} Close to the plasmon cutoff, the SPP momentum becomes increasingly large. This marks the transition towards the non-retarded regime, in which magnetic and electric fields are decoupled, so that we can approximate $\nabla\times\Eb^{\rm SP}=0$ and define a potential $\phi=\exp[\ii(x-z)\ksp]$ to write $\Eb^{\rm SP}=-\nabla\phi$. The electric field appears to be longitudinal, and the plasmons are then known as SPs rather than SPPs, signalled by the condition $\epsilon=-1$.

An aloof electron like that of Fig.\ \ref{planar-surface} (left)
couples to SPPs of momentum component along the velocity vector
given by $\hbar\omega/v$, according to Eq.\ (\ref{wqv}). Consequently, the electron
only excites parallel momenta above that value. This condition is
represented in Fig.\ \ref{SP} (left) for two different electron
energies. The corresponding electric fields are shown in the
same figure (right) within a plane that contains the velocity vector
and the surface normal. The electron is passing at a distance of 10
nm above the surface. We can see several characteristic features in
these plots: (1) the field shows a wake pattern characterized
by oscillations of wavelength $\approx 2\pi/\QQ$, with
$\QQ=\omega/v$ determined by the intersection points of Fig.\
\ref{SP} (left); (2) the electron that moves faster excites plasmons
of lower energy, thus giving rise to oscillations of longer spatial period;
(3) at variance with the continuity of the parallel electric field,
the normal component changes sign across the surface in order to preserve
the continuity of the normal displacement, because the
metal permittivity is negative; (4) the normal field
component takes large values compared to the parallel one, and
this effect is more pronounced for faster electrons, which involve lower $\omega$'s.


In a different direction, the early 1990s witnessed remarkable experiments of energy
losses in coincidence with SEE, initially collected in amorphous carbon samples \cite{PK91,MB92}.
Angle-resolved measurements were also carried out in this context to
conclude that the SEE yield is larger in more localized excitations,
involving larger-momentum transfers \cite{DS93}. Comparative studies
for amorphous carbon and silicon suggested that bulk plasmons do
not play a central role in SEE \cite{SDW93,DSL93}, possibly due to the noted begrenzungseffekt, which limits the strength of
volume plasmons in favor of SPs within the region accessible to the
escape depth of SEs. Finally, coincidence
experiments were conducted using the aloof beam geometry in diamond and
MgO to compare EELS and SEE rates and to provide direct measurements
of the probability that a surface excitation gives rise to a SE, which was found to be below $\sim 5$\% for low-energy
losses \cite{MBH93}.

\subsubsection{Guided modes in thin films} \label{guidedmodes}

\begin{figure}
\includegraphics[width=80mm,angle=0,clip]{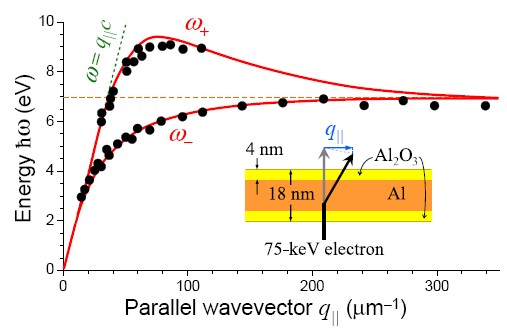}
\caption{\label{Silcox} (Color online) Experimental
determination of the SPP dispersion relation
in a thin film (adapted from \onlinecite{PSV1975}). Symbols: measured
data obtained from energy and angle distributions of fast electrons
traversing a partially oxidized Al thin film. Continuous curves:
SPP branches calculated from optical data.}
\end{figure}

The SPPs on both sides of a thin film interact to produced two
hybridized modes. One of them can travel longer distances along the
film due to exclusion of the electric field from the metal
\cite{S1981}. This mode is currently being applied to
propagate electromagnetic signals in plasmonic devices \cite{BDE03,BML07}.
The dispersion relation of coupled SPPs in thin films was obtained
from the EELS signal of transmitted electrons over 30 years ago by a series of
outstanding experiments conducted by Silcox and co-workers in Al
\cite{VS1973,PSV1975} and Si films \cite{CSV1975}.
An example of these studies is given in Fig.\ \ref{Silcox} for a partially
oxidized thin Al film. Both plasmon branches follow the light line
at low energies, and they converge to the non-retarded limit of the
Al/Al$_2$O$_3$ interface for large momentum transfer. This limit
corresponds to $\epsilon_{\rm Al}+4=1$, or equivalently,
$\hbar\omega_s\approx 6.7$ eV when Eq.\ (\ref{drude}) is used to model Al,
since the permittivity of Al$_2$O$_3$ in this energy window is $\approx 4$.
Large $\QQ$'s involve fast plasmon oscillations along the film,
so that alternating induced charges weakens the interaction between
both film sides and we recover the limit of the single interface separating
two semi-infinite media. For small $\QQ$, the interaction between plasmons in both sides of the film gives rise to two SPP branches.

Guided modes in graphite \cite{CS1975} and aluminum-oxide
\cite{CS1975_2} films were also characterized using the same
technique. In this case, the film behaved as a dielectric, and the
electron coupled to optical Fabry-Perot resonances set up by
successive total internal reflections at the film boundaries. The
experiments were in excellent agreement with theory \cite{CS1979}.

In a more recent development, a complex multilayer structure containing HfO, which is of interest in microelectronics, has been characterized by \textcite{CKS07}. The agreement with theory is excellent after retardation effects are incorporated to correctly account for interface plasmons and Cherenkov guided modes. A similar study has been recently presented by \textcite{YCM08} in slabs and stacks of Si and SiO$_2$.

Additionally, a radiative plasmon branch inside the light cone, giving rise to
losses near $\omega_p$, was predicted by \textcite{F1958}
for thin metal films, and the resulting light emission was detected later
by \textcite{S1960} in Ag. We discuss this issue in more detail
in Sec.\ \ref{Ferrellmodeetc}.

\subsection{Curved geometries}
\label{curvedgeometries}

The ability of TEMs to probe local response properties has been
extensively exploited to study non-planar sample geometries.
Particular attention has been paid to spherical and cylindrical
shapes, carbon nanotubes, composites, and other more complicated
structures. The intense valence losses can provide
spectral images, the interpretation of which is not as direct as
when core losses are employed instead. Therefore, we need to strongly rely
on theory, which is actually in a very advanced state, capable of
dealing with virtually any system. For example, small particles
like fullerenes or carbon nanotubes are reliably described by
atomistic models, although first-principles theory is also possible
\cite{MRR03}. For larger structures (e.g., nanoparticles) we have a
suite of methods based upon classical dielectric theory, which gives
excellent agreement with experiment in metallic objects of
dimensions above $\sim 10$\,nm. In this section, we explore
some of these structures possessing spherical or cylindrical
geometry.

\subsubsection{Cylinders}
\label{cylinders}

Early work on cylindrical nanocavities relied on an electron beam to drill holes in AlF$_3$ \cite{MAB1989}, which were
investigated by EELS \cite{SMI1985}. The synthesis of such types of holes has considerably improved over the last few years in two different directions. (1) Bottom-up approach: chemical methods have been developed that are capable of spontaneously forming arrays of self-organized nanoholes in materials like alumina \cite{MF95}; arrays of cylinders are also formed in eutectics \cite{PKR08}. (2) Top-down approach: advanced electron-beam and focused-ion-beam (FIB) lithographies have been extensively employed to produce hole arrays for nanophotonics studies \cite{GE07}. These and similar geometries have been investigated through TEM spectroscopic analysis \cite{paper080,DLY04}.

Likewise, self-standing Si \cite{RCM99} and Ge \cite{HK04}
nanowires have been studied to show clear evidence of the
begrenzungseffekt in the former and confinement effects in the
position of the bulk plasmons in the latter. Interface losses in Bi
wires embedded in an alumina matrix have been also observed
\cite{SGL01} and theoretically explained \cite{paper070}.

\begin{figure}
\includegraphics[width=80mm,angle=0,clip]{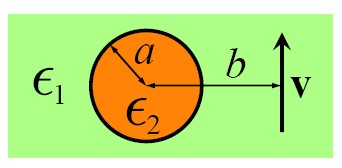}
\caption{\label{sphere} (Color online) Notation used for
homogeneous spheres and cylinders or radius $a$, with an electron
passing at a distance $b$ from the center.}
\end{figure}

The axial symmetry of a cylinder enables us to classify its modes
according to their azimuthal dependence $\exp(\ii m\varphi)$,
whereas a separation into $\exp(\ii q_zz)$ components of wavevector
$q_z$ is appropriate for the variation along the direction of
translational invariance, chosen along $\zz$. This decomposition
is employed in Sec.\ \ref{cylinder} to write the non-retarded screened interaction,
from which the cylinder modes are derived using the condition $\Delta_m=0$ [see Eq.\ (\ref{Wcylinder})].
A relatively simple expression can be obtained for these modes including retardation \cite{AE1974}:
\begin{eqnarray}
&&x_1^2x_2^2\left[\epsilon_2x_1I'_m(x_2)/I_m(x_2)-\epsilon_1x_2K'_m(x_1)/K_m(x_1)\right] \nonumber\\&&\times
\left[x_1I'_m(x_2)/I_m(x_2)-x_2K'_m(x_1)/K_m(x_1)\right] \nonumber\\
&&=m^2\left(\epsilon_1-\epsilon_2\right)^2\left(kq_za^2\right)^2,
\label{cylmodes}
\end{eqnarray}
where the notation of Fig.\ \ref{sphere} has been adopted, the subscripts 1 and 2 refer to the media outside and inside the cylinder, respectively, and $x_j=(\omega a/v)\sqrt{1-\epsilon_jv^2/c^2}$.

The interaction of the electron with a cylinder has been described using both non-retarded \cite{ZRE1989,TKT06} and fully-retarded \cite{ZRE1989,W91_2} dielectric theory. One can easily derive the loss probability using the screened interaction $W$ in the non-retarded limit [see Eq.\ (\ref{Wcylinder})]. For electrons moving parallel to a cylinder, one just needs to insert $W$ inside Eq.\ (\ref{Gfos}). The momentum transfer along the cylinder axis, is fully determined by the condition $q_z=\omega/v$, and one obtains the loss probability
\begin{eqnarray}
&&\Gamma^{\rm NR}_{\rm ref, cyl}(\Rb_0,\omega)=\frac{2e^2L}{\pi\hbar v^2}\sum_{m=-\infty}^\infty
{\rm Im}\left\{-W_m(b,b,q_z,\omega)\right\}, \nonumber
\end{eqnarray}
where $b$ is the distance from the electron beam to the cylinder axis, and different $m$'s contribute separately. Retardation effects are however tedious to deal with analytically \cite{ZRE1989,W91_2}, as compared to the planar interface. For instance, a beam moving along the center of the cylinder, which only couples to $m=0$ modes due to the symmetry of the electron-cylinder combined system, experiences a loss probability
\begin{eqnarray}
\Gamma_{\rm ref, cyl}(\omega)&=&\frac{e^2}{2\hbar v^2}\,{\rm Im}\bigg\{\left(\frac{v^2}{c^2}-\frac{1}{\epsilon_1}\right)
\nonumber \\ &\times&\,\,
\frac{\epsilon_2x_1I_0(x_1)I_1(x_2)
-\epsilon_1x_2I_1(x_1)I_0(x_2)}
{\epsilon_2x_1K_0(x_1)I_1(x_2)
+\epsilon_1x_2K_1(x_1)I_0(x_2)}\bigg\}, \nonumber
\end{eqnarray}
where we can immediately see the presence of the $m=0$ mode predicted by Eq.\ (\ref{cylmodes}) when we arrange the denominator of the last fraction as $K_0(x_1)I_0(x_2)(\epsilon_2x_1I'_0(x_2)/I_0(x_2)-\epsilon_1x_2K'_0(x_1)/K_0(x_1))$.


\subsubsection{Spheres}
\label{spheres}

The interaction of electrons with spheres has attracted considerable attention since the publication of pioneering experimental loss spectra in alkali-halide \cite{C1966} and Al \cite{FKI1967} particles, followed by their subsequent theoretical interpretation \cite{FK1968_2,CR1968}. Further EELS measurements corroborated these results and focused on the role of plasmon dispersion in small spheres \cite{B1980,OBI92} and in the effect of the interaction between neighboring particles \cite{B1982_2,B1985,UCT92}. More recently, UV surface exciton polaritons have been observed in gold nanospheres \cite{paper151} and whispering gallery modes measured in silica beads \cite{HCR08} using EELS.

A small metallic sphere embedded in a dielectric host exhibits plasmon modes at frequencies dictated by the condition \cite{K1987}
\begin{eqnarray}
(l+1)\epsilon_1+l\epsilon_2=0, \label{sphereplasmons}
\end{eqnarray}
where $l$ labels the multipolar order of the excitation, and
$\epsilon_1$ and $\epsilon_2$ are the permittivities of the
host and the metal, respectively (see Fig.\ \ref{sphere}). In
particular, for an Al sphere in which $\epsilon_2$ is described by the Drude dielectric
function of Eq.\ (\ref{drude}), the frequency of the non-retarded plasmon
oscillations becomes
\begin{eqnarray}
\omega=\omega_p\,\sqrt{\frac{l}{(l+1)\epsilon_1+l}}. \label{wlsphere}
\end{eqnarray}
For example, the dipole ($l=1$) of an Al sphere in vacuum has
energy $\hbar\omega_p/\sqrt{3}\approx 8.7$ eV. For very large $l$, the
surface-charge oscillations are so dense that they cannot resolve the particle
curvature, and therefore the mode frequency evolves to the plasmon of the
planar interface, $\omega_s=\omega/\sqrt{\epsilon_1+1}$.

The remarkable works of \textcite{FK1968_2} and \textcite{CR1968}
established the basis for the theoretical understanding of
fast-electron interaction with spherical particles. These authors
described a metallic sphere using the hydrodynamical model of
Appendix\ \ref{hydrodynamicmodel} and derived analytical expressions
for the loss probability and photon emission rate of a broad beam
neglecting retardation. Their results showed spectral features at frequencies roughly described by Eq.\ (\ref{wlsphere}).

This theoretical development was followed by studies of voids in metals \cite{AF1976}, and later by the
derivation of an analytical formula for the loss probability of a
STEM electron passing outside a sphere \cite{FE1985,FWA1987}.
Namely,
\begin{eqnarray}
\Gamma^{\rm NR}_{\rm sph}(\omega) &=& \frac{4e^2}{\pi\hbar v^2}
         \sum_{l=1}^\infty \sum_{m=-l}^l
               \frac{(\omega/v)^{2l}}{(l+m)!(l-m)!}
\label{FEfor} \\ &\times&  K_m^2\left(\frac{\omega b}{v}\right)
               {\rm Im}\left\{\alpha_l^{\rm NR}(\omega)\right\},
               \;\; kb\ll 1, \;\;  v/c\ll 1,
\nonumber
\end{eqnarray}
where
\begin{eqnarray}
\alpha_l^{\rm NR}(\omega)=a^{2l+1} \frac{l\epsilon(\omega)-l}{l\epsilon(\omega)+l+1}
\label{alphaNRvac}
\end{eqnarray}
is the non-retarded multipolar polarizability of a sphere of radius $a$ in vacuum, and $b$ is the distance from the trajectory to the sphere center, as illustrated in Fig.\ \ref{sphere}. Equation\ (\ref{FEfor}) can be derived upon insertion of the non-retarded screened interaction near a sphere [Eq.\ (\ref{Wlalpha})] into Eq.\ (\ref{Gfos}). This formula successfully explained previous measurements for aloof electrons \cite{B1982} and was soon applied to explain EELS of spheres made of tin \cite{ACK1986}, silicon \cite{UCT92}, silver \cite{OBI92}, and aluminum \cite{SBS97}, as well as corrugated metal surfaces \cite{MWF1987}, and diamond particles \cite{FB99,FB00,PFB01}. The effect of coatings, which can be important due to oxidation \cite{FWA1987,UCT92}, is described by Eq.\ (\ref{FEfor}) if we use the polarizability of a coated sphere instead of Eq.\ (\ref{alphaNRvac}).

Optical modes in nanoparticles (e.g., plasmons in metallic spheres)
are well-known to be redshifted due to retardation when the
particle size is not negligible in front of the wavelength (see Sec.\ \ref{plasmonmapping}): the
interaction between distant regions of a particle is mediated by
electromagnetic signals that propagate at the speed of light, but
the signal simply takes longer to travel over larger distances as
the particle dimensions increase, so that the period of the
oscillation modes decreases, thus producing the noted redshift. This
effect was first described by \textcite{M1908} in his analysis of
light scattering by metallic spheres. Similar phenomena
occur in particles made of polaritonic materials \cite{FK1968}.

The extension of Eq.\ (\ref{FEfor}) to incorporate retardation also leads
to an analytical formula \cite{paper030,paper041}:
\begin{eqnarray}
\Gamma_{\rm sph}(\omega) &=& \frac{e^2}{c\hbar\omega}
         \sum_{l=1}^\infty \sum_{m=-l}^l
               K_m^2\left(\frac{\omega b}{v\gamma}\right)
\label{retarded.sphere} \\ &\times&
               \left[C_{lm}^M{\rm Im}\left\{t_l^M\right\}
                +C_{lm}^E{\rm Im}\left\{t_l^E\right\}\right],
\nonumber
\end{eqnarray}
where the Lorentz factor $\gamma=1/\sqrt{1-v^2/c^2}$ accounts for the contraction of the impact parameter $b$ as seen by the moving electron. Here, $C_{lm}^M$ and $C_{lm}^E$ are coupling coefficients that depend exclusively on the $v/c$ ratio, and $t_l^M$ and $t_l^E$ are magnetic and electric Mie scattering coefficients independent of beam energy and impact parameter. Analytical expressions for these coefficients are given in Appendix\ \ref{scatteringcoefficients} for homogeneous spheres. Equation\ (\ref{retarded.sphere}) results from a multipolar expansion of the electric field induced by the electron [Eq.\ (\ref{Ehomo1})] after each multipole is scattered by the sphere to produce an outgoing wave acting back on the electron and giving rise to energy loss [see Eq.\ (\ref{Ggen})].

It should be emphasized that Eq.\ (\ref{retarded.sphere}) can be applied to any spherical object for external electron trajectories, provided $t_l^M$ and $t_l^E$ are known for that object. In particular, this has been recently done for left-handed spheres \cite{XZ08} and spherical metamaterial cloaks \cite{XDZ08}.

\begin{figure}
\includegraphics[width=80mm,angle=0,clip]{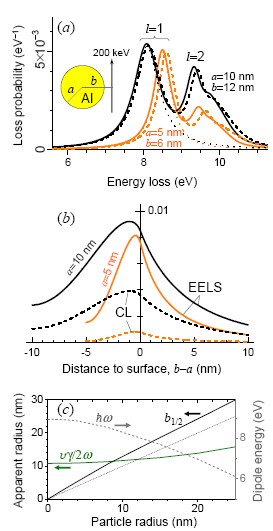}
\caption{\label{resolution} (Color online) Plasmon losses and spatial resolution in Al spheres. {\bf (a)} The loss probability is calculated for two different particle sizes, exhibiting several multipole plasmons that redshift as the particle radius increases. Solid curves: result of Eq.\ (\ref{retarded.sphere}) using local response. Dashed curves: same equation with nonlocal scattering matrices (see Appendix\ \ref{scatteringcoefficients}). Dotted curves: prediction of Eq.\ (\ref{Pdipole}). The Al is described by the dielectric function of Eq.\ (\ref{drude}) with $\hbar\omega=15\,$eV and $\hbar\eta=0.5\,$eV. {\bf (b)} Impact-parameter dependence of the dipolar mode maximum. {\bf (c)} Apparent radius of the dipole excitation in an Al sphere, modeled within local response, and defined as the impact parameter $b_{1/2}$ for which the EELS probability of a 200\,keV electron is half the maximum as a function of $b$. The dipole-energy dependence on particle size (right scale) is taken into account. The $v\gamma/2\omega$ distance of exponential decay in $\Gamma_{\rm EELS,dip}$ [Eq.\ (\ref{Pdipole})] is shown for comparison.}
\end{figure}

Equation\ (\ref{retarded.sphere}) predicts sizable redshifts even for relatively small Al spheres of radius $a=10$ nm, as shown in Fig.\ \ref{resolution}(a) by comparison to spheres with $a=5$ nm (solid curves).\footnote{Loss and CL emission probabilities are plotted per unit of lost-energy range (i.e., in eV$^{-1}$), so that the areas under the spectral curves of Fig.\ \ref{resolution}(a) yield dimensionless probabilities per electron.} Spectral features associated to dipolar ($l=1$) and quadrupolar ($l=2$) excitations are clearly discernable in the spectra. Higher-order multipoles produce a shoulder piling up near $\omega_s$ because their separation is smaller than the absorption width. Besides, the dipole feature is broader in the bigger sphere, which is again a retardation effect, since the coupling of this mode to emitted light increases with the particle dipole strength (i.e., with particle size), therefore limiting the lifetime of the dipolar plasmon due to radiative decay \cite{paper155}. This point is further discussed in Sec.\ \ref{plasmonmapping}.

For sufficiently small particles and aloof trajectories, we can just retain
dipole terms ($l=1$), in which case Eq.\ (\ref{retarded.sphere}) reduces to
\begin{eqnarray}
\Gamma_{\rm EELS,dip}(\omega)=\frac{1}{\hbar\pi} \left(\frac{2e\omega}{v^2\gamma}\right)^2f\left(\frac{\omega b}{v\gamma}\right)\;{\rm Im}\{\alpha\},
\label{Pdipole}
\end{eqnarray}
where
\begin{eqnarray}
f(\zeta)=K_1^2(\zeta)+K_0^2(\zeta)/\gamma^2. \label{fff}
\end{eqnarray}
Actually, this approximation works quite well even for relatively big spheres, as
shown by the dotted curves of Fig.\ \ref{resolution}(a), obtained
from Eq.\ (\ref{Pdipole}) with the electric polarizability
\begin{eqnarray}
\alpha=3t_1^E/2k^3 \label{alphatlE}
\end{eqnarray}
[see Eq.\ (\ref{tlE}); $t_1^E$ is a Mie scattering coefficient]. The dipole mode is given correctly, although higher multipoles are obviously missed. In contrast to the particle-size-independent dipole-plasmon energy predicted by Eq.\ (\ref{alphaNRvac}), we find that Eq.\ (\ref{alphatlE}) incorporates retardation effects leading to a sizeable dipole-plasmon redshift with increasing particle diameter.

Spheres offer an excellent playground to discuss Coulomb delocalization effects (see Sec.\ \ref{spatialresolution}). As anticipated by Eq.\ (\ref{Ehomo1}) for the external electron field, Eq.\ (\ref{retarded.sphere}) predicts an exponential decay of the EELS signal with $b$, and an attenuation distance given by $v\gamma/2\omega$. This is true when the argument of the Bessel functions is large compared to unity, a condition that is not satisfied near small spheres, in which those functions have a steeper dependence on $b$ (see Sec.\ \ref{anevanescentsource}). This is actually observed in the $b$ dependence of the loss
probability shown in Fig.\ \ref{resolution}(b) (solid curves) for the dipole peaks of the spheres considered in Fig.\ \ref{resolution}(a). The probability falls down by a factor of two at a distance of $\sim 2$\,nm away from the surface of the smaller sphere, so that its apparent radius $b_{1/2}\sim 7$\,nm is significantly smaller than $v\gamma/2\omega\sim 11$ nm. The sphere size dependence of $b_{1/2}$ [Fig.\ \ref{resolution}(c)] fully corroborates these arguments.

Small metallic particles exhibit further peak shifts originating in quantum confinement and spatial dispersion, the effect of which is to break down the hypothesis of local response. Nonlocal effects in the performance of bounded metals can be easily included using the specular-reflection-model approximation \cite{RM1966,W1966,paper162}. Then, Eq.\ (\ref{FEfor}) is still valid if we replace $\epsilon$ by \cite{RCF1988}
\begin{eqnarray}
\epsilon_l(\omega)=\frac{\pi}{2a(2l+1)}\left[\int_0^\infty\frac{dq}{\epsilon(q,\omega)}j_l^2(qa)\right]^{-1}
\label{epsl}
\end{eqnarray}
in the polarizability of Eq.\ (\ref{alphaNRvac}), where $\epsilon(q,\omega)$ is the nonlocal bulk response discussed in Secs.\ \ref{bulklosses} and \ref{LindhardMermin}, and $j_l$ is a spherical Bessel function. This level of description is similar to the work of \textcite{FK1968_2}. Under the conditions of Fig.\ \ref{resolution}(a), one needs however to simultaneously include nonlocal and retardation effects, for which the above procedure is not enough.

The hydrodynamic model described in Sec.\ \ref{hydrodynamicmodel} incorporates both spatial-dispersion and retardation effects. When applied to spherical particles, it still yields analytical expressions for the resulting nonlocal Mie scattering coefficients, which we introduce in Appendix\ \ref{scatteringcoefficients}. Using these coefficients in combination with Eq.\ (\ref{retarded.sphere}), we predict blue shifts in the plasmons of the Al particles of Fig.\ \ref{resolution}(a) (dashed curves), even for relatively large spheres. This type of dispersion operates on a length scale
comparable to the wavelength of conduction electrons at the Fermi
level, which is of the order of 1 nm for typical metals. This is
consistent with the large nonlocal effect observed for the quadrupole as
compared to the dipole in the $a=5$ nm sphere, since the $l=2$ mode involves more sign changes in the induced surface charge that increase the relative importance of short distances. It should be noted that spatial dispersion has been found to be partially compensated by the smooth electron density profile at the particle surface \cite{R1978}.

In this regard, nonlocality is important at distances $\sim v_F/\omega$
\cite{LSA04,LS05}, as derived from the minimum momentum transfer in the
e-h pair region for a given excitation frequency $\omega$
(see Fig.\ \ref{bulk}). This is the so-called Landau damping, which has
been phenomenologically accounted for through a size-dependent local
dielectric function, with the dephasing rate increased by $\sim a/v_F$
\cite{B1980,HFH93,KV95}. Nonlocal effects in noble metal particles are noticeable up to a size $\sim 10-20\,$nm at optical frequencies, but it can be important in other materials for larger sizes $\sim 100\,$nm (e.g., silver halide nanocrystals, studied with EELS by \onlinecite{O08}).

There are only scarce data on plasmon dispersion in small spheres (see, for example, \onlinecite{OBI92}), and particularly on the spatial distribution of these effects, so there is a clear need for further EELS experiments that can shed more light into this subject, which is becoming increasingly important in many nanophotonics developments that make use of metallic nanoparticles in the sub-20 nm range. We hope that these results, together with the dimer calculations of Sec.\ \ref{nonlocaleffects}, contribute to trigger future research in this direction.

\subsubsection{Coupled nanoparticles} \label{collectivemodes}

Coupling between plasmon modes in neighboring nanoparticles has
attracted considerable interest in the electron microscopy
community. Early energy-loss experiments focused on the interaction
between neighboring spheres \cite{B1982_2,B1985}, leading to
hybridization of individual particle plasmons, the energies of which were successfully explained through
analytical derivations. Further theoretical work addressed these
modes in more detail \cite{R1982,SD91}, and eventually the coupling
of the electron to two neighboring spheres was derived in the
non-retarded approximation \cite{ZRE97}.

The interaction between more than two non-aligned particles is
difficult to explore by analytical techniques. A particularly
suitable method that yields highly convergent results, including
retardation, consists in expanding the electromagnetic field in terms
of multipoles around the particles, and then allowing these
multipoles to propagate among particles, thus generating a
self-consistent field \cite{paper046}. More details on this method are
given in Sec.\ \ref{secMESME}.

Cylinder pairs have been theoretically addressed as well in the non-retarded limit by \textcite{S1989}. This study was later extended to include retardation and to cope with more than two cylinders \cite{paper070,paperxxx}.

A great deal of work has been devoted in recent years to the
interaction between metallic nanoparticles in the context of nanophotonics,
due in part to the strong enhancement of the light intensity in the
inter-particle region at close proximity \cite{XBK99}. This has
direct application to biosensing using surface-enhanced Raman
scattering \cite{TJO05}. Furthermore, the plasmon modes of
neighboring particles can be tuned to achieve on-demand frequencies
by varying their separation, as shown in optical measurements
performed on lithographically printed metal dots \cite{ASN04}.
However, there are still many issues that remain unclear in the
interaction between particles, specially when the gap distance is
of the order of a few nanometers. Among them, the role of Landau
damping and other dispersion effects. Energy loss spectroscopy turns
out again to be ideally suited for exploring these situations,
in which nanometer spatial resolution is needed to resolve optical
excitations. For example, the singular transition in the response of touching and non-touching particle dimers \cite{paper114} has already prompted EELS measurements \cite{NRM08,paper164}, although a comprehensive experimental study of this transition is still pending.

\subsection{More complex shapes} \label{morecomplex}

The real world is generally more complex than planar, cylindrical,
or spherical surfaces. There are numerous experimental studies of
EELS involving more complicated geometries, like Si tips \cite{RCM99}, metallic
particles supported by dielectric \cite{WC1987_4,WC1987} or by metal
\cite{WC1987_4,WC1987_2} substrates, MgO wedges \cite{paper025,paper048}, MoS$_2$
platelets \cite{CLI03}, 2D photonic crystals \cite{paper080}, Ag spheroids \cite{LCF1984}, Ag
nanotriangles \cite{paper125,NSK07}, Ag nanoshells \cite{EDP07}, and gold nanoparticles ranging from single nanospheres and nanorods \cite{BKW07,SHT09} to touching dimers and interacting nanorods \cite{NRM08,paper164}. The corresponding theoretical efforts needed to explain these measurements involve more elaborate analytical
developments as well as numerical simulations based upon either Poisson's equation in the non-retarded
approximation or Maxwell's equations if full retardation effects
are necessary. These methods use in general local-response functions
and assume sharp boundaries separating different dielectric phases.

\subsubsection{Analytical methods}

Several analytical studies of EELS have been carried out in the non-retarded approximation for geometries that can be naturally described using coordinate systems in which the Poisson equation is separable.

\paragraph{Ellipsoids} The electrostatic modes of ellipsoids have been known for a long time, including analytical expressions for the polarizability \cite{G1912,J1945}. In particular, the EELS probability has been worked out for oblate ellipsoids by \textcite{IAW1988}.

\paragraph{Wedges} Analytical solutions have been reported for the modes of sharp wedges \cite{DM1972,D1976}, parabolic wedges \cite{EM1976}, and hyperbolic wedges \cite{D1976}. The latter display a discrete spectrum that suffers a singular transition in the limit to a sharp edge, characterized by a continuous spectrum. The EELS probability has been calculated for parabolic wedges \cite{GGR1985} with the beam running parallel to the edge.

\paragraph{Supported particles} A sphere or a hemisphere intersected by a planar surface can be described by mode matching, consisting in expanding the potential into spherical harmonics and plane waves within different regions of the structure, thus generating a set of self-consistent linear equations upon application of the boundary conditions. This method, pioneered by \textcite{Z1985} to explore the van der Waals interaction of an atom with a surface feature, was later applied by \textcite{WC1987} to study plasmon modes of supported particles, which they measured in Al spheres sitting on AlF$_3$ \cite{WC1987} and Al surfaces \cite{WC1987_2}. These authors incorporated nonlocal effects within the hydrodynamic approximation \cite{WC1987_3}. Further work on this geometry focused on calculating the loss probability \cite{RZE92,ARA96}.


Several other geometries in which Poisson's equation is separable are still unresolved, but the list of structures that can be accessed with analytical tools is short and we need in general to rely on numerical methods for more complicated shapes. We briefly discuss two of them in this section, and a third one is presented in Sec.\ \ref{carbonstructures}.

\subsubsection{Non-retarded boundary element method}

The boundary element method (BEM) consists in expressing the fields
inside each homogeneous region of a given structure in
terms of interface sources that are then self-consistently calculated by imposing the
continuity of the potential and the electric displacement. This has the advantage that it requires a
modest numerical effort, since it involves parametrization of
boundaries rather than volumes.

The BEM can be traced back to Maxwell, who utilized it to calculate
capacitances \cite{M1891}. It was pioneered by \textcite{F1975}
in the computer age to simulate modes of ionic cubes, and later by
\textcite{LM1990} to find plasmons propagating inside channels cut
into otherwise planar surfaces. Its debut in EELS was made by
\textcite{OI1989}, who obtained loss probabilities that they applied
to study a particle supported on a substrate \cite{OI1989_2}.
Subsequent developments focused on geometries characterized by
either axial symmetry or translational invariance along a certain
direction, thus permitting us to access a large number of complex
morphologies with moderate computational effort \cite{paper025}, since
only curves rather than surfaces need to be parametrized in those
cases. This approach gives excellent results as compared to
experiment for the losses of electrons passing near Si tips
\cite{RCM99} or close to the edge of an MgO cube with different
orientations of the trajectory \cite{paper048}.

\begin{figure}
\includegraphics[width=60mm,angle=0,clip]{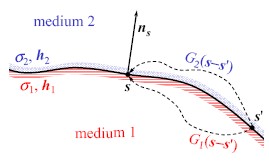}
\caption{\label{BEMparam} (Color online) Scheme of parametrization used in the boundary element method.}
\end{figure}

In the non-retarded BEM, the potential inside a given homogeneous
region of space is separated into two terms,
   \begin{eqnarray}
      \phi=\phi^{\rm ext}+\phi^{\rm ref},
   \nonumber
   \end{eqnarray}
the first of which describes the potential produced by the external
charges present in that medium, treating it as if it were infinitely
extended, while the second term accounts for reflection at the
boundary. Assuming an inhomogeneous structure composed of two
materials with dielectric functions $\epsilon_1$ and $\epsilon_2$,
respectively, the interface $S$ can be assimilated to an induced
surface charge $\sigma$, so that the reflected potential becomes
   \begin{eqnarray}
      \phi^{\rm ref}({\bf r}) = \int_S d{\bf s} \:
                           \frac{\sigma({\bf s})}
                              {|{\bf r}-{\bf s}|}.
   \nonumber
   \end{eqnarray}
Inserting these expressions into Poisson's equation, one finds the following self-consistent relation for the induced boundary charge \cite{paper025}:
   \begin{eqnarray}
      \Lambda\sigma({\bf s})=
               {\bf n}_{\bf s}\cdot\nabla\phi^{\rm ext}({\bf s})
              +\int d{\bf s}'\,F({\bf s},{\bf s}')\,\sigma({\bf s}'),
   \label{nrbem}
   \end{eqnarray}
where
   \begin{eqnarray}
      F({\bf s},{\bf s}')=\frac{-{\bf n}_{\bf s}\cdot({\bf s}-{\bf s}')}{|{\bf s}-{\bf s}'|^3},
   \nonumber
   \end{eqnarray}
\[\Lambda=2\pi(\epsilon_2+\epsilon_1)/(\epsilon_2-\epsilon_1),\]
and ${\bf n}_{\bf s}$ is the interface normal directed towards medium 2 (see Fig.\ \ref{BEMparam}).
The right-hand side of Eq.\ (\ref{nrbem}) is actually the opposite
of the electric field along the boundary normal, with the integral
describing the field produced by induced interface charges. Equation\ (\ref{nrbem}) admits a
representation in terms of the eigenvectors of $F$,
with eigenvalues denoted as $2\pi\lambda_j$. This leads to the following general
expression for the loss probability \cite{paper025}:
   \begin{eqnarray}
      &&\Gamma^{\rm NR}_{\rm ref}(\omega) \label{Pnrbem} \\
      &&=\sum_{i,j}{\rm Im}\left\{-A_{i j}\left[\frac{2}{(1+\lambda_j)\epsilon_1+(1-\lambda_j)\epsilon_2}-\frac{1}{\epsilon_i}\right]\right\},
   \nonumber
   \end{eqnarray}
where $i$ runs over the materials 1 and 2, and $A_{ij}$ are
coupling constants that depend on the actual sample geometry and
electron trajectory but not on $\epsilon_1$ or $\epsilon_2$. The second term inside the square brackets of
Eq.\ (\ref{Pnrbem}) gives a negative loss probability corresponding to
the begrenzungseffekt, the weight of which increases with $A_{ij}$ in
the same proportion as the interface mode described by the first
term.

\subsubsection{Retarded boundary element method}
\label{RBEM}

We can follow a similar procedure to incorporate retardation by
solving Maxwell's equations at the expense of a 6-fold increase in
computational cost \cite{paper030,paper070}. First of all, the electric field can
be expressed in terms of vector and scalar potentials as
   \begin{eqnarray}
      \Eb=\ii k\Ab-\nabla\phi.
   \nonumber
   \end{eqnarray}
We find it convenient to work in the Lorentz gauge, in which
$\nabla\cdot\Ab=\ii k\epsilon\phi$. Then, the potentials become
   \begin{eqnarray}
      \phi(\rb)=\phi^{\rm ext}(\rb)+\int_{S_j}  d\sb \,G_j(|\rb-\sb|) \; \sigma_j(\sb)
   \label{BEMphi}
   \end{eqnarray}
and
   \begin{eqnarray}
      \Ab(\rb)= \Ab^{\rm ext}(\rb)+ \int_{S_j} d\sb \,G_j(|\rb-\sb|) \; \hb_j(\sb)
   \label{BEMA}
   \end{eqnarray}
inside each homogeneous region $j$ delimited by a boundary ${S_j}$, where
   \begin{eqnarray}
      G_j(r) = \frac{\exp\left(\ii kr\sqrt{\epsilon_j}\right)}{r}
   \nonumber
   \end{eqnarray}
is the Green function of Helmholz equation inside that medium,
$\phi^{\rm ext}$ and $\Ab^{\rm ext}$ are the potentials produced by external sources
in an infinite homogeneous medium described by a dielectric function
$\epsilon_j$, and $\sigma_j$ and $\hb_j$ are auxiliary interface
charges and currents (see Fig.\ \ref{BEMparam}). The boundary conditions of the electromagnetic
field are then imposed and one is left with a set of self-consistent surface
integral equations that determine the boundary sources. For example, the continuity of the scalar potential leads to
   \begin{eqnarray}
      &&\int d\sb'\left[G_1(\sb-\sb')\sigma_1(\sb')-G_2(\sb-\sb')\sigma_2(\sb')\right]\nonumber\\&&=\phi^{\rm ext}_2(\sb)-\phi^{\rm ext}_1(\sb),
   \nonumber
   \end{eqnarray}
where $\phi^{\rm ext}_j$ is the external potential in medium $j$, evaluated at positions infinitesimally close to the interface. We proceed by discretizing the integrals in these equations through a finite number of representative points distributed on the interfaces. This leads to a linear set of equations that we solve by linear algebra techniques. Further details on the method and its generalization to an arbitrary number of different media have been given by \textcite{paper070}.

The BEM is particularly advantageous to simulate  axially-symmetric samples, for which different azimuthal components $m$ having $\exp(\ii m\varphi)$ dependence on azimuthal angle are uncoupled. Then, the sample surface can be parametrized with a contour line, thus effectively reducing the problem to one dimension. Similarly, only the 1D contour defining the geometry needs to be parametrized for geometries with translational invariance along a given direction. These two special cases have been discussed in detail by \textcite{paper070}. However, the method is general and can be applied to arbitrary boundaries, as recently illustrated by \textcite{paper149}.

\begin{figure}
\includegraphics[width=80mm,angle=0,clip]{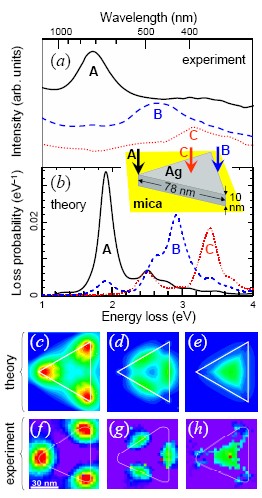}
\caption{\label{triangles} (Color online) Mapping plasmon
modes in an individual Ag triangular nanoprism supported on mica (adapted
from \onlinecite{paper125}). {\bf (a)} Loss spectra acquired at different
positions of the electron beam over the particle (see inset). {\bf (b)}
Parameter-free calculated loss spectra, given in probability per
incident electron per eV of energy-loss range. {\bf (c-e)} Simulated
energy-filtered images at the positions of absolute maxima in the
spectra A-C, respectively. {\bf (f-h)} Measured energy-filtered images.}
\end{figure}

A recent example of application of the retarded BEM is presented in
Fig.\ \ref{triangles}, adapted from \textcite{paper125}, in which
measured loss spectra for Ag nanoprisms [Fig.\ \ref{triangles}(a)] are compared to simulations
[Fig.\ \ref{triangles}(b)] when the electron beam is aimed at the corner, the edge, or the
center of the triangle. Different plasmon losses are observed at
each of these positions. Moreover, one can obtain {\it snap shots} of
these plasmons, as shown in the images of Fig.\ \ref{triangles}(c-h), by scanning the beam over the sample
area and collecting the rate of selected energy losses corresponding to the three dominant features
observed in the spectra of Fig.\ \ref{triangles}(a-b). These findings have been corroborated by tip-scattering NSOM measurement \cite{RJZ08}.

The fully retarded BEM has been used to investigate the optical response of nanoparticles \cite{paper072,HK05,paper115,paper123,paper140,paper155} and extended nanostructures \cite{paper087,paper128}, with particular emphasis placed on localized plasmon resonances, the coupling between atoms/molecules and nanoparticles \cite{paper085,TH08}, and the interaction of fast electrons with nanoparticles to explain EELS \cite{paper030,paper070,paper080,paper125,SHT09,paper164} and CL \cite{paper064,paper139,paper147,paper159,paper167} experiments.

\subsubsection{Multiple scattering approach}
\label{secMESME}

For specimens containing several neighboring particles, multiple scattering techniques provide a natural way of solving Maxwell's equations in frequency space. The field induced by each of the particles adds up to the scattering produced by the other particles in a self-consistent fashion.

In practice, we need to express the electric field around each
particle $j$ in a complete basis set, for which vector spherical
harmonics \cite{WZY93} are a convenient choice. The following
specific steps are then followed to solve the electromagnetic
problem in the presence of a passing electron \cite{paper046}: (1) the external
field is scattered by each of the particles, giving rise to
spherical outgoing waves (single-scattering contribution);
(2) every outgoing multipole centered around a given object is
expanded into propagating multipoles centered around each of
the remaining objects; (3) these propagating multipoles are scattered by the
objects around which they are centered, giving rise to outgoing waves
that add up to the ones obtained after first-order scattering of the
electron field; (4) this process is repeated iteratively, giving
rise to the self-consistent relation
   \begin{eqnarray}
      \xi_j^{\rm ind}=\xi_j^{\rm ss}+t_j \sum_{j'} H_{jj'} \xi_{j'}^{\rm ind},
   \label{MESMEeq}
   \end{eqnarray}
where $\xi_j^{\rm ss}$ and $\xi_j^{\rm ind}$ are vectors
containing the coefficients of the outgoing-wave expansion of the
single-scattering contribution and the self-consistent induced
field, respectively, $H_{jj'}$ is the propagation matrix that
expresses outgoing waves centered around object $j'$  in terms of
propagating spherical waves centered around object $j$, and $t_j$ is the
scattering matrix of object $j$, containing the scattering
outgoing-wave coefficients after interaction of propagating
spherical waves with that object. Analytical expressions of $t_j$
for homogeneous spheres are given in Appendix\
\ref{scatteringcoefficients}. The multipolar expansions are truncated
at some maximum orbital angular-momentum number $l_{\rm max}$ ($\gtrsim ka$ for particles of radius $a$), so
that the above equations contain $L=2N(l_{\rm max}+1)^2$ variables,
where $N$ is the number of particles, and the factor of 2 stands for
the fact that we have electric and magnetic multipoles, similar to
the situation encountered in Eq.\ (\ref{retarded.sphere}) for the
single sphere. Direct inversion of the full system of equations takes a prohibitive computation time proportional to $L^3$. Instead,
we can utilize a highly convergent iterative procedure based upon
Lanzcos' method (see \onlinecite{H1980_2}) to solve these equations
\cite{paper040,paper046}. These computations demand an effort proportional to
$L^2$ and yield highly converged spectra for electrons passing near
clusters of up to several hundred particles in a matter of minutes with $l_{\rm max}\sim 15$.

\begin{figure*}
\includegraphics[width=180mm,angle=0,clip]{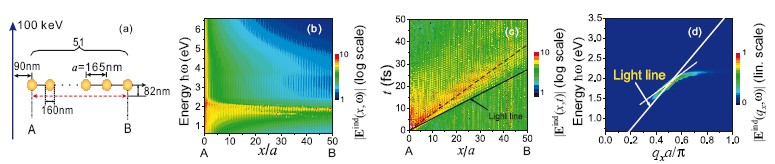}
\caption{\label{Cai} (Color online) Electron-beam-induced plasmon generation in gold nanoparticle chains (taken from \onlinecite{paper168}). {\bf (a)} Scheme of the structure and geometry under consideration. {\bf (b)} Induced electric field $|E^{\rm ind}(x,\omega)|$ (in frequency space $\omega$) produced by the electron along the points of the $\overline{AB}$ segment [see red double arrow in (a), $\parallel\xx$]. {\bf (c)} Same as (b), in the time domain: $|E^{\rm ind}(x,t)|$ [see Eq.\ (\ref{Fourierwt})]. {\bf (d)} Fourier transform of $E^{\rm ind}(x,\omega)$ with respect to $x$, showing the dispersion relation of plasmons excited by the electron and propagating along the chain. The velocity of the launched pulse is found to be $\sim 0.75 c$ [dashed line in (c)], in good agreement with the group velocity at the maximum of excitation probability along the guided-mode dispersion curve [dashed line in (d)].}
\end{figure*}

An example of application of this method is presented in Fig.\ \ref{Cai}, taken from \textcite{paper168}. A 100\,keV electron is considered to pass near the end of a linear array formed by 51 spherical gold particles [Fig.\ \ref{Cai}(a)]. The spectral distribution of the electric field [$|\Eb^{\rm ind}(\rb,\omega)|$, plotted in Fig.\ \ref{Cai}(b) along a segment parallel to the array] exhibits prominent features at the localized plasmon resonances of the spheres. The time-domain picture [Fig.\ \ref{Cai}(c), obtained from Eq.\ (\ref{Fourierwt})] shows plasmon propagation mediated by inter-particle interaction. The group velocity (dashed curve) is consistent with the slope at the intensity-maximum of the guided-mode dispersion relation derived from the space Fourier transform of $|\Eb^{\rm ind}(\rb,\omega)|$ [Fig.\ \ref{Cai}(d)].

\subsection{Composite materials}
\label{compositematerials}

A common situation arises when the sample contains composites made of a fine-scale mixture of different dielectric grains. Finding the optical response of the resulting metamaterial is not easy, specially in cases for which the actual microscopic geometry is unknown. This is even more complicated in EELS due to the local nature of the electron probe.

A first naive approach to the problem of explaining loss spectra in
a composite consists in considering the appropriately weighed bulk
probabilities derived from Eq.\ (\ref{Pbulkintegnr}) for the
constituents. The result is a reasonable approximation when the
grains are large enough so that interface losses are small in
front of volume losses. This type of analysis has been used to map
water distributions in biological specimens using valence
excitations \cite{SSL93}, which are more intense than core
losses and provide similar information in this case, accompanied by lower sample damage.

However, this approach breaks down for finer mixtures, as shown by
\textcite{HW91} through studies of losses in Al spheres created in
AlF$_3$ by the ionizing action of the electron beam. They found that
the spectra were well reproduced by an expression like Eq.\
(\ref{Pnrbem}) with $\lambda_j=1/3$ \cite{HW91,H03}, which results
from the dipolar mode of the spheres [Eq.\ (\ref{sphereplasmons})
for $l=1$], now corrected by the bulk losses due to the
begrenzungseffekt. Similar conclusions were extracted for other
systems, like Ni-Al mixed oxides \cite{CHW93}.

The \textcite{M1904_2,M1906} effective medium theory, which yields good
results in many cases, is particularly suited to describe small
spheres of a given material 2 embedded in a matrix of material 1,
and works extremely well for small volume fractions of the spheres, $f\lesssim 0.3$. This theory predicts an effective
dielectric function
\begin{eqnarray}
  \epsilon_{\rm eff}=\epsilon_1\frac{2(1-f)\epsilon_1+(1+2f)\epsilon_2}
  {(2+f)\epsilon_1+(1-f)\epsilon_2},
\label{MG}
\end{eqnarray}
accounting for the spheres through their non-retarded dipolar response [$\alpha_1^{\rm NR}$ in Eq.\ (\ref{alphaAB})]. Not surprisingly, the $l=1$ mode of Eq.\ (\ref{sphereplasmons}) ($2\epsilon_1+\epsilon_2=0$) is recovered from the zero of the denominator of Eq.\ (\ref{MG}) in the small $f$ limit. The Maxwell-Garnett theory has been successfully employed by \textcite{MH95} to explain EELS measurements in SiO$_2$ polymorphs of different density.

Corrections to the loss function of granular materials beyond the
dipolar terms captured in Maxwell-Garnett theory have been explored
by several authors \cite{FMR93,FR93}. In particular, \textcite{BF95}
obtained a nonlocal effective dielectric function for a random
system of spheres described their local bulk permittivity. This work was
followed by the derivation of a spectral representation of an
effective local response for the same system \cite{FBC96}.
More complicated situations have been considered, such as an aloof electron moving above
ordered and disordered arrays of spheres \cite{MBF98,MBF99}, and
self-similar structures formed by a continuous distribution of
sphere sizes \cite{SMB01}, all of them analyzed in the non-retarded limit.

In this context, the remarkable works of \textcite{MM1977} and \textcite{MMD1978} provide analytical expansions for the effective local response of cubic arrangements of spheres with inclusion of an arbitrarily large number of multipoles, which directly yield the loss function ${\rm Im}\{-1/\epsilon_{\rm eff}\}$. However, a similar derivation of EELS formulas for these structures is still pending, although they could shed light into the open discussion of nonlocal effects in metamaterials.

\begin{figure}
\includegraphics[width=75mm,angle=0,clip]{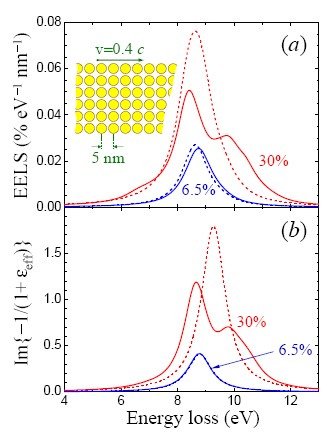}
\caption{\label{metamaterials} (Color online) {\bf (a)} Energy loss spectra for an electron moving
parallel to the $[100]$ direction of the $(100)$ surface of a
simple-cubic crystal formed by six layers of aluminum spheres in
vacuum. The lattice period is 5 nm. We consider two different filling
fractions as shown by labels. The electron is moving at a distance of 1\,nm
from the sphere surfaces with velocity $v=0.4 c$. The dashed curves
are obtained by neglecting the interaction among spheres, whereas
the solid curves correspond to the full solution of Maxwell's equations using a layer KKR approach.
{\bf (b)} Surface loss function for the same crystal as in (a), using the effective dielectric
function obtained from Maxwell-Garnett theory [broken curves, Eq.\ (\ref{MG})] and from
the layer KKR method (solid curves).}
\end{figure}

Retardation corrections to the effective dielectric function can be obtained from the photonic bands of periodic composites in the long-wavelength limit. This approach was followed by \textcite{PM94} to
explain the $\sim 8$ eV loss feature measured by \textcite{HW91} in
aggregates of small aluminum particles. Their simulations were
accompanied by a complex series of spectral features that subsequent
analysis by \textcite{paper075} demonstrated to be due to numerical inaccuracies. In fact,
the loss spectrum for small filling fraction ($f=0.06$) of an array of
equally-spaced aluminum spheres already shows a single $\sim 8.5$\,eV feature
\cite{paper075} [see Fig.\ \ref{metamaterials}(a), solid curves].
This peak splits into two for larger $f=0.3$, due to multipolar interaction between neighboring spheres, quite different from the individual-plasmon excitation predicted by Maxwell-Garnett theory [Fig.\ \ref{metamaterials}(b), dashed curves, obtained from Eq.\ (\ref{MG})]. Interestingly, the calculated EELS peaks in Fig.\ \ref{metamaterials} are displaced to the red with respect to the maxima of the surface loss function ${\rm Im}\{-1/(1+\epsilon_{\rm eff})\}$ as a result of retardation, similar to the effect discussed in Fig.\ \ref{SP}. Both the EELS probability and the effective dielectric function $\epsilon_{\rm eff}$ are calculated here using a layer Korringa-Kohn-Rostoker (KKR) method, which allows solving the multiple-scattering equation (\ref{MESMEeq}) for parallel layers of spheres \cite{SYM00,paper075}. The aluminum is described by Eq.\ (\ref{drude}) with $\hbar\omega_p=15$\,eV and $\hbar\eta=1$\,eV. A value of $l_{\rm max}=6$ is sufficient to achieve convergence. We plot for comparison the loss probability computed in the absence of multiple-scattering interaction among spheres [Fig.\ \ref{metamaterials}(a), broken curves], predicting a single peak that is roughly independent of $f$.

\subsection{Carbon molecules and low-dimensional structures: The discrete-dipole approximation}
\label{carbonstructures}


Carbon molecules and other low-dimensional structures have
fascinated chemists and physicists since the discovery of C$_{60}$
\cite{KHO1985} and carbon nanotubes \cite{I91,M06}, followed by BN
nanotubes \cite{CLC95} and other related molecules.

Plasmon modes were first observed in solid C$_{60}$ as
6.5\,eV and 28\,eV EELS features \cite{GPT91,SSO91}, corresponding to
collective oscillations of $\pi$ and $\sigma$ electrons,
respectively. The optical properties of the individual molecules is
not significantly altered in the solid, since their binding interaction is
basically due to weak van der Waals attraction. Actually, subsequent studies in gas phase
C$_{60}$ molecules corroborated these results \cite{KC92}.

The clear-cut distinction between loosely-bound $\pi$-like orbitals and
tightly-bound $\sigma$-like orbitals persisted in the loss spectra of
carbon nanotubes \cite{KTT02}, in both single-wall and multishell
configurations \cite{STK02}, in multishell fullerenes \cite{HMR99}, in BN nanotubes \cite{KHS00}, and in other similar
types of molecules \cite{KSH01}.

These results were soon related to the dielectric properties of
graphite, the anisotropic permittivity of which was used as
input to describe fullerenes and nanotubes via continuous dielectric
theory, with different $\epsilon$ for field components parallel and
perpendicular to the carbon shells \cite{LHL94}. The dielectric
approach yielded reasonable results compared to measurements of plasmons
sustained by a diverse number of geometries and chemical identities,
including multishell BN nanotubes, C nanotubes, and C fullerenes
\cite{HMR99,KHS00,STK02}, WS$_2$ hollow nanoparticles \cite{KSH01}, and WS$_2$
nanotubes \cite{TKC02}. Good agreement was obtained for nanotubes of different thicknesses, down
to the single-wall level \cite{STK02}.

Also, first-principles calculations have been carried out to describe the interaction of fast electrons with carbon nanotubes \cite{NSC06}. The agreement with experimental loss spectra of single-wall tubes is quite satisfactory \cite{MRR03}. However, this approach is currently limited to relatively small structures, due to the heavy computational demand of spectroscopy {\it ab initio} methods.

The discrete nature of these molecules suggests an atomistic
approach to describe their response, with each atom
represented by an induced dipole of different strength in the
directions out-of-plane and along the tangent plane of the atomic shells. This is the so-called
discrete-dipole approximation (DDA), in which the atomic
polarizability is obtained from the dielectric function of the bulk
material (e.g., graphite for carbon nanotubes and fullerenes) by inverting
the Clausius-Mossotti relation.\footnote{For carbon structures, the effective polarizability per atom is obtained from the dielectric function of graphite using $\alpha_i^{-1}=(4\pi n_C/3)\,[(\epsilon_i+2)/(\epsilon_i-1)-S_i]$, where $n_C=113.63\,$nm$^{-3}$ is the atomic density of this material, $i=\parallel,\perp$ refers to the response to an electric field either parallel or perpendicular to the carbon planes, and the constants $S_\perp=-2S_\parallel=2.8216$ originate in the lack of inversion symmetry of the graphite atomic lattice \cite{AM1976}.} A detailed comparison between the DDA and the continuous dielectric theory was carried out by \textcite{HL96}, who showed that both approaches are in excellent mutual agreement.

The DDA was pioneered by \textcite{PP1973} in the study of the
optical response of dielectric grains by dividing them into fine
volume elements that were described through induced dipoles. Actually, the DDA
is a versatile method for dealing with arbitrary
sample morphologies and is extensively employed to predict
optical properties of nanoparticles \cite{DF94,KCZ03,N07}. This
versatility should find application in the simulation of EELS, and
therefore, we offer a simple derivation of the method that
generalizes previous approaches to fully include retardation
effects.

In order to apply the DDA to carbon structures, each of the atoms (or alternatively, each of the bonds in a similar formulation) is assimilated to an induced point particle of polarizability tensor $\alpha(\omega)$. The dipoles induced on these points in response to an external electric field $\Eb^{\rm ext}(\rb)$ are then obtained from their self-consistent interaction, so that the dipole of the particle located at position $\rb_j$ is given by
\begin{eqnarray}
\pb_j=\alpha\cdot\left[\Eb^{\rm ext}(\rb_j)+\sum_{j'\neq j}\mathcal{G}^0_{jj'}\cdot\pb_{j'}\right], \label{DDAeq}
\end{eqnarray}
where the $3\times 3$ symmetric tensor $\mathcal{G}^0_{jj'}$ describes the coupling with the dipole at $\rb_{j'}$.\footnote{The electric field created by dipole $\pb_{j'}$ is given by \begin{eqnarray}&&\mathcal{G}^0_{jj'}\cdot\pb_{j'}=\frac{\ee^{\ii ku}}{u^3}\nonumber\\&&\times\left\{\left[(ku)^2+\ii ku-1\right]\,\pb_{j'}\,-\left[(ku)^2+3\ii ku-3\right]\,\frac{(\rb\cdot\pb_{j'})\,{\bf u}}{u^2}\right\},\nonumber\end{eqnarray} where ${\bf u}=\rb_j-\rb_{j'}$. This quantity is related to the Green tensor $G$ of Eq.\ (\ref{defgreen}) in vacuum through $\mathcal{G}^0=-4\pi\omega^2G$.} Equation\ (\ref{DDAeq}) can be solved by direct matrix inversion to yield
\begin{eqnarray}
p=\frac{1}{\alpha^{-1}-\mathcal{G}^0}\cdot E^{\rm ext}, \label{secdip}
\end{eqnarray}
where $p$ represents the vector formed by the components of all dipoles $\pb_j$, the matrix $\mathcal{G}^0$ is made of the components of $\mathcal{G}^0_{jj'}$, and so on. The electric field corresponding to a passing electron is given by Eq.\ (\ref{Ehomo1}), that is, $\Eb^{\rm ext}(\rb_j)=(2e\omega/v^2\gamma)\,\gb_j$, where $\gb_j=\gb(\rb_j)$ [see Eq.\ (\ref{Ehomo2})].

Using the same matrix notation as in Eq.\ (\ref{secdip}), the fully-retarded loss probability in the presence of our polarizable point particles representing carbon atoms reads
\begin{eqnarray}
\Gamma_{\rm EELS}(\omega)=\frac{1}{\hbar\pi} \left(\frac{2e\omega}{v^2\gamma}\right)^2
{\rm Im}\left\{g^+\cdot\frac{1}{\alpha^{-1}-\mathcal{G}^0}\cdot g\right\}. \label{DDAEELS}
\end{eqnarray}
Notice that this expression reduces to Eq.\ (\ref{Pdipole}) in the single-atom case ($\mathcal{G}^0=0$).

\begin{figure}
\includegraphics[width=80mm,angle=0,clip]{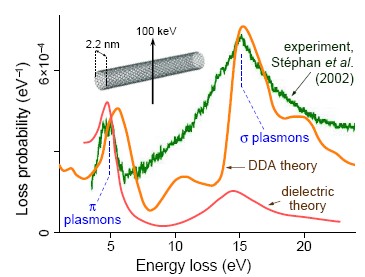}
\caption{\label{carbon} (Color online) Probing plasmons in a single-wall carbon nanotube via EELS. The experiment shows prominent $\pi$ and $\sigma$ plasmons that are relatively well described both by the discrete-dipole approximation (DDA) and by a dielectric approach. The experimental curve and the dielectric theory have been taken from \textcite{STK02}. The DDA calculation performed here for a $16\times 16$ nanotube is based upon the dielectric function of graphite, taken from \textcite{P91}.}
\end{figure}

As an example of application of Eq.\ (\ref{DDAEELS}), Fig.\ \ref{carbon} shows the loss probability for an electron passing near a single-wall carbon tube under grazing incidence conditions. The experimental data and the results of the anisotropic dielectric theory have been taken from \textcite{STK02}. The position of the plasmon features predicted by the above DDA formalism compares very well to both of them. It is remarkable that dielectric theory is still producing meaningful results for such thin nanotubes, although the DDA seems to attribute a relative weight to $\pi$ and $\sigma$ plasmons in better agreement with the observed spectrum.

\subsection{Relation to the photonic local density of states}
\label{relationtophotonic}

We can ask ourselves how exactly the loss probability is related to the photonic properties of the sample and whether it is possible to retrieve from measured data for bounded media a relevant quantity similar to the dielectric function. This question has positive answer in systems with translational invariance along the direction of electron motion: the EELS probability is proportional to the photonic local density of  states (LDOS) \cite{paper145}.\footnote{There is a univocal correspondence between the spatially-dependent local dielectric function and the Green tensor, as can be clearly deduced upon inspection of Eq.\ (\ref{defgreen}). This correspondence works frequency by frequency. Furthermore, the photonic LDOS is directly obtained from the Green tensor following Eq.\ (\ref{eeq2}). The inverse relation is less direct, although one can extend the equivalent of the \textcite{HK1964} theorem to light, thus asserting that there is a univocal correspondence between dielectric functions and LDOS, with both quantities defined within a finite frequency range above $\omega=0$ (i.e., the correspondence is not local in frequency).}

The photonic LDOS is defined by analogy to its electronic counterpart in solid state physics as the combined local intensity of all normalized eigenmodes of the system \cite{AM1976}, but unlike electronic wave functions, the electric field is a vectorial quantity, which suggests defining separate Cartesian projections of the LDOS along unit vectors $\nt$ as \cite{FMM05}
\begin{eqnarray}
\rho_\nt(\rb,\omega)=\frac{-2\omega}{\pi}\,{\rm Im}\{\nt\cdot G(\rb,\rb,\omega)\cdot\nt\} \label{eeq2}
\end{eqnarray}
in terms of the Green tensor defined in Eq.\ (\ref{defgreen}).\footnote{The definition of Eq.\ (\ref{eeq2}) assumes that the photon eigenmodes are well defined (that is, in the absence of absorption) \cite{CGW01}. An alternative interpretation, which holds even in the presence of lossy materials \cite{DMC04}, comes from the realization that $(4\pi^2\omega D^2/\hbar)\,\rho$ gives the spontaneous decay rate for an excitation dipole strength $D$ \cite{BL96,FMM04}. It should be also noted that a complete definition of the LDOS must include a magnetic part \cite{JCM03}, which is however uncoupled to our fast electrons.}
The uniform LDOS in free space is known from black-body theory:
\begin{eqnarray}
\rho^0_\nt(\rb,\omega)=\omega^2/3\pi^2c^3, \nonumber
\end{eqnarray}
which contains an additional factor of $1/3$ arising from the projection over a specific Cartesian vector $\nt$.

If we choose the direction of translation invariance along the $\zz$ axis, the Green tensor $G(\rb,\rb',\omega)$ depends on $z$ and $z'$ only via $z-z'$, and consequently we can write
   \begin{eqnarray}
      G(\rb,\rb',\omega)=\int\frac{dq_z}{2\pi}\,\tilde{G}(\Rb,\Rb',q_z,\omega)\,\ee^{\ii q_z (z-z')}.
   \label{eeq15}
   \end{eqnarray}
Using this in Eq.\ (\ref{eq14}), the loss probability can be recast
as
   \begin{eqnarray}
      \Gamma_{\rm EELS}(\Rb_0,\omega) &=&
      \frac{4e^2L}{\hbar} \, {\rm Im} \{-G_{zz}(\Rb_0,\Rb_0,q_z,\omega)\}
      \nonumber \\ &=& \frac{2\pi e^2L}{\hbar\omega}\,\tilde{\rho}_\zt(\Rb_0,q_z,\omega),
   \label{eq14bis}
   \end{eqnarray}
where $q_z=\omega/v$ is determined by energy conservation [see Eq.\ (\ref{wqv})], $L$ is the length of the trajectory, and
\begin{eqnarray}
\tilde{\rho}_\nt(\Rb,q_z,\omega)=\frac{-2\omega}{\pi}\,{\rm Im}\{\nt\cdot\tilde{G}(\Rb,\Rb,q_z,\omega)\cdot\nt\} \nonumber
\end{eqnarray}
is the photonic density of states, local in real space along the $\Rb$ directions perpendicular to the beam and local in wavevector space $q_z$ along the direction of motion, $\zz$. This relation between the photonic LDOS and the EELS probability is further illustrated in Sec.\ \ref{cherenkoveffect} for electrons moving inside 2D photonic crystals.

For a non-lossy homogeneous dielectric of real permittivity $\epsilon$, one finds
$\tilde{\rho}_{\zt}(\Rb,q_z,\omega)=(\omega/2\pi c^2)(1-q_z^2/k^2\epsilon)
\,\theta(v^2\epsilon-c^2)$, which reflects the fact that only electrons
moving faster than light in the material can lose energy [see Eq.\
(\ref{Cherenkovcond})]. This is the Cherenkov effect that we
discuss in Sec.\ \ref{cherenkovradiation}. As expected, the vacuum
density of states does not contribute to the EELS signal.

Equation\ (\ref{eq14bis}) constitutes a rigorous relation between the EELS probability and the LDOS valid for 2D systems with translational invariance along the direction of electron motion. The LDOS enters this relation through its projection over specific components $q_z$ of light wavevector along that direction. A similar relation can be obtained for fully arbitrary geometries from Eq.\ (\ref{eq14}) \cite{paper145}:
   \begin{eqnarray}
      \Gamma_{\rm EELS}(\Rb_0,\omega) &=&
      \frac{4e^2}{\hbar} \, {\rm Im} \{-G^{\rm ind}_{zz}(\Rb_0,\Rb_0,q_z,-q_z,\omega)\},
      \nonumber \\ &=& \frac{2\pi e^2}{\hbar\omega}\,\rho_\zt(\Rb_0,q_z,\omega),
   \nonumber
   \end{eqnarray}
where $G^{\rm ind}_{zz}(\Rb,\Rb',q_z,-q_z',\omega)$ is the Fourier transform
of the induced Green tensor with respect to $z$ and $z'$, $q_z=\omega/v$, and we have defined, by analogy to Eq.\ (\ref{eeq2}), the quantity
\begin{eqnarray}
\rho_\nt(\Rb,q,\omega)=\frac{-2\omega}{\pi}\,{\rm Im}\{\nt\cdot G(\Rb,\Rb,q,-q,\omega)\cdot\nt\} \nonumber
\end{eqnarray}
as a generalized density of states that is local in real space along the $\Rb$ directions and local in momentum space along the remaining $z$ direction. In 2D systems, we have $\tilde{\rho}_\nt=L\rho_\nt$.

\subsection{Electronic structure determination}
\label{electronicstructure-1}

The publications containing EELS studies of local electronic structures are so numerous that only a succinct review is possible here. A spectacular degree of spatial resolution is achieved by resorting to high-energy losses, for instance when mapping $sp^2$ and $sp^3$ carbon in a silicon-diamond interface with sub-nanometer detail \cite{MTR93}, and also when imaging K atoms in doped fullerenes \cite{GSS05}. However, valence losses can provide excellent resolution too with minimum sample damage, as illustrated by \textcite{DBB03}, who mapped carbon chemical states through characteristic bulk plasmon losses with nanometer detail. In general, comparison with {\it ab initio} theory greatly facilitates the identification of spectral features in the measured loss spectra \cite{MRR03,KB08}.

The determination of band gaps in insulators is successfully addressed by valence EELS \cite{DFM98}, although this requires special care to deal with the subtraction of the ZLP and to get rid of CR losses, which can be misleading in the assessment of the inelastic signal threshold \cite{SFS06,SS07,GSS07,SLS08,S08}.

\begin{figure}
\includegraphics[width=80mm,angle=0,clip]{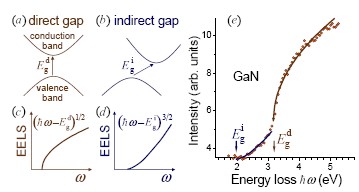}
\caption{\label{gaps} (Color online) Gap determination
using EELS. {\bf (a-b)} Direct and indirect gaps. {\bf (c-d)} Loss probability
near direct and indirect gaps (adapted from \onlinecite{RB98}). {\bf (e)}
Example of simultaneous fit of indirect and direct gaps in GaN
for $E_{\rm g}^{\rm i}=2$ eV and $E_{\rm g}^{\rm d}=3.2$ eV
(adapted from \onlinecite{LBW03}).}
\end{figure}

The loss probability has been shown to be closely related to the
so-called joint density of states (JDOS) \cite{PFW1984,PFW1985}, which is the convolution
of density of states of occupied and unoccupied electronic bands. The JDOS
gives an intuitive measure of the interband-transition probability when
more subtle effects in the excitation matrix elements and lifetime effects are
neglected. The JDOS predicts a dependence of the loss probability on the energy transfer $\hbar\omega$ as
$\sim(\hbar\omega-E_g^d)^{1/2}$ and $\sim(\hbar\omega-E_g^i)^{3/2}$ at the
edge of direct- and indirect-gap energies, $E_g^d$ and $E_g^i$, respectively
\cite{RB98}. A good example of this is found in GaN, as shown in Fig.\ \ref{gaps},
adapted from \textcite{LBW03}. A direct gap involves a more abrupt
decay of the loss probability near the gap [Fig.\
\ref{gaps}(a)-(d)]. Interestingly, both the smooth tail of an
indirect gap and the sharper edge of a higher-energy direct gap are
clearly resolved in the loss spectra of GaN [Fig.\
\ref{gaps}(e)].

In this context, Van Hove singularities \cite{V1953} appearing in the electronic
density of states at energies corresponding to maxima, minima, and
saddle points of the bands have been also resolved using
valence EELS \cite{DFM98}.

Finally, the spatial resolution of electron microscopes can be employed at
its best to study electronic bands of nanostructures. For instance, this has been
done to measure optical gaps in single-, double-, and
triple-wall BN nanotubes \cite{ASK05}, showing that the gap is
rather independent of thickness and stays near the in-plane value of
hexagonal bulk BN ($\sim 5.8$ eV). In a separate development, quantum confinement effects in the
gaps of individual CdSe quantum dots have been observed using
space-resolved EELS \cite{EB07}.

\section{Cathodoluminescence: Generation of Light by Incoming Electrons}
\label{cathodoluminescence}

Cathodoluminescence (CL) was first observed in the mid nineteenth century as light emission produced when cathode rays (electrons) struck the glass of evacuated discharge tubes. Since then, a whole industry has been generated in the search for outstanding phosphorescent materials \cite{O1990}, and in particular, CL has been extensively used in TV screens, and even TEMs have benefited from it by projecting sample images in phosphor screens.

Cathodoluminescence offers a method for characterizing microstructures causing minimum sample damage. It can be regarded as a contactless technique under the aloof configuration (see Sec.\ \ref{planarsurfaces}). CL is widely utilized in mineralogy \cite{SKR00,PGV03,P08} to resolve the composition, contamination, and defects of natural structures with sub-micron detail. It is also a common tool to study the electronic bands of insulators \cite{BP1971} and semiconductors \cite{HY1986,SIJ04}, and particularly the effect of dopants \cite{HY1990}. Recently, CL has been used as a means to read stored information by tracing phase transitions in Ga nanoparticles \cite{paper144}, which can be in turn controlled by electron excitation \cite{PMK04}. Cathodoluminescence is typically performed in adapted SEMs using 1-50 keV beams, and it is now possible to record spectra for every pixel of a given image \cite{GMT03}. In an separate development, low-energy electrons (150-300 eV) emitted from scanning tunneling microscope tips have been employed to produce surface CL, which benefits from the small inelastic mean free path of electrons at those energies \cite{MLM06}.

Here, we are interested in the application of CL to investigate the optical properties of nanostructured materials. Besides e-h pair recombination, which dominates CL in semiconductors, we are interested in other mechanisms of light emission such as excitation of leaky plasmons in metal nanoparticles. Overall, CL constitutes an excellent technique to access optical properties with nanometer resolution and without spurious effects due to modifications produced by the presence of a probe in the region near the sample, as is the case in NSOM. A close relative of CL is found in light emission accompanying inelastic electron tunneling, which can be assisted by localized plasmon excitations \cite{PB92}, although the photon yield is very low in that case.

\subsection{Mechanisms of light emission}
\label{mechanismsoflight}

We find it convenient to classify the mechanisms of CL emission
according to their degree of coherence with respect to the external
electric field of the bombarding electrons.

\subsubsection{Coherent electron-induced radiation emission}
\label{eire}

The interaction of the electron with a material gives rise to electromagnetic far-field components that are coherent with respect to the external evanescent field accompanying the moving charge, described by Eq.\ (\ref{Ehomo1}). These far-field components produce what we call electron-induced radiation emission (EIRE). There are several mechanisms of coherent light emission, the most notorious of which are transition radiation (Sec.\ \ref{transitionradiation}), Cherenkov radiation (Sec.\ \ref{cherenkovradiation}), and diffraction radiation (Sec.\ \ref{diffractionradiation}). The electric field associated to each of these mechanisms is subject to the same set of Maxwell equations, so that we must observe interference effects between them.

\begin{figure}
\includegraphics[width=80mm,angle=0,clip]{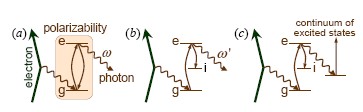}
\caption{\label{polarizability} (Color online) Some of
the processes taking place during energy transfer from a swift
electron to a sample. {\bf (a)} Excitation from the specimen ground state g
to an excited state e, and subsequent radiative decay of the latter (see Sec.\ \ref{eire}).
{\bf (b)} Excitation followed by incoherent radiative decay to an
intermediate state i (see Sec.\ \ref{incoherentluminescence}). {\bf (c)}
Excitation followed by non-radiative decay.}
\end{figure}

The interference of coherently emitted light is actually due to the
fact that the quantum-mechanical state of the sample does not change
after the emission process takes place. All the energy released by
the electron in a scattering event giving rise to EIRE is liberated
in the form of radiation, as schematically shown in Fig.\
\ref{polarizability}(a).

EIRE sources have a relatively delocalized origin, since they
involve regions of the sample that are large from the electronic
point of view, although they can be small compared to the wavelength
of the emitted light. As a consequence, the coherent photon
yield increases with electron energy: the faster the charge, the
more extended the external field (see Sec.\ \ref{anevanescentsource}) and the more efficient the coupling to delocalized excitations.

When heavy encounters with sample atoms take place, the sudden change in
the electron trajectory gives rise to radiation. This is the so-called
bremsstrahlung \cite{J99}, which can maintain certain degree of
coherence with the sources of EIRE discussed so far. In particular,
theoretical analysis has shown that bremsstrahlung produces
interference with transition radiation (TR) upon electron bombardment
\cite{RAE1965,SR1970}. Experimental evidence of this effect has
been reported under grazing incidence conditions \cite{CA1967}, so that deflections
in the electron trajectory occur close to the surface. However,
the contribution of bremsstrahlung to the total photon yield under usual TEM and SEM operation
conditions is minor, unless very grazing trajectories are considered
\cite{TLS1972,TLS1974}.

\subsubsection{Incoherent cathodoluminescence}
\label{incoherentluminescence}

The incoherent part of the emission bears close relation to
photoluminescence: the sample is first excited and then it decays
inelastically by emitting light of a different color, displaced to
the red with respect to the external illumination. The only
difference lies in the source of excitation: while external light
preferentially couples to strong-dipole modes, electrons can
efficiently excite metastable levels because they are capable of
transferring relatively large momenta to the sample (see, for example, Fig.\
\ref{bulk}) that enhances the probability of non-dipole transitions.
Furthermore, the electron field can be regarded as a super-continuum source of
evanescent light, suitable for creating energetic excitations (for
instance, in large band-gap insulators) that would be difficult to
produce using a laser instead.

Figure\ \ref{polarizability}(b) shows a typical incoherent emission
process, whereby the excited state produced by the electron decays
into an intermediate state, giving rise to light of frequency
$\omega'$ below the transferred $\omega$. The
sample changes its quantum-mechanical state after the interaction with the electron and the optical
emission take place, and consequently, the resulting light cannot produce interference with coherent EIRE.
This contribution to CL is similar to the outcome of excited-atom decay (i.e., proportional to the radiative part of the photonic LDOS).

Incoherent emission is generally associated to the creation of
e-h pairs and excitons. Actually, CL is routinely used to
investigate these types of excitations \cite{SSW00,KWH01,RSP05}, and
particularly when one is interested in resolving them in space. For
instance, to study quantum dots \cite{GCL95,RSP05},
quantum-confinement in semi-conductors \cite{CCS06,YBW06},
and optical effects driven by surface atomic steps \cite{MY06}.

We should expect the opposite dependence of the incoherent emission yield on electron velocity as compared to coherent EIRE. The excitations in this case are highly localized, so that the strength of the coupling to the more delocalized field of faster electrons turns out to be weaker. This effect is actually reflected in the energy dependence of the electron inelastic-mean-free-path (e-imfp) \cite{TPP93}, represented in Fig.\ \ref{techniques}: this quantity passes through a minimum at $\sim 50$ eV, in which inelastic excitation is more probable, and exhibits a monotonic increase with energy due to weaker Coulomb interaction when the electron moves faster. Incoherent CL can take very large values in suitably designed insulators [for example, $>5$ photons per incident 4\,keV electron in LaF$_3$ doped with rare earths \cite{WCL08}].

There are other situations in which the excited states involved in incoherent CL are not so localized. For instance, in emission assisted by nanoparticle plasmons. More precisely, $d$-band holes created by photo-excitation in Au nanocrystals have been shown to recombine non-radiatively with $sp$ electrons that give rise to excitation of intermediate particle plasmons, the decay of which results in photoluminescence intensities that are four orders of magnitude higher than in metal films \cite{DNK04}, but still insignificant compared with electronic decay channels [see Fig.\ \ref{polarizability}(c)]. Nevertheless, these phenomena are relevant to understand CL assisted by particle plasmons in metal clusters embedded in dielectrics \cite{DAN02,KCV05}.

Incoherent CL can be regarded as a way to deliver a localized source
of light (the inelastic excitations) with which one can investigate local photonic
properties. This is possible thanks to the effect predicted by
\textcite{P1946}: an increase in the radiative decay rate of excited
nuclei coupled to a resonant circuit. The Purcell effect is
quantified by a linear dependence of the light emission rate of
excited atoms on the photonic LDOS (see Sec.\ \ref{relationtophotonic}) that has been observed
in numerous experiments (see \onlinecite{AB01}, and references
therein). But of course, this decay channel must compete with
non-radiative processes and their relative magnitudes depend on the
dielectric environment. As a thumb rule, atomic radiative decay dominates in
insulators and is negligible in metals.

A microscopic quantitative analysis of incoherent CL is made more
complicated by the excitation of hot electrons, which are in turn a
source of CL. Some of these electrons
escape from the sample surface, giving rise to SEE, the yield of which can reach values above 20 electrons per
primary electron in some insulators at $\sim 1$ keV incidence energy
\cite{C06}. This high emission is due to hot charge carriers right above
the insulator gap, characterized by long mean free paths and therefore
larger probability of escaping from the solid \cite{paper031}.

Secondary electrons of significant energy above the Fermi level are less abundant in metals because they are rapidly quenched by strong dynamical screening. Moreover, the CL emission comes from the skin-depth subsurface layer ($\sim 20$\,nm in the visible and near-IR for Al and noble metals), which drastically erases any effects arising from the electron cascade that is produced by the energetic impinging probe in deeper regions.

As noted above, the incoherent CL emission must compete with other
relaxation channels, like Auger de-excitation [see Fig.\
\ref{polarizability}(c)], which are relatively inefficient in
insulators. However, radiative inelastic decay gives rise to just a minor
contribution to CL in metals because electronic relaxation channels
are several orders of magnitude faster. Therefore, the CL signal
arising from metallic nanostructures, such as the ones employed in
plasmonics, is dominated by EIRE, the details of which we discuss
in the next sections.\footnote{A word of caution should be added here: while EIRE is dominant at relatively-high electron-beam energies (for example, above 10\,keV), it decays and is overcome by incoherent CL at sufficiently-low electron energies.}

\subsection{Calculation of coherent light emission}
\label{calculationofEIRE}

The intensity of light emission produced by swift electrons interacting with a sample can be calculated from the induced far field. Using the retarded BEM to solve Maxwell's equations (Sec.\ \ref{RBEM}), we obtain the far field from the integrals of Eqs.\ (\ref{BEMphi}) and (\ref{BEMA}) in the $k r\rightarrow\infty$ limit. The radiation is commonly produced and measured in the vacuum medium $j=1$, in which $\epsilon_1=1$ and $G_1(|\rb-\sb|)\approx\exp[\ii(kr-k\rh\cdot\sb)]/r$ (for $kr\gg 1$), with $\rh=\rb/r$. Then, the induced field reduces to
\begin{equation}
\Eb^{\rm ind}\mathop{\longrightarrow}\limits_{kr\to\infty}\fb(\Omega)\;\frac{\ee^{\ii kr}}{r},
\nonumber
\end{equation}
where $\Omega$ denotes the orientation of $\rb$, and the far-field amplitude $\fb=\tilde{\fb}-(\tilde{\fb}\cdot\rh)\rh$ is expressed in terms of the boundary current on the vacuum side as
\begin{equation}
\tilde{\fb}(\Omega)=\ii k\int_{S_1}d\sb\,\ee^{-\ii k\rh\cdot\sb}\,\hb_1(\sb). \label{ftilde}
\end{equation}
Here, we are relying on the fact that the contribution of $-\nabla\phi$ to $\Eb^{\rm ind}$ only produces longitudinal components. However, the far field is transversal, and consequently, entirely coming from the $\ii k\Ab$ contribution to $\Eb$. This gives rise to $\tilde{\fb}$ [Eq.\ (\ref{ftilde})], from which the longitudinal part has been subtracted to obtain $\fb$.

The EIRE intensity can be obtained from the Poynting vector integrated over emission directions. The emitted energy per incoming electron is then given by
\begin{equation}
\Delta E=\frac{c}{4\pi}\int dt\,r^2\,\int d\Omega\,\,\rh\cdot\left[\Eb(\rb,t)\times\Hb(\rb,t)\right]. \nonumber
\end{equation}
Now, expressing the fields in time Fourier transform [Eq.\ (\ref{Fourierwt})] and using the far-field limit just discussed, $\Delta E$ can be decomposed into photon-energy components $\hbar\omega$ as
\begin{equation}
\Delta E=\int_0^\infty\hbar\omega\,d\omega\int d\Omega\; \Gamma_{\rm CL}(\Omega,\omega), \nonumber
\end{equation}
where
\begin{equation}
\Gamma_{\rm CL}(\Omega,\omega)=\frac{1}{4\pi^2\hbar k} |\fb(\Omega)|^2 \nonumber
\end{equation}
is the number of photons emitted per incoming electron, per unit of solid angle of emission $\Omega$, and per unit of photon
frequency range $\omega$.

\subsection{Transition radiation}
\label{transitionradiation}

\begin{figure*}
\includegraphics[width=120mm,angle=0,clip]{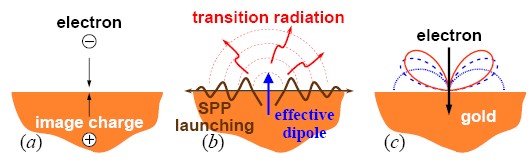}
\caption{\label{transition-radiation} (Color online)
Intuitive explanation of transition radiation. {\bf (a)} An electron cancels its image charge when it crosses a metal surface. {\bf (b)} This cancelation creates an effective dipole capable of emitting light and generating plasmons. {\bf (c)} The resulting transition radiation has indeed a dipolar angular pattern (solid curve) similar to that arising from a dipole normal to the surface (dashed curve), and vanishing for grazing emission, in contrast to the pattern of the dipole in the absence of the surface (dotted curve). The angular pattern is calculated for 200\,keV electrons impinging on a planar Au surface and for 600 nm light wavelength.}
\end{figure*}

When a fast charge impinges on a metal surface, the sudden
annihilation of its image acts like an induced dipole that
produces radiation, as illustrated in Fig.\
\ref{transition-radiation}. This effect was predicted by
\textcite{GF1946} and observed by \textcite{GJ1959} in the visible
range for protons incident on Ag, Al, and Au surfaces. In fact,
transition radiation occurs whenever a swift electron crosses
the interface between different dielectric phases.

A detailed exposition of TR was given by \textcite{T1972}, and a particularly useful result was produced by \textcite{K1968_2}, who reported a closed-form expression for the EELS probability of an electron traversing a thin film under arbitrary oblique incidence, including the contributions of Cherenkov modes and TR. This formula has been valuable in devising methods to eliminate retardation effects from measured spectra and to retrieve bulk dielectric functions \cite{S08}. In a related development, the theoretical analysis of the radiation produced by an electron passing by the center of a sphere showed an interesting evolution from CR (see Sec.\ \ref{cherenkovradiation}) to TR when considering large or small spheres compared to the wavelength \cite{PY1973}.

On the experimental side, the results of \textcite{YTA96,YST96} for CR and TR in mica, Ag, Al, and Si films, including detailed comparison with theory, constitute an outstanding reference that illustrates the complex interplay between both types of radiation emission channels in dielectric films as compared to metals (for example, Ag and Al show relatively featureless backward emission spectra for wavelengths above the bulk plasmons, in contrast to mica or Si). In a separate work, \textcite{BKK1976} reported CL spectra for 5-500 eV electrons impinging on Ag, Au, Bi, Cu, and W surfaces, and showed a dense series of features that might originate in incoherent CL, since coherent EIRE is strongly suppressed at such low energies (see Sec.\ \ref{eire}). More recently, relativistic beams (6.1\,MeV) have been utilized to produce millimeter TR that is subsequently focussed with a parabolic mirror \cite{NCK08}.

The theory of TR for an electron normally incident on a planar surface is covered by several textbooks  \cite{LLP1984,J99}. We outline it here in a form that is suitable for obtaining plasmon launching probabilities in Sec.\ \ref{plasmonemission}. We consider a swift electron moving along the $z$ axis and crossing the free surface of a metal occupying the $z<0$ region (Fig.\ \ref{planar-surface}, right).\footnote{The electron velocity varies dramatically inside the metal due to both heavy encounters with target atoms and inelastic collisions, until it is eventually stopped. However, no relevant changes in the trajectory occur with significant probability for energetic electrons ($\gtrsim 50$ keV) within a distance of the order of the skin depth underneath the surface.} The electric field produced by the moving charge in each medium $j$ (with $j=1$ and 2 representing the materials above and below the interface, respectively) can be separated into the contribution of the external electron field for an infinite bulk material plus the field reflected at the surface,
\begin{eqnarray}
   \Eb=\Eb^{\rm ext}_j+\Eb^{\rm ref}_j.\nonumber
\end{eqnarray}
In particular, the electric field inside an infinite medium $j$ obtained from Eq.\ (\ref{Eqw}) after integrating the $z$ component of $\qb$ reads
\begin{eqnarray}
   \Eb^{\rm ext}_j(\rb,\omega)=\int\frac{d^2\QQv}{(2\pi)^2}\,\Eb_j^{\rm ext}(\QQv,z,\omega)\,\ee^{\ii\QQv\cdot\Rb}
   \nonumber
\end{eqnarray}
with
\begin{eqnarray}
   \Eb_j^{\rm ext}(\QQv,z,\omega)=\frac{4\pi\ii e}{|v|\epsilon_j}\,\ee^{\ii\omega z/v}\,\frac{\qb-\vb k\epsilon_j/c}{q^2-k^2\epsilon_j},
   \nonumber
\end{eqnarray}
where $\qb=(\QQv,\omega/v)$ and $\th=\zz\times\hat{\QQv}$, so that the set $\{\hat{\QQv},\th,\zz\}$ forms a positively oriented 3D reference frame. Faraday's law permits writing the corresponding magnetic field as
\begin{eqnarray}
   \Hb_j^{\rm ext}(\QQv,z,\omega)=\frac{4\pi\ii e\QQ}{c}\,\ee^{\ii\omega z/v}\,\frac{\th}{q^2-k^2\epsilon_j}.
   \nonumber
\end{eqnarray}
It should be stressed that the bulk fields are evanescent away from the trajectory, unless $\epsilon_j$ is a positive real number and $v^2\sqrt{\epsilon_j}>c^2$, in which case CR can be produced.

The reflected component of the field finds its sources in the induced surface charges and currents. The charges can be in turn expressed in terms of the currents by using the continuity equation. After some algebra, we obtain
\begin{eqnarray}
\Eb^{\rm ref}(\QQv,z,\omega)=\nonumber
\end{eqnarray}
\begin{displaymath}
\frac{-2\pi k}{q_{zj}}\times\left\{\begin{array}{ll}
\ee^{\ii q_{z1}z}\left[\hb_1-\frac{\QQv\cdot\hb_1}{k^2\epsilon_1}(\QQv,q_{z1})\right],&\;\;\;j=1\\ \\
\ee^{-\ii q_{z2}z}\left[\hb_2-\frac{\QQv\cdot\hb_2}{k^2\epsilon_2}(\QQv,-q_{z2})\right],&\;\;\;j=2
\end{array}\right.
\end{displaymath}
and
\begin{displaymath}
\Hb_{j}^{\rm ref}(\QQv,z,\omega)=\frac{-2\pi}{q_{zj}}\times\left\{\begin{array}{ll}
\ee^{\ii q_{z1}z}(\QQv,q_{z1})\times\hb_1,&\;\;\;j=1\\ \\
\ee^{-\ii q_{z2}z}(\QQv,-q_{z2})\times\hb_2,&\;\;\;j=2
\end{array}\right.
\end{displaymath}
where $q_{zj}=\sqrt{k^2\epsilon_j-\QQ^2}$. Here, $\hb_j$ are the noted surface currents (defined on either side of the interface), which are determined from the continuity of the parallel components of $\Eb$ and $\Hb$. Finally, the reflected magnetic field reduces to
\begin{eqnarray}
   \Hb_{j}^{\rm ref}(\QQv,z,\omega)=-2\pi \ee^{\ii q_{zj} |z|}\,D\,\mu_j\,{\rm sign}(z)\,\th,
   \label{eeq7}
\end{eqnarray}
where
\begin{eqnarray}
      \mu_1&=&\frac{-(\omega/v)\epsilon_2+q_{z2}\epsilon_1}{q^2-k^2\epsilon_1}
            -\frac{-(\omega/v)\epsilon_1+q_{z2}\epsilon_1}{q^2-k^2\epsilon_2},\label{muja}\\
      \mu_2&=&\frac{(\omega/v)\epsilon_2+q_{z1}\epsilon_2}{q^2-k^2\epsilon_1}
                -\frac{(\omega/v)\epsilon_1+q_{z1}\epsilon_2}{q^2-k^2\epsilon_2},\label{mujb}
\end{eqnarray}
and
\begin{eqnarray}
   D=\frac{2\ii e\QQ/c}{q_{z1}\epsilon_2+q_{z2}\epsilon_1}.
   \label{eeq10}
\end{eqnarray}
Performing the integral over the azimuthal angle of $\QQv$, one
finds
\begin{eqnarray}
   \Hb_{j}^{\rm ref}(\rb,\omega)&=&\label{eeq12}\\
       &&-\ii s_j\,\fh\int_0^\infty\QQ d\QQ\,D\,\mu_j\,\ee^{\ii q_{zj}|z|}\,J_1(\QQ R),
   \nonumber
\end{eqnarray}
where $\fh$ is the azimuthal unit vector of $\rb$. The TR emission is obviously $p$ polarized for normal incidence, with the magnetic field pointing along the azimuthal direction, parallel to the surface. This is in agreement with experimental observations, in which residual $s$-polarized emission is claimed to originate in surface roughness \cite{H1977_2}.

The homogeneous field $\Hb^{\rm ext}_j$ decays evanescently away from the
electron trajectory (see Sec.\ \ref{anevanescentsource}). Therefore, any TR emission must arise from
$\Hb^{\rm ref}$, which behaves as
\begin{eqnarray}
   \Hb^{\rm ref}(\rb,\omega)\mathop {\longrightarrow}\limits_{kr\to\infty}\fb_H(\theta,\omega)\;\frac{\ee^{\ii kr}}{r}
   \nonumber
\end{eqnarray}
in the far field. Asymptotic analysis of Eq.\ (\ref{eeq12})
shows that
\begin{eqnarray}
   \fb_H(\theta,\omega)=\ii k\cos\theta\,D\,\mu_1\,\fh.
   \nonumber
\end{eqnarray}
This expression must be evaluated at $\QQ=k\sin\theta$, where $\theta$ is the emission angle with respect to the surface normal. Finally, proceeding in a way similar to Sec.\ \ref{calculationofEIRE}, the TR probability is given by the angular integral
\begin{eqnarray}
   \Gamma_{\rm TR}(\omega)=\frac{1}{2\pi\hbar k}\int_0^{\pi/2}d\theta\,|\fb(\theta,\omega)|^2,
   \label{eqTR}
\end{eqnarray}
performed over the upper hemisphere (vacuum).

\begin{figure*}
\includegraphics[width=180mm,angle=0,clip]{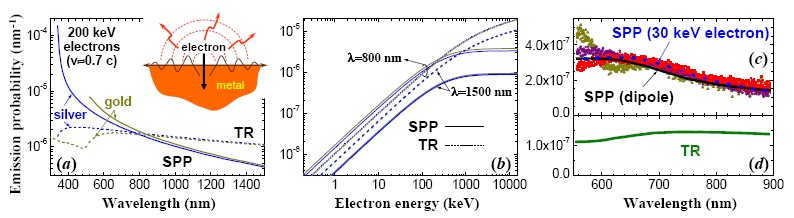}
\caption{\label{SP-launching-1} (Color online) {\bf (a)} Spectral dependence of the TR and SPP emission probabilities upon electron beam bombardment under normal incidence. A 200\,keV electron impinges on a planar metal surface, giving rise to TR [dashed curves, calculated from Eq.\ (\ref{eqTR})] and SPPs [solid curves, obtained from Eq.\ (\ref{eeq20})]. The metal dielectric function has been taken from \textcite{JC1972}. The probability is given per unit of light wavelength range. {\bf (b)} Dependence of the TR and SPP emission probabilities on electron kinetic energy for two different free-space light wavelengths, $\lambda=800\,$nm and $1500\,$nm. {\bf (c)} Measured (symbols) and simulated (dashed curve) SPP emission probability spectra for normally-incident 30 keV electrons. The SPP emission rate produced by an oscillating dipole placed at the surface is shown for comparison (solid curve), normalized to the spectral integral of TR and far-field dipole radiation. {\bf (d)} Computed TR spectrum for 30 keV electrons normally impinging on a gold surface. (c) and (d) have been taken from \textcite{paper167}, to which the reader is referred for more details on the microscope setup, experimental methods, and theoretical analysis.}
\end{figure*}

Typical TR spectra are shown in Fig.\ \ref{SP-launching-1}(a) (broken curves) for 200\,keV electrons incident on Ag and Au surfaces. The spectral emission intensity exhibits an approximately linear decrease with wavelength above the SP threshold. The integrated emission in the 600-1200 eV region yields $\sim 10^{-3}$ photons per incident electron and shows a monotonic increase with probe energy [Fig.\ \ref{SP-launching-1}(b)].

\subsection{Cherenkov radiation} \label{cherenkovradiation}

The electric field set up by a swift electron in a homogeneous
dielectric can exhibit far-field oscillatory behavior according to
Eq.\ (\ref{Ehomo2}) whenever the charged probe moves faster than light in the
medium under the condition (\ref{Cherenkovcond}).
This is reflected in the fact that the argument of the Bessel functions in Eq.\ (\ref{Ehomo2}) becomes imaginary under such conditions. Actually, these oscillations signal the emission of light, originally observed by
\textcite{C1934_2} and successfully explained soon after its discovery
\cite{FT1937_2,T1939_2}. The theory of the Cherenkov effect in
homogeneous media was later generalized to accommodate frequency
dispersion \cite{T1939_2,F1940}, and more sophisticated descriptions
have been produced since then \cite{LW1971,P1982}.

Cherenkov radiation has been extensively discussed (see
\onlinecite{G96}, and references therein), but a particularly
interesting demonstration was reported by \textcite{OOT91}, who
observed the effect for bunched ultra-relativistic electrons moving faster
than light in air.

The probability of CR emission in a homogeneous material was first
obtained by \textcite{FT1937_2}. It can be easily derived by
integrating Eq.\ (\ref{Pbulk}) over wavevector transfers, assuming real and
local $\epsilon(\omega)$. One finds
\begin{eqnarray}
\Gamma_{\rm Che}(\omega)=\frac{e^2L}{\hbar}\left[\frac{1}{c^2}-\frac{1}{v^2\epsilon(\omega)}\right],
\label{CR}
\end{eqnarray}
where the dependence on emission frequency comes exclusively from
the dielectric function. Equation\ (\ref{CR}) is a pillar in
the design of Cherenkov detectors for high-energy physics, in
which CR signals the passage of charged particles resulting from
energetic subatomic collisions \cite{J1958_2}.

Non-trivial finite-trajectory effects have been successfully
described using the field of Eq.\ (\ref{Ehomo1}) in structures
of large dimensions compared to the emission wavelength, for which
the contribution of reflection at the boundaries can be neglected
\cite{rmp-cr1,rmp-cr2,rmp-cr3}. However, this approximation is not
sufficient for small samples like the thin films theoretically
addressed by \textcite{K1968_2}. Actually, the experiments reported
by \textcite{YST96} on mica and Si films show a complex evolution of
the CR intensity with film thickness (particularly for backward
emission) that is correctly explained by Kr\"{o}ger's formula.

The Cherenkov effect has been theoretically investigated for a
charge traversing planar interfaces
\cite{K1968_2,VK1968,LK1970,T1972} or moving parallel to a surface
\cite{SS98}, for layered structures \cite{LTV00,rmp-cr4}, for an
electron moving along the axis of a cylinder \cite{OK1980}, and for
beams passing near a number of different geometries
\cite{paper084}. Other related effects have been explored, like CR due to bunched electrons moving along
waveguides \cite{OSS02}, solitons producing CR in optical fibers
\cite{AK95,SLK03}, a second narrower Cherenkov cone in strongly
dispersive active media \cite{CAR01}, THz SPPs emitted by browsing a
light spot at superluminal velocities on a planar surface
\cite{BMB05}, and inverted Cherenkov cones in materials of
negative refraction-index \cite{V1968,MG04,AY05}.

Furthermore, exotic production of CR has been observed due to superluminal vortices in Josephson junctions \cite{GWT98}, whereas subluminal CR emission has been reported assisted by phonons that supply the required extra momentum \cite{SWK01}.

\subsubsection{Cherenkov effect in photonic crystals}
\label{cherenkoveffect}

\begin{figure}
\includegraphics[width=80mm,angle=0,clip]{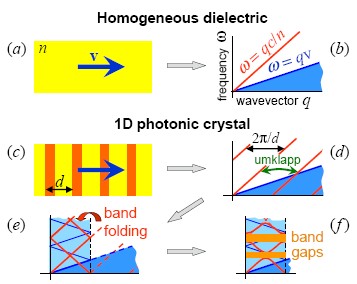}
\caption{\label{Cherenkov} (Color online) Tutorial
description of the Cherenkov effect in a photonic crystal. A charge
moving in a homogeneous dielectric with velocity below the Cherenkov
threshold $v<c/n$ (a) cannot transfer energy and momentum [shaded
region in (b)] overlapping the light dispersion line [upper line in
(b)]. However, a periodic modulation of the index of reflection (c)
allows light to undergo umklapp processes [i.e., absorption or emission
of lattice momentum, as shown in (d)], so that band folding over the first
Brillouin zone of the resulting photonic crystal (e) produces mixing
and gap openings in the photonic band structure (f). The charge can now
couple to propagating optical modes.}
\end{figure}

An interesting situation is presented when an electron moves inside a photonic crystal, consisting of a periodic modulation of the index of refraction along certain(s) spatial direction(s). The resulting CR is then affected by the photonic band structure, which can considerably differ from the vacuum light cone, as schematically illustrated in Fig.\ \ref{Cherenkov} for a 1D crystal. The loss of translational invariance allows subluminal electrons to couple to photonic bands of the crystals, thus resulting in the emission of CR.

This effect has been theoretically analyzed by \textcite{LIJ03} and \textcite{KCK09} for electrons moving perpendicularly with respect to the pores of a 2D photonic crystal. These authors found a complex evolution of the electric field, which is molded by the photonic bands.

\begin{figure}
\includegraphics[width=76mm,angle=0,clip]{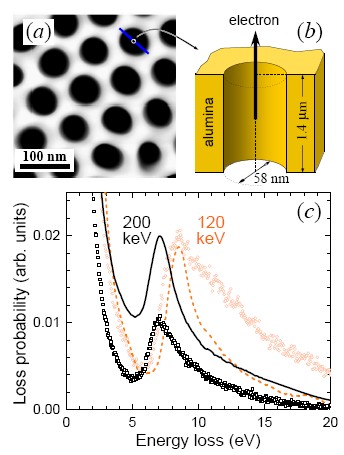}
\caption{\label{porous-alumina} (Color online)
Experimental observation of CR losses in a photonic crystal (adapted from \onlinecite{paper080}).
A 1.4\,$\mu$m thick self-standing porous-alumina film, as shown in the
TEM image of (a), is used to measure the energy loss experienced by
electrons moving inside one of the holes (b). The measured
energy-loss spectrum is represented in (c) for 120\,keV and 200\,keV
electrons (symbols) alongside theoretical calculations based upon full solution of Maxwell's
equations for an electron moving inside the 2D photonic crystal
(dashed and solid curves, respectively). The theory has been
convoluted with a measured ZLP of 0.86\,eV FWHM.}
\end{figure}

An experimental investigation of the interaction of fast electrons with a 2D photonic crystal has been reported
\cite{paper080} and is illustrated in Fig.\ \ref{porous-alumina}.
The photonic crystal consisted of a porous alumina film, grown by
electrochemical methods under the appropriate conditions to form a
locally regular array of pores \cite{MF95}. The precise control of
electron optics in a STEM allows focusing the probes so that
they pass along one of the pores of the crystal in aloof
configuration, with the beam positioned tens of nanometers away from the
pore wall. The resulting loss spectra display prominent peaks around
7-9 eV, depending on the electron energy, which were attributed to
radiative losses produced by coupling to photonic modes of
the crystal. The spectral features are of course related to the photonic
LDOS, as discussed in Sec.\ \ref{relationtophotonic}, and they
exhibit singularities similar to those explained by \textcite{V1953}, but now
arising from characteristic points in photonic rather than electronic bands. These results are indirect evidence of CR in photonic crystals. A direct experimental detection of CL in these systems has not been yet reported.

Incidentally, similar effects could be responsible for unusual spectral features in porous silicon at large beam-surface separation \cite{WLA00}.

\subsection{Diffraction radiation} \label{diffractionradiation}

We have seen that radiation can be produced when an electron crosses a
dielectric interface or when it moves faster than light in a homogeneous medium.
Another common source of light emission originates in the passage of the electron
near a structured surface (without actually crossing it), giving rise to the so-called diffraction
radiation \cite{P98}. For instance, this effect occurs if a swift charge passes near and perpendicularly to metallic slits, a geometry that has been actually used for beam diagnosis in electron accelerators \cite{KAH04}.

\begin{figure}
\includegraphics[width=80mm,angle=0,clip]{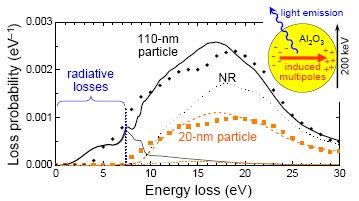}
\caption{\label{alumina-spheres} (Color online) Losses in aloof trajectories due to diffraction radiation. The
figure shows measured and simulated energy-loss spectra for 200\,keV
electrons passing just outside alumina spherical particles. Symbols:
measured data taken from \textcite{AKH00}. Thick curves: retarded
EELS calculations using Eq.\ (\ref{retarded.sphere}). Dotted curve:
non-retarded EELS calculation for the larger sphere, obtained from Eq.\
(\ref{FEfor}). Thin solid curves: CL calculations using Eq.\
(\ref{CLfor}). The computed probability is given per eV of
energy-loss range. The dielectric function of alumina is taken from
\textcite{P1985}.}
\end{figure}

We present in Fig.\ \ref{alumina-spheres} the effect of diffraction
radiation on the loss spectrum of electrons grazingly passing with
respect to alumina spheres of different size. The largest sphere
produces a sizable inelastic signal in the region below $\sim 7$ eV,
in which the material is basically transparent for the dimensions under
consideration. This is actually originating in the polarization
charges induced by the passing electron in the sphere, which oscillate giving rise to light emission. Understandably, the smaller sphere is poorly polarized, and therefore, couples weakly to propagating light, so that it does not produce observable losses within the low-energy end of the spectrum.

The emission probability admits a closed-form expression similar to Eq.\
(\ref{retarded.sphere}) \cite{paper041}:
\begin{eqnarray}
&&\Gamma_{\rm CL, sph}(\omega) \label{CLfor} \\
&&=\frac{e^2}{c\hbar\omega}
         \sum_{l=1}^\infty \sum_{m=-l}^l
               K_m^2\left(\frac{\omega b}{v\gamma}\right)
               \left[C_{lm}^M\left|t_l^M\right|^2
                +C_{lm}^E\left|t_l^E\right|^2\right].
\nonumber
\end{eqnarray}
One can easily see from Eqs.\ (\ref{tlM}) and (\ref{tlE}) that the scattering matrices satisfy the relation $|t_l^\nu|={\rm Im}\{t_l^\nu\}$ in spheres made of real dielectric function, implying that the loss probability equals the CL rate in that case, because the material is unable to absorb any energy. The CL rate calculated from Eq.\ (\ref{CLfor}) coincides with the loss probability in the region below 7 eV of Fig.\ \ref{alumina-spheres} (thin and thick continuous curves, respectively), which further supports the dominant effect of the diffraction radiation mechanism in that part of the spectrum.

Incidentally, the EELS spectra of Fig.\ \ref{alumina-spheres} show absolute maxima in the 15-20 eV range, corresponding to plasmons of the alumina spheres (for comparison, the bulk plasmon occurs at 25 eV and the planar SP at 21 eV).

\subsubsection{Smith-Purcell emission} \label{smithpurcell}

The effect discovered by \textcite{SP1953} (S-P) is a well-known example
of diffraction radiation. These authors recorded light emitted when
an electron beam passes parallel and above a metallic grating.
The direction and wavelength of emission followed a simple relation
derived from a Huygens construction. More precisely,
\begin{eqnarray}
\Delta t=d\cos\theta/c=d/v+\lambda n/c, \label{SPfor}
\end{eqnarray}
where $n$ is an integer number, $\lambda$ is the wavelength, and the
rest of the parameters is
defined in the following inset:\\
\includegraphics[width=65mm,angle=0,clip]{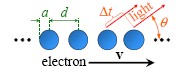}\\
Equation\ (\ref{SPfor}) arises from the condition of constructive interference in the emission originating in contiguous spheres.

Successive experimental demonstrations of this effect were carried out using 30-300 keV electrons to produce light of wavelength lying in the $\mu$m region \cite{S1970,B1972}. Also, relativistic beams have been employed for generating millimeter radiation \cite{DMO92,SHK98,BDP08}, which turns out to be very intense along the forward direction \cite{WWS95}. Furthermore, 20-40 keV dense electron beams coming from conventional electron microscopes and coupled to metallic gratings of periods in the sub-millimeter domain have been used to produce super-generation of far-infrared radiation \cite{UGK98,T98}.\footnote{Within linear response theory, valid for most situations considered in this review, the EELS and CL probabilities are proportional to the square of the particle charge, and they scale linearly with the beam intensity for moderate currents. However, bunching can produce nonlinear scaling when the electrons move closer to each other than the emission wavelength \cite{UGK98}, so that they behave like a composite charge, the square of which is obviously larger than the sum of the squares. This leads to super-radiant Smith-Purcell emission.}

Early theoretical analyses of the S-P effect relied on diffraction of the evanescent waves accompanying the moving electron \cite{T1960,V1973,VT1974,HRS94}. It was also shown that an equivalent description results from considering the currents induced by the passage of the electron on the grating surface \cite{BWD98}. Moreover, a large deal of work has been devoted to studying the conditions upon which this effect could be utilized for efficient x-ray generation \cite{M92}, for instance in free-electron lasers (FELs) \cite{LIY07,LIG07,LYL07}.

\begin{figure}
\includegraphics[width=80mm,angle=0,clip]{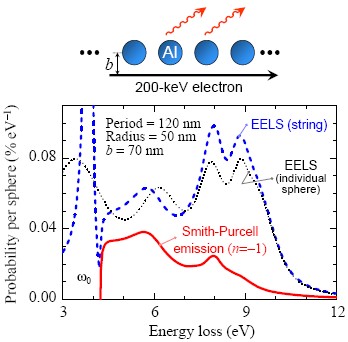}
\caption{\label{Smith-Purcell} (Color online) Energy loss
probability and Smith-Purcell radiation emission for an electron
moving parallel to an infinite periodic string of aligned Al
nanoparticles (see upper inset). The probability is normalized per
sphere and per eV of energy-loss range. The loss probability for an
individual sphere is also shown for comparison. Adapted from
\textcite{paper040}.}
\end{figure}

The excellent spatial resolution of electron microscopes could be used to direct a beam parallel to an array of
nanoparticles and produce S-P emission. This situation is considered in Fig.\
\ref{Smith-Purcell}, which predicts intense super-continuum emission in the
UV domain. This type of sub-micron light emitters could find
eventual application in integrated photonic devices. Interestingly,
Fig.\ \ref{Smith-Purcell} shows the presence of an absolute
threshold below which no emission is possible because Eq.\
(\ref{SPfor}) cannot be satisfied.

In recent works, the S-P emission from photonic crystals has been extensively investigated \cite{PM94,paper075,OO04,OO04_2,OO05} and compared with gratings \cite{OY01,YIH02} using theoretical tools. Besides, experimental observations have been reported for millimeter waves emitted by electrons passing near photonic-crystal opals \cite{YSY04}. Finally, interesting effects due to finite size of the periodic structure along the direction of the beam have been noted \cite{OO06} and theoretically analyzed \cite{HOY06}.

\subsection{Cathodoluminescence and plasmons}

Plasmons in metallic nanoparticles have finite lifetime due to
radiative decay, Landau damping (e-h pair creation), and
other absorption mechanisms. The former is rather efficient
when the particle has non-negligible size compared to the
corresponding light wavelength, as we discuss below. Actually, the CL signal resulting from plasmon decay offers an excellent opportunity to spectrally and spatially resolve plasmons.

\subsubsection{Plasmons in metallic films and gratings}
\label{Ferrellmodeetc}

An electron traversing a thin film can give rise to a peak in the CL emission at the bulk plasmon frequency $\omega_p$, as
predicted by \textcite{F1958}, who found that the emitted light was $p$ polarized (magnetic field parallel to the film) for normal incidence, and the emission intensity vanished at normal and grazing directions for a Drude metal. This prediction was soon confirmed by \textcite{S1960} and \textcite{BWT1960} in Ag films, and later by \textcite{AHB1964} in Ag and Mg films.
A study by \textcite{RE1962} generalized Ferrell's results to arbitrary metals and $v\ll c$, while the comprehensive work of \textcite{CA1967} showed excellent agreement between theoretical and experimental results in Ag films, compared on the same absolute-intensity scale. Additionally, the Ferrell mode was observed using optical excitation rather than electron beams \cite{BR1965}, and its presence was shown to affect the response properties of thin films near the bulk plasmon frequency \cite{KF1967}. Subsequent TEM experiments managed to resolve the dispersion relation of the Ferrell mode in Al \cite{VS1973}, basically following the transversal bulk plasmon curve of Fig.\ \ref{bulk}(b).

In contrast, SPPs in a thin film cannot directly couple to external light because their parallel momentum lies outside the light cone, according to Eq.\ (\ref{SPdisp}), with $\epsilon_1$ and $\epsilon_2$ having opposite sign. However, SPPs can be made to radiate if they find obstacles in their propagation along an otherwise planar surface. In particular, gratings provide a source of parallel momentum that plasmons can use to couple to external light. The optical emission resulting from excitation of SPPs by electrons impinging on a grating was first observed by \textcite{TS1967}. These authors managed to correlate SPP-related peaks in the angular distribution of the emission with dips observed in the specular reflectivity of periodically corrugated surfaces.

In a remarkable development, \textcite{H1977} used CL to map the SP dispersion in sinusoidal Ag gratings of different height. The gratings assisted SPPs to couple to freely propagating light. His analysis of extrapolated results in the limit of shallow corrugation were in excellent accord with both the dispersion relation and the momentum-energy width predicted by Eq.\ (\ref{SPdisp}). Furthermore, his results for stronger modulation of the grating profile resulted in dispersion relations further apart from the light line, so that the grating contributed to increase the binding of SPPs to the surface. A similar increase in surface-mode binding driven by periodic corrugations has been also observed in microwaves trapped by drilled metal films \cite{UT1973}. This behavior is accompanied by gaps in the boundaries of the Brillouin zones, as first discussed by \textcite{RAC1968} for $p$-polarized light incident on gratings, and later studied in 2D plasmon crystals \cite{KBS96} capable of guiding plasmons through lines of defects \cite{BEL01}.

Two recent outstanding experiments have independently supplemented Heitmann's work by out-coupling electron-generated SPPs in planar surfaces using distant gratings \cite{BJK06,VVP06}, thus suggesting the possibility of using fast electrons as plasmon sources. We address this issue in more detail in Sec.\ \ref{plasmonemission}.

\subsubsection{Plasmon mapping} \label{plasmonmapping}

Cathodoluminescence has been used to map plasmon excitations in scanning mode, by analogy to the EELS plasmon-mapping study presented in Fig.\ \ref{triangles}. The CL signal is generally much weaker that the loss intensity, which is problematic when good signal-to-noise ratio is needed to resolve neighboring spectral features. However, CL has the advantage that it can be performed in SEMs, which are more widely available than TEMs, and it works for thick samples, since there is no need to collect transmitted electrons.

Several examples of application of this technique to spatially
resolve plasmons in nanostructures have been presented by Yamomoto
and co-workers over the last few years, including studies of
plasmons in spherical Ag nanoparticles \cite{paper064}, gratings
\cite{YAT01}, holes in thin films \cite{DLY04}, and particle dimers
and nanorods \cite{YNS06}.

\begin{figure*}
\includegraphics[width=140mm,angle=0,clip]{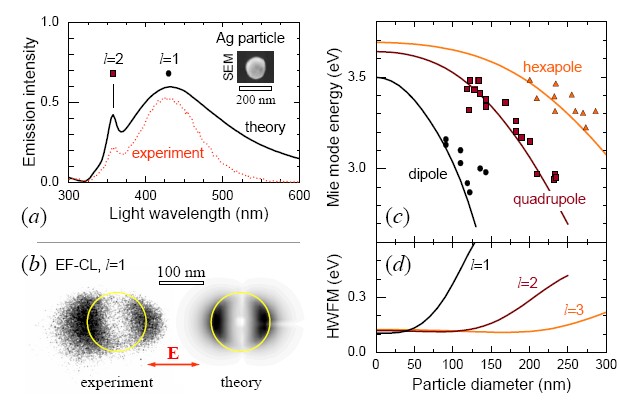}
\caption{\label{Yamamoto} (Color online) Plasmon mapping
of single metallic nanoparticles (adapted from \onlinecite{paper064}).
{\bf (a)} Measured and calculated CL spectra for a 122\,nm Ag nanoparticle
under grazing incidence, showing dipole ($l=1$) and quadrupole
($l=2$) plasmons. The dielectric function of Ag is taken from \textcite{JC1972}.
{\bf (b)} Energy-filtered CL scan corresponding to the
dipole-mode maximum of the spectra shown in (a). The collected light is passing
through a polarizer with the orientation shown by the arrow. {\bf (c)} Symbols: measured dispersion of multipole plasmons
with particle size. Curves: Mie theory. {\bf (d)} Plasmon width, obtained
from Mie theory.}
\end{figure*}

A characteristic example of these studies is shown in Fig.\
\ref{Yamamoto} for Ag particles. The spectra of Fig.\
\ref{Yamamoto}(a) unveil the excitation of a broad dipole mode and a
narrower quadrupole. The measurement is in excellent agreement with the theory of Eq.\
(\ref{CLfor}). By scanning the electron beam over the particle
and collecting polarized photons emitted at the peak wavelength, one
can form the image of the dipole plasmon shown in Fig.\
\ref{Yamamoto}(b). The intensity of each pixel represents the photon rate measured when the electron is positioned at that spot of the sample. This plasmon map displays two characteristic maxima on either
side of the particle along the direction of the polarizer. This
study was carried out for particles of various sizes, the results of
which are summarized in Fig.\ \ref{Yamamoto}(c). The plasmon modes
show a distinctive redshift with increasing particle size, in
good agreement with the theory of \textcite{M1908}, who provided the
basis for understanding light scattering by spherical particles. The
origin of these shifts is retardation, as we discuss in Sec.\
\ref{spheres}, and they occur when the diameter of the particle is a
sizable fraction of the wavelength. As an empirical rule, we can
assert that retardation effects show up for radius $a\gtrsim
\lambda l/2\pi$.\footnote{The multipolar order $l$ intervenes in determining the relevant distance to which the wavelength has
to be compared for assessing the relative importance of retardation. This must be of the order of the separation between nodes in the induced surface charge, which is $\sim a/l$, since the interaction between charges at larger distances is partially washed out by cancelation of contributions with different sign.} These effects are
accompanied by an increase in the width of the multipole plasmons
with particle size, which arises as a result of stronger coupling to
propagating light [see Fig.\ \ref{Yamamoto}(d)] \cite{paper155}.

\begin{figure}
\includegraphics[width=80mm,angle=0,clip]{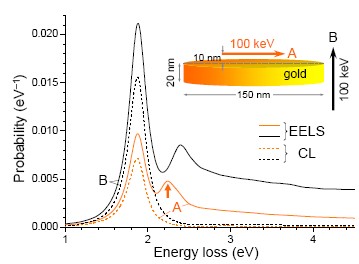}
\caption{\label{EELS-CL} (Color online) Comparison of EELS and CL intensities in a gold nanodisk. Two different electron beam trajectories are considered (see inset) separated 10\,nm from the disk in both cases. The disk edges have been rounded with a curvature radius of 3 nm. These results have been obtained using the retarded BEM with the dielectric function of gold taken from \textcite{JC1972}.}
\end{figure}

Figures\ \ref{Yamamoto}(a)-(b) clearly demonstrate that aloof
geometries (external trajectories) provide significant information
on the particle plasmons. This idea is further illustrated in Fig.\
\ref{EELS-CL}, which presents a quantitative comparison of EELS and
CL probabilities for Au disks. The difference between these two
probabilities corresponds to events in which the lost energy is absorbed by
the particle rather than being re-emitted. These calculations have
been carried out using the boundary element method described in
Sec.\ \ref{RBEM}.

The CL spectra of Fig.\ \ref{EELS-CL} (dashed curves) show just a
subset of the features observed in the loss spectra (solid curves),
as an indication that some plasmon excitations are dark modes, in
the sense that their main decay mechanism is via absorption by the
particle rather than coupling to radiation \cite{paper164}. More precisely, the
feature marked by an arrow in the loss spectrum of trajectory A is
not found in the light cross section of randomly oriented disks (not
shown), but it can be clearly resolved by EELS.

The spatial resolution of CL should be analogous to EELS for
mapping plasmons, as shown in Fig.\ \ref{resolution}(b). This figure
proves that aloof beams can yield similar emission intensities
as penetrating electrons, in agreement with the images of Fig.\
\ref{Yamamoto}(b). Finally, an additional thought on the aloof
configuration: the incoherent CL signal, which we argued in Sec.\
\ref{incoherentluminescence} to be small compared to the coherent
EIRE contribution reported in these figures, should be actually
smaller when the electron does not traverse the metal and
cannot create localized excitations, which are the main source of
incoherent CL.

\subsection{Ultrafast cathodoluminescence}

In a recent development, the CL technique has evolved to allow picosecond resolution using a microscope in which the electron source consists of a photocathode being excited by external light pulses, so that the time-resolved CL emission can be assigned an excitation time origin determined for each of the pulses, taking into account the time of flight delay of the electron in its way to the target. This technique has been recently utilized to resolve charge-carrier dynamics in quantum dots and quantum wells with 50\,nm spatial resolution and 10\,ps time resolution \cite{MSC05,SFM06}. Similarly, picosecond electron pulses have been used to develop ultrafast electron microscopy \cite{LSZ05}, so far applied just for simultaneous dynamical and structural characterization \cite{GLW06,BPK08}. These studies open an avenue towards a whole range of interesting phenomena in the domain of the ultrafast nanoworld.

\section{Related Phenomena and Suggested Experiments}
\label{othereffects}

Electron microscopy is a rather versatile discipline that finds in
the nanoworld its natural domain. We can certainly think of many new
phenomena to be explored with focused electron beams, particularly
when low-energy, valence excitations are involved. In this section, we
briefly go through some of them.

\subsection{Mechanical momentum transfer}
\label{momentumtransfer}

A fast electron beam has the potential to act as a nanotool that manipulates molecules and nanoparticles by transferring linear and angular momentum. This should become an excellent complement to optical tweezers, currently employed to trap small particles ranging in size from nanometers to several microns \cite{MQ00}, and to study their mechanical behavior by exerting electromagnetic forces \cite{KTG02,G03}. Studies based on optical tweezers range from elastic properties of deformable living cells \cite{GAM00} to fluctuation forces in biomolecules \cite{MQ00}. Likewise, an electron microscope can modify (and simultaneously image) small structures with far greater resolution than what is achieved by optical imaging. A more sophisticated setup would involve in-vacuo optical tweezers integrated in an electron microscope for observation of trapped particles. This system is experimentally challenging because it requires to trap particles in vacuum, for which pioneering results on optical levitation were reported three decades ago by \textcite{AD1976}. A solution that is perhaps more compatible with trapping in vacuum consists in relying on evanescent fields (e.g., localized SPs), as proposed by \textcite{NBX97} and recently demonstrated by \textcite{RVG08}.

Here, we estimate the magnitude of the momentum transfer imparted on small objects using the methods elaborated in previous sections \cite{paper093}. The electromagnetic force is obtained by integrating Maxwell's stress tensor over a surface surrounding the object \cite{J99}. Performing the time-integral of this force, we find
\begin{eqnarray}
  {\rm Momentum\;\,transfer} = \int \Fb(t) \, dt = \int_0^\infty \Fb(\omega) \, d\omega,
  \label{Dp}
\end{eqnarray}
where the spectral decomposition in the last identity is similar
to the one used for the EELS probability in Sec.\ \ref{classicaldielectric}.

For simplicity, we first limit our discussion to a small isotropic particle described
by its electric polarizability $\alpha(\omega)$. The force exerted by
an external field $\Eb(\rb,\omega)$ acting on such particle is given by
the expression \cite{GA1980}
\begin{eqnarray}
  \Fb(\omega)={\rm Re}\left\{\alpha\,\sum_jE_j(\rb,\omega)\,\nabla\left[E_j(\rb,\omega)\right]^*\right\},
  \label{Falpha}
\end{eqnarray}
where the sum runs over Cartesian components and the external field supplied by the electron is given by
Eq.\ (\ref{Ehomo1}). Using this in Eq.\ (\ref{Falpha}), one finds
\begin{eqnarray}
  \Fb(\omega)=\frac{2 e^2 \omega^3}{v^5\gamma^3} \bigg[&-&{\rm Re}\{\alpha\}
                      \, f^\prime\left(\frac{\omega b}{v\gamma}\right) \, \hat{\bf x}
                      \label{smallparticle}\\
                  &+&2\gamma \, {\rm Im}\{\alpha\} \, f\left(\frac{\omega b}{v\gamma}\right) \,
                  \hat{\bf z}\bigg],
  \nonumber
\end{eqnarray}
where the function $f$ is the same as in Eq.\ (\ref{fff})
and the particle is located at a distance $b$ from the electron beam.

Symmetry considerations lead to the conclusion that Rayleigh scattering by the particle produces a radiation pattern independent of the direction of $\vb$. This means that the overall transfer of momentum in the form of induced radiation is zero in the small-particle limit, so that Eq. (\ref{Dp}) accounts for all momentum transfer emanating from the moving electron along the $z$ axis, parallel to the trajectory. Then, the contribution of each $\omega$ component to the electron energy-loss probability is $v F_z(\omega)$. In fact, we find from Eqs. (\ref{smallparticle}) and (\ref{Pdipole}) the identity $vF_z(\omega)=\hbar\omega\Gamma_{\rm EELS,dip}(\omega)$: it is reassuring that the mechanical energy exchanged through momentum transfer equals the loss probability multiplied by the transferred energy for small targets.

\begin{figure}
\includegraphics[width=80mm,angle=0,clip]{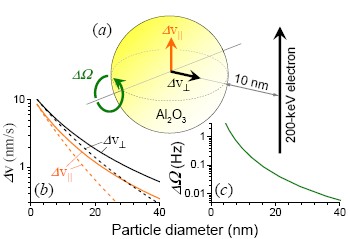}
\caption{\label{momentum} (Color online) Momentum transfer from an electron to an spherical alumina particle (taken from \onlinecite{paper093}). {\bf (a)} Sketch of a 200\,keV electron passing at a distance of 10 nm from the surface of an alumina particle. {\bf (b)} Linear momentum transfer, quantified by the velocity change of the particle along directions parallel and perpendicular to the beam, and normalized per incident electron. Solid curves: full electromagnetic calculation. Dashed curves: dipole approximation [Eq.\ (\ref{smallparticle}), with the prescription of Eq.\ (\ref{alphatlE}) for $\alpha$]. {\bf (c)} Change in rotational velocity produced by angular momentum transfer from a single electron (full electromagnetic simulation).}
\end{figure}

The mechanical effect can be significant in nanoparticles, as shown in Fig.\ \ref{momentum}(b), which compares  Eq.\ (\ref{smallparticle}) (dashed curves) with the results obtained when all multipoles are included (solid curves). For instance, a typical STEM beam with a current of 0.1\,nA passing 10\,nm away from the surface of a 40\,nm alumina particle exerts a macroscopic attractive acceleration equal to 0.13 times the gravity.

The transfer of angular momentum $\Delta\mathcal{L}$ can be written
\begin{eqnarray}
\Delta\mathcal{L}=\int_0^\infty\tau(\omega)\,d\omega, \nonumber
\end{eqnarray}
where the torque $\tau$ is also given by the integral of the Maxwell stress tensor
\cite{J99} over a surface surrounding the target. For the small particle considered above, one obtains
\begin{eqnarray}
  \tau(\omega)&=&\label{tautau}\\&&\frac{8 \sqrt{2}\omega^2}{v^4\gamma^2} K_0\left(\frac{\omega b}{v\gamma}\right)
  K_1\left(\frac{\omega b}{v\gamma}\right) \left[{\rm Im}\{\alpha\}-\frac{2k^3}{3}|\alpha|^2\right].
  \nonumber
\end{eqnarray}
Interestingly, the expression inside the square brackets of this equation vanishes for a non-dissipative particle.\footnote{This is a consequence of the optical theorem for small particles, stating that ${\rm Im}\{-1/\alpha\}\ge 2k^3/3$, with the equal sign only occurring for non-absorbing particles \cite{V1981}.} Equation (\ref{tautau}) shows that the torque produced on a small particle drops exponentially with impact parameter [see Fig.\ \ref{momentum}(c)], but it is still sufficient to cause an astonishing 3.9 MHz/s angular acceleration when a 0.1\,nA electron beam passes 20 nm away from a 40\,nm alumina particle. The type of experiment needed to demonstrate this prediction should be attainable using the combination of optical tweezers and TEM suggested above. Perhaps the reader shares the curiosity of the author to see what happens to the particle as it spins increasingly faster until centrifugal forces compete with atomic binding.

\subsection{Vicinage effects}
\label{vicinageeffects}

At half the speed of light, electrons in a typical 100\,keV, 0.1\,nA
STEM beam are separated by an average distance of 26 cm along the
propagation direction and they impact the specimen with a time delay
of 1.6\,ns. The interaction among them is therefore negligible in vacuum,
but a leading electron produces excitations in the sample that a
second trailing projectile can pick up, whenever they last more than the
delay time. This situation is discussed in Sec.\
\ref{interactionbetweentwoelectrons}.

Another type of vicinage effect involves the electron
self-interaction, taking place when its wave function extends over different
regions of the sample. Current technology makes the observation of
this effect feasible, and it should find direct application to
measure nonlocal screened interactions, as we show in Sec.
\ref{selfinteractionofasingleelectron}.

\subsubsection{Interaction between two electrons}
\label{interactionbetweentwoelectrons}

This effect bears close relation to the amply discussed topic
of swift charged-molecule dissociation after traversing thin foils,
in which the interaction among the resulting ionic fragments affects
their stopping \cite{GRP1975} and the SE
yield \cite{TYU06}. Likewise, mutual interaction between bunched electrons is
at the heart of super-radiant Smith-Purcell emission
\cite{UGK98,LIY06}, as discussed in Sec.\ \ref{smithpurcell}.

We can easily find from Eq.\ (\ref{Ggen}) the overall loss
probability for two electrons moving a distance $a$ apart from each
other along the same straight-line trajectory. The ratio of this
probability to that of a single electron is exactly given by
$4\cos(\omega a/2v)$ with independence of the nature of the sampled
excitations. When the electrons move in close proximity
as compared to $v/\omega$, a two-fold increase in the combined loss
probability is produced. An interesting situation occurs under the
condition $a=\pi v/\omega$: the energy released from the leading
probe is picked up by the trailing one and the overall
probability vanishes. This may find application to shadow intense
energy-loss or light-emission features in order to make weaker
spectral details more visible, in the spirit of the masking
interferometry techniques utilized in astronomy \cite{LMI06}.

\subsubsection{Electron self-interaction}
\label{selfinteractionofasingleelectron}

The quantum nature of the electron probe can be exploited to retrieve
nonlocal information on the dielectric response if we separate the
incident beam into two or more coherent spots at the position of the sample.\footnote{Nonlocality in this paragraph refers to $W(\rb,\rb',\omega)$ for $\rb\neq\rb'$, and not to spatial dispersion in $\epsilon$.} This
is routinely done in TEM holography by inserting a prism in the
electron optics, basically exploiting partial coherence of the energy-loss signal \cite{SW05,H08_2,VBS08}.
Inelastic holography has been performed both in the real image plane
\cite{LF00} and in the diffraction plane \cite{H05}, although the current
theoretical understanding of the obtained results suggests that
further development of the technique is required in order to extract
new information on the coherence of inelastic excitations
\cite{V06}. This technique permits addressing profound questions on quantum mechanics, such as the Feynman thought experiment \cite{PVS07}.

Here, we discuss a similar scenario, consisting of an electron beam split into two coherent beams that are focused on identical neighboring spots. This situation is well described by Eq.\ (\ref{PQM2}) if one assumes that the lateral wave function of the incident electron has the form
\begin{eqnarray}
\psi_{i\perp}(\Rb)=\frac{1}{\sqrt{2}}\left[F(\Rb-\Rb_1)+F(\Rb-\Rb_2)\,\ee^{\ii\delta}\right], \label{eq2spots}
\end{eqnarray}
where $F$ gives the spot profile (e.g., a Gaussian of width $\Delta R$),
$\Rb_j$ are the centers of the two separate spots ($|\Rb_1-\Rb_2|\gg\Delta R$),
and the phase $\delta\approx m_eva/\hbar$ accounts for the difference in path length
between the two beams ($a$), which can be controlled through the electron optics
of the microscope.

We now make two assumptions regarding the spot size: (1) the
induced part of the screened interaction $W^{\rm ind}$ is considered to be a smooth function that
does not vary significantly over the extension $\Delta R$; and (2)
the lateral variation of the detected plane wave components across a
distance $\Delta R$ is negligible, that is, $p_{f\perp}\Delta R\ll
1$ in Eq.\ (\ref{PQM2}). The first assumption limits our analysis to
excitations extended over large regions as compared to the size of each beam
spot. This is quite reasonable for a realistic value of $\Delta
R\sim 2$\,{\AA}, since most valence excitations including plasmons
are extended over much larger distances (see, for example, Fig.\ \ref{triangles}). The second assumption defines the angular region of transmitted electrons in which we are interested (for instance, $p_{f\perp}\ll 1/\Delta R \sim 0.5$\,\AA$^{-1}$, or equivalently, collection angles $\ll 15$\,mrad for 100\,keV electrons).

Inserting Eq.\ (\ref{eq2spots}) into Eq.\ (\ref{PQM2}), one finds
\begin{eqnarray}
\frac{d\Gamma^{\rm NR}(\omega)}{d\pb_{f\perp}}\propto\Gamma_{\rm EELS}^{\rm NR}(\Rb_1,\omega)+\Gamma_{\rm EELS}^{\rm NR}(\Rb_2,\omega)+\Gamma^{\rm NR}_{\rm int}(\Rb_1,\Rb_2,\omega), \nonumber
\end{eqnarray}
where
\begin{eqnarray}
&&\Gamma^{\rm NR}_{\rm int}(\Rb_1,\Rb_2,\omega)=\frac{2e^2}{\pi\hbar v^2} \int dz_1dz_2\;{\rm Im}\left\{-W(\rb_1,\rb_2,\omega)\right\}
\nonumber \\
&&\,\,\,\,\,\,\,\,\times\,\,\cos\left[\delta+\pb_{f\perp}\cdot(\Rb_1-\Rb_2)-\frac{\omega}{v}(z_1-z_2)\right] \nonumber
\end{eqnarray}
contains the interference between the two spots.

At variance with the EELS probability $\Gamma_{\rm EELS}^{\rm NR}(\Rb,\omega)$ of Eq.\
(\ref{Gfos}), obtained when the entire inelastic signal is
collected, the quantity $d\Gamma^{\rm NR}(\omega)/d\pb_{f\perp}$ offers direct
access to the nonlocal dependence of the screened interaction,
${\rm Im}\{W(\rb_1,\rb_2,\omega)\}$. This requires to have exquisite wavevector resolution, $\Delta p_{f\perp}\lesssim 1/|\Rb_1-\Rb_2|$, in order to prevent cancelations of the interference signal produced by the oscillations of the cosine function in $\Gamma^{\rm NR}_{\rm int}$. Using 100\,keV electron with an achievable angular resolution of $\varphi_{\rm out}\sim 1\,\mu$rad, one should be able to retrieve information for spots separated by a distance of $\sim 50$ nm, which is sufficient to study excitations of interest in nanoparticles and nanostructured materials.

\begin{figure}
\includegraphics[width=75mm,angle=0,clip]{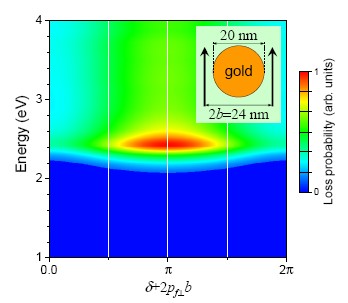}
\caption{\label{vicinage} (Color online) Loss probability according to Eq.\ (\ref{eqvicinage}) for two coherent beams passing close to opposite sides of a 20\,nm gold sphere (see inset). The gold is described with the dielectric function of \textcite{JC1972}. The electron energy is 200\,keV.}
\end{figure}

We illustrate this formalism by considering two coherent beams of impact parameters $\Rb_1=b\xx$ and $\Rb_2=-b\xx$ relative to the center of a sphere of radius $a$, described by Eq.\ (\ref{Wlalpha}). We obtain
\begin{eqnarray}
&&\frac{d\Gamma^{\rm NR}_{\rm sph}(\omega)}{d\pb_{f\perp}}\propto
         \sum_{l=1}^\infty \sum_{m=-l}^l
               \frac{(\omega/v)^{2l}}{(l+m)!(l-m)!}
\label{eqvicinage} \\ &&\;\;\;\times\;K_m^2\left(\frac{\omega b}{v}\right)
               {\rm Im}\left\{\alpha_l^{\rm NR}\right\} \left[1+(-1)^m\cos(\delta+2p_{f\perp}b)\right],
\nonumber
\end{eqnarray}
where $\alpha_l^{\rm NR}$ is defined in Eq.\ (\ref{alphaNRvac}). Figure\ \ref{vicinage} represents the results obtained from
Eq.\ (\ref{eqvicinage}) for a 200\,keV electron split into two beams that are passing at a distance of 2\,nm from a 20\,nm gold particle, as shown in the inset. Interestingly, $d\Gamma^{\rm NR}_{\rm sph}/d\pb_{f\perp}$ is periodic in $p_{f\perp}$ with period $\pi/b$ and is dominated by a dipolar feature . Furthermore, the interference disappears under the condition $\delta +2p_{f\perp}b=(2n+1)\pi/2$ for integers $n$, leading to a loss probability proportional to Eq.\ (\ref{FEfor}).

\subsection{Electron energy-gain spectroscopy}
\label{electronenergygain}

\begin{figure}
\includegraphics[width=75mm,angle=0,clip]{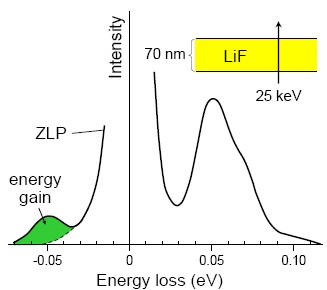}
\caption{\label{Boersch} (Color online) Pioneering evidence of
energy gain after transmission of fast electrons through a thin
foil (adapted from \onlinecite{BGS1966}). Energy gain events involve room-temperature phonon
absorption by the electron and is strongly reduced at lower
temperature.}
\end{figure}

The width of the ZLP limits the resolution of EELS and masks
low-energy features sitting in the elastic peak tail. These problems
are partially solved as we show below if one is capable of
reversing the process, so that energy gain rather loss is recorded.
Electron acceleration events were actually reported over 40 years
ago when \textcite{BGS1966} observed absorption of thermal phonons
by electrons transmitted through thin LiF films, as illustrated in
Fig.\ \ref{Boersch}. The gain signal was later shown to depended on temperature
as \[\Gamma_{\rm EEGS}^{\rm thermal}(\omega)=n(\omega)\,\Gamma_{\rm EELS}(\omega),\] where $\Gamma_{\rm EELS}(\omega)$ is the EELS probability discussed in Sec.\ \ref{interactionofelectron} and $n(\omega)=1/[\exp(\hbar\omega/k_BT)-1]$ is the Bose-Einstein distribution function at temperature $T$ \cite{paper157}. Actually, \textcite{BGS1966} managed to suppress the measured $\Gamma_{\rm EEGS}^{\rm thermal}$ by cooling their samples. Unfortunately, thermal transfers are limited to modes of energies $\lesssim k_BT\approx 26$\,meV at room temperature.

\begin{figure}
\includegraphics[width=80mm,angle=0,clip]{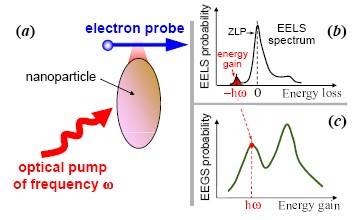}
\caption{\label{eegs} (Color online) Schematic
description of electron energy-gain spectroscopy (EEGS). {\bf (a)}
An electron passes near a nanostructure (a nanoparticle in the
illustration) that is being illuminated by light of frequency
$\omega$. {\bf (b)} A regular EELS spectrum acquired in a scanning
transmission electron microscope (STEM) shows a gain peak on the
left-hand side with respect to the zero-loss peak,
corresponding to electrons that have absorbed one photon out of the
external light. {\bf (c)} The electron energy-gain probability for
light of frequency $\omega$ is proportional to the area of the
energy-gain peak. The gain process is assisted by localized
excitations of the nanostructure. They supply the necessary
momentum to couple evanescent, scattered light to the electron, so
that the EEGS probability exhibits features as a function of photon
frequency mimicking the excitation modes of the sample.}
\end{figure}

A more flexible strategy consists in exciting through external illumination only those modes that we want to probe, as already suggested by \textcite{H99}. The proposed technique is illustrated in Fig.\ \ref{eegs} and can be termed electron energy-gain spectroscopy (EEGS) \cite{paper157}, in which the electrons absorb photons from an external light source. However, the energy-momentum mismatch between freely propagating charges and photons prevents the linear coupling of fast electrons and light plane waves in vacuum (for example, an electron moving in vacuum cannot emit CR).\footnote{In-vacuum optical acceleration of an electron moving along the axis of a Bessel beam has been recently proposed \cite{ZL08}.} Therefore, we must rely on the interaction with a material that is structured on subwavelength scales (the specimen under scrutiny) to produce evanescent optical fields carrying the necessary complex momentum to make this coupling feasible. This is already the case in the inverse Cherenkov effect, whereby an electron moving under the condition given by (\ref{Cherenkovcond}) can absorb energy from a laser aligned close to the Cherenkov cone, as proposed by \textcite{M1966_2} and observed fifteen years later by \textcite{EKP1981}. The inverse Cherenkov effect can find application to accelerate charged particles \cite{KKR95}. In a related context, the observation of the inverse Smith-Purcell effect \cite{MOS1975,MPN1987} demonstrated the coupling of free electrons to evanescent components of momentum produced by interaction of external light with a grating.

Here, we consider a more suitable scenario for spectroscopy.
Light scattered from a nanostructure, and in particular evanescent
components in its vicinity, posses the necessary
momentum to couple to the electron. Consequently, the energy-gain
process can occur near the specimen and should yield information on
its spectral response, directly mapping the evanescent light
strength. In such setup, the full width of the gain peak must
correspond to electrons that have absorbed one photon out of the
external light, and therefore the energy resolution of this
technique is only limited by the width of the external illumination,
typically in the sub-meV domain \cite{RK1988}. This is a 100-fold increase in
spectral resolution with respect to currently-available EELS
technology. [Notice that low-energy excitations (e.g., sub-meV modes) are difficult to resolve due to the ZLP background. Here we refer instead to the ability to resolve higher-energy excitation features (e.g., 1 eV) separated by minute energy differences (e.g., 1 eV versus 1.001 eV).]

In order to evaluate EEGS probabilities, we consider an energetic
electron focused near the sample and describe it through the wave function
of Eq.\ (\ref{psii}) (Sec.\ \ref{quantumeffects}). We adopt
the non-recoil approximation and assume that the normalized transversal wave function $\psi_{i\perp}(\Rb)$ is focused within a narrow spot centered around $\Rb=\Rb_0$, so that only the motion along the beam
direction $z$ needs to be considered. The electron wave function is
perturbed by interaction with the external light field and picks up
components of higher energy to become
\begin{eqnarray}
  \psi_\parallel(z) = \frac{1}{L^{1/2}}\,\ee^{\ii p_iz}\ee^{-\ii \varepsilon_it} + \delta\psi_\parallel(z)\,\ee^{-\ii \varepsilon_f t},
  \nonumber
\end{eqnarray}
where we now have $\varepsilon_f=\varepsilon_i+\omega$ and $\omega$ is the photon frequency.
Using first-order perturbation theory, we find
\begin{eqnarray}
  \delta\psi_\parallel(z)=\int dz'\,{\rm G}_f(z-z')\,H_\omega(z')\,\psi_\parallel(z'),
  \label{eq1}
\end{eqnarray}
where $H_\omega(z)=(-e\hbar/m_e\omega)\,E_z(\Rb_0,z)\,\partial_z$ is the
energy-gain part of the electron-photon coupling Hamiltonian,
$E_z$ is the $z$ component of the light electric field (parallel to the trajectory), $\partial_z$ denotes differentiation with respect to $z$, and
\begin{eqnarray}
  {\rm G}_f(z-z')&=&\int\frac{dp}{2\pi} \frac{\ee^{\ii p (z-z')}}{(m_e\varepsilon_f/\hbar-p^2/2+\ii 0^+)}
         \nonumber\\&=&\frac{-\ii\hbar^2}{m_ep_f}\,\ee^{\ii p_f |z-z'|}
  \nonumber
\end{eqnarray}
is the non-relativistic Green function of the free electron propagating along the $z$
direction with final energy $\hbar\varepsilon_f$. The integral in Eq.\ (\ref{eq1}) must be performed in
the $z\rightarrow\infty$ limit in order to ensure that the electron has already
abandoned the region of interaction with the nanostructure. Then,
$\delta\psi_\parallel$ can be expressed as a coefficient multiplying a plane wave
function of energy $\hbar\varepsilon_f$. Finally, the gain probability is
given by the square of that coefficient as the dimensionless quantity
\begin{eqnarray}
  P_{\rm EEGS}(\omega)=\left(\frac{e}{\hbar\omega}\right)^2\left|\int dz\;\ee^{-\ii \omega z/v}\,E_z(\Rb_0,z)\right|^2,
  \label{eq2}
\end{eqnarray}
where we have used the non-recoil approximation $p_f-p_i\approx\omega/v\ll p_i$ for the electron moving with high velocity $v=\hbar p_i/m_e$. If the incident field is a plane wave and
the electron moves in vacuum, only scattered components contribute
to the EEGS probability, because the momentum gained by the electron
($\hbar\omega/v$) is always larger then the free-space momentum of
light. Therefore, we can substitute the induced field for the total field in Eq.\ (\ref{eq2}).

Useful analytical formulas can be derived for small particles, in
which the dipolar response is dominant, via their electric
polarizability $\alpha$. The EEGS probability is given by
integration of the dipole scattered field,
\[E_z^{\rm ind}=\alpha[(\omega^2/c^2)E_z^{\rm ext}+(\Eb^{\rm
ext}\cdot\nabla)\partial_z]\exp(\ii\omega r/c)/r.\]
Direct application of Eq.\ (\ref{eq2}) yields
\begin{eqnarray}
  &&P_{\rm EEGS, dip}(\omega)=\left|\frac{2e\omega\alpha}{\hbar v^2\gamma}\right|^2\label{dipeegs}\\&\times&\;\;\;\;\left[|\Eb^{\rm ext}\cdot\hat{\bf{b}}|^2K_1^2\left(\frac{\omega b}{v\gamma_\epsilon}\right)+\frac{|E_z^{\rm ext}|^2}{\gamma^2}K_0^2\left(\frac{\omega b}{v\gamma_\epsilon}\right)\right],
  \nonumber
\end{eqnarray}
where $\bf{b}$ is the impact parameter of the electron relative to the particle ($\bf{b}\perp\zz$) and $\hat{{\bf b}}={\bf b}/b$. Interestingly, for an external field that is perpendicularly oriented with respect to the electron trajectory, the scattered field picks up $z$ components that couple to the electron, giving rise to the $K_1^2$ contribution in
Eq.\ (\ref{dipeegs}). This expression needs to be compared with the EELS probability for a small particle [$\Gamma_{\rm EELS,dip}$, Eq.\ (\ref{Pdipole})]. For a lossless particle, the optical theorem (${\rm
Im}\{\alpha\}=(2\omega^3/3c^3)\,|\alpha|^2$) allows us to write $\Gamma_{\rm EELS,dip}=\langle P_{\rm
EEGS,dip}\rangle\rho^0 f$, where $\rho^0=\omega^2/\pi^2c^3$ is the
free-space local density of photonic states, the factor
$f=2\pi\hbar\omega$ corrects for the number of photons contained in
a normalized external field $\Eb^{\rm ext}$ (i.e., for $|\Eb^{\rm
ext}|=1$), and the average over orientations of this field has been
performed. In other words, the EELS probability, or equivalently the
CL probability ($\Gamma_{\rm CL}$) in our lossless particle,
coincides with the EEGS probability for an incoherent source
containing one photon per photon state. For a dissipative particle,
we have $\Gamma_{\rm EELS}>\Gamma_{\rm CL}$, but the relation
$\Gamma_{\rm CL}=\langle P_{\rm EEGS,dip}\rangle\rho^0 f$ still holds.

We are now ready to make a quantitative comparison of EEGS and EELS.
Incidentally, $\Gamma_{\rm EELS}$ has units of inverse energy,
whereas $P_{\rm EEGS}$ is dimensionless, so that the
comparison requires integrating $\Gamma_{\rm EELS}$ over a
finite energy-loss range $\hbar\Delta\omega$. We have $P_{\rm
EEGS}/(\hbar\Delta\omega\Gamma_{\rm EELS})\sim\pi|E^{\rm
ext}|^2/(\hbar\Delta\omega{\rm Im}\{-1/\alpha\})$. For instance, for
a 100\,nm gold nanoparticle illuminated with a laser of moderate
intensity $\sim 10^6$\,W/cm$^2$, tuned to the particle plasmon energy
$\hbar\omega=2.5$\,eV, and taking $\hbar\Delta\omega=0.1$\,eV (the
plasmon width), we find $P_{\rm EEGS}/(\hbar\Delta\omega\Gamma_{\rm
EELS})\approx 40$. This must be compared with $\Gamma_{\rm
CL}/\Gamma_{\rm EELS}<1$.

We conclude that EEGS should provide an excellent probe of local photonic properties, combining the spatial resolution of TEMs and the energy resolution of optical sources. The EEGS signal is free from bulk features that dominate the loss spectrum as
soon as the electron trajectory intersects the sample. Control over the intensity of the pumped light adds an extra handle to facilitate the detection of the gain signal, in contrast to CL. Finally, EEGS is sensitive to light polarization and it can be applied to investigate nonlinear optical response with nanometer resolution.

\subsection{Surface plasmon launching}
\label{plasmonemission}

As microchip features continue to shrink, lithographically-patterned
metal structures are becoming natural candidates to replace
electronic microcircuits. The new devices are expected to operate
at frequencies above the THz, rather than GHz, and to encode
information in plasmon signals rather than electric currents. The
field of plasmonics is actually experiencing a tremendous expansion
in preparation for this emerging technology \cite{BDE03} but is
still facing the lack of suitable plasmon sources that provide the
versatility demanded by highly integrated plasmonic circuits, which
conventional techniques like frustrated total-internal reflection
and NSOM cannot satisfy.

Fast electrons are a promising alternative for generating plasmons with
exquisite spatial control and relatively high efficiency. Plasmon
launching by swift electrons was already demonstrated several decades
ago by \textcite{TS1967} and \textcite{H1977} in metallic gratings. These authors utilized the periodic
corrugation to couple plasmons to propagating light, which was
analyzed in energy and angle of emission.

These pioneering studies have been recently supplemented by monitoring SPP generation in metallic planar surfaces. The SPPs were eventually decoupled by means of a grating placed at some distance from the position in which they were generated \cite{BJK06,VVP06}, thus yielding information on SPP propagation lengths through the dependence of the resulting emission intensity on the distance between the beam and the grating, and conclusively demonstrating the excitation of SPPs by fast electrons. Subsequent studies have added more insight into the spectrally- and spatially-resolved propagation of SPPs \cite{BJM07,paper167,YS08} and the increase in plasmon losses produced by grain boundaries in polycrystalline gold compared to single-crystal gold \cite{KVV08}. The excitation near the plasmon threshold $\omega_s$ has been also addressed \cite{PLC08}.

Plasmons can be identified in Eq.\ (\ref{eeq7}) with the pole in the denominator of $D$ [Eq.\ (\ref{eeq10})], which yields the dispersion relation given by Eq.\ (\ref{SPdisp}). The $\QQ$ integral of Eq.\ (\ref{eeq12}) can be then performed in the $R\gg\lambda$ limit by only retaining the contribution of the $\QQ=\ksp$ pole. This is the so-called plasmon-pole approximation \cite{FW1984}, which enables us to approximate
\begin{eqnarray}
   D\approx\frac{C}{\QQ-\ksp}
   \label{ppa}
\end{eqnarray}
with
\begin{eqnarray}
   C=\frac{2\ii eq_{z1}}{c}\frac{\epsilon_2}{\epsilon_1^2-\epsilon_2^2}.
   \nonumber
\end{eqnarray}
[This comes from the Taylor expansion of $\epsilon_2\;q_{z1}+\epsilon_1\;q_{z2}$ around the $\QQ=\ksp$ pole.]
Now, we proceed by inserting the plasmon-pole expression back into the reflected components of the fields [Eq.\ (\ref{eeq12})] and by approximating the integral over $\QQ$ using the $\QQ=\ksp$ pole of Eq.\ (\ref{ppa}). Finally, Eq.\ (\ref{eeq12}) reduces to
\begin{eqnarray}
   \Hb^{\rm ref}(\rb,\omega)\approx A_j\,\ee^{\ii q_{zj} |z|}\,H^{(1)}_1(\ksp R)\,\fh,
   \label{eq18}
\end{eqnarray}
where
\begin{eqnarray}
   A_j=\pi\ksp\,{\rm sign}(z)\,C\,\mu_j\left(\ksp\right),
   \nonumber
\end{eqnarray}
and $\mu_j$ is defined in Eqs.\ (\ref{muja}) and (\ref{mujb}). Here, $H^{(1)}_1$ is a Hankel function, dying off as $\approx(2/\pi\ksp R)^{1/2}\,\exp(\ii\ksp R)$ at large distance $R$ from the electron trajectory\footnote{The $1/\sqrt{R}$ dependence of the field is the signature of a surface mode, since the resulting Poynting vector scales as $1/R$, so that its integral over a circle of radius $R$ centered at the electron source stays independent of $R$, except for the exponential attenuation due to non-vanishing SPP absorption, $\exp(-2{\rm Im}\{\ksp\}R)$.} \cite{AS1972}. Equation\ (\ref{eq18}) indicates that the induced plasmon field is a cylindrical wave of $m=0$ symmetry and parallel wavevector $\ksp$.

\begin{figure*}
\includegraphics[width=180mm,angle=0,clip]{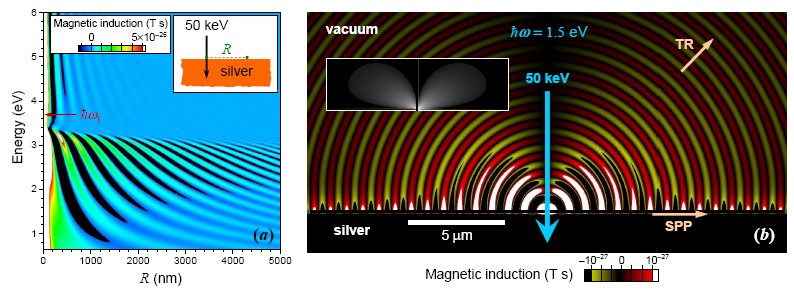}
\caption{\label{SP-launching-2} (Color online) Spatial visualization of plasmon launching. {\bf (a)} The magnetic induction $\Bb(\rb,\omega)$ set up by an electron impinging on a semi-infinite silver sample is represented as a function of distance to the trajectory along the surface and is resolved in energy components $\hbar\omega$. Plasmon generation leads to oscillations of increasing period with decreasing plasmon energy, in agreement with the dispersion relation of Fig.\ \ref{SP}. The induction is calculated at a distance of 10\,nm from the surface. {\bf (b)} Full spatial dependence of the $\hbar\omega=1.5\,$eV component of the magnetic induction under the same conditions as in (a). The left inset shows a log-scale overview stretching $1.2\,$mm along the surface, in which SPP and TR components are clearly separated. The dielectric function of Ag has been taken from \textcite{JC1972}.}
\end{figure*}

Figure\ \ref{SP-launching-2}(a) shows the real part of the reflected magnetic induction ($\Bb^{\rm
ref}(\rb,\omega)$), calculated from Eqs.\  (\ref{eeq10}) and (\ref{eeq12}) as a function of $R$ and $\omega$ for $\rb=(R,0,10\,{\rm nm})$ along a silver surface bombarded by a 50\,keV electron. Surface plasmons produce strong modulations of the field at energies below Ritchie's non-retarded limit ($\hbar\omega_s\approx 3.7$ eV for Ag). The
period of these oscillations increases with decreasing $\omega$ in
accordance with the $H^{(1)}_1$ behavior noted above. The surface
field exhibits faster oscillations for $\omega>\omega_s$, which rapidly decay away from the trajectory. The full spatial dependence of $\Bb^{\rm ref}(\rb,\omega)$ is represented in Fig.\ \ref{SP-launching-2}(b) for $\hbar\omega=1.5\,$eV, clearly showing the excitation of a cylindrical SPP wave (see intensity maxima near the surface) and TR (oscillatory wave components evolving away from the surface). The TR region is dominated by oscillations of short wavelength ($2\pi/k$) compared to the SPP wavelength ($2\pi/{\rm Re}\{\ksp\}$), in consonance with the larger momentum of SPPs (${\rm Re}\{\ksp\}>k$).

We are ultimately interested in determining how many plasmons are launched per
incident electron, which can be calculated from the integral of the
Poynting vector derived from Eq.\ (\ref{eq18}) over a cylindrical
surface centered around the electron trajectory. This leads to
\begin{eqnarray}
   \Gamma_{\rm SP}(\omega) = \frac{1}{2\hbar k^2} \sum_{j=1,2}
         \frac{|A_j|^2}{{\rm Im}\{q_{zj}\}}\,{\rm Re}\left\{\frac{\QQ
         |\QQ|}{\epsilon_j}\right\}
   \label{eeq20}
\end{eqnarray}
for the probability normalized per incident electron and per unit of plasmon frequency range.\footnote{It should be noted that the approximation of Eq.\ (\ref{ppa}) is fully justified for $R\gg\lambda$, that is, in the limit that we have examined to derive the plasmon energy flux far away from the electron trajectory \cite{paper130}.}

The prediction of Eq.\ (\ref{eeq20}) is represented in Fig.\ \ref{SP-launching-1}(a) (solid curves) for Au and Ag surfaces bombarded by 200\,keV electrons, and normalized per unit of emission wavelength $\lambda$ [i.e., the plotted quantity is $(\omega/\lambda)\Gamma_{\rm SP}(\omega)$]. Plasmon production is boosted near $\omega_s$ in both cases, and it actually surpasses the production of TR in that region. Roughly speaking, $\sim 10^{-3}$ gold plasmons are produced for each normally-incident electron within the 600-1200 nm spectral range, a figure that reduces to $\sim 1.7\times 10^{-4}$ for 20\,keV incident energy and monotonically grows with electron velocity until it levels up at a value of $\sim 2\times 10^{-3}$ around 2 MeV [Fig.\ \ref{SP-launching-1}(b)]. This behavior is qualitatively consistent with the spatial extension of the external electron field $\sim v\gamma/\omega$ [see Eq.\ (\ref{Ehomo2})], and one should expect maximum coupling to the plasmon field when that extension is of the order of the SPP wavelength. The dependence of the emission on primary electron-energy, plotted in Fig.\ \ref{SP-launching-1}(b), confirms this explanation.

Good agreement between the above theory and experiment has been recently obtained by \textcite{paper167}, who inferred the absolute SPP yield by first normalizing the measured CL signal to the calculated TR in a flat surface, and then analyzing the interference between SPPs decoupled by a grating and TR. The results are shown in Fig.\ \ref{SP-launching-1}(c).

Propagating plasmons can be generally excited with electron beams in any plasmon-supporting system, including metal films, nanowires, and particle arrays \cite{paper168}. An example of plasmon excitation in a linear particle array is discussed in Fig.\ \ref{Cai}, whereas experimental observation of electron coupling to plasmons in nanowires was obtained by \textcite{YNS06} and \textcite{VDK07}.

The emission mechanism so far considered is coherent in the sense explained in Sec.\ \ref{eire}. However, there are at least two other incoherent mechanisms of emission involving plasmons: (1) plasmon-assisted decay of
localized excitations produced by the impinging electron close to
its trajectory and (2) SEs, which can in turn couple to
the plasmon, acting as primary electrons. These contributions are
only effective within a distance of the order of the skin depth
inside the metal. We can invoke Bethe's theory of charge stopping \cite{B1930} to conclude
that the energy deposited within that region scales as $\sim 1/E$ with the primary electron energy $E$. Consequently, one should observe relatively more incoherent emission at low electron energies and an increase in the degree of coherence with faster electrons.

\subsection{Nonlocal effects in nanostructured metals}
\label{nonlocaleffects}

The interaction between closely spaced particles has recently attracted much interest because they produce large electric field enhancement in the gap region, which is beneficial for molecular sensing applications \cite{XBK99}. Gap distances in the range of one to a few nanometers have actually been used to measure surface-enhanced Raman scattering (SERS) signals, but nonlocal effects have been largely neglected in this context until very recently \cite{paper162,ZPN09}. Spatially-resolved EELS is well suited to experimentally study these unexplored nonlocal effects in dimers, and therefore we discuss this problem in this section.

Spatial dispersion in the material response generally produces blue shifts and decreasing intensity of plasmon features. Since the early work of \textcite{FK1968_2} on spheres, nonlocal effects in EELS have been theoretically addressed for several geometries in the non-retarded limit, and they have been predicted to show up for electron impact parameters in the nanometer range \cite{ZE1990,AR08}.

Nonlocal effects in the optical response of small metal spheres ($\lesssim 10\,$nm) have been known for a long time [see, for example, \textcite{OBI92} for an EELS experimental study of small silver particles], and they are well described by a phenomenological model incorporating conduction-electron scattering at the particle boundary as a source of plasmon decoherence \cite{KV95}. This model consists in supplementing the damping rate $\eta$ in Eq.\ (\ref{drude}) with a size-dependent term, $v_F/a$. However, the $v_F/a$ quantum-confinement model is inconsistent with the fact that an electron state trapped inside a finite cavity (mimicking a conduction electron in a nanoparticle) has infinite lifetime. Therefore, many-body effects rather than reflection at the boundaries seem to be the source of dephasing in small particles, facilitated by spatial dispersion,
which is responsible for the breakdown of the local response approximation at distances below a few nanometers \cite{paper162}.

The hydrodynamic model offers an alternative way of accounting for nonlocal effects, as discussed in Appendix.\ \ref{hydrodynamicmodel} for a general case and in Appendix\ \ref{scatteringcoefficients} for spheres. The results of this model for single aluminum spheres are shown in Fig.\ \ref{resolution}(a) (dashed curves), compared with a local description (solid curves). As expected, nonlocal calculations are blue shifted with respect to the local theory.

\begin{figure}
\includegraphics[width=80mm,angle=0,clip]{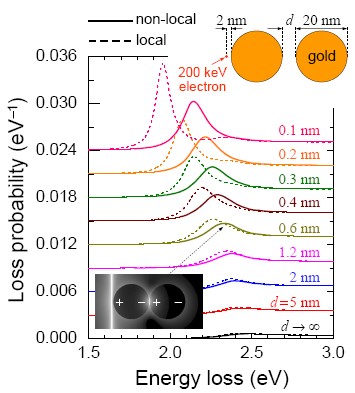}
\caption{\label{nonlocal} (Color online) Loss probability of a 200\,keV electron passing near dimers formed by spherical gold particles of 20 nm in diameter. The loss spectra are dominated by the longitudinal dipole mode. Local (dashed curves) and nonlocal (solid curves) calculations are compared for several separations between the particle surfaces, $d$. Consecutive curves are vertically offset by $0.003\,$eV$^{-1}$ for clarity. The electron trajectory is perpendicular to the dimer axis and located as indicated in the upper-right inset. The lower-left inset shows the $\hbar\omega=2.26\,$eV component of the total-field intensity in logarithmic scale for $d=0.6\,$nm, calculated from the BEM; the field diverges along the electron trajectory (vertical bright line) and it has a local maximum in the dimer-gap region.}
\end{figure}

For small gold dimers at optical frequencies, it is reasonable to describe the particles using Eq.\ (\ref{epsl}), and to plug the resulting multipolar polarizability into the multiple-scattering formalism of Sec.\ \ref{secMESME}. We have used Eq.\ (\ref{eqwjpcc}) for $\epsilon(q,\omega)$. The calculated loss probability is represented in Fig.\ \ref{nonlocal} for 20\,nm particles (see \onlinecite{paper162}, for more details on the model). The loss spectra show a dominant feature that corresponds to the lowest-order longitudinal dipole mode of the structure (i.e., this mode corresponds to the co-axial dipole-dipole mode characterized by a charge pattern "$(+-)\cdot(+-)$" at large separations). Significant enhancement of the electric field is produced near the gap region at the plasmon mode frequency (Fig.\ \ref{nonlocal}, lower-left inset), so that the details of the dielectric response at small distances become critical. The local description of the sphere (dashed curves) is clearly insufficient at distances below 2\,nm, since they produce redshift and increase of intensity as compared to the nonlocal results (solid curves). The response of the dimer is extremely sensitive to the gap distance, but electron microscopes are ideally suited to simultaneously determine this parameter and the plasmon energy.

\section{Prospects for Plasmonics}
\label{prospectsforplasmonics}

Browsing the figures of this review, we observe that the historical
development of plasmons has been intimately correlated with advances
in electron microscopy and spectroscopy. Since the early days of
SPs, which \textcite{R1957} predicted to show up as
characteristic energy losses in electrons traversing metallic thin foils (see
Fig.\ \ref{Swan}), up until now that particle plasmons can be mapped
with nanometer precision (see Fig.\ \ref{triangles}), these
collective excitations have captured the imagination of electron
microscopists and material scientists alike.

Plasmons have now taken off as excellent tools in ultrafast technology for processing information encoded as plasmonic signals
\cite{BDE03,O06,ZSC06} and for applications in biology \cite{SSA93} and biomedicine \cite{BCN05}. In particular, localized plasmon resonances (LPRs) play a relevant role because they are typically accompanied by large enhancement of the electromagnetic field that is useful in molecular sensing [for example, using SERS \cite{XBK99}].\footnote{Localized plasmons can be confined down to a few nanometers in small particles, for which retardation effects are negligible, so that LPRs are safely described by Poisson's equation. Actually, the minimum size of well-defined LPRs is only limited by broadening due to nonlocal effects (see Sec.\ \ref{nonlocaleffects}). Now, the question arises, are metals (in the optical sense) necessary to produce localized electromagnetic resonances? In other words, do we need to have negative $\epsilon$ materials? A positive answer to this question comes from examining the eigenvalues $\Lambda=2\pi(\epsilon_2+\epsilon_1)/(\epsilon_2-\epsilon_1)$ that appear in the solution of Poisson's equation, recast into Eq.\ (\ref{nrbem}) (i.e., the solution corresponding to vanishing external field): one finds that the eigenvalues are real and they must satisfy $|\Lambda|<2\pi$, which is only possible if $\epsilon_1$ and $\epsilon_2$ have opposite sign. An alternative explanation based upon physical intuition can be found in the formal equivalence between Poisson's equation and the heat transport equation under stationary conditions, with $\epsilon$ and $\phi$ playing the role of the thermal conductivity and the temperature, respectively: real materials have positive thermal conductivity, which guarantees that the second law of thermodynamics if fulfilled (heat does flow towards regions of lower temperature); but we know that heat cannot be spatially confined for ever using passive materials, so that $\epsilon>0$ media are unable to trap the nonuniform $\phi$ associated to a localized resonance.} It is of course important to have experimental tools that permit us to visualize LPRs in patterned metallic surfaces and nanoparticles. One of these tools is NSOM (see \onlinecite{KS07}, and references therein), which has been successfully applied to map plasmons \cite{BVD06,DHK07}, although it is limited to around several tens of nanometers in spatial resolution. In this context, absorption and scattering spectroscopy has been extensively used to characterize the optical properties of metallic nanoparticles in colloids \cite{BCN05}, but it requires narrow distributions of particle size and morphology, and it can only access the average properties of particle assemblies. Similarly, dark-field microscopy (DFM) is becoming a popular technique in the colloidal chemistry community, allowing localized plasmons to be studied in individual nanoparticles \cite{MBS02}. However, unlike EELS, the DFM technique cannot provide details on the local distribution of these modes.

\begin{figure*}
\includegraphics[width=160mm,angle=0,clip]{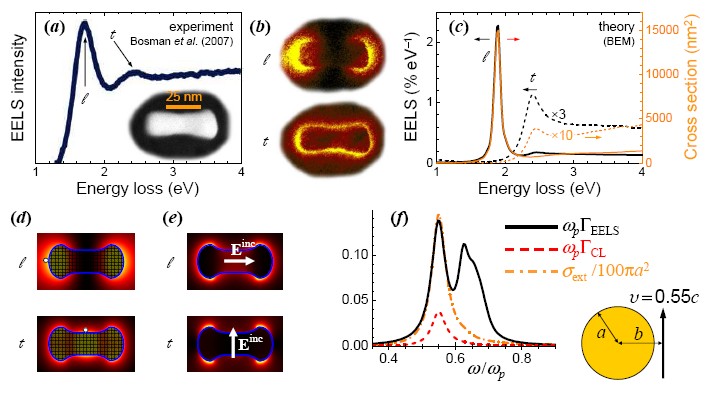}
\caption{\label{vsoptics} (Color online) Comparison between EELS, CL, and optical extinction spectroscopy. {\bf (a)} Electron energy-loss spectrum (EELS) of the dumbbell gold nanoparticle shown in the dark-field TEM-image inset. {\bf (b)} Plasmon maps obtained from the intensity of electrons that have lost an energy corresponding to the dipolar longitudinal mode (1.7 eV) and the dipolar transversal mode (2.4 eV). {\bf (c)} Calculated EELS probability (left scale) and extinction cross section (right scale) of a nanoparticle mimicking the dumbbell of (a). The cross section is obtained for external electric field orientations both parallel and perpendicular to the particle axis, in order to excite longitudinal ({\it l}) and transversal ({\it t}) plasmons, respectively. The EELS spectra correspond to positions of the electron beam as shown in (d) by white circles. The longitudinal mode is red-shifted in the experiment, as compared to the calculation, presumably due to the effect of a mica substrate, which is not accounted for here. {\bf (d)} Calculated EELS plasmon maps, with the particle contour shown as a shaded area. Notice that the experimental images of (b) exhibit depletion of electron intensity for trajectories passing through the particle, due to elastic scattering of the electrons with gold atoms. {\bf (e)} Near-field plots for longitudinal and transversal polarization of the incident electric field under external optical illumination. {\bf (f)} Comparison of EELS and CL probabilities (solid and dashed curves, respectively) with the optical extinction cross section $\sigma_{\rm ext}$ (dash-dotted curve) for a metal sphere described by the dielectric function of Eq.\ (\ref{drude}). The damping rate is $\eta=0.04\omega_p$ (similar to Al), the sphere radius is a tenth of the free-space wavelength at frequency $\omega_p$, and the impact parameter is $b=1.1a$. The experimental data of (a) and (b) are taken from \textcite{BKW07}. The calculations of (c)-(e) are adapted from \textcite{paper155}.}
\end{figure*}

Spatially-resolved EELS and CL offer alternative methods to
characterize plasmons with unmatched resolution. Actually,
no other available technique has achieved the degree of detail
obtained by the plasmon studies that we review in Figs.\
\ref{triangles}, \ref{Yamamoto}, and \ref{vsoptics}. Furthermore, the theoretical
tools that are necessary for interpreting EELS and CL spectral images
have already reached a mature status that allows addressing arbitrarily
complex geometries (see Sec.\ \ref{morecomplex}).

The plasmon energies and widths obtained from EELS and optical extinction are in excellent agreement [Fig.\ \ref{vsoptics}]. In particular, the spectra of dumbbells that we show in Fig.\ \ref{vsoptics}(c) provide similar information in both techniques. However, EELS adds the spatial resolution that allows mapping plasmons in real space [see Figs.\ \ref{triangles} and \ref{vsoptics}(d)]. The resulting EELS images compare reasonably well with the near-field enhancement that is expected upon conventional optical illumination [Fig.\ \ref{vsoptics}(e)]. Moreover, EELS can access plasmon modes that are not resolvable with light \cite{paper164}. For example, multipoles of order $l>1$ in small metal spheres [Fig.\ \ref{vsoptics}(f)].

Like far-field optical spectroscopy, the EELS and CL techniques benefit from
the use of relatively non-invasive probes (the electrons) that produce negligible changes in
the optical properties of what is being observed. In contrast, the NSOM tip exerts a strong perturbation on the sample.
The spatial resolution of EELS and CL is now pushed down to the sub-{\AA} level,
but even a modest SEM setup equipped with CL can focus an electron
beam down to $<2$ nm spots. The photon-energy resolution of CL is similar
to that of other optical techniques, although the signal-to-noise
ratio becomes annoying, particularly when the
electron does not penetrate the specimen but simply interacts
through electromagnetic fields at a short distance from it (aloof geometry). Regarding EELS, the
0.1\,eV resolution achieved by forefront commercial STEMs is already sufficient
for most plasmonic structures, and this situation should improve in
coming years with further advances in electron monochromators and spectrometers.

Finally, it would be unfair to conclude this section without a summary of the impressive amount of information that has been gathered on both localized and extended plasmons using electron microscopes. We cannot review the vast literature here, but we present a succinct overview with some relevant references in Table\ \ref{table}.

\begin{table*}
\begin{center}
\begin{tabular}{|l|l|l|c|r|l|l|}
\hline & Structure & Technique & $\hbar\omega\,$(eV) & Reference & Comments\\ \hline
SP  & Al planar surfaces  & REELS & $5-100$ & \textcite{PS1959} & first observation of planar SPs \\
SPP & Al grating          & CL & $2.3$ & \textcite{TS1967} & SPPs signature in CL \\
LPR & Al nanoparticles    & CL & $5-100$ & \textcite{FKI1967} & Mie modes \\
SPP & Al films & EELS     & $3-10$ & \textcite{PSV1975} & SPPs dispersion relation \\
SP  & Al and In surfaces  & EELS & $10-12$ & \textcite{KR1976} & large-momentum dispersion \\
SPP & Ag shallow grating  & CL & $2-3$ & \textcite{H1977} & SPP dispersion relation \\
LPR & Al nanospheres      & EELS & $1-20$ & \textcite{B1980} & \\
LPR & Al nanospheres on AlO$_3$ & EELS & $1-40$ & \textcite{WC1987} & plasmon maps \\
LPR & Al nanospheres on Al & EELS & $1-40$ & \textcite{WC1987_2} & plasmon maps \\
LPR & Ag nanospheres      & EELS & $2-35$ & \textcite{OBI92} & nonlocal effects \\
LPR & Si nanospheres      & EELS & $2-35$ & \textcite{UCT92} & plasmon maps \\
LPR & Al nanospheres      & EELS & $2-40$ & \textcite{SBS97} & \\
SP  & Al$_2$O$_3$ nanospheres & EELS & $2-30$ & \textcite{AKH00} & diffraction radiation losses\\
SPP & Ag gratings         & CL & $1-4$ & \textcite{YAT01} & plasmon standing waves \\
LPR & Ag nanospheres      & CL & $2-4$ & \textcite{paper064} & plasmon maps of Mie modes \\
LPR & WS$_3$ nanotubes    & EELS & $3-30$ & \textcite{KSH01} & \\
LPR & C nanotubes & EELS  & $3-25$ & \textcite{STK02} & $\pi$ and $\sigma$ plasmons \\
LPR & Ag rough surfaces   & PEEM & $3$ & \textcite{KOP05} & femtosecond dynamics \\
SPP & Au planar surfaces  & CL & $1-3$ & \textcite{BJK06} & SPP propagation \\
    &&&& \textcite{VVP06} &  \\
LPR & Ag sphere dimer     & CL & $2-4$ & \textcite{YNS06} & plasmon maps \\
LPR & Ag nanowires        & CL & $1-4$ & \textcite{YNS06} & plasmon maps \\
SPP & Au atom wire array  & LEEM & $1-3$ & \textcite{NYI06} & 1D plasmon dispersion \\
LPR & Ag triangular nanoprisms & EELS & $1-4$ & \textcite{paper125} & plasmon maps \\
LPR & Ag nanowires        & CL & $1-4$ & \textcite{VDK07} & plasmon maps \\
SPP & Au planar surfaces  & CL & $1-3$ & \textcite{BJM07} & hyper-spectral imaging \\
SPP & Ag planar surfaces  & PEEM & $3$ & \textcite{KPP07} & femtosecond dynamics \\
LPR & Au annular grating  & CL & $1-3$ & \textcite{paper139} & plasmon maps \\
SP  & HfO$_2$ in stacks   & EELS & $2-30$ & \textcite{CKS07} & \\
SP  & Si/SiO$_2$/Si stacks & EELS & $2-30$ & \textcite{CYM08} & \\
SPP & Au planar surfaces  & CL & $1-3$ & \textcite{KVV08} & SPP propagation \\
    &&&& \textcite{PLC08} &  \\
    &&&& \textcite{YS08}  &  \\
LPR & Au ridge            & CL & $1-3$ & \textcite{paper147} & plasmon maps \\
LPR & Au nanorods, nanospheres          & EELS & $1-4$ & \textcite{BKW07} & plasmon maps and spectra \\
    & ellipsoids, dimers, and &&& \textcite{NRM08} & \\
    & touching dimers         &&& \textcite{paper164} &  \\
    &                         &&& \textcite{SHT09} &  \\
\hline
\end{tabular}
\end{center}
\caption{Historical overview of experimental surface plasmon studies making use of electron beams.}
\label{table}
\end{table*}

\section{Conclusion}
\label{conclusion}

Overall, valence excitations, probed by electron microscopes, have been a continuous source of information on electronic structure and optical response of materials on the nanometer scale over the last six decades.
The energetic electrons employed in these microscopes permit obtaining structural and spectroscopic details of the investigated samples with sub-nanometer resolution, and they can be regarded as relatively contactless probes compared to other techniques such as STM of NSOM, particularly when the aloof configuration is employed, in which the electron beam passes outside the targeted materials and interacts with them via electromagnetic coupling. Optical excitations are the most intense in the energy-loss spectra, so that they can be collected using low beam-currents, and therefore, sample damage is reduced to small levels, even for penetrating trajectories. However, electron microscopes are versatile instruments in which the beam current can reach doses that cause substantial changes in the specimen, including reshaping of the morphology of metallic nanoparticles \cite{B08}.

In the context of electron microscopy, CL constitutes an inexpensive addition to electron microscopes that can be very valuable when combined with appropriate theoretical tools, the versatility of which we have tried to illustrate in this review. Electron microscopes in scanning mode of operation, combined with either EELS or CL analysis, are suitable tools for studying the optical response of nanostructures, and no other existing technique can match them in combined space-energy resolution. Furthermore, these instruments should facilitate the eventual application of swift electrons to light and plasmon sources for nanoscale devices. Finally, there is a huge background of both experimental and theoretical expertise that constitutes a valuable guide for future research making use of electron microscopes in nanophotonics studies.

\section*{Acknowledgments}

It is a pleasure to express my gratitude to Archie Howie for countless suggestions, and to Benito Hombrados Madrid and Carmen Mart\'{\i}n Nieto for their generous technical assistance. I would also like to acknowledge many enjoyable and stimulating discussions with Harry A. Atwater, Cheng-Hsuan Chen, Ming-Wen Chu, Christian Colliex, Pedro M. Echenique, Archie Howie, Mathieu Kociak, Martin Kuttge, Viktor Myroshnychenko, Albert Polman, Alberto Rivacoba, Peter Schattschneider, Naoki Yamamoto, Nerea Zabala, and Nikolay Zheludev. This work was supported in part by the Spanish MICyT (MAT2007-66050 and Consolider NanoLight.es) and by the EU (NMP4-2006-016881 "SPANS" and NMP4-SL-2008-213669 "ENSEMBLE").

\appendix

\section{Dielectric response of homogeneous media}
\label{homogeneousmedia}

The local response approximation yields excellent results when comparing theory and experiment both in the context of nanophotonics and in numerous EELS and CL studies that focus on the role of sample geometry. The different materials forming a structure are then described by frequency-dependent dielectric functions, usually taken from tabulated optical data (see, for example, \onlinecite{P1985}).

Several metals, including some alkali elements and aluminum, are
remarkably well described by a free-electron gas model, in which the
local dielectric function takes the form\footnote{Equation
(\ref{drude}) describes the correct way of incorporating damping,
via $\omega(\omega+\ii\eta)$ rather than $(\omega+\ii\eta)^2$. The
latter is inconsistent with conservation of the number of electrons
in the metal \cite{M1970}.}
\begin{eqnarray}
\epsilon^{\rm D}(\omega)=1-\frac{\omega_p^2}{\omega(\omega+\ii\eta)}, \label{drude}
\end{eqnarray}
where
\begin{eqnarray}
\omega_p=(4\pi e^2n_0/m_e)^{1/2} \label{wpfor}
\end{eqnarray}
is the bulk plasmon frequency, determined by the conduction electron density $n_0$,\footnote{The dependence of $\omega_p$ on $n_0$ has been experimentally monitored by \textcite{IN1970} using EELS and changing $n_0$ through thermal expansion in the $3-295\,$K range.} and $\eta$ is a small phenomenological relaxation rate that accounts for electromagnetic damping in the material (e.g., $\hbar\eta=0.6\,$eV in Al). This model gives $\hbar\omega_p=15.8\,$eV for Al, in good agreement with the value of $15\,$eV obtained from EELS measurements (see Fig.\ \ref{bulk} and Table\ \ref{table2}).

Some ionic materials, and alkali-halide crystals in particular, are
better described by assuming localized ionic oscillation modes,
which result in the dielectric function \cite{AM1976}
\begin{eqnarray}
\epsilon(\omega)=\epsilon_\infty\left[1-\frac{\omega_{\rm lon}^2-\omega_{\rm tr}^2}{\omega(\omega+\ii\eta)-\omega_{\rm tr}^2}\right], \label{epspol}
\end{eqnarray}
where $\omega_{\rm tr}$ is the frequency of transverse vibrations and $\omega_{\rm lon}$ signals a
plasmon-like longitudinal mode characterized by $\epsilon\approx 0$.
Retardation produces dispersion of these modes for momentum transfers
comparable to $\hbar k$, as shown in Fig.\ \ref{homogeneous}.

Neglecting the influence of the atomic lattice, the ratio between components of the external potential and the total potential (resolved in momentum and frequency) defines in general a dispersive dielectric function $\epsilon(q,\omega)$. Spatial dispersion becomes important for wavevector transfers $q$ of the order of $\omega/v_F$, where the Fermi velocity $v_F$ is typically two or three orders of magnitude smaller than the speed of light $c$. Consequently, the dependence on $q$ is typically important for $q\gg k$, away from the region showing retardation [see Fig.\ \ref{bulk}(b)].

\begin{table*}
\begin{center}
\begin{tabular}{|lcccccc|}
\hline \;\; Material \;\;&\;\; Equation \;\;&\;\; $\hbar\omega_p$ or $\hbar\omega_{\rm lon}$ (eV) \;\;&\;\; $\hbar\eta$ (eV) \;\;&\;\; $\epsilon_b$ or $\epsilon_\infty$ \;\;&\;\; $\hbar\omega_{\rm tr}$ (eV) \;\;&\;\; Energy range (eV)\;\;\\ \hline
Ag  (a) & (\ref{epst})   & 9.17$\pm$0.19   & 0.025$\pm$0.01  & $\approx$4    & $-$             & $<1.8$     \\
Al  (b) & (\ref{drude})  & $\approx$15     & $\approx$0.6    & $-$           & $-$             & $8-25$     \\
Au  (a) & (\ref{epst})   & 9.06$\pm$0.18   & 0.071$\pm$0.007 & $\approx$9.5  & $-$             & $<1.8$     \\
Cu  (a) & (\ref{epst})   & 8.88$\pm$0.18   & 0.096$\pm$0.01  & $\approx$8    & $-$             & $<1.8$     \\
KCl (c) & (\ref{epspol}) & $\approx$0.0274 & $\approx$0.0014 & $\approx$2.19 & $\approx$0.0185 & $0.01-0.025$ \\
SiC (b) & (\ref{epspol}) & $\approx$0.075  & $\approx$0.012  & $\approx$6.15 & $\approx$0.099  & $0.01-0.2$ \\
\hline
\end{tabular}
\end{center}
\caption{Dielectric parameters of selected materials, described by Eqs.\ (\ref{drude}), (\ref{epspol}), and (\ref{epst}), and chosen to fit the measured dielectric functions reported in the following references within the specified light-energy range: (a) \textcite{JC1972}; (b) \textcite{P1985}; (c) \textcite{AM1976}. The reader is referred to the book of \textcite{R1980} for comprehensive tables of bulk and surface plasmon energies and widths.}
\label{table2}
\end{table*}

\subsection{Lindhard and Mermin dielectric functions}
\label{LindhardMermin}

Spatial dispersion effects in good metals such as aluminum and silver can be accounted for by assimilating their conduction band to a gas of non-interacting free electrons \cite{AM1976}. An analytical formula for $\epsilon(q,\omega)$ in the free-electron gas was derived by \textcite{L1954}. The Lindhard formula, which predicts a continuum of e-h pair excitations and collective plasmon modes as illustrated in Fig.\ \ref{bulk}(a), is obtained by perturbing the plane-wave representing each electron within the Fermi sphere by means of an external potential of the form $\phi^{\rm ext}=\exp(\ii\qb\cdot\rb-\ii\omega t)$. The sum of these perturbations for all conduction electrons produces an induced charge $\rho^{\rm ind}$ proportional to $\phi^{\rm ext}$. The system is then assumed to be perturbed by the self-consistent total potential rather than the external potential in what has been termed the random-phase approximation (RPA). This leads to \cite{AM1976}
\begin{eqnarray}
\epsilon(q,\omega)&=&\left[1+\frac{4\pi}{q^2}\frac{\rho^{\rm ind}}{\phi^{\rm ext}}\right]^{-1}\nonumber\\&=&1+\frac{e^2}{\pi^2\hbar q^2}\int d^3\kb\,\frac{f_{\kb-\qb/2}-f_{\kb+\qb/2}}{\hbar\,\kb\cdot\qb/m_e+\omega},\nonumber
\end{eqnarray}
where the integral extends over 3D electron wavevectors $\kb$, $f_k$ is the Fermi distribution (approximately 1 for $k\le k_F$ and 0 otherwise at room temperature), and $k_F=m_ev_F/\hbar$ is the Fermi wavevector. This integral admits the closed-form expression \cite{L1954}
\begin{eqnarray}
\epsilon^{\rm L}(q,\omega)=1+\frac{2m_ee^2k_F}{\pi\hbar^2q^2}\left[1+R(x,y)+R(x,-y)\right],\nonumber
\end{eqnarray}
where $x=q/k_F$, $y=\omega/\varepsilon_F$, $\hbar\varepsilon_F=\hbar^2k_F^2/2m_e$ is the Fermi energy, and
\begin{eqnarray}
R(x,y)=\frac{1}{2x}\left[1-\left(\frac{x^2+y}{2x}\right)^2\right]\ln\left(\frac{x^2+2x+y}{x^2-2x+y}\right).\nonumber
\end{eqnarray}
Here, $\omega$ is assumed to have a positive imaginary part. If ${\rm Im}\{\omega\}$ is infinitesimally small
then ${\rm Im}\{\epsilon^{\rm L}(q,\omega)\}$ takes nonzero values only for $x(x-2)\le y\le x(x+2)$ (i.e., in the so-called
e-h region), which corresponds to the shaded area of Fig.\ \ref{bulk}(a).

A naive way of introducing a damping rate $\eta$ in the electron-gas motion consists in adding a finite
imaginary part to $\omega$, but this results in a net loss of the number of electrons with time. The Lindhard formula was
extended by \textcite{M1970} to include finite damping while maintaining the number of electrons constant. He found
the expression
\begin{eqnarray}
&&\epsilon^{\rm M}(q,\omega)\label{epsMermin}\\&&=1+\frac{(\omega+\ii\eta)\left[\epsilon^{\rm L}(q,\omega+\ii\eta)-1\right]}
{\omega+\ii\eta\left[\epsilon^{\rm L}(q,\omega+\ii\eta)-1\right]/\left[\epsilon^{\rm L}(q,0)-1\right]},\nonumber
\end{eqnarray}
where
\begin{eqnarray}
\epsilon^{\rm L}(q,0)=1+\frac{2me^2k_F}{\pi\hbar^2q^2}\left[1+\frac{4-x^2}{4x}\ln\left|\frac{x+2}{x-2}\right|\right]\nonumber
\end{eqnarray}
is the static limit of $\epsilon^{\rm L}$.

\subsection{Hydrodynamic model}
\label{hydrodynamicmodel}

A simpler approach, which can be extended to nonlocal structured metals, consists in describing the conduction band as a classical electron plasma \cite{B1933}. We generalize this approach here to incorporate retardation effects. The force equation for the hydrodynamic flow takes the
form\footnote{The linearized Bloch hydrodynamic equations constitute
a suitable approximation within linear response under the assumption
$|\rho^{\rm ind}|\ll en_0$.}
\begin{eqnarray}
\left(\partial/\partial t+\eta\right)\jb^{\rm ind}&=&(e^2/m_e)\,n_0\Eb\label{hyd}\\&+&(\hbar/m_e)^2\left[\beta^2-(1/4)\nabla^2\right]\nabla\rho^{\rm ind}, \nonumber
\end{eqnarray}
where $\rho^{\rm ind}$ and $\jb^{\rm ind}$ are the charge density and current density
perturbations induced in the electron gas, $\Eb$ is the
self-consistent electric field, $n_0$ is the unperturbed electron
density, and the hydrodynamic pressure is related to the Fermi
wavevector $k_F$ through the coefficient $\beta=(3/5)^{1/2}k_F$ \cite{R1957}.
We have supplemented this equation with the cutoff term
$-(\hbar/m_e)^2(1/4)\nabla^2\nabla\rho^{\rm ind}$, which accounts for e-h pair
excitations in an effective way. The response of the system is then
obtained by combining Eq.\ (\ref{hyd}) with Maxwell's equations for
the total charge and current ($\rho^{\rm ext}+\rho^{\rm ind}$ and $\jb^{\rm ext}+\jb^{\rm
ind}$, respectively) in the presence of the ionic background, the
response of which is accounted for through a permittivity $\epsilon_b$.\footnote{Interband transitions in aluminum give the main contribution to $\eta$, but $\epsilon_b=1$ works well for that material, indicating that they are not strongly polarized near the bulk plasmon frequency \cite{R1980}. In contrast, interband transitions play a dominant role in noble metals such as Ag, Au, and Cu \cite{JC1972}, where $\epsilon_b$ differs considerably from 1 (see Table\ \ref{table2}).}$^,$\footnote{Notice that the continuity equation $\nabla\cdot\jb=\ii\omega\rho$ is implied by Maxwell's equations.}
This formalism is a direct generalization of
the common non-retarded approach that is based upon Poisson's equation
and expresses the current density through a velocity potential
$\xi$ as $\jb^{\rm ind}=n_0\nabla\xi$ (see, for instance, \onlinecite{R1957},
for application to planar surfaces, \onlinecite{FK1968_2},
\onlinecite{CR1968}, and \onlinecite{BB1985}, for spheres, and
\onlinecite{WC1987_3}, for supported particles).
The customary boundary conditions of the electromagnetic field at
the metal surface must be now supplemented by the condition that the
normal component of $\jb^{\rm ind}$ is zero. The results of this formalism for
planar surfaces and spheres are presented in Appendices\
\ref{screenedinteractionplanar} and \ref{scatteringcoefficients}, respectively.

In homogeneous media, it is convenient to project the hydrodynamic and Maxwell equations in the space of wavevector $\qb$ and frequency $\omega$, from which we can work out the longitudinal and transversal dielectric functions $\epsilon_{\rm lon}$ and $\epsilon_{\rm tr}$ using the definition of electric displacement
\[\Db=\epsilon_{\rm lon}\Eb_{\rm lon}+\epsilon_{\rm tr}\Eb_{\rm tr}\]
in terms of electric field components parallel and perpendicular to $\qb$, respectively. We obtain
\begin{eqnarray}
\epsilon_{\rm lon}(q,\omega)&=&\epsilon_b\label{epsPP}\\&+&\frac{\omega_p^2}{(\hbar/m_e)^2\left[\beta^2
q^2+q^4/4\right]-\omega(\omega+\ii\eta)}\nonumber
\end{eqnarray}
and
\begin{eqnarray}
\epsilon_{\rm tr}(q,\omega)=\epsilon_b-\frac{\omega_p^2}{\omega(\omega+\ii\eta)}. \label{epst}
\end{eqnarray}
Eq.\ (\ref{epsPP}) was derived by \textcite{R1957} from an expansion of Lindhard's function for large $\omega$. The dispersion relation of longitudinal and transversal bulk plasmons is represented in Fig.\ \ref{bulk}(b) for aluminum.

\section{Non-retarded screened interaction in simple geometries}
\label{screenedinteraction}

The electrostatic screened interaction $W(\rb,\rb',\omega)$ is
defined as the amplitude of the potential created at $\rb$ by a
charge of magnitude $\exp(-\ii\omega t)$ placed at $\rb'$. For
materials described by a local, frequency-dependent dielectric
function $\epsilon(\rb,\omega)$, Poisson's equation leads to
\begin{eqnarray}
\nabla\cdot\epsilon(\rb,\omega)\nabla W(\rb,\rb',\omega)=-4\pi\delta(\rb-\rb'). \label{Poissoneq}
\end{eqnarray}
In free space, one has $W^0(\rb-\rb')=|\rb-\rb'|^{-1}$. While Eq.\ (\ref{Poissoneq}) is valid for arbitrary, smooth spatial dependence of the permittivity, most situations encountered in practice are reliably
modeled by sharp boundaries. We can then obtain $W$ from the Coulomb
interaction in each homogeneous region after imposing the continuity of
both the potential and the normal displacement at the interfaces. Next, we offer analytical expressions for $W$ in the simple cases of planes, cylinders, and spheres.

\subsection{Planar surface}
\label{screenedinteractionplanar}

A planar interface at $z=0$ is best described by expressing $W$ in
the space of parallel wavevector $\QQv$. In particular, the bare
Coulomb interaction can be written
\begin{eqnarray}
\frac{1}{|\rb-\rb'|}=\int \frac{d^2\QQv}{(2\pi)^2}\,\ee^{\ii\QQv\cdot(\Rb-\Rb')}\,W^0(\QQ,z,z'), \label{planarsurfaceeq}
\end{eqnarray}
where $\rb=(\Rb,z)$ and
\begin{eqnarray}
W^0(\QQ,z,z')=\frac{2\pi}{\QQ}\,\ee^{-\QQ|z-z'|}. \nonumber
\end{eqnarray}
The full screened interaction $W$ admits an expansion similar to Eq.\ (\ref{planarsurfaceeq}), in which
$W^0(\QQ,z,z')$ must be replaced by
\begin{widetext}
\begin{eqnarray}
   W(\QQ,z,z',\omega)=\frac{2\pi}{\QQ}\times
   \begin{cases}
   \frac{1}{\epsilon_1}\left(\ee^{-\QQ|z-z'|}+\frac{\epsilon_1-\epsilon_2}{\epsilon_1+\epsilon_2}\ee^{-\QQ(z+z')}\right),&
      \text{$z,z'>0$}, \\
   \frac{1}{\epsilon_2}\left(\ee^{-\QQ|z-z'|}-\frac{\epsilon_1-\epsilon_2}{\epsilon_1+\epsilon_2}\ee^{\QQ(z+z')}\right),&
      \text{$z,z'<0$}, \\
                                     \frac{2}{\epsilon_1+\epsilon_2}\ee^{-\QQ(|z|+|z'|)},&
      \text{otherwise},
   \end{cases}
   \label{Wplane}
\end{eqnarray}
using the notation of Fig.\ \ref{planar-surface}.

Nonlocal effects can be incorporated through the hydrodynamic model
outlined in Appendix\ \ref{hydrodynamicmodel}. The result in the
non-retarded approximation reduces to
\begin{eqnarray}
   W(\QQ,z,z',\omega)=\frac{2\pi}{\QQ}\times
   \begin{cases}
   \ee^{-\QQ|z-z'|}+\frac{\epsilon_s(0)-1}{\epsilon_s(0)+1}\ee^{-\QQ(z+z')},&
      \text{$z,z'\geq 0$}, \\
   \epsilon_s(z-z')+\epsilon_s(z+z')-\frac{2\epsilon_s(z)\epsilon_s(z')}{\epsilon_s(0)+1},&
      \text{$z,z'<0$}, \\
                                     \frac{2\epsilon_s(z')}{\epsilon_s(0)+1}\ee^{-\QQ z},&
      \text{$z\geq 0$, $z'<0$}, \\
                                     \frac{2\epsilon_s(z)}{\epsilon_s(0)+1}\ee^{-\QQ z'},&
      \text{$z<0$, $z'\geq 0$},
   \end{cases}
   \nonumber
\end{eqnarray}
where
\[\epsilon_s(z)=\int \frac{dq_z}{q^2} \frac{\ee^{\ii q_zz}}{\epsilon_{\rm lon}(q,\omega)},\]
and $q^2=\QQ^2+q_z^2$. An analytical expression for $\epsilon_s(z)$ can be obtained using Eq. (\ref{epsPP}) \cite{paper004}. This approach is formally equivalent to the specular reflection model \cite{RM1966,W1966}, which applies to any $\epsilon_{\rm lon}(q,\omega)$.

\subsection{Cylinder}
\label{cylinder}

For axially-symmetric problems, it is convenient to write the bare
Coulomb interaction as
\begin{eqnarray}
   \frac{1}{|\rb-\rb'|}&=&\frac{1}{(2\pi)^2}\int_{-\infty}^\infty dq_z\,\ee^{\ii q_z(z-z')}\sum_{m=-\infty}^\infty \ee^{\ii m(\varphi-\varphi')}
   \,\, W^0_m(R,R',q_z,\omega),
   \nonumber
\end{eqnarray}
where
\[W^0_m(R,R',q_z)=4\pi I_m(q_zR_<)\,K_m(q_zR_>),\]
$R_<=\min\{R,R'\}$, and $R_>=\max\{R,R'\}$.
In a homogeneous cylinder centered around the $z$ axis (see Fig.\
\ref{sphere}), the screened interaction is found to be
\begin{eqnarray}
   W_m(R,R',q_z,\omega)=4\pi\times
   \begin{cases}
   \frac{1}{\epsilon_1}\left[I_m(q_zR_<)+\frac{1}{\Delta_m}(\epsilon_1-\epsilon_2)I_m(x)I_m'(x)K_m(q_zR_<)\right]K_m(q_zR_>),&
      \text{if $R,R'>a$}, \\
   \frac{1}{\epsilon_2}\left[K_m(q_zR_>)+\frac{1}{\Delta_m}(\epsilon_1-\epsilon_2)K_m(x)K_m'(x)I_m(q_zR_>)\right]I_m(q_zR_<),&
      \text{if $R,R'<a$}, \\
                                     \frac{1}{\Delta_mx}I_m(q_zR_<)K_m(q_zR_>),&
      \text{otherwise},
   \end{cases}
   \label{Wcylinder}
\end{eqnarray}
where $\Delta_m=\epsilon_2I_m'(x)K_m(x)-\epsilon_1I_m(x)K_m'(x)$ and $x=q_za$.

\subsection{Sphere}

The bare Coulomb interaction can be expanded in spherical harmonics as
\[\frac{1}{|\rb-\rb'|}=\sum_{lm} \frac{4\pi}{2l+1}\,W^0_l(r,r')\,Y_{lm}(\Omega)Y_{lm}^*(\Omega'),\]
where
\[W^0_l(r,r')=\frac{r_<^l}{r_>^{l+1}},\]
$r_<=\min\{r,r'\}$, $r_>=\max\{r,r'\}$, and $\Omega$ denotes the direction of $\rb$.
Using the notation of Fig.\ \ref{sphere}, the interaction near a homogeneous sphere
is given by the above expansion with
\begin{eqnarray}
   W_l(r,r',\omega)=
   \begin{cases}
   \frac{1}{\epsilon_1}W_l^0(r,r')+\left(\frac{2l+1}{\eta_l}-\frac{1}{\epsilon_1}\right)\frac{a^{2l+1}}{(rr')^{l+1}},&
      \text{$r,r'>a$}, \\
   \frac{1}{\epsilon_2}W_l^0(r,r')+\left(\frac{2l+1}{\eta_l}-\frac{1}{\epsilon_2}\right)\frac{(rr')^l}{a^{2l+1}},&
      \text{$r,r'<a$}, \\
                                     \frac{2l+1}{\eta_l}W_l^0(r,r'),&
      \text{otherwise},
   \end{cases}
   \nonumber
\end{eqnarray}
\end{widetext}
substituted for $W_l^0(r,r')$. Here, $\eta_l=(l+1)\epsilon_1+l\epsilon_2$. It is also useful to write the $r,r'>a$ part in terms of the multipolar polarizability
\begin{eqnarray}
   \alpha_l^{\rm NR}(\omega)=a^{2l+1}\frac{l(\epsilon_2-\epsilon_1)}{l\epsilon_2+(l+1)\epsilon_1} \label{alphaAB}
\end{eqnarray}
as
\begin{eqnarray}
   W_l(r,r',\omega)=\frac{1}{\epsilon_1}\left[W_l^0(r,r')-\frac{\alpha_l^{\rm NR}}{(rr')^{l+1}}\right]. \label{Wlalpha}
\end{eqnarray}

\section{Retarded scattering matrices and coupling coefficients in a sphere including nonlocal effects}
\label{scatteringcoefficients}

The coupling coefficients of Eqs.\ (\ref{retarded.sphere}) and (\ref{CLfor}) are
dimensionless functions of $v/c$ that admit the following
closed-form expressions:
\begin{eqnarray}
C_{lm}^M=\frac{1}{l(l+1)}\left|\frac{2 m v}{c}N_{lm}\right|^2
\nonumber
\end{eqnarray}
and
\begin{eqnarray}
C_{lm}^E=\frac{1}{l(l+1)}\left|\frac{c}{v\gamma}M_{lm}\right|^2,
\nonumber
\end{eqnarray}
where
\begin{eqnarray}
N_{lm}=\sqrt{\frac{2l+1}{\pi} \frac{(l-|m|)!}{(l+|m|)!}} \,
\,\frac{\left(2|m|-1\right)!!}{(v\gamma/c)^{|m|}} \,\,
C^{(|m|+1/2)}_{l-|m|}\!\!\left(\frac{c}{v}\right), \nonumber
\end{eqnarray}
\begin{eqnarray}
M_{lm}&=& N_{lm+1}\sqrt{(l+m-1)(l-m)}\nonumber\\
&+& N_{lm-1}\sqrt{(l-m-1)(l+m)},\nonumber
\end{eqnarray}
and $C^{(\nu)}_m$ are Gegenbauer polynomials \cite{AS1972}. Similar formulas are encountered to describe the Coulomb excitation of atomic nuclei \cite{WK1979}.

The scattering matrix elements of Eqs.\ (\ref{retarded.sphere}) and (\ref{CLfor})
depend on the composition and structure of the spheres.
For the homogeneous particle of Fig.\ \ref{sphere},
Mie theory leads to \cite{M1908,J99}
\begin{eqnarray}
t_l^M = \frac{- j_l(x_1) x_2 j_l^\prime(x_2) + x_1
j_l^\prime(x_1) j_l(x_2)} { h_l^{(+)}(x_1) x_2
j_l^\prime(x_2)-x_1  [h_l^{(+)}(x_1)]^\prime j_l(x_2)}
\label{tlM}
\end{eqnarray}
and
\begin{eqnarray}
&&t_l^E \label{tlE} \\
&&=\frac{- \epsilon_1 j_l(x_1) [x_2 j_l(x_2)]^\prime +
\epsilon_2 [x_1 j_l(x_1)]^\prime j_l(x_2)}{\epsilon_1
h_l^{(+)}(x_1) [x_2 j_l(x_2)]^\prime - \epsilon_2 [x_1
h_l^{(+)}(x_1)]^\prime j_l(x_2)}, \nonumber
\end{eqnarray}
where $x_1=ka\sqrt{\epsilon_1}$, $x_2=ka\sqrt{\epsilon_2}$,
the prime denotes differentiation with respect to the argument, and
$j_l$ and $h^{(+)}_l=\ii j_l-y_l$ are spherical Bessel and Hankel
functions, respectively.

We can incorporate nonlocal effects in the sphere response using
the formalism of Sec.\ \ref{hydrodynamicmodel}. This allows us to generalize the
non-retarded study of dispersive spheres reported by
\textcite{FK1968_2}. Considering a metallic sphere in vacuum [i.e., $\epsilon_1=1$, and with medium 2 inside the sphere described by $\epsilon_{\rm lon}(q,\omega)$ and $\epsilon_{\rm tr}(\omega)$], a detailed analysis leads to
\begin{eqnarray}
t_l^E = \frac{-j_l(x) \left\{[x_2 j_l(x_2)]^\prime+g_l\right\} + \epsilon_2 [x j_l(x)]^\prime j_l(x_2)}
{h_l^{(+)}(x) \left\{[x_2 j_l(x_2)]^\prime+g_l\right\} - \epsilon_2 [x  h_l^{(+)}(x)]^\prime j_l(x_2)}, \nonumber
\end{eqnarray}
where
\[g_l=\frac{l(l+1)j_l(x_2)j_l(q_0a)}{q_0aj'_l(q_0a)}\left(\frac{\epsilon_{\rm tr}}{\epsilon_b}-1\right),\]
$x=ka$, $x_2=ka\sqrt{\epsilon_{\rm tr}}$, and $q_0$ is given by the
condition $\epsilon_{\rm lon}(q_0,\omega)=0$ [see Eq.\ (\ref{epsPP})]. The
matrix elements $t_l^M$ are still given by Eq.\ (\ref{tlM}) with
$\epsilon_{\rm tr}$ [Eq.\ (\ref{epst})] substituted for $\epsilon_2$.

\section{List of selected symbols and acronyms}
\label{listofsymbols}

\noindent
\begin{tabular}{lll}
$\Ab$               && vector potential \\
$a$                 && sphere or cylinder radius \\
$\Bb$               && magnetic induction \\
$b$                 && electron impact parameter \\
$c$                 && speed of light \\
$C_{lm}^E$, $C_{lm}^M$ && electron-sphere coupling coefficients \\
$e$                 && minus the electron charge \\
$\Eb$               && electric field \\
$\Fb$               && electromagnetic force \\
$\fb$               && far field amplitude \\
$G$                 && electromagnetic Green tensor \\
$\mathcal{G}^0$     && dipole-dipole interaction in vacuum \\
$\Hb$               && magnetic field \\
$\hb$               && boundary current \\
$\hbar$             && Planck's constant \\
$h^{(+)}_l$         && spherical Hankel function of order $l$ \\
$I_m$               && modified Bessel function of order $m$ \\
$\jb$               && electron current density \\
$j_l$               && spherical Bessel function of order $l$ \\
$k=\omega/c$        && free-space light wavevector \\
$k_B$               && Boltzmann's constant \\
$k_F$               && Fermi wavevector \\
$K_m$               && modified Bessel function of order $m$ \\
$L$                 && length of the electron trajectory \\
$l,m$               && orbital and azimuthal numbers \\
$l_{\rm max}$       && maximum orbital angular-momentum number \\
$m_e$               && electron mass \\
$\pb_i$, $\pb_f$    && initial and final electron wavevectors \\
$\qb$               && wavevector transferred from the electron \\
$\qb_\parallel=(q_x,q_y)$ && wavevector parallel to a planar surface \\
$\qb_\perp$         && wavevector transfer perpendicular to $\vb$ \\
$R$                 && radial distance in cylindrical coordinates \\
$\Rb=(x,y)$         && in-plane or $\perp\zz$ component of $\rb$ \\
$\Rh$               && cylindrical radial unit vector \\
$r$                 && radial distance in spherical coordinates \\
$\rb=(\Rb,z)$       && position coordinate vector \\
$\rb_e(t)$          && electron trajectory \\
$\rb_0=(\Rb_0,z_0)$ && electron impact parameter at time $t=0$ \\
$t$                 && time \\
$t_l^E$, $t_l^M$    && sphere scattering matrices \\
$\vb$               && electron velocity vector \\
$v$                 && electron velocity \\
$v_F$               && Fermi velocity \\
$W$                 && scalar screened interaction \\
$\xx$, $\yy$, $\zz$ && Cartesian unit vectors \\
\end{tabular}

\begin{tabular}{lll}
$\alpha$            && particle polarizability \\
$\beta$             && hydrodynamic wavevector \\
$\Gamma(\omega)$    && probability per unit frequency $\omega$ \\
$\gamma$            && $\gamma_\epsilon$ in vacuum ($\epsilon=1$)\\
$\gamma_\epsilon$   && Lorentz's contraction factor in a medium \\&& of permittivity $\epsilon$ \\
$\epsilon$          && dielectric function \\
$\epsilon^{\rm D}$, $\epsilon^{\rm L}$, $\epsilon^{\rm M}$
                    && Drude, Lindhard, and Mermin $\epsilon$ \\
$\epsilon_{\rm lon}$, $\epsilon_{\rm tr}$
                    && longitudinal and transversal $\epsilon$ \\
$\varepsilon$       && $\hbar\varepsilon$ is an electron energy \\
$\eta$              && bulk plasmon damping rate \\
$\lambda$           && free-space light wavelength \\
$\mu_j$             && coefficients for TR and SPP emission\\
$\rho$              && electron charge density \\
$\rho_\nt$          && photonic local density of states (LDOS) \\
$\sigma$            && boundary charge \\
$\phi$              && scalar potential \\
$\fh$               && azimuthal unit vector \\
$\fh_\qb$           && azimuthal unit vector in $\qb$ space, $\perp\qb_\parallel,\,\zz$ \\
$\varphi$           && azimuthal angle \\
$\varphi_{\rm in}$  && electron-beam divergence half-angle \\
$\varphi_{\rm out}$ && electron-spectrometer aperture half-angle \\
$\chi$              && electric susceptibility \\
$\psi$              && electron wave function \\
$\Omega$            && solid angle \\
$\omega$            && frequency \\
$\omega_p$          && classical bulk plasma frequency \\
$\omega_s$          && Ritchie's surface plasma frequency \\
\end{tabular}

\begin{tabular}{lll}
1D,2D,3D          && one-, two-, three-dimensional \\
BEM               && boundary element method \\
CL                && cathodoluminescence \\
CR                && Cherenkov radiation \\
DDA               && discrete-dipole approximation \\
DFM               && dark-field microscopy \\
EEGS              && electron energy-gain spectroscopy \\
EELS              && electron energy-loss spectroscopy \\
EFTEM             && energy-filtered TEM \\
e-h               && electron-hole \\
EIRE              && electron-induced radiation emission \\
FWHM              && full width at half maximum \\
HREELS            && high-resolution EELS \\
JDOS              && joint density of states \\
KK                && Kramers-Kronig \\
KKR               && Korringa-Kohn-Rostoker \\
LDOS              && local density of states \\
LEEM              && low-energy electron microscopy \\
LPR               && localized plasmon resonance \\
NIR               && near infrared \\
NSOM              && near-field scanning optical microscopy \\
PEEM              && photo-electron emission microscopy \\
REELS             && reflection EELS \\
SE                && secondary electron \\
SEE               && secondary electron emission \\
SEM               && scanning electron microscopy/microscope \\
SERS              && surface-enhanced Raman scattering \\
SP                && surface plasmon \\
SPP               && surface-plasmon polariton \\
S-P               && Smith-Purcell \\
STEM              && scanning transmission electron microscopy \\
STM               && scanning tunneling microscopy \\
STS               && scanning tunneling spectroscopy \\
TEM               && transmission electron\\&& microscopy/microscope \\
TR                && transition radiation \\
UV                && ultraviolet \\
XRM               && x-ray microscopy \\
ZLP               && zero-loss peak \\
\end{tabular}


\begin{thebibliography}{511}
\expandafter\ifx\csname natexlab\endcsname\relax\def\natexlab#1{#1}\fi
\expandafter\ifx\csname bibnamefont\endcsname\relax
  \def\bibnamefont#1{#1}\fi
\expandafter\ifx\csname bibfnamefont\endcsname\relax
  \def\bibfnamefont#1{#1}\fi
\expandafter\ifx\csname citenamefont\endcsname\relax
  \def\citenamefont#1{#1}\fi
\expandafter\ifx\csname url\endcsname\relax
  \def\url#1{\texttt{#1}}\fi
\expandafter\ifx\csname urlprefix\endcsname\relax\def\urlprefix{URL }\fi
\providecommand{\bibinfo}[2]{#2}
\providecommand{\eprint}[2][]{\url{#2}}

\bibitem[{\citenamefont{Abe} \emph{et~al.}(2000)\citenamefont{Abe, Kurata, and
  Hojou}}]{AKH00}
\bibinfo{author}{\bibnamefont{Abe}, \bibfnamefont{H.}},
  \bibinfo{author}{\bibfnamefont{H.}~\bibnamefont{Kurata}}, and
  \bibinfo{author}{\bibfnamefont{K.}~\bibnamefont{Hojou}},
  \bibinfo{year}{2000}, \bibinfo{journal}{J.\ Phys.\ Soc.\ Jpn.}
  \textbf{\bibinfo{volume}{69}}, \bibinfo{pages}{1553}.

\bibitem[{\citenamefont{Abramowitz and Stegun}(1972)}]{AS1972}
\bibinfo{author}{\bibnamefont{Abramowitz}, \bibfnamefont{M.}}, and
  \bibinfo{author}{\bibfnamefont{I.~A.} \bibnamefont{Stegun}},
  \bibinfo{year}{1972}, \emph{\bibinfo{title}{Handbook of Mathematical
  Functions}} (\bibinfo{publisher}{Dover}, \bibinfo{address}{New York}).

\bibitem[{\citenamefont{{Ach\`{e}che}}
  \emph{et~al.}(1986)\citenamefont{{Ach\`{e}che}, Colliex, Kohl, Nourtier, and
  Trebbia}}]{ACK1986}
\bibinfo{author}{\bibnamefont{{Ach\`{e}che}}, \bibfnamefont{M.}},
  \bibinfo{author}{\bibfnamefont{C.}~\bibnamefont{Colliex}},
  \bibinfo{author}{\bibfnamefont{H.}~\bibnamefont{Kohl}},
  \bibinfo{author}{\bibfnamefont{A.}~\bibnamefont{Nourtier}}, and
  \bibinfo{author}{\bibfnamefont{P.}~\bibnamefont{Trebbia}},
  \bibinfo{year}{1986}, \bibinfo{journal}{Ultramicroscopy}
  \textbf{\bibinfo{volume}{20}}, \bibinfo{pages}{99}.

\bibitem[{\citenamefont{Aeschlimann}
  \emph{et~al.}(2007)\citenamefont{Aeschlimann, Bauer, Bayer, Brixner,
  {Garc\'{\i}a de Abajo}, Pfeiffer, Rohmer, Spindler, and Steeb}}]{paper120}
\bibinfo{author}{\bibnamefont{Aeschlimann}, \bibfnamefont{M.}},
  \bibinfo{author}{\bibfnamefont{M.}~\bibnamefont{Bauer}},
  \bibinfo{author}{\bibfnamefont{D.}~\bibnamefont{Bayer}},
  \bibinfo{author}{\bibfnamefont{T.}~\bibnamefont{Brixner}},
  \bibinfo{author}{\bibfnamefont{F.~J.} \bibnamefont{{Garc\'{\i}a de Abajo}}},
  \bibinfo{author}{\bibfnamefont{W.}~\bibnamefont{Pfeiffer}},
  \bibinfo{author}{\bibfnamefont{M.}~\bibnamefont{Rohmer}},
  \bibinfo{author}{\bibfnamefont{C.}~\bibnamefont{Spindler}}, and
  \bibinfo{author}{\bibfnamefont{F.}~\bibnamefont{Steeb}},
  \bibinfo{year}{2007}, \bibinfo{journal}{Nature}
  \textbf{\bibinfo{volume}{446}}, \bibinfo{pages}{301}.

\bibitem[{\citenamefont{Ahn}(2004)}]{A04}
\bibinfo{editor}{\bibnamefont{Ahn}, \bibfnamefont{C.~C.}} (ed.),
  \bibinfo{year}{2004}, \emph{\bibinfo{title}{Transmission Electron Energy Loss
  Spectrometry in Materials Science and the EELS Atlas}}
  (\bibinfo{publisher}{Wiley-VCH}, \bibinfo{address}{Weinheim}).

\bibitem[{\citenamefont{Aizpurua} \emph{et~al.}(2003)\citenamefont{Aizpurua,
  Hanarp, Sutherland, K\"{a}ll, Bryant, and {Garc\'{\i}a de Abajo}}}]{paper072}
\bibinfo{author}{\bibnamefont{Aizpurua}, \bibfnamefont{J.}},
  \bibinfo{author}{\bibfnamefont{P.}~\bibnamefont{Hanarp}},
  \bibinfo{author}{\bibfnamefont{D.~S.} \bibnamefont{Sutherland}},
  \bibinfo{author}{\bibfnamefont{M.}~\bibnamefont{K\"{a}ll}},
  \bibinfo{author}{\bibfnamefont{G.~W.} \bibnamefont{Bryant}}, and
  \bibinfo{author}{\bibfnamefont{F.~J.} \bibnamefont{{Garc\'{\i}a de Abajo}}},
  \bibinfo{year}{2003}, \bibinfo{journal}{Phys.\ Rev.\ Lett.}
  \textbf{\bibinfo{volume}{90}}, \bibinfo{pages}{057401}.

\bibitem[{\citenamefont{Aizpurua} \emph{et~al.}(1999)\citenamefont{Aizpurua,
  Howie, and {Garc\'{\i}a de Abajo}}}]{paper048}
\bibinfo{author}{\bibnamefont{Aizpurua}, \bibfnamefont{J.}},
  \bibinfo{author}{\bibfnamefont{A.}~\bibnamefont{Howie}}, and
  \bibinfo{author}{\bibfnamefont{F.~J.} \bibnamefont{{Garc\'{\i}a de Abajo}}},
  \bibinfo{year}{1999}, \bibinfo{journal}{Phys.\ Rev.\ B}
  \textbf{\bibinfo{volume}{60}}, \bibinfo{pages}{11149}.

\bibitem[{\citenamefont{Aizpurua and Rivacoba}(2008)}]{AR08}
\bibinfo{author}{\bibnamefont{Aizpurua}, \bibfnamefont{J.}}, and
  \bibinfo{author}{\bibfnamefont{A.}~\bibnamefont{Rivacoba}},
  \bibinfo{year}{2008}, \bibinfo{journal}{Phys.\ Rev.\ B}
  \textbf{\bibinfo{volume}{78}}, \bibinfo{pages}{035404}.

\bibitem[{\citenamefont{Aizpurua} \emph{et~al.}(1996)\citenamefont{Aizpurua,
  Rivacoba, and Apell}}]{ARA96}
\bibinfo{author}{\bibnamefont{Aizpurua}, \bibfnamefont{J.}},
  \bibinfo{author}{\bibfnamefont{A.}~\bibnamefont{Rivacoba}}, and
  \bibinfo{author}{\bibfnamefont{S.~P.} \bibnamefont{Apell}},
  \bibinfo{year}{1996}, \bibinfo{journal}{Phys.\ Rev.\ B}
  \textbf{\bibinfo{volume}{54}}, \bibinfo{pages}{2901}.

\bibitem[{\citenamefont{Akhmediev and Karlsson}(1995)}]{AK95}
\bibinfo{author}{\bibnamefont{Akhmediev}, \bibfnamefont{N.}}, and
  \bibinfo{author}{\bibfnamefont{M.}~\bibnamefont{Karlsson}},
  \bibinfo{year}{1995}, \bibinfo{journal}{Phys.\ Rev.\ A}
  \textbf{\bibinfo{volume}{51}}, \bibinfo{pages}{2602}.

\bibitem[{\citenamefont{Allen} \emph{et~al.}(2003)\citenamefont{Allen, Findlay,
  Lupini, Oxley, and Pennycook}}]{AFL03}
\bibinfo{author}{\bibnamefont{Allen}, \bibfnamefont{L.~J.}},
  \bibinfo{author}{\bibfnamefont{S.~D.} \bibnamefont{Findlay}},
  \bibinfo{author}{\bibfnamefont{A.~R.} \bibnamefont{Lupini}},
  \bibinfo{author}{\bibfnamefont{M.~P.} \bibnamefont{Oxley}}, and
  \bibinfo{author}{\bibfnamefont{S.~J.} \bibnamefont{Pennycook}},
  \bibinfo{year}{2003}, \bibinfo{journal}{Phys.\ Rev.\ Lett.}
  \textbf{\bibinfo{volume}{91}}, \bibinfo{pages}{105503}.

\bibitem[{\citenamefont{Andrew and Barnes}(2001)}]{AB01}
\bibinfo{author}{\bibnamefont{Andrew}, \bibfnamefont{P.}}, and
  \bibinfo{author}{\bibfnamefont{W.~L.} \bibnamefont{Barnes}},
  \bibinfo{year}{2001}, \bibinfo{journal}{Phys.\ Rev.\ B}
  \textbf{\bibinfo{volume}{64}}, \bibinfo{pages}{125405}.

\bibitem[{\citenamefont{Arakawa} \emph{et~al.}(1964)\citenamefont{Arakawa,
  Herickhoff, and Birkhoff}}]{AHB1964}
\bibinfo{author}{\bibnamefont{Arakawa}, \bibfnamefont{E.~T.}},
  \bibinfo{author}{\bibfnamefont{R.~J.} \bibnamefont{Herickhoff}}, and
  \bibinfo{author}{\bibfnamefont{R.~D.} \bibnamefont{Birkhoff}},
  \bibinfo{year}{1964}, \bibinfo{journal}{Phys.\ Rev.\ Lett.}
  \textbf{\bibinfo{volume}{12}}, \bibinfo{pages}{319}.

\bibitem[{\citenamefont{Arenal} \emph{et~al.}(2005)\citenamefont{Arenal, {O.
  St\'{e}phan}, Kociak, Taverna, Loiseau, and Colliex}}]{ASK05}
\bibinfo{author}{\bibnamefont{Arenal}, \bibfnamefont{R.}},
  \bibinfo{author}{\bibnamefont{{O. St\'{e}phan}}},
  \bibinfo{author}{\bibfnamefont{M.}~\bibnamefont{Kociak}},
  \bibinfo{author}{\bibfnamefont{D.}~\bibnamefont{Taverna}},
  \bibinfo{author}{\bibfnamefont{A.}~\bibnamefont{Loiseau}}, and
  \bibinfo{author}{\bibfnamefont{C.}~\bibnamefont{Colliex}},
  \bibinfo{year}{2005}, \bibinfo{journal}{Phys.\ Rev.\ Lett.}
  \textbf{\bibinfo{volume}{95}}, \bibinfo{pages}{127601}.

\bibitem[{\citenamefont{Ashcroft and Mermin}(1976)}]{AM1976}
\bibinfo{author}{\bibnamefont{Ashcroft}, \bibfnamefont{N.~W.}}, and
  \bibinfo{author}{\bibfnamefont{N.~D.} \bibnamefont{Mermin}},
  \bibinfo{year}{1976}, \emph{\bibinfo{title}{Solid State Physics}}
  (\bibinfo{publisher}{Harcourt College Publishers}, \bibinfo{address}{New
  York}).

\bibitem[{\citenamefont{Ashkin and Dziedzic}(1976)}]{AD1976}
\bibinfo{author}{\bibnamefont{Ashkin}, \bibfnamefont{A.}}, and
  \bibinfo{author}{\bibfnamefont{J.~M.} \bibnamefont{Dziedzic}},
  \bibinfo{year}{1976}, \bibinfo{journal}{Appl.\ Phys.\ Lett.}
  \textbf{\bibinfo{volume}{28}}, \bibinfo{pages}{333}.

\bibitem[{\citenamefont{Ashley and Emerson}(1974)}]{AE1974}
\bibinfo{author}{\bibnamefont{Ashley}, \bibfnamefont{J.~C.}}, and
  \bibinfo{author}{\bibfnamefont{L.~C.} \bibnamefont{Emerson}},
  \bibinfo{year}{1974}, \bibinfo{journal}{Surf.\ Sci.}
  \textbf{\bibinfo{volume}{41}}, \bibinfo{pages}{615}.

\bibitem[{\citenamefont{Ashley and Ferrell}(1976)}]{AF1976}
\bibinfo{author}{\bibnamefont{Ashley}, \bibfnamefont{J.~C.}}, and
  \bibinfo{author}{\bibfnamefont{T.~L.} \bibnamefont{Ferrell}},
  \bibinfo{year}{1976}, \bibinfo{journal}{Phys.\ Rev.\ B}
  \textbf{\bibinfo{volume}{14}}, \bibinfo{pages}{3277}.

\bibitem[{\citenamefont{Atay} \emph{et~al.}(2004)\citenamefont{Atay, Song, and
  Nurmikko}}]{ASN04}
\bibinfo{author}{\bibnamefont{Atay}, \bibfnamefont{T.}},
  \bibinfo{author}{\bibfnamefont{J.~H.} \bibnamefont{Song}}, and
  \bibinfo{author}{\bibfnamefont{A.~V.} \bibnamefont{Nurmikko}},
  \bibinfo{year}{2004}, \bibinfo{journal}{Nano\ Lett.}
  \textbf{\bibinfo{volume}{4}}, \bibinfo{pages}{1627}.

\bibitem[{\citenamefont{Averkov and Yakovenko}(2005)}]{AY05}
\bibinfo{author}{\bibnamefont{Averkov}, \bibfnamefont{Y.~O.}}, and
  \bibinfo{author}{\bibfnamefont{V.~M.} \bibnamefont{Yakovenko}},
  \bibinfo{year}{2005}, \bibinfo{journal}{Phys.\ Rev.\ B}
  \textbf{\bibinfo{volume}{72}}, \bibinfo{pages}{205110}.

\bibitem[{\citenamefont{Bachheimer}(1972)}]{B1972}
\bibinfo{author}{\bibnamefont{Bachheimer}, \bibfnamefont{J.~P.}},
  \bibinfo{year}{1972}, \bibinfo{journal}{Phys.\ Rev.\ B}
  \textbf{\bibinfo{volume}{6}}, \bibinfo{pages}{2985}.

\bibitem[{\citenamefont{Bakunov} \emph{et~al.}(2005)\citenamefont{Bakunov,
  Maslov, and Bodrov}}]{BMB05}
\bibinfo{author}{\bibnamefont{Bakunov}, \bibfnamefont{M.~I.}},
  \bibinfo{author}{\bibfnamefont{A.~V.} \bibnamefont{Maslov}}, and
  \bibinfo{author}{\bibfnamefont{S.~B.} \bibnamefont{Bodrov}},
  \bibinfo{year}{2005}, \bibinfo{journal}{Phys.\ Rev.\ B}
  \textbf{\bibinfo{volume}{72}}, \bibinfo{pages}{195336}.

\bibitem[{\citenamefont{Balberg and Pankove}(1971)}]{BP1971}
\bibinfo{author}{\bibnamefont{Balberg}, \bibfnamefont{I.}}, and
  \bibinfo{author}{\bibfnamefont{J.~I.} \bibnamefont{Pankove}},
  \bibinfo{year}{1971}, \bibinfo{journal}{Phys.\ Rev.\ Lett.}
  \textbf{\bibinfo{volume}{27}}, \bibinfo{pages}{1371}.

\bibitem[{\citenamefont{Barber\'an and Bausells}(1985)}]{BB1985}
\bibinfo{author}{\bibnamefont{Barber\'an}, \bibfnamefont{N.}}, and
  \bibinfo{author}{\bibfnamefont{J.}~\bibnamefont{Bausells}},
  \bibinfo{year}{1985}, \bibinfo{journal}{Phys.\ Rev.\ B}
  \textbf{\bibinfo{volume}{31}}, \bibinfo{pages}{6354}.

\bibitem[{\citenamefont{Barnes} \emph{et~al.}(2003)\citenamefont{Barnes,
  Dereux, and Ebbesen}}]{BDE03}
\bibinfo{author}{\bibnamefont{Barnes}, \bibfnamefont{W.~L.}},
  \bibinfo{author}{\bibfnamefont{A.}~\bibnamefont{Dereux}}, and
  \bibinfo{author}{\bibfnamefont{T.~W.} \bibnamefont{Ebbesen}},
  \bibinfo{year}{2003}, \bibinfo{journal}{Nature}
  \textbf{\bibinfo{volume}{424}}, \bibinfo{pages}{824}.

\bibitem[{\citenamefont{Barnett and Loudon}(1996)}]{BL96}
\bibinfo{author}{\bibnamefont{Barnett}, \bibfnamefont{S.~M.}}, and
  \bibinfo{author}{\bibfnamefont{R.}~\bibnamefont{Loudon}},
  \bibinfo{year}{1996}, \bibinfo{journal}{Phys.\ Rev.\ Lett.}
  \textbf{\bibinfo{volume}{77}}, \bibinfo{pages}{2444}.

\bibitem[{\citenamefont{Barrera and Fuchs}(1995)}]{BF95}
\bibinfo{author}{\bibnamefont{Barrera}, \bibfnamefont{R.~G.}}, and
  \bibinfo{author}{\bibfnamefont{R.}~\bibnamefont{Fuchs}},
  \bibinfo{year}{1995}, \bibinfo{journal}{Phys.\ Rev.\ B}
  \textbf{\bibinfo{volume}{52}}, \bibinfo{pages}{3256}.

\bibitem[{\citenamefont{Barwick} \emph{et~al.}(2008)\citenamefont{Barwick,
  Park, Kwon, Baskin, and Zewail}}]{BPK08}
\bibinfo{author}{\bibnamefont{Barwick}, \bibfnamefont{B.}},
  \bibinfo{author}{\bibfnamefont{H.~S.} \bibnamefont{Park}},
  \bibinfo{author}{\bibfnamefont{O.~H.} \bibnamefont{Kwon}},
  \bibinfo{author}{\bibfnamefont{J.~S.} \bibnamefont{Baskin}}, and
  \bibinfo{author}{\bibfnamefont{A.~H.} \bibnamefont{Zewail}},
  \bibinfo{year}{2008}, \bibinfo{journal}{Science}
  \textbf{\bibinfo{volume}{322}}, \bibinfo{pages}{1227}.

\bibitem[{\citenamefont{Bashevoy} \emph{et~al.}(2006)\citenamefont{Bashevoy,
  Jonsson, Krasavin, Zheludev, Chen, and Stockman}}]{BJK06}
\bibinfo{author}{\bibnamefont{Bashevoy}, \bibfnamefont{M.~V.}},
  \bibinfo{author}{\bibfnamefont{F.}~\bibnamefont{Jonsson}},
  \bibinfo{author}{\bibfnamefont{A.~V.} \bibnamefont{Krasavin}},
  \bibinfo{author}{\bibfnamefont{N.~I.} \bibnamefont{Zheludev}},
  \bibinfo{author}{\bibfnamefont{Y.}~\bibnamefont{Chen}}, and
  \bibinfo{author}{\bibfnamefont{M.~I.} \bibnamefont{Stockman}},
  \bibinfo{year}{2006}, \bibinfo{journal}{Nano\ Lett.}
  \textbf{\bibinfo{volume}{6}}, \bibinfo{pages}{1113}.

\bibitem[{\citenamefont{Bashevoy} \emph{et~al.}(2007)\citenamefont{Bashevoy,
  Jonsson, MacDonald, Chen, and Zheludev}}]{BJM07}
\bibinfo{author}{\bibnamefont{Bashevoy}, \bibfnamefont{M.~V.}},
  \bibinfo{author}{\bibfnamefont{F.}~\bibnamefont{Jonsson}},
  \bibinfo{author}{\bibfnamefont{K.~F.} \bibnamefont{MacDonald}},
  \bibinfo{author}{\bibfnamefont{Y.}~\bibnamefont{Chen}}, and
  \bibinfo{author}{\bibfnamefont{N.~I.} \bibnamefont{Zheludev}},
  \bibinfo{year}{2007}, \bibinfo{journal}{Opt.\ Express}
  \textbf{\bibinfo{volume}{15}}, \bibinfo{pages}{11313}.

\bibitem[{\citenamefont{Batson}(1980)}]{B1980}
\bibinfo{author}{\bibnamefont{Batson}, \bibfnamefont{P.~E.}},
  \bibinfo{year}{1980}, \bibinfo{journal}{Solid\ State\ Commun.}
  \textbf{\bibinfo{volume}{34}}, \bibinfo{pages}{477}.

\bibitem[{\citenamefont{Batson}(1982{\natexlab{a}})}]{B1982}
\bibinfo{author}{\bibnamefont{Batson}, \bibfnamefont{P.~E.}},
  \bibinfo{year}{1982}{\natexlab{a}}, \bibinfo{journal}{Ultramicroscopy}
  \textbf{\bibinfo{volume}{9}}, \bibinfo{pages}{277}.

\bibitem[{\citenamefont{Batson}(1982{\natexlab{b}})}]{B1982_2}
\bibinfo{author}{\bibnamefont{Batson}, \bibfnamefont{P.~E.}},
  \bibinfo{year}{1982}{\natexlab{b}}, \bibinfo{journal}{Phys.\ Rev.\ Lett.}
  \textbf{\bibinfo{volume}{49}}, \bibinfo{pages}{936}.

\bibitem[{\citenamefont{Batson}(1985)}]{B1985}
\bibinfo{author}{\bibnamefont{Batson}, \bibfnamefont{P.~E.}},
  \bibinfo{year}{1985}, \bibinfo{journal}{Surf.\ Sci.}
  \textbf{\bibinfo{volume}{156}}, \bibinfo{pages}{720}.

\bibitem[{\citenamefont{Batson}(1993)}]{B93}
\bibinfo{author}{\bibnamefont{Batson}, \bibfnamefont{P.~E.}},
  \bibinfo{year}{1993}, \bibinfo{journal}{Nature}
  \textbf{\bibinfo{volume}{366}}, \bibinfo{pages}{727}.

\bibitem[{\citenamefont{Batson}(2008)}]{B08}
\bibinfo{author}{\bibnamefont{Batson}, \bibfnamefont{P.~E.}},
  \bibinfo{year}{2008}, \bibinfo{journal}{Microsc.\ Microanal.}
  \textbf{\bibinfo{volume}{14}}, \bibinfo{pages}{89}.

\bibitem[{\citenamefont{Batson} \emph{et~al.}(1976)\citenamefont{Batson, Chen,
  and Silcox}}]{BCS1976}
\bibinfo{author}{\bibnamefont{Batson}, \bibfnamefont{P.~E.}},
  \bibinfo{author}{\bibfnamefont{C.~H.} \bibnamefont{Chen}}, and
  \bibinfo{author}{\bibfnamefont{J.}~\bibnamefont{Silcox}},
  \bibinfo{year}{1976}, \bibinfo{journal}{Phys.\ Rev.\ Lett.}
  \textbf{\bibinfo{volume}{37}}, \bibinfo{pages}{937}.

\bibitem[{\citenamefont{Batson} \emph{et~al.}(2002)\citenamefont{Batson,
  Dellby, and Krivanek}}]{BDK02}
\bibinfo{author}{\bibnamefont{Batson}, \bibfnamefont{P.~E.}},
  \bibinfo{author}{\bibfnamefont{N.}~\bibnamefont{Dellby}}, and
  \bibinfo{author}{\bibfnamefont{O.~L.} \bibnamefont{Krivanek}},
  \bibinfo{year}{2002}, \bibinfo{journal}{Nature}
  \textbf{\bibinfo{volume}{418}}, \bibinfo{pages}{617}.

\bibitem[{\citenamefont{Batson and Silcox}(1983)}]{BS1983}
\bibinfo{author}{\bibnamefont{Batson}, \bibfnamefont{P.~E.}}, and
  \bibinfo{author}{\bibfnamefont{J.}~\bibnamefont{Silcox}},
  \bibinfo{year}{1983}, \bibinfo{journal}{Phys.\ Rev.\ B}
  \textbf{\bibinfo{volume}{27}}, \bibinfo{pages}{5224}.

\bibitem[{\citenamefont{Bauer}(1994)}]{B94}
\bibinfo{author}{\bibnamefont{Bauer}, \bibfnamefont{E.}}, \bibinfo{year}{1994},
  \bibinfo{journal}{Rep.\ Prog.\ Phys.} \textbf{\bibinfo{volume}{57}},
  \bibinfo{pages}{895}.

\bibitem[{\citenamefont{Berini} \emph{et~al.}(2007)\citenamefont{Berini,
  Mattiussi, Lahoud, and Charbonneau}}]{BML07}
\bibinfo{author}{\bibnamefont{Berini}, \bibfnamefont{P.}},
  \bibinfo{author}{\bibfnamefont{G.}~\bibnamefont{Mattiussi}},
  \bibinfo{author}{\bibfnamefont{N.}~\bibnamefont{Lahoud}}, and
  \bibinfo{author}{\bibfnamefont{R.}~\bibnamefont{Charbonneau}},
  \bibinfo{year}{2007}, \bibinfo{journal}{Appl.\ Phys.\ Lett.}
  \textbf{\bibinfo{volume}{90}}, \bibinfo{pages}{061108}.

\bibitem[{\citenamefont{Bertoni and Verbeeck}(2008)}]{BV08}
\bibinfo{author}{\bibnamefont{Bertoni}, \bibfnamefont{G.}}, and
  \bibinfo{author}{\bibfnamefont{J.}~\bibnamefont{Verbeeck}},
  \bibinfo{year}{2008}, \bibinfo{journal}{Ultramicroscopy}
  \textbf{\bibinfo{volume}{108}}, \bibinfo{pages}{782}.

\bibitem[{\citenamefont{Bethe}(1930)}]{B1930}
\bibinfo{author}{\bibnamefont{Bethe}, \bibfnamefont{H.~A.}},
  \bibinfo{year}{1930}, \bibinfo{journal}{Ann.\ Phys.\ (Leipzig)}
  \textbf{\bibinfo{volume}{5}}, \bibinfo{pages}{325}.

\bibitem[{\citenamefont{Betzig} \emph{et~al.}(1992)\citenamefont{Betzig, Finn,
  and Weiner}}]{BFW92}
\bibinfo{author}{\bibnamefont{Betzig}, \bibfnamefont{E.}},
  \bibinfo{author}{\bibfnamefont{P.~L.} \bibnamefont{Finn}}, and
  \bibinfo{author}{\bibfnamefont{J.~S.} \bibnamefont{Weiner}},
  \bibinfo{year}{1992}, \bibinfo{journal}{Appl.\ Phys.\ Lett.}
  \textbf{\bibinfo{volume}{60}}, \bibinfo{pages}{2484}.

\bibitem[{\citenamefont{Blackmore} \emph{et~al.}(2008)\citenamefont{Blackmore,
  Doucas, Perry, and Kimmitt}}]{BDP08}
\bibinfo{author}{\bibnamefont{Blackmore}, \bibfnamefont{V.}},
  \bibinfo{author}{\bibfnamefont{G.}~\bibnamefont{Doucas}},
  \bibinfo{author}{\bibfnamefont{C.}~\bibnamefont{Perry}}, and
  \bibinfo{author}{\bibfnamefont{M.~F.} \bibnamefont{Kimmitt}},
  \bibinfo{year}{2008}, \bibinfo{journal}{Nucl.\ Instrum.\ Methods\ Phys.\
  Res.\ B} \textbf{\bibinfo{volume}{266}}, \bibinfo{pages}{3803}.

\bibitem[{\citenamefont{Blanco and {Garc\'{\i}a de
  Abajo}}(2004{\natexlab{a}})}]{paper087}
\bibinfo{author}{\bibnamefont{Blanco}, \bibfnamefont{L.~A.}}, and
  \bibinfo{author}{\bibfnamefont{F.~J.} \bibnamefont{{Garc\'{\i}a de Abajo}}},
  \bibinfo{year}{2004}{\natexlab{a}}, \bibinfo{journal}{Opt.\ Lett.}
  \textbf{\bibinfo{volume}{29}}, \bibinfo{pages}{1494}.

\bibitem[{\citenamefont{Blanco and {Garc\'{\i}a de
  Abajo}}(2004{\natexlab{b}})}]{paper085}
\bibinfo{author}{\bibnamefont{Blanco}, \bibfnamefont{L.~A.}}, and
  \bibinfo{author}{\bibfnamefont{F.~J.} \bibnamefont{{Garc\'{\i}a de Abajo}}},
  \bibinfo{year}{2004}{\natexlab{b}}, \bibinfo{journal}{Phys.\ Rev.\ B}
  \textbf{\bibinfo{volume}{69}}, \bibinfo{pages}{205414}.

\bibitem[{\citenamefont{Bloch}(1933)}]{B1933}
\bibinfo{author}{\bibnamefont{Bloch}, \bibfnamefont{F.}}, \bibinfo{year}{1933},
  \bibinfo{journal}{Z.\ Phys.} \textbf{\bibinfo{volume}{81}},
  \bibinfo{pages}{363}.

\bibitem[{\citenamefont{Boersch}
  \emph{et~al.}(1966{\natexlab{a}})\citenamefont{Boersch, Geiger, Imbusch, and
  Niedrig}}]{BGI1966}
\bibinfo{author}{\bibnamefont{Boersch}, \bibfnamefont{H.}},
  \bibinfo{author}{\bibfnamefont{J.}~\bibnamefont{Geiger}},
  \bibinfo{author}{\bibfnamefont{A.}~\bibnamefont{Imbusch}}, and
  \bibinfo{author}{\bibfnamefont{N.}~\bibnamefont{Niedrig}},
  \bibinfo{year}{1966}{\natexlab{a}}, \bibinfo{journal}{Phys.\ Lett.}
  \textbf{\bibinfo{volume}{22}}, \bibinfo{pages}{146}.

\bibitem[{\citenamefont{Boersch}
  \emph{et~al.}(1966{\natexlab{b}})\citenamefont{Boersch, Geiger, and
  Stickel}}]{BGS1966}
\bibinfo{author}{\bibnamefont{Boersch}, \bibfnamefont{H.}},
  \bibinfo{author}{\bibfnamefont{J.}~\bibnamefont{Geiger}}, and
  \bibinfo{author}{\bibfnamefont{W.}~\bibnamefont{Stickel}},
  \bibinfo{year}{1966}{\natexlab{b}}, \bibinfo{journal}{Phys.\ Rev.\ Lett.}
  \textbf{\bibinfo{volume}{17}}, \bibinfo{pages}{379}.

\bibitem[{\citenamefont{Boltasseva and Shalaev}(2008)}]{BS08}
\bibinfo{author}{\bibnamefont{Boltasseva}, \bibfnamefont{A.}}, and
  \bibinfo{author}{\bibfnamefont{V.~M.} \bibnamefont{Shalaev}},
  \bibinfo{year}{2008}, \bibinfo{journal}{Metamaterials}
  \textbf{\bibinfo{volume}{2}}, \bibinfo{pages}{1}.

\bibitem[{\citenamefont{Bolton and Chen}(1995)}]{rmp-cr4}
\bibinfo{author}{\bibnamefont{Bolton}, \bibfnamefont{J.~P.~R.}}, and
  \bibinfo{author}{\bibfnamefont{M.}~\bibnamefont{Chen}}, \bibinfo{year}{1995},
  \bibinfo{journal}{J.\ Phys.\ Condens.\ Matter} \textbf{\bibinfo{volume}{7}},
  \bibinfo{pages}{3373}.

\bibitem[{\citenamefont{Borziak} \emph{et~al.}(1976)\citenamefont{Borziak,
  Konovalov, Kulyupin, and Pilipchak}}]{BKK1976}
\bibinfo{author}{\bibnamefont{Borziak}, \bibfnamefont{P.}},
  \bibinfo{author}{\bibfnamefont{I.}~\bibnamefont{Konovalov}},
  \bibinfo{author}{\bibfnamefont{Y.}~\bibnamefont{Kulyupin}}, and
  \bibinfo{author}{\bibfnamefont{K.}~\bibnamefont{Pilipchak}},
  \bibinfo{year}{1976}, \bibinfo{journal}{Thin Solid Films}
  \textbf{\bibinfo{volume}{35}}, \bibinfo{pages}{L9}.

\bibitem[{\citenamefont{Bosman} \emph{et~al.}(2007)\citenamefont{Bosman, Keast,
  Watanabe, Maaroof, and Cortie}}]{BKW07}
\bibinfo{author}{\bibnamefont{Bosman}, \bibfnamefont{M.}},
  \bibinfo{author}{\bibfnamefont{V.~J.} \bibnamefont{Keast}},
  \bibinfo{author}{\bibfnamefont{M.}~\bibnamefont{Watanabe}},
  \bibinfo{author}{\bibfnamefont{A.~I.} \bibnamefont{Maaroof}}, and
  \bibinfo{author}{\bibfnamefont{M.~B.} \bibnamefont{Cortie}},
  \bibinfo{year}{2007}, \bibinfo{journal}{Nanotechnology}
  \textbf{\bibinfo{volume}{18}}, \bibinfo{pages}{165505}.

\bibitem[{\citenamefont{Bozhevolnyi}
  \emph{et~al.}(2001)\citenamefont{Bozhevolnyi, Erland, Leosson, Skovgaard, and
  Hvam}}]{BEL01}
\bibinfo{author}{\bibnamefont{Bozhevolnyi}, \bibfnamefont{S.~I.}},
  \bibinfo{author}{\bibfnamefont{J.}~\bibnamefont{Erland}},
  \bibinfo{author}{\bibfnamefont{K.}~\bibnamefont{Leosson}},
  \bibinfo{author}{\bibfnamefont{P.~M.~W.} \bibnamefont{Skovgaard}}, and
  \bibinfo{author}{\bibfnamefont{J.~M.} \bibnamefont{Hvam}},
  \bibinfo{year}{2001}, \bibinfo{journal}{Phys.\ Rev.\ Lett.}
  \textbf{\bibinfo{volume}{86}}, \bibinfo{pages}{3008}.

\bibitem[{\citenamefont{Bozhevolnyi}
  \emph{et~al.}(2006)\citenamefont{Bozhevolnyi, Volkov, Devaux, Laluet, and
  Ebbesen}}]{BVD06}
\bibinfo{author}{\bibnamefont{Bozhevolnyi}, \bibfnamefont{S.~I.}},
  \bibinfo{author}{\bibfnamefont{V.~S.} \bibnamefont{Volkov}},
  \bibinfo{author}{\bibfnamefont{E.}~\bibnamefont{Devaux}},
  \bibinfo{author}{\bibfnamefont{J.~Y.} \bibnamefont{Laluet}}, and
  \bibinfo{author}{\bibfnamefont{T.~W.} \bibnamefont{Ebbesen}},
  \bibinfo{year}{2006}, \bibinfo{journal}{Nature}
  \textbf{\bibinfo{volume}{440}}, \bibinfo{pages}{508}.

\bibitem[{\citenamefont{Bradley} \emph{et~al.}(2005)\citenamefont{Bradley, Dai,
  Erni, Browning, Graham, Weber, Smith, Hutcheon, Ishii, Bajt, Floss,
  Stadermann} \emph{et~al.}}]{BDE05}
\bibinfo{author}{\bibnamefont{Bradley}, \bibfnamefont{J.}},
  \bibinfo{author}{\bibfnamefont{Z.~R.} \bibnamefont{Dai}},
  \bibinfo{author}{\bibfnamefont{R.}~\bibnamefont{Erni}},
  \bibinfo{author}{\bibfnamefont{N.}~\bibnamefont{Browning}},
  \bibinfo{author}{\bibfnamefont{G.}~\bibnamefont{Graham}},
  \bibinfo{author}{\bibfnamefont{P.}~\bibnamefont{Weber}},
  \bibinfo{author}{\bibfnamefont{J.}~\bibnamefont{Smith}},
  \bibinfo{author}{\bibfnamefont{I.}~\bibnamefont{Hutcheon}},
  \bibinfo{author}{\bibfnamefont{H.}~\bibnamefont{Ishii}},
  \bibinfo{author}{\bibfnamefont{S.}~\bibnamefont{Bajt}},
  \bibinfo{author}{\bibfnamefont{C.}~\bibnamefont{Floss}},
  \bibinfo{author}{\bibfnamefont{F.}~\bibnamefont{Stadermann}}, \emph{et~al.},
  \bibinfo{year}{2005}, \bibinfo{journal}{Science}
  \textbf{\bibinfo{volume}{307}}, \bibinfo{pages}{244}.

\bibitem[{\citenamefont{Brambring and Raether}(1965)}]{BR1965}
\bibinfo{author}{\bibnamefont{Brambring}, \bibfnamefont{J.}}, and
  \bibinfo{author}{\bibfnamefont{H.}~\bibnamefont{Raether}},
  \bibinfo{year}{1965}, \bibinfo{journal}{Phys.\ Rev.\ Lett.}
  \textbf{\bibinfo{volume}{15}}, \bibinfo{pages}{882}.

\bibitem[{\citenamefont{Brink} \emph{et~al.}(2003)\citenamefont{Brink, Barfels,
  Burgner, and Edwards}}]{BBB03}
\bibinfo{author}{\bibnamefont{Brink}, \bibfnamefont{H.~A.}},
  \bibinfo{author}{\bibfnamefont{M.~M.~G.} \bibnamefont{Barfels}},
  \bibinfo{author}{\bibfnamefont{R.~P.} \bibnamefont{Burgner}}, and
  \bibinfo{author}{\bibfnamefont{B.~N.} \bibnamefont{Edwards}},
  \bibinfo{year}{2003}, \bibinfo{journal}{Ultramicroscopy}
  \textbf{\bibinfo{volume}{96}}, \bibinfo{pages}{367–}.

\bibitem[{\citenamefont{Brown} \emph{et~al.}(1960)\citenamefont{Brown, Wessel,
  and Trounson}}]{BWT1960}
\bibinfo{author}{\bibnamefont{Brown}, \bibfnamefont{R.~W.}},
  \bibinfo{author}{\bibfnamefont{P.}~\bibnamefont{Wessel}}, and
  \bibinfo{author}{\bibfnamefont{E.~P.} \bibnamefont{Trounson}},
  \bibinfo{year}{1960}, \bibinfo{journal}{Phys.\ Rev.\ Lett.}
  \textbf{\bibinfo{volume}{5}}, \bibinfo{pages}{472}.

\bibitem[{\citenamefont{Brownell} \emph{et~al.}(1998)\citenamefont{Brownell,
  Walsh, and Doucas}}]{BWD98}
\bibinfo{author}{\bibnamefont{Brownell}, \bibfnamefont{J.~H.}},
  \bibinfo{author}{\bibfnamefont{J.}~\bibnamefont{Walsh}}, and
  \bibinfo{author}{\bibfnamefont{G.}~\bibnamefont{Doucas}},
  \bibinfo{year}{1998}, \bibinfo{journal}{Phys.\ Rev.\ E}
  \textbf{\bibinfo{volume}{57}}, \bibinfo{pages}{1075}.

\bibitem[{\citenamefont{Browning} \emph{et~al.}(1993)\citenamefont{Browning,
  Chisholm, and Pennycook}}]{BCP93}
\bibinfo{author}{\bibnamefont{Browning}, \bibfnamefont{N.~D.}},
  \bibinfo{author}{\bibfnamefont{M.~F.} \bibnamefont{Chisholm}}, and
  \bibinfo{author}{\bibfnamefont{S.~J.} \bibnamefont{Pennycook}},
  \bibinfo{year}{1993}, \bibinfo{journal}{Nature}
  \textbf{\bibinfo{volume}{366}}, \bibinfo{pages}{143}.

\bibitem[{\citenamefont{Brydson}(2001)}]{B06}
\bibinfo{author}{\bibnamefont{Brydson}, \bibfnamefont{R.}},
  \bibinfo{year}{2001}, \emph{\bibinfo{title}{Electron Energy Loss
  Spectroscopy}} (\bibinfo{publisher}{BIOS Scientific Publishers},
  \bibinfo{address}{Oxford}).

\bibitem[{\citenamefont{Burda} \emph{et~al.}(2005)\citenamefont{Burda, Chen,
  Narayanan, and {El-Sayed}}}]{BCN05}
\bibinfo{author}{\bibnamefont{Burda}, \bibfnamefont{C.}},
  \bibinfo{author}{\bibfnamefont{X.}~\bibnamefont{Chen}},
  \bibinfo{author}{\bibfnamefont{R.}~\bibnamefont{Narayanan}}, and
  \bibinfo{author}{\bibfnamefont{M.~A.} \bibnamefont{{El-Sayed}}},
  \bibinfo{year}{2005}, \bibinfo{journal}{Chem.\ Rev.}
  \textbf{\bibinfo{volume}{105}}, \bibinfo{pages}{1025}.

\bibitem[{\citenamefont{{Cadete Santos Aires}}
  \emph{et~al.}(1993)\citenamefont{{Cadete Santos Aires}, Howie, and
  Walsh}}]{CHW93}
\bibinfo{author}{\bibnamefont{{Cadete Santos Aires}}, \bibfnamefont{F.~J.}},
  \bibinfo{author}{\bibfnamefont{A.}~\bibnamefont{Howie}}, and
  \bibinfo{author}{\bibfnamefont{C.~A.} \bibnamefont{Walsh}},
  \bibinfo{year}{1993}, \bibinfo{journal}{J.\ Sol.\ State\ Chem.}
  \textbf{\bibinfo{volume}{106}}, \bibinfo{pages}{48}.

\bibitem[{\citenamefont{Cai} \emph{et~al.}(2009)\citenamefont{Cai, Sainidou,
  Xu, Polman, and {Garc\'{\i}a de Abajo}}}]{paper168}
\bibinfo{author}{\bibnamefont{Cai}, \bibfnamefont{W.}},
  \bibinfo{author}{\bibfnamefont{R.}~\bibnamefont{Sainidou}},
  \bibinfo{author}{\bibfnamefont{J.}~\bibnamefont{Xu}},
  \bibinfo{author}{\bibfnamefont{A.}~\bibnamefont{Polman}}, and
  \bibinfo{author}{\bibfnamefont{F.~J.} \bibnamefont{{Garc\'{\i}a de Abajo}}},
  \bibinfo{year}{2009}, \bibinfo{journal}{Nano\ Lett.} , \bibinfo{pages}{DOI:
  10.1021/nl803825n}.

\bibitem[{\citenamefont{Calliari} \emph{et~al.}(2008)\citenamefont{Calliari,
  Fanchenko, and Filippi}}]{CFF08}
\bibinfo{author}{\bibnamefont{Calliari}, \bibfnamefont{L.}},
  \bibinfo{author}{\bibfnamefont{S.}~\bibnamefont{Fanchenko}}, and
  \bibinfo{author}{\bibfnamefont{M.}~\bibnamefont{Filippi}},
  \bibinfo{year}{2008}, \bibinfo{journal}{Surf.\ Interface\ Anal.}
  \textbf{\bibinfo{volume}{40}}, \bibinfo{pages}{814}.

\bibitem[{\citenamefont{Carusotto} \emph{et~al.}(2001)\citenamefont{Carusotto,
  Artoni, {G. C. La Rocca}, and Bassani}}]{CAR01}
\bibinfo{author}{\bibnamefont{Carusotto}, \bibfnamefont{I.}},
  \bibinfo{author}{\bibfnamefont{M.}~\bibnamefont{Artoni}},
  \bibinfo{author}{\bibnamefont{{G. C. La Rocca}}}, and
  \bibinfo{author}{\bibfnamefont{F.}~\bibnamefont{Bassani}},
  \bibinfo{year}{2001}, \bibinfo{journal}{Phys.\ Rev.\ Lett.}
  \textbf{\bibinfo{volume}{87}}, \bibinfo{pages}{064801}.

\bibitem[{\citenamefont{Cazaux}(2006)}]{C06}
\bibinfo{author}{\bibnamefont{Cazaux}, \bibfnamefont{J.}},
  \bibinfo{year}{2006}, \bibinfo{journal}{Nucl.\ Instrum.\ Methods\ Phys.\
  Res.\ B} \textbf{\bibinfo{volume}{244}}, \bibinfo{pages}{307}.

\bibitem[{\citenamefont{Chao} \emph{et~al.}(2005)\citenamefont{Chao, Harteneck,
  Liddle, Anderson, and Attwood}}]{CHL05}
\bibinfo{author}{\bibnamefont{Chao}, \bibfnamefont{W.}},
  \bibinfo{author}{\bibfnamefont{B.~D.} \bibnamefont{Harteneck}},
  \bibinfo{author}{\bibfnamefont{J.~A.} \bibnamefont{Liddle}},
  \bibinfo{author}{\bibfnamefont{E.~H.} \bibnamefont{Anderson}}, and
  \bibinfo{author}{\bibfnamefont{D.~T.} \bibnamefont{Attwood}},
  \bibinfo{year}{2005}, \bibinfo{journal}{Nature}
  \textbf{\bibinfo{volume}{435}}, \bibinfo{pages}{1210}.

\bibitem[{\citenamefont{Chen} \emph{et~al.}(1980)\citenamefont{Chen, Meixner,
  and Kincaid}}]{CMK1980}
\bibinfo{author}{\bibnamefont{Chen}, \bibfnamefont{C.~H.}},
  \bibinfo{author}{\bibfnamefont{A.~E.} \bibnamefont{Meixner}}, and
  \bibinfo{author}{\bibfnamefont{B.~M.} \bibnamefont{Kincaid}},
  \bibinfo{year}{1980}, \bibinfo{journal}{Phys.\ Rev.\ Lett.}
  \textbf{\bibinfo{volume}{44}}, \bibinfo{pages}{951}.

\bibitem[{\citenamefont{Chen and Silcox}(1975{\natexlab{a}})}]{CS1975}
\bibinfo{author}{\bibnamefont{Chen}, \bibfnamefont{C.~H.}}, and
  \bibinfo{author}{\bibfnamefont{J.}~\bibnamefont{Silcox}},
  \bibinfo{year}{1975}{\natexlab{a}}, \bibinfo{journal}{Phys.\ Rev.\ Lett.}
  \textbf{\bibinfo{volume}{35}}, \bibinfo{pages}{390}.

\bibitem[{\citenamefont{Chen and Silcox}(1975{\natexlab{b}})}]{CS1975_2}
\bibinfo{author}{\bibnamefont{Chen}, \bibfnamefont{C.~H.}}, and
  \bibinfo{author}{\bibfnamefont{J.}~\bibnamefont{Silcox}},
  \bibinfo{year}{1975}{\natexlab{b}}, \bibinfo{journal}{Solid\ State\ Commun.}
  \textbf{\bibinfo{volume}{17}}, \bibinfo{pages}{273}.

\bibitem[{\citenamefont{Chen and Silcox}(1979)}]{CS1979}
\bibinfo{author}{\bibnamefont{Chen}, \bibfnamefont{C.~H.}}, and
  \bibinfo{author}{\bibfnamefont{J.}~\bibnamefont{Silcox}},
  \bibinfo{year}{1979}, \bibinfo{journal}{Phys.\ Rev.\ B}
  \textbf{\bibinfo{volume}{20}}, \bibinfo{pages}{3605}.

\bibitem[{\citenamefont{Chen} \emph{et~al.}(1976)\citenamefont{Chen, Silcox,
  Garito, Heeger, and MacDiarmid}}]{CSG1976}
\bibinfo{author}{\bibnamefont{Chen}, \bibfnamefont{C.~H.}},
  \bibinfo{author}{\bibfnamefont{J.}~\bibnamefont{Silcox}},
  \bibinfo{author}{\bibfnamefont{A.~F.} \bibnamefont{Garito}},
  \bibinfo{author}{\bibfnamefont{A.~J.} \bibnamefont{Heeger}}, and
  \bibinfo{author}{\bibfnamefont{A.~G.} \bibnamefont{MacDiarmid}},
  \bibinfo{year}{1976}, \bibinfo{journal}{Phys.\ Rev.\ Lett.}
  \textbf{\bibinfo{volume}{36}}, \bibinfo{pages}{525}.

\bibitem[{\citenamefont{Chen} \emph{et~al.}(1975)\citenamefont{Chen, Silcox,
  and Vincent}}]{CSV1975}
\bibinfo{author}{\bibnamefont{Chen}, \bibfnamefont{C.~H.}},
  \bibinfo{author}{\bibfnamefont{J.}~\bibnamefont{Silcox}}, and
  \bibinfo{author}{\bibfnamefont{R.}~\bibnamefont{Vincent}},
  \bibinfo{year}{1975}, \bibinfo{journal}{Phys.\ Rev.\ B}
  \textbf{\bibinfo{volume}{12}}, \bibinfo{pages}{64}.

\bibitem[{\citenamefont{Chen} \emph{et~al.}(2006)\citenamefont{Chen, Chen,
  Shen, Ganguly, Chen, Wu, Wen, and Pong}}]{CCS06}
\bibinfo{author}{\bibnamefont{Chen}, \bibfnamefont{C.~W.}},
  \bibinfo{author}{\bibfnamefont{K.~H.} \bibnamefont{Chen}},
  \bibinfo{author}{\bibfnamefont{C.~H.} \bibnamefont{Shen}},
  \bibinfo{author}{\bibfnamefont{A.}~\bibnamefont{Ganguly}},
  \bibinfo{author}{\bibfnamefont{L.~C.} \bibnamefont{Chen}},
  \bibinfo{author}{\bibfnamefont{J.~J.} \bibnamefont{Wu}},
  \bibinfo{author}{\bibfnamefont{H.~I.} \bibnamefont{Wen}}, and
  \bibinfo{author}{\bibfnamefont{W.~F.} \bibnamefont{Pong}},
  \bibinfo{year}{2006}, \bibinfo{journal}{Appl.\ Phys.\ Lett.}
  \textbf{\bibinfo{volume}{88}}, \bibinfo{pages}{241905}.

\bibitem[{\citenamefont{Cherenkov}(1934)}]{C1934_2}
\bibinfo{author}{\bibnamefont{Cherenkov}, \bibfnamefont{P.~A.}},
  \bibinfo{year}{1934}, \bibinfo{journal}{Dokl.\ Akad.\ Nauk\ SSSR}
  \textbf{\bibinfo{volume}{2}}, \bibinfo{pages}{451}.

\bibitem[{\citenamefont{Chopra} \emph{et~al.}(1995)\citenamefont{Chopra,
  Luyken, Cherrey, Crespi, Cohen, Louie, and Zettl}}]{CLC95}
\bibinfo{author}{\bibnamefont{Chopra}, \bibfnamefont{N.~G.}},
  \bibinfo{author}{\bibfnamefont{R.~J.} \bibnamefont{Luyken}},
  \bibinfo{author}{\bibfnamefont{K.}~\bibnamefont{Cherrey}},
  \bibinfo{author}{\bibfnamefont{V.~H.} \bibnamefont{Crespi}},
  \bibinfo{author}{\bibfnamefont{M.~L.} \bibnamefont{Cohen}},
  \bibinfo{author}{\bibfnamefont{S.~G.} \bibnamefont{Louie}}, and
  \bibinfo{author}{\bibfnamefont{A.}~\bibnamefont{Zettl}},
  \bibinfo{year}{1995}, \bibinfo{journal}{Science}
  \textbf{\bibinfo{volume}{269}}, \bibinfo{pages}{966}.

\bibitem[{\citenamefont{Chu} \emph{et~al.}(2008)\citenamefont{Chu, Chen,
  {Garc\'{\i}a de Abajo}, Deng, and Mou}}]{paper151}
\bibinfo{author}{\bibnamefont{Chu}, \bibfnamefont{M.~W.}},
  \bibinfo{author}{\bibfnamefont{C.~H.} \bibnamefont{Chen}},
  \bibinfo{author}{\bibfnamefont{F.~J.} \bibnamefont{{Garc\'{\i}a de Abajo}}},
  \bibinfo{author}{\bibfnamefont{J.~P.} \bibnamefont{Deng}}, and
  \bibinfo{author}{\bibfnamefont{C.~Y.} \bibnamefont{Mou}},
  \bibinfo{year}{2008}, \bibinfo{journal}{Phys.\ Rev.\ B}
  \textbf{\bibinfo{volume}{77}}, \bibinfo{pages}{245402}.

\bibitem[{\citenamefont{Chu} \emph{et~al.}(2009)\citenamefont{Chu,
  Myroshnychenko, Chen, Deng, Mou, and {Garc\'{\i}a de Abajo}}}]{paper164}
\bibinfo{author}{\bibnamefont{Chu}, \bibfnamefont{M.~W.}},
  \bibinfo{author}{\bibfnamefont{V.}~\bibnamefont{Myroshnychenko}},
  \bibinfo{author}{\bibfnamefont{C.~H.} \bibnamefont{Chen}},
  \bibinfo{author}{\bibfnamefont{J.~P.} \bibnamefont{Deng}},
  \bibinfo{author}{\bibfnamefont{C.~Y.} \bibnamefont{Mou}}, and
  \bibinfo{author}{\bibfnamefont{F.~J.} \bibnamefont{{Garc\'{\i}a de Abajo}}},
  \bibinfo{year}{2009}, \bibinfo{journal}{Nano\ Lett.}
  \textbf{\bibinfo{volume}{9}}, \bibinfo{pages}{399}.

\bibitem[{\citenamefont{{\v{C}iljak}}
  \emph{et~al.}(2003)\citenamefont{{\v{C}iljak}, {Ru\v{z}i\v{c}ka}, Vodopianov,
  Ivanshin, Tyapkin, Tyapkin, Zinchenko, and Zrelov}}]{rmp-cr3}
\bibinfo{author}{\bibnamefont{{\v{C}iljak}}, \bibfnamefont{M.}},
  \bibinfo{author}{\bibfnamefont{J.}~\bibnamefont{{Ru\v{z}i\v{c}ka}}},
  \bibinfo{author}{\bibfnamefont{A.~S.} \bibnamefont{Vodopianov}},
  \bibinfo{author}{\bibfnamefont{Y.~I.} \bibnamefont{Ivanshin}},
  \bibinfo{author}{\bibfnamefont{A.~A.} \bibnamefont{Tyapkin}},
  \bibinfo{author}{\bibfnamefont{I.~A.} \bibnamefont{Tyapkin}},
  \bibinfo{author}{\bibfnamefont{A.~I.} \bibnamefont{Zinchenko}}, and
  \bibinfo{author}{\bibfnamefont{V.~P.} \bibnamefont{Zrelov}},
  \bibinfo{year}{2003}, \bibinfo{journal}{Nucl.\ Instrum.\ Methods\ Phys.\
  Res.\ A} \textbf{\bibinfo{volume}{498}}, \bibinfo{pages}{126–}.

\bibitem[{\citenamefont{Cinchetti} \emph{et~al.}(2005)\citenamefont{Cinchetti,
  Gloskovskii, Nepjiko, {Sch\"onhense}, Rochholz, and Kreiter}}]{CGN05}
\bibinfo{author}{\bibnamefont{Cinchetti}, \bibfnamefont{M.}},
  \bibinfo{author}{\bibfnamefont{A.}~\bibnamefont{Gloskovskii}},
  \bibinfo{author}{\bibfnamefont{S.~A.} \bibnamefont{Nepjiko}},
  \bibinfo{author}{\bibfnamefont{G.}~\bibnamefont{{Sch\"onhense}}},
  \bibinfo{author}{\bibfnamefont{H.}~\bibnamefont{Rochholz}}, and
  \bibinfo{author}{\bibfnamefont{M.}~\bibnamefont{Kreiter}},
  \bibinfo{year}{2005}, \bibinfo{journal}{Phys.\ Rev.\ Lett.}
  \textbf{\bibinfo{volume}{95}}, \bibinfo{pages}{047601}.

\bibitem[{\citenamefont{Cohen} \emph{et~al.}(2003)\citenamefont{Cohen,
  Lembrikov, Itskovsky, and Maniv}}]{CLI03}
\bibinfo{author}{\bibnamefont{Cohen}, \bibfnamefont{H.}},
  \bibinfo{author}{\bibfnamefont{B.~I.} \bibnamefont{Lembrikov}},
  \bibinfo{author}{\bibfnamefont{M.~A.} \bibnamefont{Itskovsky}}, and
  \bibinfo{author}{\bibfnamefont{T.}~\bibnamefont{Maniv}},
  \bibinfo{year}{2003}, \bibinfo{journal}{Nano\ Lett.}
  \textbf{\bibinfo{volume}{3}}, \bibinfo{pages}{203}.

\bibitem[{\citenamefont{Cohen} \emph{et~al.}(1998)\citenamefont{Cohen, Maniv,
  Tenne, Hacohen, {O. St\'{e}phan}, and Colliex}}]{CMT98}
\bibinfo{author}{\bibnamefont{Cohen}, \bibfnamefont{H.}},
  \bibinfo{author}{\bibfnamefont{T.}~\bibnamefont{Maniv}},
  \bibinfo{author}{\bibfnamefont{R.}~\bibnamefont{Tenne}},
  \bibinfo{author}{\bibfnamefont{Y.~R.} \bibnamefont{Hacohen}},
  \bibinfo{author}{\bibnamefont{{O. St\'{e}phan}}}, and
  \bibinfo{author}{\bibfnamefont{C.}~\bibnamefont{Colliex}},
  \bibinfo{year}{1998}, \bibinfo{journal}{Phys.\ Rev.\ Lett.}
  \textbf{\bibinfo{volume}{80}}, \bibinfo{pages}{782}.

\bibitem[{\citenamefont{Cohen} \emph{et~al.}(1999)\citenamefont{Cohen, Maniv,
  Tenne, Hacohen, {O. St\'{e}phan}, and Colliex}}]{CMT99}
\bibinfo{author}{\bibnamefont{Cohen}, \bibfnamefont{H.}},
  \bibinfo{author}{\bibfnamefont{T.}~\bibnamefont{Maniv}},
  \bibinfo{author}{\bibfnamefont{R.}~\bibnamefont{Tenne}},
  \bibinfo{author}{\bibfnamefont{Y.~R.} \bibnamefont{Hacohen}},
  \bibinfo{author}{\bibnamefont{{O. St\'{e}phan}}}, and
  \bibinfo{author}{\bibfnamefont{C.}~\bibnamefont{Colliex}},
  \bibinfo{year}{1999}, \bibinfo{journal}{Phys.\ Rev.\ Lett.}
  \textbf{\bibinfo{volume}{83}}, \bibinfo{pages}{659}.

\bibitem[{\citenamefont{{Colas des Francs}}
  \emph{et~al.}(2001)\citenamefont{{Colas des Francs}, Girard, Weeber, Chicane,
  David, Dereux, and Peyrade}}]{CGW01}
\bibinfo{author}{\bibnamefont{{Colas des Francs}}, \bibfnamefont{G.}},
  \bibinfo{author}{\bibfnamefont{C.}~\bibnamefont{Girard}},
  \bibinfo{author}{\bibfnamefont{J.~C.} \bibnamefont{Weeber}},
  \bibinfo{author}{\bibfnamefont{C.}~\bibnamefont{Chicane}},
  \bibinfo{author}{\bibfnamefont{T.}~\bibnamefont{David}},
  \bibinfo{author}{\bibfnamefont{A.}~\bibnamefont{Dereux}}, and
  \bibinfo{author}{\bibfnamefont{D.}~\bibnamefont{Peyrade}},
  \bibinfo{year}{2001}, \bibinfo{journal}{Phys.\ Rev.\ Lett.}
  \textbf{\bibinfo{volume}{86}}, \bibinfo{pages}{4950}.

\bibitem[{\citenamefont{Cole} \emph{et~al.}(2007)\citenamefont{Cole, Baumberg,
  {Garc\'{\i}a de Abajo}, Mahajan, Abdelsalam, and Bartlett}}]{paper128}
\bibinfo{author}{\bibnamefont{Cole}, \bibfnamefont{R.~M.}},
  \bibinfo{author}{\bibfnamefont{J.~J.} \bibnamefont{Baumberg}},
  \bibinfo{author}{\bibfnamefont{F.~J.} \bibnamefont{{Garc\'{\i}a de Abajo}}},
  \bibinfo{author}{\bibfnamefont{S.}~\bibnamefont{Mahajan}},
  \bibinfo{author}{\bibfnamefont{M.}~\bibnamefont{Abdelsalam}}, and
  \bibinfo{author}{\bibfnamefont{P.~N.} \bibnamefont{Bartlett}},
  \bibinfo{year}{2007}, \bibinfo{journal}{Nano\ Lett.}
  \textbf{\bibinfo{volume}{7}}, \bibinfo{pages}{2094}.

\bibitem[{\citenamefont{Couillard} \emph{et~al.}(2007)\citenamefont{Couillard,
  Kociak, {O. St\'{e}phan}, Botton, and Colliex}}]{CKS07}
\bibinfo{author}{\bibnamefont{Couillard}, \bibfnamefont{M.}},
  \bibinfo{author}{\bibfnamefont{M.}~\bibnamefont{Kociak}},
  \bibinfo{author}{\bibnamefont{{O. St\'{e}phan}}},
  \bibinfo{author}{\bibfnamefont{G.~A.} \bibnamefont{Botton}}, and
  \bibinfo{author}{\bibfnamefont{C.}~\bibnamefont{Colliex}},
  \bibinfo{year}{2007}, \bibinfo{journal}{Phys.\ Rev.\ B}
  \textbf{\bibinfo{volume}{76}}, \bibinfo{pages}{165131}.

\bibitem[{\citenamefont{Couillard} \emph{et~al.}(2008)\citenamefont{Couillard,
  Yurtsever, and Muller}}]{CYM08}
\bibinfo{author}{\bibnamefont{Couillard}, \bibfnamefont{M.}},
  \bibinfo{author}{\bibfnamefont{A.}~\bibnamefont{Yurtsever}}, and
  \bibinfo{author}{\bibfnamefont{D.~A.} \bibnamefont{Muller}},
  \bibinfo{year}{2008}, \bibinfo{journal}{Phys.\ Rev.\ B}
  \textbf{\bibinfo{volume}{77}}, \bibinfo{pages}{085318}.

\bibitem[{\citenamefont{Cowley}(1982{\natexlab{a}})}]{C1982}
\bibinfo{author}{\bibnamefont{Cowley}, \bibfnamefont{J.~M.}},
  \bibinfo{year}{1982}{\natexlab{a}}, \bibinfo{journal}{Phys.\ Rev.\ B}
  \textbf{\bibinfo{volume}{25}}, \bibinfo{pages}{1401}.

\bibitem[{\citenamefont{Cowley}(1982{\natexlab{b}})}]{C1982_2}
\bibinfo{author}{\bibnamefont{Cowley}, \bibfnamefont{J.~M.}},
  \bibinfo{year}{1982}{\natexlab{b}}, \bibinfo{journal}{Surf.\ Sci.}
  \textbf{\bibinfo{volume}{114}}, \bibinfo{pages}{587}.

\bibitem[{\citenamefont{Coyle} \emph{et~al.}(2001)\citenamefont{Coyle, Netti,
  Baumberg, Ghanem, Birkin, Bartlett, and Whittaker}}]{CNB01}
\bibinfo{author}{\bibnamefont{Coyle}, \bibfnamefont{S.}},
  \bibinfo{author}{\bibfnamefont{M.~C.} \bibnamefont{Netti}},
  \bibinfo{author}{\bibfnamefont{J.~J.} \bibnamefont{Baumberg}},
  \bibinfo{author}{\bibfnamefont{M.~A.} \bibnamefont{Ghanem}},
  \bibinfo{author}{\bibfnamefont{P.~R.} \bibnamefont{Birkin}},
  \bibinfo{author}{\bibfnamefont{P.~N.} \bibnamefont{Bartlett}}, and
  \bibinfo{author}{\bibfnamefont{D.~M.} \bibnamefont{Whittaker}},
  \bibinfo{year}{2001}, \bibinfo{journal}{Phys.\ Rev.\ Lett.}
  \textbf{\bibinfo{volume}{87}}, \bibinfo{pages}{176801}.

\bibitem[{\citenamefont{Cram and Arakawa}(1967)}]{CA1967}
\bibinfo{author}{\bibnamefont{Cram}, \bibfnamefont{L.~S.}}, and
  \bibinfo{author}{\bibfnamefont{E.~T.} \bibnamefont{Arakawa}},
  \bibinfo{year}{1967}, \bibinfo{journal}{Phys.\ Rev.}
  \textbf{\bibinfo{volume}{153}}, \bibinfo{pages}{455}.

\bibitem[{\citenamefont{Creuzburg}(1966)}]{C1966}
\bibinfo{author}{\bibnamefont{Creuzburg}, \bibfnamefont{M.}},
  \bibinfo{year}{1966}, \bibinfo{journal}{Z.\ Phys.}
  \textbf{\bibinfo{volume}{194}}, \bibinfo{pages}{211}.

\bibitem[{\citenamefont{Crewe} \emph{et~al.}(1971)\citenamefont{Crewe,
  Isaacson, and Johnson}}]{CIJ1971}
\bibinfo{author}{\bibnamefont{Crewe}, \bibfnamefont{A.~V.}},
  \bibinfo{author}{\bibfnamefont{M.}~\bibnamefont{Isaacson}}, and
  \bibinfo{author}{\bibfnamefont{D.}~\bibnamefont{Johnson}},
  \bibinfo{year}{1971}, \bibinfo{journal}{Nature}
  \textbf{\bibinfo{volume}{231}}, \bibinfo{pages}{262}.

\bibitem[{\citenamefont{Crowell and Ritchie}(1968)}]{CR1968}
\bibinfo{author}{\bibnamefont{Crowell}, \bibfnamefont{J.}}, and
  \bibinfo{author}{\bibfnamefont{R.~H.} \bibnamefont{Ritchie}},
  \bibinfo{year}{1968}, \bibinfo{journal}{Phys.\ Rev.}
  \textbf{\bibinfo{volume}{172}}, \bibinfo{pages}{436}.

\bibitem[{\citenamefont{{D'Aguanno}}
  \emph{et~al.}(2004)\citenamefont{{D'Aguanno}, Mattiucci, Centini, Scalora,
  and Bloemer}}]{DMC04}
\bibinfo{author}{\bibnamefont{{D'Aguanno}}, \bibfnamefont{G.}},
  \bibinfo{author}{\bibfnamefont{N.}~\bibnamefont{Mattiucci}},
  \bibinfo{author}{\bibfnamefont{M.}~\bibnamefont{Centini}},
  \bibinfo{author}{\bibfnamefont{M.}~\bibnamefont{Scalora}}, and
  \bibinfo{author}{\bibfnamefont{M.~J.} \bibnamefont{Bloemer}},
  \bibinfo{year}{2004}, \bibinfo{journal}{Phys.\ Rev.\ E}
  \textbf{\bibinfo{volume}{69}}, \bibinfo{pages}{057601}.

\bibitem[{\citenamefont{Daniels} \emph{et~al.}(2003)\citenamefont{Daniels,
  Brydson, Brown, and Rand}}]{DBB03}
\bibinfo{author}{\bibnamefont{Daniels}, \bibfnamefont{H.~R.}},
  \bibinfo{author}{\bibfnamefont{R.}~\bibnamefont{Brydson}},
  \bibinfo{author}{\bibfnamefont{A.}~\bibnamefont{Brown}}, and
  \bibinfo{author}{\bibfnamefont{B.}~\bibnamefont{Rand}}, \bibinfo{year}{2003},
  \bibinfo{journal}{Ultramicroscopy} \textbf{\bibinfo{volume}{96}},
  \bibinfo{pages}{547}.

\bibitem[{\citenamefont{Davis}(1976)}]{D1976}
\bibinfo{author}{\bibnamefont{Davis}, \bibfnamefont{L.~C.}},
  \bibinfo{year}{1976}, \bibinfo{journal}{Phys.\ Rev.\ B}
  \textbf{\bibinfo{volume}{14}}, \bibinfo{pages}{5523}.

\bibitem[{\citenamefont{{de Broglie}}(1925)}]{L1925}
\bibinfo{author}{\bibnamefont{{de Broglie}}, \bibfnamefont{L.}},
  \bibinfo{year}{1925}, \bibinfo{journal}{Ann.\ Phys.\ (Paris)}
  \textbf{\bibinfo{volume}{3}}, \bibinfo{pages}{22}.

\bibitem[{\citenamefont{Degiron} \emph{et~al.}(2004)\citenamefont{Degiron,
  Lezec, Yamamoto, and Ebbesen}}]{DLY04}
\bibinfo{author}{\bibnamefont{Degiron}, \bibfnamefont{A.}},
  \bibinfo{author}{\bibfnamefont{H.~J.} \bibnamefont{Lezec}},
  \bibinfo{author}{\bibfnamefont{N.}~\bibnamefont{Yamamoto}}, and
  \bibinfo{author}{\bibfnamefont{T.~W.} \bibnamefont{Ebbesen}},
  \bibinfo{year}{2004}, \bibinfo{journal}{Opt.\ Commun.}
  \textbf{\bibinfo{volume}{239}}, \bibinfo{pages}{61–}.

\bibitem[{\citenamefont{Demkov and Meyer}(2004)}]{DM04}
\bibinfo{author}{\bibnamefont{Demkov}, \bibfnamefont{Y.~N.}}, and
  \bibinfo{author}{\bibfnamefont{J.~D.} \bibnamefont{Meyer}},
  \bibinfo{year}{2004}, \bibinfo{journal}{Eur.\ Phys.\ J.\ B}
  \textbf{\bibinfo{volume}{42}}, \bibinfo{pages}{361–}.

\bibitem[{\citenamefont{Denisyuk} \emph{et~al.}(2008)\citenamefont{Denisyuk,
  Jonsson, MacDonald, Zheludev, and {Garc\'{\i}a de Abajo}}}]{paper144}
\bibinfo{author}{\bibnamefont{Denisyuk}, \bibfnamefont{A.~I.}},
  \bibinfo{author}{\bibfnamefont{F.}~\bibnamefont{Jonsson}},
  \bibinfo{author}{\bibfnamefont{K.~F.} \bibnamefont{MacDonald}},
  \bibinfo{author}{\bibfnamefont{N.~I.} \bibnamefont{Zheludev}}, and
  \bibinfo{author}{\bibfnamefont{F.~J.} \bibnamefont{{Garc\'{\i}a de Abajo}}},
  \bibinfo{year}{2008}, \bibinfo{journal}{Appl.\ Phys.\ Lett.}
  \textbf{\bibinfo{volume}{92}}, \bibinfo{pages}{093112}.

\bibitem[{\citenamefont{Diaconescu}
  \emph{et~al.}(2007)\citenamefont{Diaconescu, Pohl, Vattuone, Savio, Hofmann,
  Silkin, Pitarke, Chulkov., Echenique, Far\'{\i}as, and Rocca}}]{DPV07}
\bibinfo{author}{\bibnamefont{Diaconescu}, \bibfnamefont{B.}},
  \bibinfo{author}{\bibfnamefont{K.}~\bibnamefont{Pohl}},
  \bibinfo{author}{\bibfnamefont{L.}~\bibnamefont{Vattuone}},
  \bibinfo{author}{\bibfnamefont{L.}~\bibnamefont{Savio}},
  \bibinfo{author}{\bibfnamefont{P.}~\bibnamefont{Hofmann}},
  \bibinfo{author}{\bibfnamefont{V.~M.} \bibnamefont{Silkin}},
  \bibinfo{author}{\bibfnamefont{J.~M.} \bibnamefont{Pitarke}},
  \bibinfo{author}{\bibfnamefont{E.~V.} \bibnamefont{Chulkov.}},
  \bibinfo{author}{\bibfnamefont{P.~M.} \bibnamefont{Echenique}},
  \bibinfo{author}{\bibfnamefont{D.}~\bibnamefont{Far\'{\i}as}}, and
  \bibinfo{author}{\bibfnamefont{M.}~\bibnamefont{Rocca}},
  \bibinfo{year}{2007}, \bibinfo{journal}{Nature}
  \textbf{\bibinfo{volume}{448}}, \bibinfo{pages}{57}.

\bibitem[{\citenamefont{Dobrzynski and Maradudin}(1972)}]{DM1972}
\bibinfo{author}{\bibnamefont{Dobrzynski}, \bibfnamefont{L.}}, and
  \bibinfo{author}{\bibfnamefont{A.~A.} \bibnamefont{Maradudin}},
  \bibinfo{year}{1972}, \bibinfo{journal}{Phys.\ Rev.\ B}
  \textbf{\bibinfo{volume}{6}}, \bibinfo{pages}{3810}.

\bibitem[{\citenamefont{Dorneich} \emph{et~al.}(1998)\citenamefont{Dorneich,
  French, {M\"ullejans}, Loughin, and {R\"uhle}}}]{DFM98}
\bibinfo{author}{\bibnamefont{Dorneich}, \bibfnamefont{A.~D.}},
  \bibinfo{author}{\bibfnamefont{R.~H.} \bibnamefont{French}},
  \bibinfo{author}{\bibfnamefont{H.}~\bibnamefont{{M\"ullejans}}},
  \bibinfo{author}{\bibfnamefont{S.}~\bibnamefont{Loughin}}, and
  \bibinfo{author}{\bibfnamefont{M.}~\bibnamefont{{R\"uhle}}},
  \bibinfo{year}{1998}, \bibinfo{journal}{J.\ Microsc.}
  \textbf{\bibinfo{volume}{191}}, \bibinfo{pages}{286–}.

\bibitem[{\citenamefont{Doucas} \emph{et~al.}(1992)\citenamefont{Doucas,
  Mulvey, Omori, Walsh, and Kimmitt}}]{DMO92}
\bibinfo{author}{\bibnamefont{Doucas}, \bibfnamefont{G.}},
  \bibinfo{author}{\bibfnamefont{J.~H.} \bibnamefont{Mulvey}},
  \bibinfo{author}{\bibfnamefont{M.}~\bibnamefont{Omori}},
  \bibinfo{author}{\bibfnamefont{J.}~\bibnamefont{Walsh}}, and
  \bibinfo{author}{\bibfnamefont{M.~F.} \bibnamefont{Kimmitt}},
  \bibinfo{year}{1992}, \bibinfo{journal}{Phys.\ Rev.\ Lett.}
  \textbf{\bibinfo{volume}{69}}, \bibinfo{pages}{1761}.

\bibitem[{\citenamefont{Drachsel} \emph{et~al.}(2002)\citenamefont{Drachsel,
  Adelt, Nilius, and Freund}}]{DAN02}
\bibinfo{author}{\bibnamefont{Drachsel}, \bibfnamefont{W.}},
  \bibinfo{author}{\bibfnamefont{M.}~\bibnamefont{Adelt}},
  \bibinfo{author}{\bibfnamefont{N.}~\bibnamefont{Nilius}}, and
  \bibinfo{author}{\bibfnamefont{H.~J.} \bibnamefont{Freund}},
  \bibinfo{year}{2002}, \bibinfo{journal}{J.\ Electron\ Spectrosc.\ Relat.\
  Phenom.} \textbf{\bibinfo{volume}{122}}, \bibinfo{pages}{239–}.

\bibitem[{\citenamefont{Draine and Flatau}(1994)}]{DF94}
\bibinfo{author}{\bibnamefont{Draine}, \bibfnamefont{B.~T.}}, and
  \bibinfo{author}{\bibfnamefont{P.~J.} \bibnamefont{Flatau}},
  \bibinfo{year}{1994}, \bibinfo{journal}{J.\ Opt.\ Soc.\ Am.\ A}
  \textbf{\bibinfo{volume}{11}}, \bibinfo{pages}{1491}.

\bibitem[{\citenamefont{Drezet} \emph{et~al.}(2007)\citenamefont{Drezet,
  Hohenau, Krenn, Brun, and Huant}}]{DHK07}
\bibinfo{author}{\bibnamefont{Drezet}, \bibfnamefont{A.}},
  \bibinfo{author}{\bibfnamefont{A.}~\bibnamefont{Hohenau}},
  \bibinfo{author}{\bibfnamefont{J.~R.} \bibnamefont{Krenn}},
  \bibinfo{author}{\bibfnamefont{M.}~\bibnamefont{Brun}}, and
  \bibinfo{author}{\bibfnamefont{S.}~\bibnamefont{Huant}},
  \bibinfo{year}{2007}, \bibinfo{journal}{Micron}
  \textbf{\bibinfo{volume}{38}}, \bibinfo{pages}{427–}.

\bibitem[{\citenamefont{Drucker and Scheinfein}(1993)}]{DS93}
\bibinfo{author}{\bibnamefont{Drucker}, \bibfnamefont{J.}}, and
  \bibinfo{author}{\bibfnamefont{M.~R.} \bibnamefont{Scheinfein}},
  \bibinfo{year}{1993}, \bibinfo{journal}{Phys.\ Rev.\ B}
  \textbf{\bibinfo{volume}{47}}, \bibinfo{pages}{15973}.

\bibitem[{\citenamefont{Drucker} \emph{et~al.}(1993)\citenamefont{Drucker,
  Scheinfein, Liu, and Weiss}}]{DSL93}
\bibinfo{author}{\bibnamefont{Drucker}, \bibfnamefont{J.}},
  \bibinfo{author}{\bibfnamefont{M.~R.} \bibnamefont{Scheinfein}},
  \bibinfo{author}{\bibfnamefont{J.}~\bibnamefont{Liu}}, and
  \bibinfo{author}{\bibfnamefont{J.~K.} \bibnamefont{Weiss}},
  \bibinfo{year}{1993}, \bibinfo{journal}{J.\ Appl.\ Phys.}
  \textbf{\bibinfo{volume}{74}}, \bibinfo{pages}{7329}.

\bibitem[{\citenamefont{Dulkeith} \emph{et~al.}(2004)\citenamefont{Dulkeith,
  Niedereichholz, Klar, Feldmann, {von Plessen}, Gittins, Mayya, and
  Caruso}}]{DNK04}
\bibinfo{author}{\bibnamefont{Dulkeith}, \bibfnamefont{E.}},
  \bibinfo{author}{\bibfnamefont{T.}~\bibnamefont{Niedereichholz}},
  \bibinfo{author}{\bibfnamefont{T.~A.} \bibnamefont{Klar}},
  \bibinfo{author}{\bibfnamefont{J.}~\bibnamefont{Feldmann}},
  \bibinfo{author}{\bibfnamefont{G.}~\bibnamefont{{von Plessen}}},
  \bibinfo{author}{\bibfnamefont{D.~I.} \bibnamefont{Gittins}},
  \bibinfo{author}{\bibfnamefont{K.~S.} \bibnamefont{Mayya}}, and
  \bibinfo{author}{\bibfnamefont{F.}~\bibnamefont{Caruso}},
  \bibinfo{year}{2004}, \bibinfo{journal}{Phys.\ Rev.\ B}
  \textbf{\bibinfo{volume}{70}}, \bibinfo{pages}{205424}.

\bibitem[{\citenamefont{Dwyer} \emph{et~al.}(2008)\citenamefont{Dwyer, Erni,
  and Etheridge}}]{DEE08}
\bibinfo{author}{\bibnamefont{Dwyer}, \bibfnamefont{C.}},
  \bibinfo{author}{\bibfnamefont{R.}~\bibnamefont{Erni}}, and
  \bibinfo{author}{\bibfnamefont{J.}~\bibnamefont{Etheridge}},
  \bibinfo{year}{2008}, \bibinfo{journal}{Appl.\ Phys.\ Lett.}
  \textbf{\bibinfo{volume}{93}}, \bibinfo{pages}{021115}.

\bibitem[{\citenamefont{Echenique} \emph{et~al.}(1987)\citenamefont{Echenique,
  Bausells, and Rivacoba}}]{EBR1987}
\bibinfo{author}{\bibnamefont{Echenique}, \bibfnamefont{P.~M.}},
  \bibinfo{author}{\bibfnamefont{J.}~\bibnamefont{Bausells}}, and
  \bibinfo{author}{\bibfnamefont{A.}~\bibnamefont{Rivacoba}},
  \bibinfo{year}{1987}, \bibinfo{journal}{Phys.\ Rev.\ B}
  \textbf{\bibinfo{volume}{35}}, \bibinfo{pages}{1521}.

\bibitem[{\citenamefont{Echenique} \emph{et~al.}(1990)\citenamefont{Echenique,
  Flores, and Ritchie}}]{EFR1990}
\bibinfo{author}{\bibnamefont{Echenique}, \bibfnamefont{P.~M.}},
  \bibinfo{author}{\bibfnamefont{F.}~\bibnamefont{Flores}}, and
  \bibinfo{author}{\bibfnamefont{R.~H.} \bibnamefont{Ritchie}},
  \bibinfo{year}{1990}, \bibinfo{journal}{Solid\ State\ Physics}
  \textbf{\bibinfo{volume}{43}}, \bibinfo{pages}{229}.

\bibitem[{\citenamefont{Echenique} \emph{et~al.}(1999)\citenamefont{Echenique,
  Howie, and Ritchie}}]{EHR99}
\bibinfo{author}{\bibnamefont{Echenique}, \bibfnamefont{P.~M.}},
  \bibinfo{author}{\bibfnamefont{A.}~\bibnamefont{Howie}}, and
  \bibinfo{author}{\bibfnamefont{R.~H.} \bibnamefont{Ritchie}},
  \bibinfo{year}{1999}, \bibinfo{journal}{Phys.\ Rev.\ Lett.}
  \textbf{\bibinfo{volume}{83}}, \bibinfo{pages}{658}.

\bibitem[{\citenamefont{Echenique and Pendry}(1975)}]{EP1975}
\bibinfo{author}{\bibnamefont{Echenique}, \bibfnamefont{P.~M.}}, and
  \bibinfo{author}{\bibfnamefont{J.~B.} \bibnamefont{Pendry}},
  \bibinfo{year}{1975}, \bibinfo{journal}{J.\ Phys.\ C}
  \textbf{\bibinfo{volume}{8}}, \bibinfo{pages}{2936}.

\bibitem[{\citenamefont{Edighoffer}
  \emph{et~al.}(1981)\citenamefont{Edighoffer, Kimura, Pantell, Piestrup, and
  Wang}}]{EKP1981}
\bibinfo{author}{\bibnamefont{Edighoffer}, \bibfnamefont{J.~A.}},
  \bibinfo{author}{\bibfnamefont{W.~D.} \bibnamefont{Kimura}},
  \bibinfo{author}{\bibfnamefont{R.~H.} \bibnamefont{Pantell}},
  \bibinfo{author}{\bibfnamefont{M.~A.} \bibnamefont{Piestrup}}, and
  \bibinfo{author}{\bibfnamefont{D.~Y.} \bibnamefont{Wang}},
  \bibinfo{year}{1981}, \bibinfo{journal}{Phys.\ Rev.\ A}
  \textbf{\bibinfo{volume}{23}}, \bibinfo{pages}{1848}.

\bibitem[{\citenamefont{Egerton}(1996)}]{E96}
\bibinfo{author}{\bibnamefont{Egerton}, \bibfnamefont{R.~F.}},
  \bibinfo{year}{1996}, \emph{\bibinfo{title}{Electron Energy-Loss Spectroscopy
  in the Electron Microscope}} (\bibinfo{publisher}{Plenum Press},
  \bibinfo{address}{New York}).

\bibitem[{\citenamefont{Egerton}(2003)}]{E03}
\bibinfo{author}{\bibnamefont{Egerton}, \bibfnamefont{R.~F.}},
  \bibinfo{year}{2003}, \bibinfo{journal}{Micron}
  \textbf{\bibinfo{volume}{34}}, \bibinfo{pages}{127–}.

\bibitem[{\citenamefont{Egerton}(2007)}]{E07}
\bibinfo{author}{\bibnamefont{Egerton}, \bibfnamefont{R.~F.}},
  \bibinfo{year}{2007}, \bibinfo{journal}{Ultramicroscopy}
  \textbf{\bibinfo{volume}{107}}, \bibinfo{pages}{575–}.

\bibitem[{\citenamefont{Eggeman} \emph{et~al.}(2007)\citenamefont{Eggeman,
  Dobson, and Petford-Long}}]{EDP07}
\bibinfo{author}{\bibnamefont{Eggeman}, \bibfnamefont{A.~S.}},
  \bibinfo{author}{\bibfnamefont{P.~J.} \bibnamefont{Dobson}}, and
  \bibinfo{author}{\bibfnamefont{A.~K.} \bibnamefont{Petford-Long}},
  \bibinfo{year}{2007}, \bibinfo{journal}{J.\ Appl.\ Phys.}
  \textbf{\bibinfo{volume}{101}}, \bibinfo{pages}{024307}.

\bibitem[{\citenamefont{Eguiluz and Maradudin}(1976)}]{EM1976}
\bibinfo{author}{\bibnamefont{Eguiluz}, \bibfnamefont{A.}}, and
  \bibinfo{author}{\bibfnamefont{A.~A.} \bibnamefont{Maradudin}},
  \bibinfo{year}{1976}, \bibinfo{journal}{Phys.\ Rev.\ B}
  \textbf{\bibinfo{volume}{14}}, \bibinfo{pages}{5526}.

\bibitem[{\citenamefont{Erni and Browning}(2005)}]{EB05}
\bibinfo{author}{\bibnamefont{Erni}, \bibfnamefont{R.}}, and
  \bibinfo{author}{\bibfnamefont{N.~D.} \bibnamefont{Browning}},
  \bibinfo{year}{2005}, \bibinfo{journal}{Ultramicroscopy}
  \textbf{\bibinfo{volume}{104}}, \bibinfo{pages}{176–}.

\bibitem[{\citenamefont{Erni and Browning}(2007)}]{EB07}
\bibinfo{author}{\bibnamefont{Erni}, \bibfnamefont{R.}}, and
  \bibinfo{author}{\bibfnamefont{N.~D.} \bibnamefont{Browning}},
  \bibinfo{year}{2007}, \bibinfo{journal}{Ultramicroscopy}
  \textbf{\bibinfo{volume}{107}}, \bibinfo{pages}{267–}.

\bibitem[{\citenamefont{Erni and Browning}(2008)}]{EB08}
\bibinfo{author}{\bibnamefont{Erni}, \bibfnamefont{R.}}, and
  \bibinfo{author}{\bibfnamefont{N.~D.} \bibnamefont{Browning}},
  \bibinfo{year}{2008}, \bibinfo{journal}{Ultramicroscopy}
  \textbf{\bibinfo{volume}{108}}, \bibinfo{pages}{84}.

\bibitem[{\citenamefont{Erni} \emph{et~al.}(2008)\citenamefont{Erni, Lazar, and
  Browning}}]{ELB08}
\bibinfo{author}{\bibnamefont{Erni}, \bibfnamefont{R.}},
  \bibinfo{author}{\bibfnamefont{S.}~\bibnamefont{Lazar}}, and
  \bibinfo{author}{\bibfnamefont{N.~D.} \bibnamefont{Browning}},
  \bibinfo{year}{2008}, \bibinfo{journal}{Ultramicroscopy}
  \textbf{\bibinfo{volume}{108}}, \bibinfo{pages}{270}.

\bibitem[{\citenamefont{Fehlhaber and Bursill}(1999)}]{FB99}
\bibinfo{author}{\bibnamefont{Fehlhaber}, \bibfnamefont{R.~P.}}, and
  \bibinfo{author}{\bibfnamefont{L.~A.} \bibnamefont{Bursill}},
  \bibinfo{year}{1999}, \bibinfo{journal}{Phys.\ Rev.\ B}
  \textbf{\bibinfo{volume}{60}}, \bibinfo{pages}{14147}.

\bibitem[{\citenamefont{Fehlhaber and Bursill}(2000)}]{FB00}
\bibinfo{author}{\bibnamefont{Fehlhaber}, \bibfnamefont{R.~P.}}, and
  \bibinfo{author}{\bibfnamefont{L.~A.} \bibnamefont{Bursill}},
  \bibinfo{year}{2000}, \bibinfo{journal}{Phys.\ Rev.\ B}
  \textbf{\bibinfo{volume}{62}}, \bibinfo{pages}{17094}.

\bibitem[{\citenamefont{Fermi}(1940)}]{F1940}
\bibinfo{author}{\bibnamefont{Fermi}, \bibfnamefont{E.}}, \bibinfo{year}{1940},
  \bibinfo{journal}{Phys.\ Rev.} \textbf{\bibinfo{volume}{57}},
  \bibinfo{pages}{485}.

\bibitem[{\citenamefont{Ferrell}(1958)}]{F1958}
\bibinfo{author}{\bibnamefont{Ferrell}, \bibfnamefont{R.~A.}},
  \bibinfo{year}{1958}, \bibinfo{journal}{Phys.\ Rev.}
  \textbf{\bibinfo{volume}{111}}, \bibinfo{pages}{1214}.

\bibitem[{\citenamefont{Ferrell and Echenique}(1985)}]{FE1985}
\bibinfo{author}{\bibnamefont{Ferrell}, \bibfnamefont{T.~L.}}, and
  \bibinfo{author}{\bibfnamefont{P.~M.} \bibnamefont{Echenique}},
  \bibinfo{year}{1985}, \bibinfo{journal}{Phys.\ Rev.\ Lett.}
  \textbf{\bibinfo{volume}{55}}, \bibinfo{pages}{1526}.

\bibitem[{\citenamefont{Ferrell} \emph{et~al.}(1987)\citenamefont{Ferrell,
  Warmack, Anderson, and Echenique}}]{FWA1987}
\bibinfo{author}{\bibnamefont{Ferrell}, \bibfnamefont{T.~L.}},
  \bibinfo{author}{\bibfnamefont{R.~J.} \bibnamefont{Warmack}},
  \bibinfo{author}{\bibfnamefont{V.~E.} \bibnamefont{Anderson}}, and
  \bibinfo{author}{\bibfnamefont{P.~M.} \bibnamefont{Echenique}},
  \bibinfo{year}{1987}, \bibinfo{journal}{Phys.\ Rev.\ B}
  \textbf{\bibinfo{volume}{35}}, \bibinfo{pages}{7365}.

\bibitem[{\citenamefont{Ford and Weber}(1984)}]{FW1984}
\bibinfo{author}{\bibnamefont{Ford}, \bibfnamefont{G.~W.}}, and
  \bibinfo{author}{\bibfnamefont{W.~H.} \bibnamefont{Weber}},
  \bibinfo{year}{1984}, \bibinfo{journal}{Phys.\ Rep.}
  \textbf{\bibinfo{volume}{113}}, \bibinfo{pages}{195}.

\bibitem[{\citenamefont{Forstmann} \emph{et~al.}(1991)\citenamefont{Forstmann,
  {A. Gras-Mart\'{\i}}, Ferrell, Warmack, and Mamola}}]{FGF91}
\bibinfo{author}{\bibnamefont{Forstmann}, \bibfnamefont{F.}},
  \bibinfo{author}{\bibnamefont{{A. Gras-Mart\'{\i}}}},
  \bibinfo{author}{\bibfnamefont{T.~L.} \bibnamefont{Ferrell}},
  \bibinfo{author}{\bibfnamefont{R.~J.} \bibnamefont{Warmack}}, and
  \bibinfo{author}{\bibfnamefont{K.~C.} \bibnamefont{Mamola}},
  \bibinfo{year}{1991}, \bibinfo{journal}{Phys.\ Rev.\ B}
  \textbf{\bibinfo{volume}{44}}, \bibinfo{pages}{4884}.

\bibitem[{\citenamefont{Frank and Tamm}(1937)}]{FT1937_2}
\bibinfo{author}{\bibnamefont{Frank}, \bibfnamefont{I.~M.}}, and
  \bibinfo{author}{\bibfnamefont{I.}~\bibnamefont{Tamm}}, \bibinfo{year}{1937},
  \bibinfo{journal}{Dokl.\ Akad.\ Nauk\ SSSR} \textbf{\bibinfo{volume}{14}},
  \bibinfo{pages}{109}.

\bibitem[{\citenamefont{French} \emph{et~al.}(1998)\citenamefont{French,
  M\"ullejans, and Jones}}]{FMJ98}
\bibinfo{author}{\bibnamefont{French}, \bibfnamefont{R.~H.}},
  \bibinfo{author}{\bibfnamefont{H.}~\bibnamefont{M\"ullejans}}, and
  \bibinfo{author}{\bibfnamefont{D.~J.} \bibnamefont{Jones}},
  \bibinfo{year}{1998}, \bibinfo{journal}{J.\ Am.\ Ceram.\ Soc.}
  \textbf{\bibinfo{volume}{81}}, \bibinfo{pages}{2459}.

\bibitem[{\citenamefont{Fu} \emph{et~al.}(1993)\citenamefont{Fu, Macedo, and
  Resca}}]{FMR93}
\bibinfo{author}{\bibnamefont{Fu}, \bibfnamefont{L.}},
  \bibinfo{author}{\bibfnamefont{P.~B.} \bibnamefont{Macedo}}, and
  \bibinfo{author}{\bibfnamefont{L.}~\bibnamefont{Resca}},
  \bibinfo{year}{1993}, \bibinfo{journal}{Phys.\ Rev.\ B}
  \textbf{\bibinfo{volume}{47}}, \bibinfo{pages}{13818}.

\bibitem[{\citenamefont{Fu and Resca}(1993)}]{FR93}
\bibinfo{author}{\bibnamefont{Fu}, \bibfnamefont{L.}}, and
  \bibinfo{author}{\bibfnamefont{L.}~\bibnamefont{Resca}},
  \bibinfo{year}{1993}, \bibinfo{journal}{Phys.\ Rev.\ B}
  \textbf{\bibinfo{volume}{47}}, \bibinfo{pages}{16194}.

\bibitem[{\citenamefont{Fuchs}(1975)}]{F1975}
\bibinfo{author}{\bibnamefont{Fuchs}, \bibfnamefont{R.}}, \bibinfo{year}{1975},
  \bibinfo{journal}{Phys.\ Rev.\ B} \textbf{\bibinfo{volume}{11}},
  \bibinfo{pages}{1732}.

\bibitem[{\citenamefont{Fuchs} \emph{et~al.}(1996)\citenamefont{Fuchs, Barrera,
  and Carrillo}}]{FBC96}
\bibinfo{author}{\bibnamefont{Fuchs}, \bibfnamefont{R.}},
  \bibinfo{author}{\bibfnamefont{R.~G.} \bibnamefont{Barrera}}, and
  \bibinfo{author}{\bibfnamefont{J.~L.} \bibnamefont{Carrillo}},
  \bibinfo{year}{1996}, \bibinfo{journal}{Phys.\ Rev.\ B}
  \textbf{\bibinfo{volume}{54}}, \bibinfo{pages}{12824}.

\bibitem[{\citenamefont{Fuchs and Kliewer}(1968)}]{FK1968}
\bibinfo{author}{\bibnamefont{Fuchs}, \bibfnamefont{R.}}, and
  \bibinfo{author}{\bibfnamefont{K.~L.} \bibnamefont{Kliewer}},
  \bibinfo{year}{1968}, \bibinfo{journal}{J.\ Opt.\ Soc.\ Am.}
  \textbf{\bibinfo{volume}{58}}, \bibinfo{pages}{319}.

\bibitem[{\citenamefont{Fujimoto and Komaki}(1968)}]{FK1968_2}
\bibinfo{author}{\bibnamefont{Fujimoto}, \bibfnamefont{F.}}, and
  \bibinfo{author}{\bibfnamefont{K.}~\bibnamefont{Komaki}},
  \bibinfo{year}{1968}, \bibinfo{journal}{J.\ Phys.\ Soc.\ Jpn.}
  \textbf{\bibinfo{volume}{25}}, \bibinfo{pages}{1679}.

\bibitem[{\citenamefont{Fujimoto} \emph{et~al.}(1967)\citenamefont{Fujimoto,
  Komaki, and Ishida}}]{FKI1967}
\bibinfo{author}{\bibnamefont{Fujimoto}, \bibfnamefont{F.}},
  \bibinfo{author}{\bibfnamefont{K.}~\bibnamefont{Komaki}}, and
  \bibinfo{author}{\bibfnamefont{K.}~\bibnamefont{Ishida}},
  \bibinfo{year}{1967}, \bibinfo{journal}{J.\ Phys.\ Soc.\ Jpn.}
  \textbf{\bibinfo{volume}{23}}, \bibinfo{pages}{1186}.

\bibitem[{\citenamefont{Fussell} \emph{et~al.}(2004)\citenamefont{Fussell,
  McPhedran, and {Martijn de Sterke}}}]{FMM04}
\bibinfo{author}{\bibnamefont{Fussell}, \bibfnamefont{D.~P.}},
  \bibinfo{author}{\bibfnamefont{R.~C.} \bibnamefont{McPhedran}}, and
  \bibinfo{author}{\bibfnamefont{C.}~\bibnamefont{{Martijn de Sterke}}},
  \bibinfo{year}{2004}, \bibinfo{journal}{Phys.\ Rev.\ E}
  \textbf{\bibinfo{volume}{70}}, \bibinfo{pages}{066608}.

\bibitem[{\citenamefont{Fussell} \emph{et~al.}(2005)\citenamefont{Fussell,
  McPhedran, and {Martijn de Sterke}}}]{FMM05}
\bibinfo{author}{\bibnamefont{Fussell}, \bibfnamefont{D.~P.}},
  \bibinfo{author}{\bibfnamefont{R.~C.} \bibnamefont{McPhedran}}, and
  \bibinfo{author}{\bibfnamefont{C.}~\bibnamefont{{Martijn de Sterke}}},
  \bibinfo{year}{2005}, \bibinfo{journal}{Phys.\ Rev.\ A}
  \textbf{\bibinfo{volume}{71}}, \bibinfo{pages}{013815}.

\bibitem[{\citenamefont{Galloway} \emph{et~al.}(2003)\citenamefont{Galloway,
  Miller, Thomas, and Harmon}}]{GMT03}
\bibinfo{author}{\bibnamefont{Galloway}, \bibfnamefont{S.~A.}},
  \bibinfo{author}{\bibfnamefont{P.}~\bibnamefont{Miller}},
  \bibinfo{author}{\bibfnamefont{P.}~\bibnamefont{Thomas}}, and
  \bibinfo{author}{\bibfnamefont{R.}~\bibnamefont{Harmon}},
  \bibinfo{year}{2003}, \bibinfo{journal}{phys.\ stat.\ sol.\ (c)}
  \textbf{\bibinfo{volume}{0}}, \bibinfo{pages}{1028–}.

\bibitem[{\citenamefont{Gans}(1912)}]{G1912}
\bibinfo{author}{\bibnamefont{Gans}, \bibfnamefont{R.}}, \bibinfo{year}{1912},
  \bibinfo{journal}{Ann.\ Phys.\ (Leipzig)} \textbf{\bibinfo{volume}{37}},
  \bibinfo{pages}{881}.

\bibitem[{\citenamefont{{Garc\'{\i}a de Abajo}}(1999{\natexlab{a}})}]{paper040}
\bibinfo{author}{\bibnamefont{{Garc\'{\i}a de Abajo}}, \bibfnamefont{F.~J.}},
  \bibinfo{year}{1999}{\natexlab{a}}, \bibinfo{journal}{Phys.\ Rev.\ Lett.}
  \textbf{\bibinfo{volume}{82}}, \bibinfo{pages}{2776}.

\bibitem[{\citenamefont{{Garc\'{\i}a de Abajo}}(1999{\natexlab{b}})}]{paper046}
\bibinfo{author}{\bibnamefont{{Garc\'{\i}a de Abajo}}, \bibfnamefont{F.~J.}},
  \bibinfo{year}{1999}{\natexlab{b}}, \bibinfo{journal}{Phys.\ Rev.\ B}
  \textbf{\bibinfo{volume}{60}}, \bibinfo{pages}{6103}.

\bibitem[{\citenamefont{{Garc\'{\i}a de Abajo}}(1999{\natexlab{c}})}]{paper041}
\bibinfo{author}{\bibnamefont{{Garc\'{\i}a de Abajo}}, \bibfnamefont{F.~J.}},
  \bibinfo{year}{1999}{\natexlab{c}}, \bibinfo{journal}{Phys.\ Rev.\ B}
  \textbf{\bibinfo{volume}{59}}, \bibinfo{pages}{3095}.

\bibitem[{\citenamefont{{Garc\'{\i}a de Abajo}}(2004)}]{paper093}
\bibinfo{author}{\bibnamefont{{Garc\'{\i}a de Abajo}}, \bibfnamefont{F.~J.}},
  \bibinfo{year}{2004}, \bibinfo{journal}{Phys.\ Rev.\ B}
  \textbf{\bibinfo{volume}{70}}, \bibinfo{pages}{115422}.

\bibitem[{\citenamefont{{Garc\'{\i}a de Abajo}}(2007)}]{paper130}
\bibinfo{author}{\bibnamefont{{Garc\'{\i}a de Abajo}}, \bibfnamefont{F.~J.}},
  \bibinfo{year}{2007}, \bibinfo{journal}{Rev.\ Mod.\ Phys.}
  \textbf{\bibinfo{volume}{79}}, \bibinfo{pages}{1267}.

\bibitem[{\citenamefont{{Garc\'{\i}a de Abajo}}(2008)}]{paper162}
\bibinfo{author}{\bibnamefont{{Garc\'{\i}a de Abajo}}, \bibfnamefont{F.~J.}},
  \bibinfo{year}{2008}, \bibinfo{journal}{J.\ Phys.\ Chem.\ C}
  \textbf{\bibinfo{volume}{112}}, \bibinfo{pages}{17983}.

\bibitem[{\citenamefont{{Garc\'{\i}a de Abajo}}
  \emph{et~al.}(2003)\citenamefont{{Garc\'{\i}a de Abajo}, {A. G.
  Pattantyus-Abraham}, Zabala, Rivacoba, Wolf, and Echenique}}]{paper080}
\bibinfo{author}{\bibnamefont{{Garc\'{\i}a de Abajo}}, \bibfnamefont{F.~J.}},
  \bibinfo{author}{\bibnamefont{{A. G. Pattantyus-Abraham}}},
  \bibinfo{author}{\bibfnamefont{N.}~\bibnamefont{Zabala}},
  \bibinfo{author}{\bibfnamefont{A.}~\bibnamefont{Rivacoba}},
  \bibinfo{author}{\bibfnamefont{M.~O.} \bibnamefont{Wolf}}, and
  \bibinfo{author}{\bibfnamefont{P.~M.} \bibnamefont{Echenique}},
  \bibinfo{year}{2003}, \bibinfo{journal}{Phys.\ Rev.\ Lett.}
  \textbf{\bibinfo{volume}{91}}, \bibinfo{pages}{143902}.

\bibitem[{\citenamefont{{Garc\'{\i}a de Abajo} and Aizpurua}(1997)}]{paper025}
\bibinfo{author}{\bibnamefont{{Garc\'{\i}a de Abajo}}, \bibfnamefont{F.~J.}},
  and \bibinfo{author}{\bibfnamefont{J.}~\bibnamefont{Aizpurua}},
  \bibinfo{year}{1997}, \bibinfo{journal}{Phys.\ Rev.\ B}
  \textbf{\bibinfo{volume}{56}}, \bibinfo{pages}{15873}.

\bibitem[{\citenamefont{{Garc\'{\i}a de Abajo} and Blanco}(2003)}]{paper075}
\bibinfo{author}{\bibnamefont{{Garc\'{\i}a de Abajo}}, \bibfnamefont{F.~J.}},
  and \bibinfo{author}{\bibfnamefont{L.~A.} \bibnamefont{Blanco}},
  \bibinfo{year}{2003}, \bibinfo{journal}{Phys.\ Rev.\ B}
  \textbf{\bibinfo{volume}{67}}, \bibinfo{pages}{125108}.

\bibitem[{\citenamefont{{Garc\'{\i}a de Abajo} and Echenique}(1992)}]{paper004}
\bibinfo{author}{\bibnamefont{{Garc\'{\i}a de Abajo}}, \bibfnamefont{F.~J.}},
  and \bibinfo{author}{\bibfnamefont{P.~M.} \bibnamefont{Echenique}},
  \bibinfo{year}{1992}, \bibinfo{journal}{Phys.\ Rev.\ B}
  \textbf{\bibinfo{volume}{46}}, \bibinfo{pages}{2663}.

\bibitem[{\citenamefont{{Garc\'{\i}a de Abajo} and Howie}(1998)}]{paper030}
\bibinfo{author}{\bibnamefont{{Garc\'{\i}a de Abajo}}, \bibfnamefont{F.~J.}},
  and \bibinfo{author}{\bibfnamefont{A.}~\bibnamefont{Howie}},
  \bibinfo{year}{1998}, \bibinfo{journal}{Phys.\ Rev.\ Lett.}
  \textbf{\bibinfo{volume}{80}}, \bibinfo{pages}{5180}.

\bibitem[{\citenamefont{{Garc\'{\i}a de Abajo} and Howie}(2002)}]{paper070}
\bibinfo{author}{\bibnamefont{{Garc\'{\i}a de Abajo}}, \bibfnamefont{F.~J.}},
  and \bibinfo{author}{\bibfnamefont{A.}~\bibnamefont{Howie}},
  \bibinfo{year}{2002}, \bibinfo{journal}{Phys.\ Rev.\ B}
  \textbf{\bibinfo{volume}{65}}, \bibinfo{pages}{115418}.

\bibitem[{\citenamefont{{Garc\'{\i}a de Abajo} and
  Kociak}(2008{\natexlab{a}})}]{paper157}
\bibinfo{author}{\bibnamefont{{Garc\'{\i}a de Abajo}}, \bibfnamefont{F.~J.}},
  and \bibinfo{author}{\bibfnamefont{M.}~\bibnamefont{Kociak}},
  \bibinfo{year}{2008}{\natexlab{a}}, \bibinfo{journal}{New\ J.\ Phys.}
  \textbf{\bibinfo{volume}{10}}, \bibinfo{pages}{073035}.

\bibitem[{\citenamefont{{Garc\'{\i}a de Abajo} and
  Kociak}(2008{\natexlab{b}})}]{paper145}
\bibinfo{author}{\bibnamefont{{Garc\'{\i}a de Abajo}}, \bibfnamefont{F.~J.}},
  and \bibinfo{author}{\bibfnamefont{M.}~\bibnamefont{Kociak}},
  \bibinfo{year}{2008}{\natexlab{b}}, \bibinfo{journal}{Phys.\ Rev.\ Lett.}
  \textbf{\bibinfo{volume}{100}}, \bibinfo{pages}{106804}.

\bibitem[{\citenamefont{{Garc\'{\i}a de Abajo}}
  \emph{et~al.}(2004)\citenamefont{{Garc\'{\i}a de Abajo}, Rivacoba, Zabala,
  and Yamamoto}}]{paper084}
\bibinfo{author}{\bibnamefont{{Garc\'{\i}a de Abajo}}, \bibfnamefont{F.~J.}},
  \bibinfo{author}{\bibfnamefont{A.}~\bibnamefont{Rivacoba}},
  \bibinfo{author}{\bibfnamefont{N.}~\bibnamefont{Zabala}}, and
  \bibinfo{author}{\bibfnamefont{N.}~\bibnamefont{Yamamoto}},
  \bibinfo{year}{2004}, \bibinfo{journal}{Phys.\ Rev.\ B}
  \textbf{\bibinfo{volume}{69}}, \bibinfo{pages}{155420}.

\bibitem[{\citenamefont{{Garc\'{\i}a-Molina}}
  \emph{et~al.}(1985{\natexlab{a}})\citenamefont{{Garc\'{\i}a-Molina}, {A.
  Gras-Mart\'{\i}}, Howie, and Ritchie}}]{GGH1985}
\bibinfo{author}{\bibnamefont{{Garc\'{\i}a-Molina}}, \bibfnamefont{R.}},
  \bibinfo{author}{\bibnamefont{{A. Gras-Mart\'{\i}}}},
  \bibinfo{author}{\bibfnamefont{A.}~\bibnamefont{Howie}}, and
  \bibinfo{author}{\bibfnamefont{R.~H.} \bibnamefont{Ritchie}},
  \bibinfo{year}{1985}{\natexlab{a}}, \bibinfo{journal}{J.\ Phys.\ C}
  \textbf{\bibinfo{volume}{18}}, \bibinfo{pages}{5335}.

\bibitem[{\citenamefont{{Garc\'{\i}a-Molina}}
  \emph{et~al.}(1985{\natexlab{b}})\citenamefont{{Garc\'{\i}a-Molina}, {A.
  Gras-Mart\'{\i}}, and Ritchie}}]{GGR1985}
\bibinfo{author}{\bibnamefont{{Garc\'{\i}a-Molina}}, \bibfnamefont{R.}},
  \bibinfo{author}{\bibnamefont{{A. Gras-Mart\'{\i}}}}, and
  \bibinfo{author}{\bibfnamefont{R.~H.} \bibnamefont{Ritchie}},
  \bibinfo{year}{1985}{\natexlab{b}}, \bibinfo{journal}{Phys.\ Rev.\ B}
  \textbf{\bibinfo{volume}{31}}, \bibinfo{pages}{121}.

\bibitem[{\citenamefont{Gemmell} \emph{et~al.}(1975)\citenamefont{Gemmell,
  Remillieux, Poizat, Gaillard, Holland, and Vager}}]{GRP1975}
\bibinfo{author}{\bibnamefont{Gemmell}, \bibfnamefont{D.~S.}},
  \bibinfo{author}{\bibfnamefont{J.}~\bibnamefont{Remillieux}},
  \bibinfo{author}{\bibfnamefont{J.~C.} \bibnamefont{Poizat}},
  \bibinfo{author}{\bibfnamefont{M.~J.} \bibnamefont{Gaillard}},
  \bibinfo{author}{\bibfnamefont{R.~E.} \bibnamefont{Holland}}, and
  \bibinfo{author}{\bibfnamefont{Z.}~\bibnamefont{Vager}},
  \bibinfo{year}{1975}, \bibinfo{journal}{Phys.\ Rev.\ Lett.}
  \textbf{\bibinfo{volume}{34}}, \bibinfo{pages}{1420}.

\bibitem[{\citenamefont{Genet and Ebbesen}(2007)}]{GE07}
\bibinfo{author}{\bibnamefont{Genet}, \bibfnamefont{C.}}, and
  \bibinfo{author}{\bibfnamefont{T.~W.} \bibnamefont{Ebbesen}},
  \bibinfo{year}{2007}, \bibinfo{journal}{Nature}
  \textbf{\bibinfo{volume}{445}}, \bibinfo{pages}{39}.

\bibitem[{\citenamefont{Gensterblum}
  \emph{et~al.}(1991)\citenamefont{Gensterblum, Pireaux, Thiry, Caudano,
  Vigneron, Lambin, Lucas, and {Kr\"atschmer}}}]{GPT91}
\bibinfo{author}{\bibnamefont{Gensterblum}, \bibfnamefont{G.}},
  \bibinfo{author}{\bibfnamefont{J.~J.} \bibnamefont{Pireaux}},
  \bibinfo{author}{\bibfnamefont{P.~A.} \bibnamefont{Thiry}},
  \bibinfo{author}{\bibfnamefont{R.}~\bibnamefont{Caudano}},
  \bibinfo{author}{\bibfnamefont{J.~P.} \bibnamefont{Vigneron}},
  \bibinfo{author}{\bibfnamefont{P.}~\bibnamefont{Lambin}},
  \bibinfo{author}{\bibfnamefont{A.~A.} \bibnamefont{Lucas}}, and
  \bibinfo{author}{\bibfnamefont{W.}~\bibnamefont{{Kr\"atschmer}}},
  \bibinfo{year}{1991}, \bibinfo{journal}{Phys.\ Rev.\ Lett.}
  \textbf{\bibinfo{volume}{67}}, \bibinfo{pages}{2171}.

\bibitem[{\citenamefont{Ginzburg}(1996)}]{G96}
\bibinfo{author}{\bibnamefont{Ginzburg}, \bibfnamefont{V.~L.}},
  \bibinfo{year}{1996}, \bibinfo{journal}{Phys.\ Usp.}
  \textbf{\bibinfo{volume}{39}}, \bibinfo{pages}{973}.

\bibitem[{\citenamefont{Ginzburg and Frank}(1946)}]{GF1946}
\bibinfo{author}{\bibnamefont{Ginzburg}, \bibfnamefont{V.~L.}}, and
  \bibinfo{author}{\bibfnamefont{I.~M.} \bibnamefont{Frank}},
  \bibinfo{year}{1946}, \bibinfo{journal}{Zh.\ Eksp.\ Theor.\ Fiz. [Sov.\
  Phys.\ JETP]} \textbf{\bibinfo{volume}{16}}, \bibinfo{pages}{15}.

\bibitem[{\citenamefont{Goldobin} \emph{et~al.}(1998)\citenamefont{Goldobin,
  Wallraff, Thyssen, and Ustinov}}]{GWT98}
\bibinfo{author}{\bibnamefont{Goldobin}, \bibfnamefont{E.}},
  \bibinfo{author}{\bibfnamefont{A.}~\bibnamefont{Wallraff}},
  \bibinfo{author}{\bibfnamefont{N.}~\bibnamefont{Thyssen}}, and
  \bibinfo{author}{\bibfnamefont{A.~V.} \bibnamefont{Ustinov}},
  \bibinfo{year}{1998}, \bibinfo{journal}{Phys.\ Rev.\ B}
  \textbf{\bibinfo{volume}{57}}, \bibinfo{pages}{130}.

\bibitem[{\citenamefont{Goldsmith and Jelley}(1959)}]{GJ1959}
\bibinfo{author}{\bibnamefont{Goldsmith}, \bibfnamefont{P.}}, and
  \bibinfo{author}{\bibfnamefont{J.~V.} \bibnamefont{Jelley}},
  \bibinfo{year}{1959}, \bibinfo{journal}{Philos.\ Mag.}
  \textbf{\bibinfo{volume}{4}}, \bibinfo{pages}{836}.

\bibitem[{\citenamefont{{G\'omez-Medina}}
  \emph{et~al.}(2008)\citenamefont{{G\'omez-Medina}, Yamamoto, Nakano, and
  {Garc\'{\i}a de Abajo}}}]{paper159}
\bibinfo{author}{\bibnamefont{{G\'omez-Medina}}, \bibfnamefont{R.}},
  \bibinfo{author}{\bibfnamefont{N.}~\bibnamefont{Yamamoto}},
  \bibinfo{author}{\bibfnamefont{M.}~\bibnamefont{Nakano}}, and
  \bibinfo{author}{\bibfnamefont{F.~J.} \bibnamefont{{Garc\'{\i}a de Abajo}}},
  \bibinfo{year}{2008}, \bibinfo{journal}{New\ J.\ Phys.}
  \textbf{\bibinfo{volume}{10}}, \bibinfo{pages}{105009}.

\bibitem[{\citenamefont{Gordon and Ashkin}(1980)}]{GA1980}
\bibinfo{author}{\bibnamefont{Gordon}, \bibfnamefont{J.~P.}}, and
  \bibinfo{author}{\bibfnamefont{A.}~\bibnamefont{Ashkin}},
  \bibinfo{year}{1980}, \bibinfo{journal}{Phys.\ Rev.\ A}
  \textbf{\bibinfo{volume}{21}}, \bibinfo{pages}{1606}.

\bibitem[{\citenamefont{Grier}(2003)}]{G03}
\bibinfo{author}{\bibnamefont{Grier}, \bibfnamefont{D.~G.}},
  \bibinfo{year}{2003}, \bibinfo{journal}{Nature}
  \textbf{\bibinfo{volume}{424}}, \bibinfo{pages}{810}.

\bibitem[{\citenamefont{Grinolds} \emph{et~al.}(2006)\citenamefont{Grinolds,
  Lobastov, Weissenrieder, and Zewail}}]{GLW06}
\bibinfo{author}{\bibnamefont{Grinolds}, \bibfnamefont{M.~S.}},
  \bibinfo{author}{\bibfnamefont{V.~A.} \bibnamefont{Lobastov}},
  \bibinfo{author}{\bibfnamefont{J.}~\bibnamefont{Weissenrieder}}, and
  \bibinfo{author}{\bibfnamefont{A.~H.} \bibnamefont{Zewail}},
  \bibinfo{year}{2006}, \bibinfo{journal}{Proc.\ Natl.\ Academ.\ Sci.}
  \textbf{\bibinfo{volume}{103}}, \bibinfo{pages}{18427}.

\bibitem[{\citenamefont{Grundmann} \emph{et~al.}(1995)\citenamefont{Grundmann,
  Christen, Ledentsov, B\"{o}hrer, Bimberg, Ruvimov, Werner, Richter,
  G\"{o}sele, Heydenreich, Ustinov, Egorov} \emph{et~al.}}]{GCL95}
\bibinfo{author}{\bibnamefont{Grundmann}, \bibfnamefont{M.}},
  \bibinfo{author}{\bibfnamefont{J.}~\bibnamefont{Christen}},
  \bibinfo{author}{\bibfnamefont{N.~N.} \bibnamefont{Ledentsov}},
  \bibinfo{author}{\bibfnamefont{J.}~\bibnamefont{B\"{o}hrer}},
  \bibinfo{author}{\bibfnamefont{D.}~\bibnamefont{Bimberg}},
  \bibinfo{author}{\bibfnamefont{S.~S.} \bibnamefont{Ruvimov}},
  \bibinfo{author}{\bibfnamefont{P.}~\bibnamefont{Werner}},
  \bibinfo{author}{\bibfnamefont{U.}~\bibnamefont{Richter}},
  \bibinfo{author}{\bibfnamefont{U.}~\bibnamefont{G\"{o}sele}},
  \bibinfo{author}{\bibfnamefont{J.}~\bibnamefont{Heydenreich}},
  \bibinfo{author}{\bibfnamefont{V.~M.} \bibnamefont{Ustinov}},
  \bibinfo{author}{\bibfnamefont{A.~Y.} \bibnamefont{Egorov}}, \emph{et~al.},
  \bibinfo{year}{1995}, \bibinfo{journal}{Phys.\ Rev.\ Lett.}
  \textbf{\bibinfo{volume}{74}}, \bibinfo{pages}{4043}.

\bibitem[{\citenamefont{Grzelczak} \emph{et~al.}(2007)\citenamefont{Grzelczak,
  P\'{e}rez-Juste, {Garc\'{\i}a de Abajo}, and Liz-Marz\'{a}n}}]{paper123}
\bibinfo{author}{\bibnamefont{Grzelczak}, \bibfnamefont{M.}},
  \bibinfo{author}{\bibfnamefont{J.}~\bibnamefont{P\'{e}rez-Juste}},
  \bibinfo{author}{\bibfnamefont{F.~J.} \bibnamefont{{Garc\'{\i}a de Abajo}}},
  and \bibinfo{author}{\bibfnamefont{L.}~\bibnamefont{Liz-Marz\'{a}n}},
  \bibinfo{year}{2007}, \bibinfo{journal}{J.\ Phys.\ Chem.\ C}
  \textbf{\bibinfo{volume}{111}}, \bibinfo{pages}{6183}.

\bibitem[{\citenamefont{Gu} \emph{et~al.}(2007)\citenamefont{Gu, Srot, Sigle,
  Koch, {van Aken}, Scholz, Thapa, Kirchner, Jetter, and R\"uhle}}]{GSS07}
\bibinfo{author}{\bibnamefont{Gu}, \bibfnamefont{L.}},
  \bibinfo{author}{\bibfnamefont{V.}~\bibnamefont{Srot}},
  \bibinfo{author}{\bibfnamefont{W.}~\bibnamefont{Sigle}},
  \bibinfo{author}{\bibfnamefont{C.}~\bibnamefont{Koch}},
  \bibinfo{author}{\bibfnamefont{P.}~\bibnamefont{{van Aken}}},
  \bibinfo{author}{\bibfnamefont{F.}~\bibnamefont{Scholz}},
  \bibinfo{author}{\bibfnamefont{S.~B.} \bibnamefont{Thapa}},
  \bibinfo{author}{\bibfnamefont{C.}~\bibnamefont{Kirchner}},
  \bibinfo{author}{\bibfnamefont{M.}~\bibnamefont{Jetter}}, and
  \bibinfo{author}{\bibfnamefont{M.}~\bibnamefont{R\"uhle}},
  \bibinfo{year}{2007}, \bibinfo{journal}{Phys.\ Rev.\ B}
  \textbf{\bibinfo{volume}{75}}, \bibinfo{pages}{195214}.

\bibitem[{\citenamefont{Guan} \emph{et~al.}(2005)\citenamefont{Guan, Suenaga,
  Shi, Gu, and Iijima}}]{GSS05}
\bibinfo{author}{\bibnamefont{Guan}, \bibfnamefont{L.}},
  \bibinfo{author}{\bibfnamefont{K.}~\bibnamefont{Suenaga}},
  \bibinfo{author}{\bibfnamefont{Z.}~\bibnamefont{Shi}},
  \bibinfo{author}{\bibfnamefont{Z.}~\bibnamefont{Gu}}, and
  \bibinfo{author}{\bibfnamefont{S.}~\bibnamefont{Iijima}},
  \bibinfo{year}{2005}, \bibinfo{journal}{Phys.\ Rev.\ Lett.}
  \textbf{\bibinfo{volume}{94}}, \bibinfo{pages}{045502}.

\bibitem[{\citenamefont{Guck} \emph{et~al.}(2000)\citenamefont{Guck,
  Ananthakrishnan, Moon, Cunningham, and {J. K\"{a}s}}}]{GAM00}
\bibinfo{author}{\bibnamefont{Guck}, \bibfnamefont{J.}},
  \bibinfo{author}{\bibfnamefont{R.}~\bibnamefont{Ananthakrishnan}},
  \bibinfo{author}{\bibfnamefont{T.~J.} \bibnamefont{Moon}},
  \bibinfo{author}{\bibfnamefont{C.~C.} \bibnamefont{Cunningham}}, and
  \bibinfo{author}{\bibnamefont{{J. K\"{a}s}}}, \bibinfo{year}{2000},
  \bibinfo{journal}{Phys.\ Rev.\ Lett.} \textbf{\bibinfo{volume}{84}},
  \bibinfo{pages}{5451}.

\bibitem[{\citenamefont{{Haeberl\'{e}}}
  \emph{et~al.}(1994)\citenamefont{{Haeberl\'{e}}, Rullhusen, {Salom\'{e}}, and
  Maene}}]{HRS94}
\bibinfo{author}{\bibnamefont{{Haeberl\'{e}}}, \bibfnamefont{O.}},
  \bibinfo{author}{\bibfnamefont{P.}~\bibnamefont{Rullhusen}},
  \bibinfo{author}{\bibfnamefont{J.~M.} \bibnamefont{{Salom\'{e}}}}, and
  \bibinfo{author}{\bibfnamefont{N.}~\bibnamefont{Maene}},
  \bibinfo{year}{1994}, \bibinfo{journal}{Phys.\ Rev.\ E}
  \textbf{\bibinfo{volume}{49}}, \bibinfo{pages}{3340}.

\bibitem[{\citenamefont{Hanrath and Korgel}(2004)}]{HK04}
\bibinfo{author}{\bibnamefont{Hanrath}, \bibfnamefont{T.}}, and
  \bibinfo{author}{\bibfnamefont{B.~A.} \bibnamefont{Korgel}},
  \bibinfo{year}{2004}, \bibinfo{journal}{Nano\ Lett.}
  \textbf{\bibinfo{volume}{4}}, \bibinfo{pages}{1455}.

\bibitem[{\citenamefont{Hartschuh}(2008)}]{H08}
\bibinfo{author}{\bibnamefont{Hartschuh}, \bibfnamefont{A.}},
  \bibinfo{year}{2008}, \bibinfo{journal}{Ang.\ Chem.\ Intl.\ Ed.}
  \textbf{\bibinfo{volume}{47}}, \bibinfo{pages}{8178}.

\bibitem[{\citenamefont{Hattendorff}(1977)}]{H1977_2}
\bibinfo{author}{\bibnamefont{Hattendorff}, \bibfnamefont{H.~D.}},
  \bibinfo{year}{1977}, \bibinfo{journal}{Phys.\ Stat.\ Sol.\ (a)}
  \textbf{\bibinfo{volume}{42}}, \bibinfo{pages}{489}.

\bibitem[{\citenamefont{Haydock}(1980)}]{H1980_2}
\bibinfo{author}{\bibnamefont{Haydock}, \bibfnamefont{R.}},
  \bibinfo{year}{1980}, \bibinfo{journal}{Solid\ State\ Physics}
  \textbf{\bibinfo{volume}{35}}, \bibinfo{pages}{215}.

\bibitem[{\citenamefont{Heitmann}(1977)}]{H1977}
\bibinfo{author}{\bibnamefont{Heitmann}, \bibfnamefont{D.}},
  \bibinfo{year}{1977}, \bibinfo{journal}{J.\ Phys.\ C}
  \textbf{\bibinfo{volume}{10}}, \bibinfo{pages}{397}.

\bibitem[{\citenamefont{Henrard and Lambin}(1996)}]{HL96}
\bibinfo{author}{\bibnamefont{Henrard}, \bibfnamefont{L.}}, and
  \bibinfo{author}{\bibfnamefont{P.}~\bibnamefont{Lambin}},
  \bibinfo{year}{1996}, \bibinfo{journal}{J.\ Phys.\ B}
  \textbf{\bibinfo{volume}{29}}, \bibinfo{pages}{5127}.

\bibitem[{\citenamefont{Henrard} \emph{et~al.}(1999)\citenamefont{Henrard,
  Malengreau, Rudolf, Hevesi, Caudano, Lambin, and {Cabioc'h}}}]{HMR99}
\bibinfo{author}{\bibnamefont{Henrard}, \bibfnamefont{L.}},
  \bibinfo{author}{\bibfnamefont{F.}~\bibnamefont{Malengreau}},
  \bibinfo{author}{\bibfnamefont{P.}~\bibnamefont{Rudolf}},
  \bibinfo{author}{\bibfnamefont{K.}~\bibnamefont{Hevesi}},
  \bibinfo{author}{\bibfnamefont{R.}~\bibnamefont{Caudano}},
  \bibinfo{author}{\bibfnamefont{P.}~\bibnamefont{Lambin}}, and
  \bibinfo{author}{\bibfnamefont{T.}~\bibnamefont{{Cabioc'h}}},
  \bibinfo{year}{1999}, \bibinfo{journal}{Phys.\ Rev.\ B}
  \textbf{\bibinfo{volume}{59}}, \bibinfo{pages}{5832}.

\bibitem[{\citenamefont{Herring}(2005)}]{H05}
\bibinfo{author}{\bibnamefont{Herring}, \bibfnamefont{R.~A.}},
  \bibinfo{year}{2005}, \bibinfo{journal}{Ultramicroscopy}
  \textbf{\bibinfo{volume}{104}}, \bibinfo{pages}{261–}.

\bibitem[{\citenamefont{Herring}(2008)}]{H08_2}
\bibinfo{author}{\bibnamefont{Herring}, \bibfnamefont{R.~A.}},
  \bibinfo{year}{2008}, \bibinfo{journal}{Ultramicroscopy}
  \textbf{\bibinfo{volume}{108}}, \bibinfo{pages}{688}.

\bibitem[{\citenamefont{Hillier and Baker}(1944)}]{HB1944}
\bibinfo{author}{\bibnamefont{Hillier}, \bibfnamefont{J.}}, and
  \bibinfo{author}{\bibfnamefont{R.~F.} \bibnamefont{Baker}},
  \bibinfo{year}{1944}, \bibinfo{journal}{J.\ Appl.\ Phys.}
  \textbf{\bibinfo{volume}{15}}, \bibinfo{pages}{663}.

\bibitem[{\citenamefont{Hofmann} \emph{et~al.}(2007)\citenamefont{Hofmann,
  Vesseur, Sweatlock, Lezec, {Garc\'{\i}a de Abajo}, Polman, and
  Atwater}}]{paper139}
\bibinfo{author}{\bibnamefont{Hofmann}, \bibfnamefont{C.~E.}},
  \bibinfo{author}{\bibfnamefont{E.~J.~R.} \bibnamefont{Vesseur}},
  \bibinfo{author}{\bibfnamefont{L.~A.} \bibnamefont{Sweatlock}},
  \bibinfo{author}{\bibfnamefont{H.~J.} \bibnamefont{Lezec}},
  \bibinfo{author}{\bibfnamefont{F.~J.} \bibnamefont{{Garc\'{\i}a de Abajo}}},
  \bibinfo{author}{\bibfnamefont{A.}~\bibnamefont{Polman}}, and
  \bibinfo{author}{\bibfnamefont{H.~A.} \bibnamefont{Atwater}},
  \bibinfo{year}{2007}, \bibinfo{journal}{Nano\ Lett.}
  \textbf{\bibinfo{volume}{7}}, \bibinfo{pages}{3612}.

\bibitem[{\citenamefont{Hohenberg and Kohn}(1964)}]{HK1964}
\bibinfo{author}{\bibnamefont{Hohenberg}, \bibfnamefont{P.}}, and
  \bibinfo{author}{\bibfnamefont{W.}~\bibnamefont{Kohn}}, \bibinfo{year}{1964},
  \bibinfo{journal}{Phys.\ Rev.} \textbf{\bibinfo{volume}{136}},
  \bibinfo{pages}{B864}.

\bibitem[{\citenamefont{Hohenester and Krenn}(2005)}]{HK05}
\bibinfo{author}{\bibnamefont{Hohenester}, \bibfnamefont{U.}}, and
  \bibinfo{author}{\bibfnamefont{J.}~\bibnamefont{Krenn}},
  \bibinfo{year}{2005}, \bibinfo{journal}{Phys.\ Rev.\ B}
  \textbf{\bibinfo{volume}{72}}, \bibinfo{pages}{195429}.

\bibitem[{\citenamefont{Holt and Yacobi}(1990)}]{HY1990}
\bibinfo{author}{\bibnamefont{Holt}, \bibfnamefont{D.~B.}}, and
  \bibinfo{author}{\bibfnamefont{B.~G.} \bibnamefont{Yacobi}},
  \bibinfo{year}{1990}, \emph{\bibinfo{title}{Cathodoluminescence Microscopy of
  Inorganic Solids}} (\bibinfo{publisher}{Plenum}, \bibinfo{address}{New
  York}).

\bibitem[{\citenamefont{Horiuchi} \emph{et~al.}(2006)\citenamefont{Horiuchi,
  Ochiai, Inoue, Segawa, Shibata, Ishi, Kondo, Kanbe, Miyazaki, Hinode,
  Yamaguti, and Ohtaka}}]{HOY06}
\bibinfo{author}{\bibnamefont{Horiuchi}, \bibfnamefont{N.}},
  \bibinfo{author}{\bibfnamefont{T.}~\bibnamefont{Ochiai}},
  \bibinfo{author}{\bibfnamefont{J.}~\bibnamefont{Inoue}},
  \bibinfo{author}{\bibfnamefont{Y.}~\bibnamefont{Segawa}},
  \bibinfo{author}{\bibfnamefont{Y.}~\bibnamefont{Shibata}},
  \bibinfo{author}{\bibfnamefont{K.}~\bibnamefont{Ishi}},
  \bibinfo{author}{\bibfnamefont{Y.}~\bibnamefont{Kondo}},
  \bibinfo{author}{\bibfnamefont{M.}~\bibnamefont{Kanbe}},
  \bibinfo{author}{\bibfnamefont{H.}~\bibnamefont{Miyazaki}},
  \bibinfo{author}{\bibfnamefont{F.}~\bibnamefont{Hinode}},
  \bibinfo{author}{\bibfnamefont{S.}~\bibnamefont{Yamaguti}}, and
  \bibinfo{author}{\bibfnamefont{K.}~\bibnamefont{Ohtaka}},
  \bibinfo{year}{2006}, \bibinfo{journal}{Phys.\ Rev.\ E}
  \textbf{\bibinfo{volume}{74}}, \bibinfo{pages}{056601}.

\bibitem[{\citenamefont{{H\"ormandinger}}(1994)}]{H94}
\bibinfo{author}{\bibnamefont{{H\"ormandinger}}, \bibfnamefont{G.}},
  \bibinfo{year}{1994}, \bibinfo{journal}{Phys.\ Rev.\ B}
  \textbf{\bibinfo{volume}{49}}, \bibinfo{pages}{13897}.

\bibitem[{\citenamefont{H\"ovel} \emph{et~al.}(1993)\citenamefont{H\"ovel,
  Fritz, Hilger, Kreibig, and Vollmer}}]{HFH93}
\bibinfo{author}{\bibnamefont{H\"ovel}, \bibfnamefont{H.}},
  \bibinfo{author}{\bibfnamefont{S.}~\bibnamefont{Fritz}},
  \bibinfo{author}{\bibfnamefont{A.}~\bibnamefont{Hilger}},
  \bibinfo{author}{\bibfnamefont{U.}~\bibnamefont{Kreibig}}, and
  \bibinfo{author}{\bibfnamefont{M.}~\bibnamefont{Vollmer}},
  \bibinfo{year}{1993}, \bibinfo{journal}{Phys.\ Rev.\ B}
  \textbf{\bibinfo{volume}{48}}, \bibinfo{pages}{18178}.

\bibitem[{\citenamefont{Howie}(1999)}]{H99}
\bibinfo{author}{\bibnamefont{Howie}, \bibfnamefont{A.}}, \bibinfo{year}{1999},
  \bibinfo{journal}{Inst. Phys. Conf. Ser.} \textbf{\bibinfo{volume}{161}},
  \bibinfo{pages}{311}.

\bibitem[{\citenamefont{Howie}(2003)}]{H03}
\bibinfo{author}{\bibnamefont{Howie}, \bibfnamefont{A.}}, \bibinfo{year}{2003},
  \bibinfo{journal}{Micron} \textbf{\bibinfo{volume}{34}},
  \bibinfo{pages}{121–}.

\bibitem[{\citenamefont{Howie and Milne}(1985)}]{HM1985}
\bibinfo{author}{\bibnamefont{Howie}, \bibfnamefont{A.}}, and
  \bibinfo{author}{\bibfnamefont{R.~H.} \bibnamefont{Milne}},
  \bibinfo{year}{1985}, \bibinfo{journal}{Ultramicroscopy}
  \textbf{\bibinfo{volume}{18}}, \bibinfo{pages}{427}.

\bibitem[{\citenamefont{Howie and Walsh}(1991)}]{HW91}
\bibinfo{author}{\bibnamefont{Howie}, \bibfnamefont{A.}}, and
  \bibinfo{author}{\bibfnamefont{C.}~\bibnamefont{Walsh}},
  \bibinfo{year}{1991}, \bibinfo{journal}{Microsc.\ Microanal.\ Microstruct.}
  \textbf{\bibinfo{volume}{2}}, \bibinfo{pages}{171}.

\bibitem[{\citenamefont{Hyun} \emph{et~al.}(2008)\citenamefont{Hyun, Couillard,
  Rajendran, Liddell, and Muller}}]{HCR08}
\bibinfo{author}{\bibnamefont{Hyun}, \bibfnamefont{J.~K.}},
  \bibinfo{author}{\bibfnamefont{M.}~\bibnamefont{Couillard}},
  \bibinfo{author}{\bibfnamefont{P.}~\bibnamefont{Rajendran}},
  \bibinfo{author}{\bibfnamefont{C.~M.} \bibnamefont{Liddell}}, and
  \bibinfo{author}{\bibfnamefont{D.~A.} \bibnamefont{Muller}},
  \bibinfo{year}{2008}, \bibinfo{journal}{Appl.\ Phys.\ Lett.}
  \textbf{\bibinfo{volume}{93}}, \bibinfo{pages}{243106}.

\bibitem[{\citenamefont{Iijima}(1991)}]{I91}
\bibinfo{author}{\bibnamefont{Iijima}, \bibfnamefont{S.}},
  \bibinfo{year}{1991}, \bibinfo{journal}{Nature}
  \textbf{\bibinfo{volume}{354}}, \bibinfo{pages}{56}.

\bibitem[{\citenamefont{Illman} \emph{et~al.}(1988)\citenamefont{Illman,
  Anderson, Warmack, and Ferrell}}]{IAW1988}
\bibinfo{author}{\bibnamefont{Illman}, \bibfnamefont{B.~L.}},
  \bibinfo{author}{\bibfnamefont{V.~E.} \bibnamefont{Anderson}},
  \bibinfo{author}{\bibfnamefont{R.~J.} \bibnamefont{Warmack}}, and
  \bibinfo{author}{\bibfnamefont{T.~L.} \bibnamefont{Ferrell}},
  \bibinfo{year}{1988}, \bibinfo{journal}{Phys.\ Rev.\ B}
  \textbf{\bibinfo{volume}{38}}, \bibinfo{pages}{3045}.

\bibitem[{\citenamefont{Imbusch and Niedrig}(1970)}]{IN1970}
\bibinfo{author}{\bibnamefont{Imbusch}, \bibfnamefont{A.}}, and
  \bibinfo{author}{\bibfnamefont{H.}~\bibnamefont{Niedrig}},
  \bibinfo{year}{1970}, \bibinfo{journal}{Phys.\ Lett.\ A}
  \textbf{\bibinfo{volume}{32}}, \bibinfo{pages}{375}.

\bibitem[{\citenamefont{Intaraprasonk}
  \emph{et~al.}(2008)\citenamefont{Intaraprasonk, Xin, and Muller}}]{IXM08}
\bibinfo{author}{\bibnamefont{Intaraprasonk}, \bibfnamefont{V.}},
  \bibinfo{author}{\bibfnamefont{H.~L.} \bibnamefont{Xin}}, and
  \bibinfo{author}{\bibfnamefont{D.~A.} \bibnamefont{Muller}},
  \bibinfo{year}{2008}, \bibinfo{journal}{Ultramicroscopy}
  \textbf{\bibinfo{volume}{108}}, \bibinfo{pages}{1454}.

\bibitem[{\citenamefont{Ireland} \emph{et~al.}(2008)\citenamefont{Ireland,
  Kraus, Martinache, Lloyd, and Tuthill}}]{LMI06}
\bibinfo{author}{\bibnamefont{Ireland}, \bibfnamefont{M.~J.}},
  \bibinfo{author}{\bibfnamefont{A.}~\bibnamefont{Kraus}},
  \bibinfo{author}{\bibfnamefont{F.}~\bibnamefont{Martinache}},
  \bibinfo{author}{\bibfnamefont{J.~P.} \bibnamefont{Lloyd}}, and
  \bibinfo{author}{\bibfnamefont{P.~G.} \bibnamefont{Tuthill}},
  \bibinfo{year}{2008}, \bibinfo{journal}{Astrophys.\ J.}
  \textbf{\bibinfo{volume}{678}}, \bibinfo{pages}{463–}.

\bibitem[{\citenamefont{Itskovsky} \emph{et~al.}(2008)\citenamefont{Itskovsky,
  Cohen, and Maniv}}]{ICM08}
\bibinfo{author}{\bibnamefont{Itskovsky}, \bibfnamefont{M.~A.}},
  \bibinfo{author}{\bibfnamefont{H.}~\bibnamefont{Cohen}}, and
  \bibinfo{author}{\bibfnamefont{T.}~\bibnamefont{Maniv}},
  \bibinfo{year}{2008}, \bibinfo{journal}{Phys.\ Rev.\ B}
  \textbf{\bibinfo{volume}{78}}, \bibinfo{pages}{045419}.

\bibitem[{\citenamefont{Jackson}(1999)}]{J99}
\bibinfo{author}{\bibnamefont{Jackson}, \bibfnamefont{J.~D.}},
  \bibinfo{year}{1999}, \emph{\bibinfo{title}{Classical Electrodynamics}}
  (\bibinfo{publisher}{Wiley}, \bibinfo{address}{New York}).

\bibitem[{\citenamefont{Jiang} \emph{et~al.}(2008)\citenamefont{Jiang, Su,
  Spence, Zhou, and Qiu}}]{JSS08}
\bibinfo{author}{\bibnamefont{Jiang}, \bibfnamefont{N.}},
  \bibinfo{author}{\bibfnamefont{D.}~\bibnamefont{Su}},
  \bibinfo{author}{\bibfnamefont{J.~C.~H.} \bibnamefont{Spence}},
  \bibinfo{author}{\bibfnamefont{S.}~\bibnamefont{Zhou}}, and
  \bibinfo{author}{\bibfnamefont{J.}~\bibnamefont{Qiu}}, \bibinfo{year}{2008},
  \bibinfo{journal}{J.\ Mater.\ Res.} \textbf{\bibinfo{volume}{23}},
  \bibinfo{pages}{2467}.

\bibitem[{\citenamefont{Johnson and Christy}(1972)}]{JC1972}
\bibinfo{author}{\bibnamefont{Johnson}, \bibfnamefont{P.~B.}}, and
  \bibinfo{author}{\bibfnamefont{R.~W.} \bibnamefont{Christy}},
  \bibinfo{year}{1972}, \bibinfo{journal}{Phys.\ Rev.\ B}
  \textbf{\bibinfo{volume}{6}}, \bibinfo{pages}{4370}.

\bibitem[{\citenamefont{Jones}(1945)}]{J1945}
\bibinfo{author}{\bibnamefont{Jones}, \bibfnamefont{R.~C.}},
  \bibinfo{year}{1945}, \bibinfo{journal}{Phys.\ Rev.}
  \textbf{\bibinfo{volume}{68}}, \bibinfo{pages}{93}.

\bibitem[{\citenamefont{Jouffrey} \emph{et~al.}(2004)\citenamefont{Jouffrey,
  Schattschneider, and {H\'ebert}}}]{JSH04}
\bibinfo{author}{\bibnamefont{Jouffrey}, \bibfnamefont{B.}},
  \bibinfo{author}{\bibfnamefont{P.}~\bibnamefont{Schattschneider}}, and
  \bibinfo{author}{\bibfnamefont{C.}~\bibnamefont{{H\'ebert}}},
  \bibinfo{year}{2004}, \bibinfo{journal}{Ultramicroscopy}
  \textbf{\bibinfo{volume}{102}}, \bibinfo{pages}{61}.

\bibitem[{\citenamefont{Joulain} \emph{et~al.}(2003)\citenamefont{Joulain,
  Carminati, Mulet, and Greffet}}]{JCM03}
\bibinfo{author}{\bibnamefont{Joulain}, \bibfnamefont{K.}},
  \bibinfo{author}{\bibfnamefont{R.}~\bibnamefont{Carminati}},
  \bibinfo{author}{\bibfnamefont{J.~P.} \bibnamefont{Mulet}}, and
  \bibinfo{author}{\bibfnamefont{J.~J.} \bibnamefont{Greffet}},
  \bibinfo{year}{2003}, \bibinfo{journal}{Phys.\ Rev.\ B}
  \textbf{\bibinfo{volume}{68}}, \bibinfo{pages}{245405}.

\bibitem[{\citenamefont{Julley}(1958)}]{J1958_2}
\bibinfo{author}{\bibnamefont{Julley}, \bibfnamefont{J.~V.}},
  \bibinfo{year}{1958}, \emph{\bibinfo{title}{Cherenkov Radiation and its
  Application}} (\bibinfo{publisher}{Pergamon, New York}).

\bibitem[{\citenamefont{Karali} \emph{et~al.}(2005)\citenamefont{Karali, Can,
  Valberg, Stepanov, Townsend, Buchal, Ganeev, Ryasnyansky, Belik, Jessett, and
  Ong}}]{KCV05}
\bibinfo{author}{\bibnamefont{Karali}, \bibfnamefont{T.}},
  \bibinfo{author}{\bibfnamefont{N.}~\bibnamefont{Can}},
  \bibinfo{author}{\bibfnamefont{L.}~\bibnamefont{Valberg}},
  \bibinfo{author}{\bibfnamefont{A.~L.} \bibnamefont{Stepanov}},
  \bibinfo{author}{\bibfnamefont{P.~D.} \bibnamefont{Townsend}},
  \bibinfo{author}{\bibfnamefont{C.}~\bibnamefont{Buchal}},
  \bibinfo{author}{\bibfnamefont{R.~A.} \bibnamefont{Ganeev}},
  \bibinfo{author}{\bibfnamefont{A.~I.} \bibnamefont{Ryasnyansky}},
  \bibinfo{author}{\bibfnamefont{H.~G.} \bibnamefont{Belik}},
  \bibinfo{author}{\bibfnamefont{M.~L.} \bibnamefont{Jessett}}, and
  \bibinfo{author}{\bibfnamefont{C.}~\bibnamefont{Ong}}, \bibinfo{year}{2005},
  \bibinfo{journal}{Physica B} \textbf{\bibinfo{volume}{363}},
  \bibinfo{pages}{88–}.

\bibitem[{\citenamefont{Karataev} \emph{et~al.}(2004)\citenamefont{Karataev,
  Araki, Hamatsu, Hayano, Muto, Naumenko, Potylitsyn, Terunuma, and
  Urakawa}}]{KAH04}
\bibinfo{author}{\bibnamefont{Karataev}, \bibfnamefont{P.}},
  \bibinfo{author}{\bibfnamefont{S.}~\bibnamefont{Araki}},
  \bibinfo{author}{\bibfnamefont{R.}~\bibnamefont{Hamatsu}},
  \bibinfo{author}{\bibfnamefont{H.}~\bibnamefont{Hayano}},
  \bibinfo{author}{\bibfnamefont{T.}~\bibnamefont{Muto}},
  \bibinfo{author}{\bibfnamefont{G.}~\bibnamefont{Naumenko}},
  \bibinfo{author}{\bibfnamefont{A.}~\bibnamefont{Potylitsyn}},
  \bibinfo{author}{\bibfnamefont{N.}~\bibnamefont{Terunuma}}, and
  \bibinfo{author}{\bibfnamefont{J.}~\bibnamefont{Urakawa}},
  \bibinfo{year}{2004}, \bibinfo{journal}{Phys.\ Rev.\ Lett.}
  \textbf{\bibinfo{volume}{93}}, \bibinfo{pages}{244802}.

\bibitem[{\citenamefont{Keast and Bosman}(2008)}]{KB08}
\bibinfo{author}{\bibnamefont{Keast}, \bibfnamefont{V.~J.}}, and
  \bibinfo{author}{\bibfnamefont{M.}~\bibnamefont{Bosman}},
  \bibinfo{year}{2008}, \bibinfo{journal}{Mater.\ Sci.\ Tech.}
  \textbf{\bibinfo{volume}{24}}, \bibinfo{pages}{651}.

\bibitem[{\citenamefont{Keller and Coplan}(1992)}]{KC92}
\bibinfo{author}{\bibnamefont{Keller}, \bibfnamefont{J.~W.}}, and
  \bibinfo{author}{\bibfnamefont{M.~A.} \bibnamefont{Coplan}},
  \bibinfo{year}{1992}, \bibinfo{journal}{Chem.\ Phys.\ Lett.}
  \textbf{\bibinfo{volume}{193}}, \bibinfo{pages}{89}.

\bibitem[{\citenamefont{Kelly} \emph{et~al.}(2003)\citenamefont{Kelly,
  Coronado, Zhao, and Schatz}}]{KCZ03}
\bibinfo{author}{\bibnamefont{Kelly}, \bibfnamefont{K.~L.}},
  \bibinfo{author}{\bibfnamefont{E.}~\bibnamefont{Coronado}},
  \bibinfo{author}{\bibfnamefont{L.~L.} \bibnamefont{Zhao}}, and
  \bibinfo{author}{\bibfnamefont{G.~C.} \bibnamefont{Schatz}},
  \bibinfo{year}{2003}, \bibinfo{journal}{J.\ Phys.\ Chem.\ B}
  \textbf{\bibinfo{volume}{107}}, \bibinfo{pages}{668}.

\bibitem[{\citenamefont{Kim and Song}(2007)}]{KS07}
\bibinfo{author}{\bibnamefont{Kim}, \bibfnamefont{J.}}, and
  \bibinfo{author}{\bibfnamefont{K.~B.} \bibnamefont{Song}},
  \bibinfo{year}{2007}, \bibinfo{journal}{Micron}
  \textbf{\bibinfo{volume}{38}}, \bibinfo{pages}{409–}.

\bibitem[{\citenamefont{Kimura} \emph{et~al.}(1995)\citenamefont{Kimura, Kim,
  Romea, Steinhauer, Pogorelsky, Kusche, Fernow, Wang, and Liu}}]{KKR95}
\bibinfo{author}{\bibnamefont{Kimura}, \bibfnamefont{W.~D.}},
  \bibinfo{author}{\bibfnamefont{G.~H.} \bibnamefont{Kim}},
  \bibinfo{author}{\bibfnamefont{R.~D.} \bibnamefont{Romea}},
  \bibinfo{author}{\bibfnamefont{L.~C.} \bibnamefont{Steinhauer}},
  \bibinfo{author}{\bibfnamefont{I.~V.} \bibnamefont{Pogorelsky}},
  \bibinfo{author}{\bibfnamefont{K.~P.} \bibnamefont{Kusche}},
  \bibinfo{author}{\bibfnamefont{R.~C.} \bibnamefont{Fernow}},
  \bibinfo{author}{\bibfnamefont{X.}~\bibnamefont{Wang}}, and
  \bibinfo{author}{\bibfnamefont{Y.}~\bibnamefont{Liu}}, \bibinfo{year}{1995},
  \bibinfo{journal}{Phys.\ Rev.\ Lett.} \textbf{\bibinfo{volume}{74}},
  \bibinfo{pages}{546}.

\bibitem[{\citenamefont{Kitson} \emph{et~al.}(1996)\citenamefont{Kitson,
  Barnes, and Sambles}}]{KBS96}
\bibinfo{author}{\bibnamefont{Kitson}, \bibfnamefont{S.~C.}},
  \bibinfo{author}{\bibfnamefont{W.~L.} \bibnamefont{Barnes}}, and
  \bibinfo{author}{\bibfnamefont{J.~R.} \bibnamefont{Sambles}},
  \bibinfo{year}{1996}, \bibinfo{journal}{Phys.\ Rev.\ Lett.}
  \textbf{\bibinfo{volume}{77}}, \bibinfo{pages}{2670}.

\bibitem[{\citenamefont{Kittle}(1987)}]{K1987}
\bibinfo{author}{\bibnamefont{Kittle}, \bibfnamefont{C.}},
  \bibinfo{year}{1987}, \emph{\bibinfo{title}{Quantum theory of solids}}
  (\bibinfo{publisher}{Wiley}, \bibinfo{address}{New York}).

\bibitem[{\citenamefont{Kliewer and Fuchs}(1967)}]{KF1967}
\bibinfo{author}{\bibnamefont{Kliewer}, \bibfnamefont{K.~L.}}, and
  \bibinfo{author}{\bibfnamefont{R.}~\bibnamefont{Fuchs}},
  \bibinfo{year}{1967}, \bibinfo{journal}{Phys.\ Rev.}
  \textbf{\bibinfo{volume}{153}}, \bibinfo{pages}{498}.

\bibitem[{\citenamefont{Kociak} \emph{et~al.}(2000)\citenamefont{Kociak,
  Henrard, {O. St\'{e}phan}, Suenaga, and Colliex}}]{KHS00}
\bibinfo{author}{\bibnamefont{Kociak}, \bibfnamefont{M.}},
  \bibinfo{author}{\bibfnamefont{L.}~\bibnamefont{Henrard}},
  \bibinfo{author}{\bibnamefont{{O. St\'{e}phan}}},
  \bibinfo{author}{\bibfnamefont{K.}~\bibnamefont{Suenaga}}, and
  \bibinfo{author}{\bibfnamefont{C.}~\bibnamefont{Colliex}},
  \bibinfo{year}{2000}, \bibinfo{journal}{Phys.\ Rev.\ B}
  \textbf{\bibinfo{volume}{61}}, \bibinfo{pages}{13936}.

\bibitem[{\citenamefont{Kociak} \emph{et~al.}(2001)\citenamefont{Kociak, {O.
  St\'{e}phan}, Henrard, Charbois, Rothschild, Tenne, and Colliex}}]{KSH01}
\bibinfo{author}{\bibnamefont{Kociak}, \bibfnamefont{M.}},
  \bibinfo{author}{\bibnamefont{{O. St\'{e}phan}}},
  \bibinfo{author}{\bibfnamefont{L.}~\bibnamefont{Henrard}},
  \bibinfo{author}{\bibfnamefont{V.}~\bibnamefont{Charbois}},
  \bibinfo{author}{\bibfnamefont{A.}~\bibnamefont{Rothschild}},
  \bibinfo{author}{\bibfnamefont{R.}~\bibnamefont{Tenne}}, and
  \bibinfo{author}{\bibfnamefont{C.}~\bibnamefont{Colliex}},
  \bibinfo{year}{2001}, \bibinfo{journal}{Phys.\ Rev.\ Lett.}
  \textbf{\bibinfo{volume}{87}}, \bibinfo{pages}{075501}.

\bibitem[{\citenamefont{Kohl}(1983)}]{K1983}
\bibinfo{author}{\bibnamefont{Kohl}, \bibfnamefont{H.}}, \bibinfo{year}{1983},
  \bibinfo{journal}{Ultramicroscopy} \textbf{\bibinfo{volume}{11}},
  \bibinfo{pages}{53}.

\bibitem[{\citenamefont{Koizumi} \emph{et~al.}(2001)\citenamefont{Koizumi,
  Watanabe, Hasegawa, and Kanda}}]{KWH01}
\bibinfo{author}{\bibnamefont{Koizumi}, \bibfnamefont{S.}},
  \bibinfo{author}{\bibfnamefont{K.}~\bibnamefont{Watanabe}},
  \bibinfo{author}{\bibfnamefont{M.}~\bibnamefont{Hasegawa}}, and
  \bibinfo{author}{\bibfnamefont{H.}~\bibnamefont{Kanda}},
  \bibinfo{year}{2001}, \bibinfo{journal}{Science}
  \textbf{\bibinfo{volume}{292}}, \bibinfo{pages}{1899}.

\bibitem[{\citenamefont{Korda} \emph{et~al.}(2002)\citenamefont{Korda, Taylor,
  and Grier}}]{KTG02}
\bibinfo{author}{\bibnamefont{Korda}, \bibfnamefont{P.~T.}},
  \bibinfo{author}{\bibfnamefont{M.~B.} \bibnamefont{Taylor}}, and
  \bibinfo{author}{\bibfnamefont{D.~G.} \bibnamefont{Grier}},
  \bibinfo{year}{2002}, \bibinfo{journal}{Phys.\ Rev.\ Lett.}
  \textbf{\bibinfo{volume}{89}}, \bibinfo{pages}{128301}.

\bibitem[{\citenamefont{Krane and Raether}(1976)}]{KR1976}
\bibinfo{author}{\bibnamefont{Krane}, \bibfnamefont{K.~J.}}, and
  \bibinfo{author}{\bibfnamefont{H.}~\bibnamefont{Raether}},
  \bibinfo{year}{1976}, \bibinfo{journal}{Phys.\ Rev.\ Lett.}
  \textbf{\bibinfo{volume}{37}}, \bibinfo{pages}{1355}.

\bibitem[{\citenamefont{Kreibig and Vollmer}(1995)}]{KV95}
\bibinfo{author}{\bibnamefont{Kreibig}, \bibfnamefont{U.}}, and
  \bibinfo{author}{\bibfnamefont{M.}~\bibnamefont{Vollmer}},
  \bibinfo{year}{1995}, \emph{\bibinfo{title}{Optical Properties of Metal
  Clusters}} (\bibinfo{publisher}{Springer-Verlag}, \bibinfo{address}{Berlin}).

\bibitem[{\citenamefont{Kremers} \emph{et~al.}(2009)\citenamefont{Kremers,
  Chigrin, and Kroha}}]{KCK09}
\bibinfo{author}{\bibnamefont{Kremers}, \bibfnamefont{C.}},
  \bibinfo{author}{\bibfnamefont{D.~N.} \bibnamefont{Chigrin}}, and
  \bibinfo{author}{\bibfnamefont{J.}~\bibnamefont{Kroha}},
  \bibinfo{year}{2009}, \bibinfo{journal}{Phys.\ Rev.\ A}
  \textbf{\bibinfo{volume}{79}}, \bibinfo{pages}{013829}.

\bibitem[{\citenamefont{Krivanek} \emph{et~al.}(1999)\citenamefont{Krivanek,
  Dellby, and Lupini}}]{KDL99}
\bibinfo{author}{\bibnamefont{Krivanek}, \bibfnamefont{O.~L.}},
  \bibinfo{author}{\bibfnamefont{N.}~\bibnamefont{Dellby}}, and
  \bibinfo{author}{\bibfnamefont{A.~R.} \bibnamefont{Lupini}},
  \bibinfo{year}{1999}, \bibinfo{journal}{Ultramicroscopy}
  \textbf{\bibinfo{volume}{78}}, \bibinfo{pages}{1}.

\bibitem[{\citenamefont{Kr\"{o}ger}(1968)}]{K1968_2}
\bibinfo{author}{\bibnamefont{Kr\"{o}ger}, \bibfnamefont{E.}},
  \bibinfo{year}{1968}, \bibinfo{journal}{Z.\ Phys.}
  \textbf{\bibinfo{volume}{216}}, \bibinfo{pages}{115}.

\bibitem[{\citenamefont{Kroto} \emph{et~al.}(1985)\citenamefont{Kroto, Heath,
  {O'Brien}, Curl, and Smalley}}]{KHO1985}
\bibinfo{author}{\bibnamefont{Kroto}, \bibfnamefont{H.~W.}},
  \bibinfo{author}{\bibfnamefont{J.~R.} \bibnamefont{Heath}},
  \bibinfo{author}{\bibfnamefont{S.~C.} \bibnamefont{{O'Brien}}},
  \bibinfo{author}{\bibfnamefont{R.~F.} \bibnamefont{Curl}}, and
  \bibinfo{author}{\bibfnamefont{R.~E.} \bibnamefont{Smalley}},
  \bibinfo{year}{1985}, \bibinfo{journal}{Nature}
  \textbf{\bibinfo{volume}{318}}, \bibinfo{pages}{162}.

\bibitem[{\citenamefont{Kubo} \emph{et~al.}(2005)\citenamefont{Kubo, Onda,
  Petek, Sun, Jung, and Kim}}]{KOP05}
\bibinfo{author}{\bibnamefont{Kubo}, \bibfnamefont{A.}},
  \bibinfo{author}{\bibfnamefont{K.}~\bibnamefont{Onda}},
  \bibinfo{author}{\bibfnamefont{H.}~\bibnamefont{Petek}},
  \bibinfo{author}{\bibfnamefont{Z.}~\bibnamefont{Sun}},
  \bibinfo{author}{\bibfnamefont{Y.~S.} \bibnamefont{Jung}}, and
  \bibinfo{author}{\bibfnamefont{H.~K.} \bibnamefont{Kim}},
  \bibinfo{year}{2005}, \bibinfo{journal}{Nano\ Lett.}
  \textbf{\bibinfo{volume}{5}}, \bibinfo{pages}{1123}.

\bibitem[{\citenamefont{Kubo} \emph{et~al.}(2007)\citenamefont{Kubo, Pontius,
  and Petek}}]{KPP07}
\bibinfo{author}{\bibnamefont{Kubo}, \bibfnamefont{A.}},
  \bibinfo{author}{\bibfnamefont{N.}~\bibnamefont{Pontius}}, and
  \bibinfo{author}{\bibfnamefont{H.}~\bibnamefont{Petek}},
  \bibinfo{year}{2007}, \bibinfo{journal}{Nano\ Lett.}
  \textbf{\bibinfo{volume}{7}}, \bibinfo{pages}{470}.

\bibitem[{\citenamefont{Kumar} \emph{et~al.}(2008)\citenamefont{Kumar,
  Pastoriza-Santos, {Rodr\'{\i}guez-Gonz\'alez}, {Garc\'{\i}a de Abajo}, and
  {Liz-Marz\'{a}n}}}]{paper140}
\bibinfo{author}{\bibnamefont{Kumar}, \bibfnamefont{P.~S.}},
  \bibinfo{author}{\bibfnamefont{I.}~\bibnamefont{Pastoriza-Santos}},
  \bibinfo{author}{\bibfnamefont{B.}~\bibnamefont{{Rodr\'{\i}guez-Gonz\'alez}}%
}, \bibinfo{author}{\bibfnamefont{F.~J.} \bibnamefont{{Garc\'{\i}a de Abajo}}},
  and \bibinfo{author}{\bibfnamefont{L.~M.} \bibnamefont{{Liz-Marz\'{a}n}}},
  \bibinfo{year}{2008}, \bibinfo{journal}{Nanotechnology}
  \textbf{\bibinfo{volume}{19}}, \bibinfo{pages}{015606}.

\bibitem[{\citenamefont{Kuttge} \emph{et~al.}(2009)\citenamefont{Kuttge,
  Vesseur, Koenderink, Lezec, Atwater, {Garc\'{\i}a de Abajo}, and
  Polman}}]{paper167}
\bibinfo{author}{\bibnamefont{Kuttge}, \bibfnamefont{M.}},
  \bibinfo{author}{\bibfnamefont{E.~J.~R.} \bibnamefont{Vesseur}},
  \bibinfo{author}{\bibfnamefont{A.~F.} \bibnamefont{Koenderink}},
  \bibinfo{author}{\bibfnamefont{H.~J.} \bibnamefont{Lezec}},
  \bibinfo{author}{\bibfnamefont{H.~A.} \bibnamefont{Atwater}},
  \bibinfo{author}{\bibfnamefont{F.~J.} \bibnamefont{{Garc\'{\i}a de Abajo}}},
  and \bibinfo{author}{\bibfnamefont{A.}~\bibnamefont{Polman}},
  \bibinfo{year}{2009}, \bibinfo{journal}{Phys.\ Rev.\ B} , \bibinfo{pages}{(in
  press)}.

\bibitem[{\citenamefont{Kuttge} \emph{et~al.}(2008)\citenamefont{Kuttge,
  Vesseur, Verhoeven, Lezec, Atwater, and Polman}}]{KVV08}
\bibinfo{author}{\bibnamefont{Kuttge}, \bibfnamefont{M.}},
  \bibinfo{author}{\bibfnamefont{E.~J.~R.} \bibnamefont{Vesseur}},
  \bibinfo{author}{\bibfnamefont{J.}~\bibnamefont{Verhoeven}},
  \bibinfo{author}{\bibfnamefont{H.~J.} \bibnamefont{Lezec}},
  \bibinfo{author}{\bibfnamefont{H.~A.} \bibnamefont{Atwater}}, and
  \bibinfo{author}{\bibfnamefont{A.}~\bibnamefont{Polman}},
  \bibinfo{year}{2008}, \bibinfo{journal}{Appl.\ Phys.\ Lett.}
  \textbf{\bibinfo{volume}{93}}, \bibinfo{pages}{113110}.

\bibitem[{\citenamefont{Kuzuo} \emph{et~al.}(1992)\citenamefont{Kuzuo,
  Terauchi, and Tanaka}}]{KTT02}
\bibinfo{author}{\bibnamefont{Kuzuo}, \bibfnamefont{R.}},
  \bibinfo{author}{\bibfnamefont{M.}~\bibnamefont{Terauchi}}, and
  \bibinfo{author}{\bibfnamefont{M.}~\bibnamefont{Tanaka}},
  \bibinfo{year}{1992}, \bibinfo{journal}{Jpn.\ J.\ Appl.\ Phys.\ Part\
  2-Lett.} \textbf{\bibinfo{volume}{31}}, \bibinfo{pages}{L1484}.

\bibitem[{\citenamefont{Lalor and Wolf}(1971)}]{LW1971}
\bibinfo{author}{\bibnamefont{Lalor}, \bibfnamefont{E.}}, and
  \bibinfo{author}{\bibfnamefont{E.}~\bibnamefont{Wolf}}, \bibinfo{year}{1971},
  \bibinfo{journal}{Phys.\ Rev.\ Lett.} \textbf{\bibinfo{volume}{26}},
  \bibinfo{pages}{1274}.

\bibitem[{\citenamefont{Landau} \emph{et~al.}(1984)\citenamefont{Landau,
  Lifshitz, and Pitaevskii}}]{LLP1984}
\bibinfo{author}{\bibnamefont{Landau}, \bibfnamefont{L.~D.}},
  \bibinfo{author}{\bibfnamefont{E.~M.} \bibnamefont{Lifshitz}}, and
  \bibinfo{author}{\bibfnamefont{L.~P.} \bibnamefont{Pitaevskii}},
  \bibinfo{year}{1984}, \emph{\bibinfo{title}{Electrodynamics of Continuous
  Media}} (\bibinfo{publisher}{Pergamon Press}, \bibinfo{address}{Oxford}).

\bibitem[{\citenamefont{Lang}(1948)}]{L1948}
\bibinfo{author}{\bibnamefont{Lang}, \bibfnamefont{W.}}, \bibinfo{year}{1948},
  \bibinfo{journal}{Optik} \textbf{\bibinfo{volume}{3}}, \bibinfo{pages}{233}.

\bibitem[{\citenamefont{Larkin and Stockman}(2005)}]{LS05}
\bibinfo{author}{\bibnamefont{Larkin}, \bibfnamefont{I.~A.}}, and
  \bibinfo{author}{\bibfnamefont{M.~I.} \bibnamefont{Stockman}},
  \bibinfo{year}{2005}, \bibinfo{journal}{Nano\ Lett.}
  \textbf{\bibinfo{volume}{5}}, \bibinfo{pages}{339}.

\bibitem[{\citenamefont{Larkin} \emph{et~al.}(2004)\citenamefont{Larkin,
  Stockman, Achermann, and Klimov}}]{LSA04}
\bibinfo{author}{\bibnamefont{Larkin}, \bibfnamefont{I.~A.}},
  \bibinfo{author}{\bibfnamefont{M.~I.} \bibnamefont{Stockman}},
  \bibinfo{author}{\bibfnamefont{M.}~\bibnamefont{Achermann}}, and
  \bibinfo{author}{\bibfnamefont{V.~I.} \bibnamefont{Klimov}},
  \bibinfo{year}{2004}, \bibinfo{journal}{Phys.\ Rev.\ B}
  \textbf{\bibinfo{volume}{69}}, \bibinfo{pages}{121403(R)}.

\bibitem[{\citenamefont{Lastdrager}
  \emph{et~al.}(2000)\citenamefont{Lastdrager, Tip, and Verhoeven}}]{LTV00}
\bibinfo{author}{\bibnamefont{Lastdrager}, \bibfnamefont{B.}},
  \bibinfo{author}{\bibfnamefont{A.}~\bibnamefont{Tip}}, and
  \bibinfo{author}{\bibfnamefont{J.}~\bibnamefont{Verhoeven}},
  \bibinfo{year}{2000}, \bibinfo{journal}{Phys.\ Rev.\ E}
  \textbf{\bibinfo{volume}{61}}, \bibinfo{pages}{5767}.

\bibitem[{\citenamefont{Lazar} \emph{et~al.}(2003)\citenamefont{Lazar, Botton,
  Wu, Tichelaar, and Zandbergen}}]{LBW03}
\bibinfo{author}{\bibnamefont{Lazar}, \bibfnamefont{S.}},
  \bibinfo{author}{\bibfnamefont{G.~A.} \bibnamefont{Botton}},
  \bibinfo{author}{\bibfnamefont{M.~Y.} \bibnamefont{Wu}},
  \bibinfo{author}{\bibfnamefont{F.~D.} \bibnamefont{Tichelaar}}, and
  \bibinfo{author}{\bibfnamefont{H.~W.} \bibnamefont{Zandbergen}},
  \bibinfo{year}{2003}, \bibinfo{journal}{Ultramicroscopy}
  \textbf{\bibinfo{volume}{96}}, \bibinfo{pages}{535}.

\bibitem[{\citenamefont{Lazar} \emph{et~al.}(2006)\citenamefont{Lazar, Botton,
  and Zandbergen}}]{LBZ06}
\bibinfo{author}{\bibnamefont{Lazar}, \bibfnamefont{S.}},
  \bibinfo{author}{\bibfnamefont{G.~A.} \bibnamefont{Botton}}, and
  \bibinfo{author}{\bibfnamefont{H.~W.} \bibnamefont{Zandbergen}},
  \bibinfo{year}{2006}, \bibinfo{journal}{Ultramicroscopy}
  \textbf{\bibinfo{volume}{106}}, \bibinfo{pages}{1091}.

\bibitem[{\citenamefont{Lecante} \emph{et~al.}(1977)\citenamefont{Lecante,
  Ballu, and Newns}}]{LBN1977}
\bibinfo{author}{\bibnamefont{Lecante}, \bibfnamefont{J.}},
  \bibinfo{author}{\bibfnamefont{Y.}~\bibnamefont{Ballu}}, and
  \bibinfo{author}{\bibfnamefont{D.~M.} \bibnamefont{Newns}},
  \bibinfo{year}{1977}, \bibinfo{journal}{Phys.\ Rev.\ Lett.}
  \textbf{\bibinfo{volume}{38}}, \bibinfo{pages}{36}.

\bibitem[{\citenamefont{Leith\"auser}(1904)}]{L1904}
\bibinfo{author}{\bibnamefont{Leith\"auser}, \bibfnamefont{G.~E.}},
  \bibinfo{year}{1904}, \bibinfo{journal}{Ann.\ Phys.\ (Leipzig)}
  \textbf{\bibinfo{volume}{15}}, \bibinfo{pages}{283}.

\bibitem[{\citenamefont{Li} \emph{et~al.}(2007{\natexlab{a}})\citenamefont{Li,
  Imasaki, Gao, Yang, and Park}}]{LIG07}
\bibinfo{author}{\bibnamefont{Li}, \bibfnamefont{D.}},
  \bibinfo{author}{\bibfnamefont{K.}~\bibnamefont{Imasaki}},
  \bibinfo{author}{\bibfnamefont{X.}~\bibnamefont{Gao}},
  \bibinfo{author}{\bibfnamefont{Z.}~\bibnamefont{Yang}}, and
  \bibinfo{author}{\bibfnamefont{G.~S.} \bibnamefont{Park}},
  \bibinfo{year}{2007}{\natexlab{a}}, \bibinfo{journal}{Appl.\ Phys.\ Lett.}
  \textbf{\bibinfo{volume}{91}}, \bibinfo{pages}{221506}.

\bibitem[{\citenamefont{Li} \emph{et~al.}(2006)\citenamefont{Li, Imasaki, Yang,
  and Park}}]{LIY06}
\bibinfo{author}{\bibnamefont{Li}, \bibfnamefont{D.}},
  \bibinfo{author}{\bibfnamefont{K.}~\bibnamefont{Imasaki}},
  \bibinfo{author}{\bibfnamefont{Z.}~\bibnamefont{Yang}}, and
  \bibinfo{author}{\bibfnamefont{G.~S.} \bibnamefont{Park}},
  \bibinfo{year}{2006}, \bibinfo{journal}{Appl.\ Phys.\ Lett.}
  \textbf{\bibinfo{volume}{88}}, \bibinfo{pages}{201501}.

\bibitem[{\citenamefont{Li} \emph{et~al.}(2007{\natexlab{b}})\citenamefont{Li,
  Imasaki, Yang, Park, Miyamoto, Amano, and Mochizuki}}]{LIY07}
\bibinfo{author}{\bibnamefont{Li}, \bibfnamefont{D.}},
  \bibinfo{author}{\bibfnamefont{K.}~\bibnamefont{Imasaki}},
  \bibinfo{author}{\bibfnamefont{Z.}~\bibnamefont{Yang}},
  \bibinfo{author}{\bibfnamefont{G.~S.} \bibnamefont{Park}},
  \bibinfo{author}{\bibfnamefont{S.}~\bibnamefont{Miyamoto}},
  \bibinfo{author}{\bibfnamefont{S.}~\bibnamefont{Amano}}, and
  \bibinfo{author}{\bibfnamefont{T.}~\bibnamefont{Mochizuki}},
  \bibinfo{year}{2007}{\natexlab{b}}, \bibinfo{journal}{Nucl.\ Instrum.\
  Methods\ Phys.\ Res.\ A} \textbf{\bibinfo{volume}{572}},
  \bibinfo{pages}{948}.

\bibitem[{\citenamefont{Lichte and Freitag}(2000)}]{LF00}
\bibinfo{author}{\bibnamefont{Lichte}, \bibfnamefont{H.}}, and
  \bibinfo{author}{\bibfnamefont{B.}~\bibnamefont{Freitag}},
  \bibinfo{year}{2000}, \bibinfo{journal}{Ultramicroscopy}
  \textbf{\bibinfo{volume}{81}}, \bibinfo{pages}{177}.

\bibitem[{\citenamefont{Lindhard}(1954)}]{L1954}
\bibinfo{author}{\bibnamefont{Lindhard}, \bibfnamefont{J.}},
  \bibinfo{year}{1954}, \bibinfo{journal}{K.\ Dan.\ Vidensk.\ Selsk.\ Mat.\
  Fys.\ Medd.} \textbf{\bibinfo{volume}{28}}, \bibinfo{pages}{no. 8}.

\bibitem[{\citenamefont{Little} \emph{et~al.}(1984)\citenamefont{Little,
  Callcott, Ferrell, and Arakawa}}]{LCF1984}
\bibinfo{author}{\bibnamefont{Little}, \bibfnamefont{J.~W.}},
  \bibinfo{author}{\bibfnamefont{T.~A.} \bibnamefont{Callcott}},
  \bibinfo{author}{\bibfnamefont{T.~L.} \bibnamefont{Ferrell}}, and
  \bibinfo{author}{\bibfnamefont{E.~T.} \bibnamefont{Arakawa}},
  \bibinfo{year}{1984}, \bibinfo{journal}{Phys.\ Rev.\ B}
  \textbf{\bibinfo{volume}{29}}, \bibinfo{pages}{1606}.

\bibitem[{\citenamefont{Liu} \emph{et~al.}(2007)\citenamefont{Liu, Yang, Liang,
  Li, Imasaki, Shi, Lan, and Park}}]{LYL07}
\bibinfo{author}{\bibnamefont{Liu}, \bibfnamefont{W.}},
  \bibinfo{author}{\bibfnamefont{Z.}~\bibnamefont{Yang}},
  \bibinfo{author}{\bibfnamefont{Z.}~\bibnamefont{Liang}},
  \bibinfo{author}{\bibfnamefont{D.}~\bibnamefont{Li}},
  \bibinfo{author}{\bibfnamefont{K.}~\bibnamefont{Imasaki}},
  \bibinfo{author}{\bibfnamefont{Z.}~\bibnamefont{Shi}},
  \bibinfo{author}{\bibfnamefont{F.}~\bibnamefont{Lan}}, and
  \bibinfo{author}{\bibfnamefont{G.~S.} \bibnamefont{Park}},
  \bibinfo{year}{2007}, \bibinfo{journal}{Nucl.\ Instrum.\ Methods\ Phys.\
  Res.\ A} \textbf{\bibinfo{volume}{580}}, \bibinfo{pages}{1552}.

\bibitem[{\citenamefont{Lobastov} \emph{et~al.}(2005)\citenamefont{Lobastov,
  Srinivasan, and Zewail}}]{LSZ05}
\bibinfo{author}{\bibnamefont{Lobastov}, \bibfnamefont{V.~A.}},
  \bibinfo{author}{\bibfnamefont{R.}~\bibnamefont{Srinivasan}}, and
  \bibinfo{author}{\bibfnamefont{A.~H.} \bibnamefont{Zewail}},
  \bibinfo{year}{2005}, \bibinfo{journal}{Proc.\ Natl.\ Academ.\ Sci.}
  \textbf{\bibinfo{volume}{102}}, \bibinfo{pages}{7069}.

\bibitem[{\citenamefont{Lu and Maradudin}(1990)}]{LM1990}
\bibinfo{author}{\bibnamefont{Lu}, \bibfnamefont{J.~Q.}}, and
  \bibinfo{author}{\bibfnamefont{A.~A.} \bibnamefont{Maradudin}},
  \bibinfo{year}{1990}, \bibinfo{journal}{Phys.\ Rev.\ B}
  \textbf{\bibinfo{volume}{42}}, \bibinfo{pages}{11159}.

\bibitem[{\citenamefont{Lucas} \emph{et~al.}(1994)\citenamefont{Lucas, Henrard,
  and Lambin}}]{LHL94}
\bibinfo{author}{\bibnamefont{Lucas}, \bibfnamefont{A.~A.}},
  \bibinfo{author}{\bibfnamefont{L.}~\bibnamefont{Henrard}}, and
  \bibinfo{author}{\bibfnamefont{P.}~\bibnamefont{Lambin}},
  \bibinfo{year}{1994}, \bibinfo{journal}{Phys.\ Rev.\ B}
  \textbf{\bibinfo{volume}{49}}, \bibinfo{pages}{2888}.

\bibitem[{\citenamefont{Lucas and Kartheuser}(1970)}]{LK1970}
\bibinfo{author}{\bibnamefont{Lucas}, \bibfnamefont{A.~A.}}, and
  \bibinfo{author}{\bibfnamefont{E.}~\bibnamefont{Kartheuser}},
  \bibinfo{year}{1970}, \bibinfo{journal}{Phys.\ Rev.\ B}
  \textbf{\bibinfo{volume}{1}}, \bibinfo{pages}{3588}.

\bibitem[{\citenamefont{Lucas and {\v{S}unji\'{c}}}(1971)}]{LS1971}
\bibinfo{author}{\bibnamefont{Lucas}, \bibfnamefont{A.~A.}}, and
  \bibinfo{author}{\bibfnamefont{M.}~\bibnamefont{{\v{S}unji\'{c}}}},
  \bibinfo{year}{1971}, \bibinfo{journal}{Phys.\ Rev.\ Lett.}
  \textbf{\bibinfo{volume}{26}}, \bibinfo{pages}{229}.

\bibitem[{\citenamefont{Luo} \emph{et~al.}(2003)\citenamefont{Luo, Ibanescu,
  Johnson, and Joannopoulos}}]{LIJ03}
\bibinfo{author}{\bibnamefont{Luo}, \bibfnamefont{C.}},
  \bibinfo{author}{\bibfnamefont{M.}~\bibnamefont{Ibanescu}},
  \bibinfo{author}{\bibfnamefont{S.~G.} \bibnamefont{Johnson}}, and
  \bibinfo{author}{\bibfnamefont{J.~D.} \bibnamefont{Joannopoulos}},
  \bibinfo{year}{2003}, \bibinfo{journal}{Science}
  \textbf{\bibinfo{volume}{299}}, \bibinfo{pages}{368}.

\bibitem[{\citenamefont{Ma} \emph{et~al.}(2006)\citenamefont{Ma, Lee, Mueller,
  and Alvarado}}]{MLM06}
\bibinfo{author}{\bibnamefont{Ma}, \bibfnamefont{D.~D.~D.}},
  \bibinfo{author}{\bibfnamefont{S.~T.} \bibnamefont{Lee}},
  \bibinfo{author}{\bibfnamefont{P.}~\bibnamefont{Mueller}}, and
  \bibinfo{author}{\bibfnamefont{S.~F.} \bibnamefont{Alvarado}},
  \bibinfo{year}{2006}, \bibinfo{journal}{Nano\ Lett.}
  \textbf{\bibinfo{volume}{6}}, \bibinfo{pages}{926}.

\bibitem[{\citenamefont{Macaulay} \emph{et~al.}(1989)\citenamefont{Macaulay,
  Allen, Brown, and Berger}}]{MAB1989}
\bibinfo{author}{\bibnamefont{Macaulay}, \bibfnamefont{J.~M.}},
  \bibinfo{author}{\bibfnamefont{R.~M.} \bibnamefont{Allen}},
  \bibinfo{author}{\bibfnamefont{L.~M.} \bibnamefont{Brown}}, and
  \bibinfo{author}{\bibfnamefont{S.~D.} \bibnamefont{Berger}},
  \bibinfo{year}{1989}, \bibinfo{journal}{Microelectron.\ Eng.}
  \textbf{\bibinfo{volume}{9}}, \bibinfo{pages}{557}.

\bibitem[{\citenamefont{Mamola} \emph{et~al.}(1987)\citenamefont{Mamola,
  Warmack, and Ferrell}}]{MWF1987}
\bibinfo{author}{\bibnamefont{Mamola}, \bibfnamefont{K.~C.}},
  \bibinfo{author}{\bibfnamefont{R.~J.} \bibnamefont{Warmack}}, and
  \bibinfo{author}{\bibfnamefont{T.~L.} \bibnamefont{Ferrell}},
  \bibinfo{year}{1987}, \bibinfo{journal}{Phys.\ Rev.\ B}
  \textbf{\bibinfo{volume}{35}}, \bibinfo{pages}{2682}.

\bibitem[{\citenamefont{Manjavacas and {Garc\'{\i}a de
  Abajo}}(2008)}]{paperxxx}
\bibinfo{author}{\bibnamefont{Manjavacas}, \bibfnamefont{A.}}, and
  \bibinfo{author}{\bibfnamefont{F.~J.} \bibnamefont{{Garc\'{\i}a de Abajo}}},
  \bibinfo{year}{2008}, \bibinfo{journal}{Nano\ Lett.} , \bibinfo{pages}{DOI:
  10.1021/nl802044t}.

\bibitem[{\citenamefont{Marinopoulos}
  \emph{et~al.}(2003)\citenamefont{Marinopoulos, Reining, Rubio, and
  Vast}}]{MRR03}
\bibinfo{author}{\bibnamefont{Marinopoulos}, \bibfnamefont{A.~G.}},
  \bibinfo{author}{\bibfnamefont{L.}~\bibnamefont{Reining}},
  \bibinfo{author}{\bibfnamefont{A.}~\bibnamefont{Rubio}}, and
  \bibinfo{author}{\bibfnamefont{N.}~\bibnamefont{Vast}}, \bibinfo{year}{2003},
  \bibinfo{journal}{Phys.\ Rev.\ Lett.} \textbf{\bibinfo{volume}{91}},
  \bibinfo{pages}{046402}.

\bibitem[{\citenamefont{Marks}(1982)}]{M1982}
\bibinfo{author}{\bibnamefont{Marks}, \bibfnamefont{L.~D.}},
  \bibinfo{year}{1982}, \bibinfo{journal}{Solid\ State\ Commun.}
  \textbf{\bibinfo{volume}{43}}, \bibinfo{pages}{727}.

\bibitem[{\citenamefont{Masuda and Fukuda}(1995)}]{MF95}
\bibinfo{author}{\bibnamefont{Masuda}, \bibfnamefont{H.}}, and
  \bibinfo{author}{\bibfnamefont{K.}~\bibnamefont{Fukuda}},
  \bibinfo{year}{1995}, \bibinfo{journal}{Science}
  \textbf{\bibinfo{volume}{268}}, \bibinfo{pages}{1466}.

\bibitem[{\citenamefont{Matloob and Ghaffari}(2004)}]{MG04}
\bibinfo{author}{\bibnamefont{Matloob}, \bibfnamefont{R.}}, and
  \bibinfo{author}{\bibfnamefont{A.}~\bibnamefont{Ghaffari}},
  \bibinfo{year}{2004}, \bibinfo{journal}{Phys.\ Rev.\ A}
  \textbf{\bibinfo{volume}{70}}, \bibinfo{pages}{052116}.

\bibitem[{\citenamefont{Maxwell}(1891)}]{M1891}
\bibinfo{author}{\bibnamefont{Maxwell}, \bibfnamefont{J.~C.}},
  \bibinfo{year}{1891}, \emph{\bibinfo{title}{Treatise on Electricity and
  Magnetism}} (\bibinfo{publisher}{Dover}, \bibinfo{address}{New York}).

\bibitem[{\citenamefont{Maxwell-Garnett}(1904)}]{M1904_2}
\bibinfo{author}{\bibnamefont{Maxwell-Garnett}, \bibfnamefont{J.~C.}},
  \bibinfo{year}{1904}, \bibinfo{journal}{Philos.\ Trans.\ R.\ Soc.\ London,\
  Ser.\ A} \textbf{\bibinfo{volume}{203}}, \bibinfo{pages}{385}.

\bibitem[{\citenamefont{Maxwell-Garnett}(1906)}]{M1906}
\bibinfo{author}{\bibnamefont{Maxwell-Garnett}, \bibfnamefont{J.~C.}},
  \bibinfo{year}{1906}, \bibinfo{journal}{Philos.\ Trans.\ R.\ Soc.\ London,\
  Ser.\ A} \textbf{\bibinfo{volume}{205}}, \bibinfo{pages}{237}.

\bibitem[{\citenamefont{McComb and Howie}(1995)}]{MH95}
\bibinfo{author}{\bibnamefont{McComb}, \bibfnamefont{D.~W.}}, and
  \bibinfo{author}{\bibfnamefont{A.}~\bibnamefont{Howie}},
  \bibinfo{year}{1995}, \bibinfo{journal}{Nucl.\ Instrum.\ Methods\ Phys.\
  Res.\ B} \textbf{\bibinfo{volume}{96}}, \bibinfo{pages}{569}.

\bibitem[{\citenamefont{McKenzie and McPhedran}(1977)}]{MM1977}
\bibinfo{author}{\bibnamefont{McKenzie}, \bibfnamefont{D.~R.}}, and
  \bibinfo{author}{\bibfnamefont{R.~C.} \bibnamefont{McPhedran}},
  \bibinfo{year}{1977}, \bibinfo{journal}{Nature}
  \textbf{\bibinfo{volume}{265}}, \bibinfo{pages}{128}.

\bibitem[{\citenamefont{McKenzie} \emph{et~al.}(1978)\citenamefont{McKenzie,
  McPhedran, and Derrick}}]{MMD1978}
\bibinfo{author}{\bibnamefont{McKenzie}, \bibfnamefont{D.~R.}},
  \bibinfo{author}{\bibfnamefont{R.~C.} \bibnamefont{McPhedran}}, and
  \bibinfo{author}{\bibfnamefont{G.~H.} \bibnamefont{Derrick}},
  \bibinfo{year}{1978}, \bibinfo{journal}{Proc.\ R.\ Soc.\ London,\ Ser.\ A}
  \textbf{\bibinfo{volume}{362}}, \bibinfo{pages}{211}.

\bibitem[{\citenamefont{{Meiners} and Quake}(2000)}]{MQ00}
\bibinfo{author}{\bibnamefont{{Meiners}}, \bibfnamefont{J.~C.}}, and
  \bibinfo{author}{\bibfnamefont{S.~R.} \bibnamefont{Quake}},
  \bibinfo{year}{2000}, \bibinfo{journal}{Phys.\ Rev.\ Lett.}
  \textbf{\bibinfo{volume}{84}}, \bibinfo{pages}{5014}.

\bibitem[{\citenamefont{Mendoza} \emph{et~al.}(1998)\citenamefont{Mendoza,
  Barrera, and Fuchs}}]{MBF98}
\bibinfo{author}{\bibnamefont{Mendoza}, \bibfnamefont{C.~I.}},
  \bibinfo{author}{\bibfnamefont{R.~G.} \bibnamefont{Barrera}}, and
  \bibinfo{author}{\bibfnamefont{R.}~\bibnamefont{Fuchs}},
  \bibinfo{year}{1998}, \bibinfo{journal}{Phys.\ Rev.\ B}
  \textbf{\bibinfo{volume}{57}}, \bibinfo{pages}{11193}.

\bibitem[{\citenamefont{Mendoza} \emph{et~al.}(1999)\citenamefont{Mendoza,
  Barrera, and Fuchs}}]{MBF99}
\bibinfo{author}{\bibnamefont{Mendoza}, \bibfnamefont{C.~I.}},
  \bibinfo{author}{\bibfnamefont{R.~G.} \bibnamefont{Barrera}}, and
  \bibinfo{author}{\bibfnamefont{R.}~\bibnamefont{Fuchs}},
  \bibinfo{year}{1999}, \bibinfo{journal}{Phys.\ Rev.\ B}
  \textbf{\bibinfo{volume}{60}}, \bibinfo{pages}{13831}.

\bibitem[{\citenamefont{Merano} \emph{et~al.}(2005)\citenamefont{Merano,
  Sonderegger, Crottini, Collin, Renucci, Pelucchi, Malko, Baier, Kapon,
  Deveaud, and Gani\`{e}re}}]{MSC05}
\bibinfo{author}{\bibnamefont{Merano}, \bibfnamefont{M.}},
  \bibinfo{author}{\bibfnamefont{S.}~\bibnamefont{Sonderegger}},
  \bibinfo{author}{\bibfnamefont{A.}~\bibnamefont{Crottini}},
  \bibinfo{author}{\bibfnamefont{S.}~\bibnamefont{Collin}},
  \bibinfo{author}{\bibfnamefont{P.}~\bibnamefont{Renucci}},
  \bibinfo{author}{\bibfnamefont{E.}~\bibnamefont{Pelucchi}},
  \bibinfo{author}{\bibfnamefont{A.}~\bibnamefont{Malko}},
  \bibinfo{author}{\bibfnamefont{M.~H.} \bibnamefont{Baier}},
  \bibinfo{author}{\bibfnamefont{E.}~\bibnamefont{Kapon}},
  \bibinfo{author}{\bibfnamefont{B.}~\bibnamefont{Deveaud}}, and
  \bibinfo{author}{\bibfnamefont{J.~D.} \bibnamefont{Gani\`{e}re}},
  \bibinfo{year}{2005}, \bibinfo{journal}{Nature}
  \textbf{\bibinfo{volume}{438}}, \bibinfo{pages}{479}.

\bibitem[{\citenamefont{Mermin}(1970)}]{M1970}
\bibinfo{author}{\bibnamefont{Mermin}, \bibfnamefont{N.~D.}},
  \bibinfo{year}{1970}, \bibinfo{journal}{Phys.\ Rev.\ B}
  \textbf{\bibinfo{volume}{1}}, \bibinfo{pages}{2362}.

\bibitem[{\citenamefont{Mie}(1908)}]{M1908}
\bibinfo{author}{\bibnamefont{Mie}, \bibfnamefont{G.}}, \bibinfo{year}{1908},
  \bibinfo{journal}{Ann.\ Phys.\ (Leipzig)} \textbf{\bibinfo{volume}{25}},
  \bibinfo{pages}{377}.

\bibitem[{\citenamefont{Minoda and Yamamoto}(2006)}]{MY06}
\bibinfo{author}{\bibnamefont{Minoda}, \bibfnamefont{H.}}, and
  \bibinfo{author}{\bibfnamefont{N.}~\bibnamefont{Yamamoto}},
  \bibinfo{year}{2006}, \bibinfo{journal}{Surf.\ Interface\ Anal.}
  \textbf{\bibinfo{volume}{38}}, \bibinfo{pages}{1666}.

\bibitem[{\citenamefont{Mizuno} \emph{et~al.}(1975)\citenamefont{Mizuno, Ono,
  and Shimoe}}]{MOS1975}
\bibinfo{author}{\bibnamefont{Mizuno}, \bibfnamefont{K.}},
  \bibinfo{author}{\bibfnamefont{S.}~\bibnamefont{Ono}}, and
  \bibinfo{author}{\bibfnamefont{O.}~\bibnamefont{Shimoe}},
  \bibinfo{year}{1975}, \bibinfo{journal}{Nature}
  \textbf{\bibinfo{volume}{253}}, \bibinfo{pages}{184}.

\bibitem[{\citenamefont{Mizuno} \emph{et~al.}(1987)\citenamefont{Mizuno, Pae,
  Nozokido, and Furuya}}]{MPN1987}
\bibinfo{author}{\bibnamefont{Mizuno}, \bibfnamefont{K.}},
  \bibinfo{author}{\bibfnamefont{J.}~\bibnamefont{Pae}},
  \bibinfo{author}{\bibfnamefont{T.}~\bibnamefont{Nozokido}}, and
  \bibinfo{author}{\bibfnamefont{K.}~\bibnamefont{Furuya}},
  \bibinfo{year}{1987}, \bibinfo{journal}{Nature}
  \textbf{\bibinfo{volume}{328}}, \bibinfo{pages}{45}.

\bibitem[{\citenamefont{Mkhoyan} \emph{et~al.}(2007)\citenamefont{Mkhoyan,
  Babinec, Maccagnano, Kirkland, and Silcox}}]{MBM07}
\bibinfo{author}{\bibnamefont{Mkhoyan}, \bibfnamefont{K.~A.}},
  \bibinfo{author}{\bibfnamefont{T.}~\bibnamefont{Babinec}},
  \bibinfo{author}{\bibfnamefont{S.~E.} \bibnamefont{Maccagnano}},
  \bibinfo{author}{\bibfnamefont{E.~J.} \bibnamefont{Kirkland}}, and
  \bibinfo{author}{\bibfnamefont{J.}~\bibnamefont{Silcox}},
  \bibinfo{year}{2007}, \bibinfo{journal}{Ultramicroscopy}
  \textbf{\bibinfo{volume}{107}}, \bibinfo{pages}{345}.

\bibitem[{\citenamefont{Mock} \emph{et~al.}(2002)\citenamefont{Mock, Barbic,
  Smith, Schultz, , and Schultz}}]{MBS02}
\bibinfo{author}{\bibnamefont{Mock}, \bibfnamefont{J.~J.}},
  \bibinfo{author}{\bibfnamefont{M.}~\bibnamefont{Barbic}},
  \bibinfo{author}{\bibfnamefont{D.~R.} \bibnamefont{Smith}},
  \bibinfo{author}{\bibfnamefont{D.~A.} \bibnamefont{Schultz}}, , and
  \bibinfo{author}{\bibfnamefont{S.}~\bibnamefont{Schultz}},
  \bibinfo{year}{2002}, \bibinfo{journal}{J.\ Chem.\ Phys.}
  \textbf{\bibinfo{volume}{116}}, \bibinfo{pages}{6755}.

\bibitem[{\citenamefont{Monthioux and Kuznetsov}(2006)}]{M06}
\bibinfo{author}{\bibnamefont{Monthioux}, \bibfnamefont{M.}}, and
  \bibinfo{author}{\bibfnamefont{V.~L.} \bibnamefont{Kuznetsov}},
  \bibinfo{year}{2006}, \bibinfo{journal}{Carbon}
  \textbf{\bibinfo{volume}{44}}, \bibinfo{pages}{1621}.

\bibitem[{\citenamefont{Moran}(1992)}]{M92}
\bibinfo{author}{\bibnamefont{Moran}, \bibfnamefont{M.~J.}},
  \bibinfo{year}{1992}, \bibinfo{journal}{Phys.\ Rev.\ Lett.}
  \textbf{\bibinfo{volume}{69}}, \bibinfo{pages}{2523}.

\bibitem[{\citenamefont{More}(1966)}]{M1966_2}
\bibinfo{author}{\bibnamefont{More}, \bibfnamefont{R.~M.}},
  \bibinfo{year}{1966}, \bibinfo{journal}{Phys.\ Rev.\ Lett.}
  \textbf{\bibinfo{volume}{16}}, \bibinfo{pages}{781}.

\bibitem[{\citenamefont{Moreau} \emph{et~al.}(1997)\citenamefont{Moreau, Brun,
  Walsh, Colliex, and Howie}}]{MBW97}
\bibinfo{author}{\bibnamefont{Moreau}, \bibfnamefont{P.}},
  \bibinfo{author}{\bibfnamefont{N.}~\bibnamefont{Brun}},
  \bibinfo{author}{\bibfnamefont{C.~A.} \bibnamefont{Walsh}},
  \bibinfo{author}{\bibfnamefont{C.}~\bibnamefont{Colliex}}, and
  \bibinfo{author}{\bibfnamefont{A.}~\bibnamefont{Howie}},
  \bibinfo{year}{1997}, \bibinfo{journal}{Phys.\ Rev.\ B}
  \textbf{\bibinfo{volume}{56}}, \bibinfo{pages}{6774}.

\bibitem[{\citenamefont{M\"{u}llejans}
  \emph{et~al.}(1993)\citenamefont{M\"{u}llejans, Beloch, Howie, and
  Tomita}}]{MBH93}
\bibinfo{author}{\bibnamefont{M\"{u}llejans}, \bibfnamefont{H.}},
  \bibinfo{author}{\bibfnamefont{A.~L.} \bibnamefont{Beloch}},
  \bibinfo{author}{\bibfnamefont{A.}~\bibnamefont{Howie}}, and
  \bibinfo{author}{\bibfnamefont{M.}~\bibnamefont{Tomita}},
  \bibinfo{year}{1993}, \bibinfo{journal}{Ultramicroscopy}
  \textbf{\bibinfo{volume}{52}}, \bibinfo{pages}{360}.

\bibitem[{\citenamefont{{M\"ullejans} and Bleloch}(1992)}]{MB92}
\bibinfo{author}{\bibnamefont{{M\"ullejans}}, \bibfnamefont{H.}}, and
  \bibinfo{author}{\bibfnamefont{A.~L.} \bibnamefont{Bleloch}},
  \bibinfo{year}{1992}, \bibinfo{journal}{Phys.\ Rev.\ B}
  \textbf{\bibinfo{volume}{46}}, \bibinfo{pages}{8597}.

\bibitem[{\citenamefont{Muller} \emph{et~al.}(2008)\citenamefont{Muller,
  {Fitting Kourkoutis}, Murfitt, Song, Hwang, Silcox, Dellby, and
  Krivanek}}]{MKM08}
\bibinfo{author}{\bibnamefont{Muller}, \bibfnamefont{D.~A.}},
  \bibinfo{author}{\bibfnamefont{L.}~\bibnamefont{{Fitting Kourkoutis}}},
  \bibinfo{author}{\bibfnamefont{M.}~\bibnamefont{Murfitt}},
  \bibinfo{author}{\bibfnamefont{J.~H.} \bibnamefont{Song}},
  \bibinfo{author}{\bibfnamefont{H.~Y.} \bibnamefont{Hwang}},
  \bibinfo{author}{\bibfnamefont{J.}~\bibnamefont{Silcox}},
  \bibinfo{author}{\bibfnamefont{N.}~\bibnamefont{Dellby}}, and
  \bibinfo{author}{\bibfnamefont{O.~L.} \bibnamefont{Krivanek}},
  \bibinfo{year}{2008}, \bibinfo{journal}{Science}
  \textbf{\bibinfo{volume}{319}}, \bibinfo{pages}{1073}.

\bibitem[{\citenamefont{Muller and Silcox}(1995)}]{MS95}
\bibinfo{author}{\bibnamefont{Muller}, \bibfnamefont{D.~A.}}, and
  \bibinfo{author}{\bibfnamefont{J.}~\bibnamefont{Silcox}},
  \bibinfo{year}{1995}, \bibinfo{journal}{Ultramicroscopy}
  \textbf{\bibinfo{volume}{59}}, \bibinfo{pages}{195}.

\bibitem[{\citenamefont{Muller} \emph{et~al.}(1993)\citenamefont{Muller, Tzou,
  Raj, and Silcox}}]{MTR93}
\bibinfo{author}{\bibnamefont{Muller}, \bibfnamefont{D.~A.}},
  \bibinfo{author}{\bibfnamefont{Y.}~\bibnamefont{Tzou}},
  \bibinfo{author}{\bibfnamefont{R.}~\bibnamefont{Raj}}, and
  \bibinfo{author}{\bibfnamefont{J.}~\bibnamefont{Silcox}},
  \bibinfo{year}{1993}, \bibinfo{journal}{Nature}
  \textbf{\bibinfo{volume}{366}}, \bibinfo{pages}{725}.

\bibitem[{\citenamefont{Myroshnychenko}
  \emph{et~al.}(2008{\natexlab{a}})\citenamefont{Myroshnychenko,
  {Carb\'o-Argibay}, Pastoriza-Santos, {P\'erez-Juste}, {Liz-Marz\'an}, and
  {Garc\'{\i}a de Abajo}}}]{paper149}
\bibinfo{author}{\bibnamefont{Myroshnychenko}, \bibfnamefont{V.}},
  \bibinfo{author}{\bibfnamefont{E.}~\bibnamefont{{Carb\'o-Argibay}}},
  \bibinfo{author}{\bibfnamefont{I.}~\bibnamefont{Pastoriza-Santos}},
  \bibinfo{author}{\bibfnamefont{J.}~\bibnamefont{{P\'erez-Juste}}},
  \bibinfo{author}{\bibfnamefont{L.~M.} \bibnamefont{{Liz-Marz\'an}}}, and
  \bibinfo{author}{\bibfnamefont{F.~J.} \bibnamefont{{Garc\'{\i}a de Abajo}}},
  \bibinfo{year}{2008}{\natexlab{a}}, \bibinfo{journal}{Adv.\ Mater.}
  \textbf{\bibinfo{volume}{20}}, \bibinfo{pages}{4288–}.

\bibitem[{\citenamefont{Myroshnychenko}
  \emph{et~al.}(2008{\natexlab{b}})\citenamefont{Myroshnychenko,
  {Rodr\'{\i}guez-Fern\'andez}, Pastoriza-Santos, Funston, Novo, Mulvaney,
  {Liz-Marz\'an}, and {Garc\'{\i}a de Abajo}}}]{paper155}
\bibinfo{author}{\bibnamefont{Myroshnychenko}, \bibfnamefont{V.}},
  \bibinfo{author}{\bibfnamefont{J.}~\bibnamefont{{Rodr\'{\i}guez-Fern\'andez}%
}}, \bibinfo{author}{\bibfnamefont{I.}~\bibnamefont{Pastoriza-Santos}},
  \bibinfo{author}{\bibfnamefont{A.~M.} \bibnamefont{Funston}},
  \bibinfo{author}{\bibfnamefont{C.}~\bibnamefont{Novo}},
  \bibinfo{author}{\bibfnamefont{P.}~\bibnamefont{Mulvaney}},
  \bibinfo{author}{\bibfnamefont{L.~M.} \bibnamefont{{Liz-Marz\'an}}}, and
  \bibinfo{author}{\bibfnamefont{F.~J.} \bibnamefont{{Garc\'{\i}a de Abajo}}},
  \bibinfo{year}{2008}{\natexlab{b}}, \bibinfo{journal}{Chem.\ Soc.\ Rev.}
  \textbf{\bibinfo{volume}{37}}, \bibinfo{pages}{1792}.

\bibitem[{\citenamefont{Nagao} \emph{et~al.}(2006)\citenamefont{Nagao,
  Yaginuma, Inaoka, and Sakurai}}]{NYI06}
\bibinfo{author}{\bibnamefont{Nagao}, \bibfnamefont{T.}},
  \bibinfo{author}{\bibfnamefont{S.}~\bibnamefont{Yaginuma}},
  \bibinfo{author}{\bibfnamefont{T.}~\bibnamefont{Inaoka}}, and
  \bibinfo{author}{\bibfnamefont{T.}~\bibnamefont{Sakurai}},
  \bibinfo{year}{2006}, \bibinfo{journal}{Phys.\ Rev.\ Lett.}
  \textbf{\bibinfo{volume}{97}}, \bibinfo{pages}{116802}.

\bibitem[{\citenamefont{Naumenko} \emph{et~al.}(2008)\citenamefont{Naumenko,
  Cha, Kalinin, Popov, Potylitsyn, Saruev, and Sukhikh}}]{NCK08}
\bibinfo{author}{\bibnamefont{Naumenko}, \bibfnamefont{G.~A.}},
  \bibinfo{author}{\bibfnamefont{V.~A.} \bibnamefont{Cha}},
  \bibinfo{author}{\bibfnamefont{B.~N.} \bibnamefont{Kalinin}},
  \bibinfo{author}{\bibfnamefont{Y.~A.} \bibnamefont{Popov}},
  \bibinfo{author}{\bibfnamefont{A.~P.} \bibnamefont{Potylitsyn}},
  \bibinfo{author}{\bibfnamefont{G.~A.} \bibnamefont{Saruev}}, and
  \bibinfo{author}{\bibfnamefont{L.~G.} \bibnamefont{Sukhikh}},
  \bibinfo{year}{2008}, \bibinfo{journal}{Nucl.\ Instrum.\ Methods\ Phys.\
  Res.\ B} \textbf{\bibinfo{volume}{266}}, \bibinfo{pages}{3733}.

\bibitem[{\citenamefont{Nelayah}
  \emph{et~al.}(2007{\natexlab{a}})\citenamefont{Nelayah, Kociak, {O.
  St\'{e}phan}, {Garc\'{\i}a de Abajo}, Tenc\'e, Henrard, Taverna,
  Pastoriza-Santos, Liz-Marz\'{a}n, and Colliex}}]{paper125}
\bibinfo{author}{\bibnamefont{Nelayah}, \bibfnamefont{J.}},
  \bibinfo{author}{\bibfnamefont{M.}~\bibnamefont{Kociak}},
  \bibinfo{author}{\bibnamefont{{O. St\'{e}phan}}},
  \bibinfo{author}{\bibfnamefont{F.~J.} \bibnamefont{{Garc\'{\i}a de Abajo}}},
  \bibinfo{author}{\bibfnamefont{M.}~\bibnamefont{Tenc\'e}},
  \bibinfo{author}{\bibfnamefont{L.}~\bibnamefont{Henrard}},
  \bibinfo{author}{\bibfnamefont{D.}~\bibnamefont{Taverna}},
  \bibinfo{author}{\bibfnamefont{I.}~\bibnamefont{Pastoriza-Santos}},
  \bibinfo{author}{\bibfnamefont{L.~M.} \bibnamefont{Liz-Marz\'{a}n}}, and
  \bibinfo{author}{\bibfnamefont{C.}~\bibnamefont{Colliex}},
  \bibinfo{year}{2007}{\natexlab{a}}, \bibinfo{journal}{Nat.\ Phys.}
  \textbf{\bibinfo{volume}{3}}, \bibinfo{pages}{348}.

\bibitem[{\citenamefont{Nelayah}
  \emph{et~al.}(2007{\natexlab{b}})\citenamefont{Nelayah, {O. St\'{e}phan},
  Kociak, {Garc\'{\i}a de Abajo}, Henrard, Pastoriza-Santos, Liz-Marz\'{a}n,
  and Colliex}}]{NSK07}
\bibinfo{author}{\bibnamefont{Nelayah}, \bibfnamefont{J.}},
  \bibinfo{author}{\bibnamefont{{O. St\'{e}phan}}},
  \bibinfo{author}{\bibfnamefont{M.}~\bibnamefont{Kociak}},
  \bibinfo{author}{\bibfnamefont{F.~J.} \bibnamefont{{Garc\'{\i}a de Abajo}}},
  \bibinfo{author}{\bibfnamefont{L.}~\bibnamefont{Henrard}},
  \bibinfo{author}{\bibfnamefont{I.}~\bibnamefont{Pastoriza-Santos}},
  \bibinfo{author}{\bibfnamefont{L.~M.} \bibnamefont{Liz-Marz\'{a}n}}, and
  \bibinfo{author}{\bibfnamefont{C.}~\bibnamefont{Colliex}},
  \bibinfo{year}{2007}{\natexlab{b}}, \bibinfo{journal}{Microsc.\ Microanal.}
  \textbf{\bibinfo{volume}{13}}, \bibinfo{pages}{144}.

\bibitem[{\citenamefont{Nellist} \emph{et~al.}(2004)\citenamefont{Nellist,
  Chisholm, Dellby, Krivanek, Murfitt, Szilagyi, Lupini, Borisevich, {Sides
  Jr.}, and Pennycook}}]{NCD04}
\bibinfo{author}{\bibnamefont{Nellist}, \bibfnamefont{P.~D.}},
  \bibinfo{author}{\bibfnamefont{M.~F.} \bibnamefont{Chisholm}},
  \bibinfo{author}{\bibfnamefont{N.}~\bibnamefont{Dellby}},
  \bibinfo{author}{\bibfnamefont{O.~L.} \bibnamefont{Krivanek}},
  \bibinfo{author}{\bibfnamefont{M.~F.} \bibnamefont{Murfitt}},
  \bibinfo{author}{\bibfnamefont{Z.~S.} \bibnamefont{Szilagyi}},
  \bibinfo{author}{\bibfnamefont{A.~R.} \bibnamefont{Lupini}},
  \bibinfo{author}{\bibfnamefont{A.}~\bibnamefont{Borisevich}},
  \bibinfo{author}{\bibfnamefont{W.~H.} \bibnamefont{{Sides Jr.}}}, and
  \bibinfo{author}{\bibfnamefont{S.~J.} \bibnamefont{Pennycook}},
  \bibinfo{year}{2004}, \bibinfo{journal}{Science}
  \textbf{\bibinfo{volume}{305}}, \bibinfo{pages}{1741}.

\bibitem[{\citenamefont{{N'Gom}} \emph{et~al.}(2008)\citenamefont{{N'Gom},
  Ringnalda, Mansfield, Agarwal, Kotov, Zaluzec, and Norris}}]{NRM08}
\bibinfo{author}{\bibnamefont{{N'Gom}}, \bibfnamefont{M.}},
  \bibinfo{author}{\bibfnamefont{J.}~\bibnamefont{Ringnalda}},
  \bibinfo{author}{\bibfnamefont{J.~F.} \bibnamefont{Mansfield}},
  \bibinfo{author}{\bibfnamefont{A.}~\bibnamefont{Agarwal}},
  \bibinfo{author}{\bibfnamefont{N.}~\bibnamefont{Kotov}},
  \bibinfo{author}{\bibfnamefont{N.~J.} \bibnamefont{Zaluzec}}, and
  \bibinfo{author}{\bibfnamefont{T.~B.} \bibnamefont{Norris}},
  \bibinfo{year}{2008}, \bibinfo{journal}{Nano\ Lett.}
  \textbf{\bibinfo{volume}{8}}, \bibinfo{pages}{3200}.

\bibitem[{\citenamefont{Noguez}(2007)}]{N07}
\bibinfo{author}{\bibnamefont{Noguez}, \bibfnamefont{C.}},
  \bibinfo{year}{2007}, \bibinfo{journal}{J.\ Phys.\ Chem.\ C}
  \textbf{\bibinfo{volume}{111}}, \bibinfo{pages}{3806}.

\bibitem[{\citenamefont{Nojeh} \emph{et~al.}(2006)\citenamefont{Nojeh, Shan,
  Cho, and Pease}}]{NSC06}
\bibinfo{author}{\bibnamefont{Nojeh}, \bibfnamefont{A.}},
  \bibinfo{author}{\bibfnamefont{B.}~\bibnamefont{Shan}},
  \bibinfo{author}{\bibfnamefont{K.}~\bibnamefont{Cho}}, and
  \bibinfo{author}{\bibfnamefont{R.~F.~W.} \bibnamefont{Pease}},
  \bibinfo{year}{2006}, \bibinfo{journal}{Phys.\ Rev.\ Lett.}
  \textbf{\bibinfo{volume}{96}}, \bibinfo{pages}{056802}.

\bibitem[{\citenamefont{Novotny} \emph{et~al.}(1997)\citenamefont{Novotny,
  Bian, and Xie}}]{NBX97}
\bibinfo{author}{\bibnamefont{Novotny}, \bibfnamefont{L.}},
  \bibinfo{author}{\bibfnamefont{R.~X.} \bibnamefont{Bian}}, and
  \bibinfo{author}{\bibfnamefont{X.~S.} \bibnamefont{Xie}},
  \bibinfo{year}{1997}, \bibinfo{journal}{Phys.\ Rev.\ Lett.}
  \textbf{\bibinfo{volume}{79}}, \bibinfo{pages}{645}.

\bibitem[{\citenamefont{Ochiai and Ohtaka}(2004{\natexlab{a}})}]{OO04}
\bibinfo{author}{\bibnamefont{Ochiai}, \bibfnamefont{T.}}, and
  \bibinfo{author}{\bibfnamefont{K.}~\bibnamefont{Ohtaka}},
  \bibinfo{year}{2004}{\natexlab{a}}, \bibinfo{journal}{Phys.\ Rev.\ B}
  \textbf{\bibinfo{volume}{69}}, \bibinfo{pages}{125106}.

\bibitem[{\citenamefont{Ochiai and Ohtaka}(2004{\natexlab{b}})}]{OO04_2}
\bibinfo{author}{\bibnamefont{Ochiai}, \bibfnamefont{T.}}, and
  \bibinfo{author}{\bibfnamefont{K.}~\bibnamefont{Ohtaka}},
  \bibinfo{year}{2004}{\natexlab{b}}, \bibinfo{journal}{Phys.\ Rev.\ B}
  \textbf{\bibinfo{volume}{69}}, \bibinfo{pages}{125107}.

\bibitem[{\citenamefont{Ochiai and Ohtaka}(2005)}]{OO05}
\bibinfo{author}{\bibnamefont{Ochiai}, \bibfnamefont{T.}}, and
  \bibinfo{author}{\bibfnamefont{K.}~\bibnamefont{Ohtaka}},
  \bibinfo{year}{2005}, \bibinfo{journal}{Opt.\ Express}
  \textbf{\bibinfo{volume}{13}}, \bibinfo{pages}{7683}.

\bibitem[{\citenamefont{Ochiai and Ohtaka}(2006)}]{OO06}
\bibinfo{author}{\bibnamefont{Ochiai}, \bibfnamefont{T.}}, and
  \bibinfo{author}{\bibfnamefont{K.}~\bibnamefont{Ohtaka}},
  \bibinfo{year}{2006}, \bibinfo{journal}{Opt.\ Express}
  \textbf{\bibinfo{volume}{14}}, \bibinfo{pages}{7378}.

\bibitem[{\citenamefont{Ohkuma} \emph{et~al.}(1991)\citenamefont{Ohkuma, Okuda,
  and Tsumori}}]{OOT91}
\bibinfo{author}{\bibnamefont{Ohkuma}, \bibfnamefont{J.}},
  \bibinfo{author}{\bibfnamefont{S.}~\bibnamefont{Okuda}}, and
  \bibinfo{author}{\bibfnamefont{K.}~\bibnamefont{Tsumori}},
  \bibinfo{year}{1991}, \bibinfo{journal}{Phys.\ Rev.\ Lett.}
  \textbf{\bibinfo{volume}{66}}, \bibinfo{pages}{1967}.

\bibitem[{\citenamefont{Ohtaka and Yamaguti}(2001)}]{OY01}
\bibinfo{author}{\bibnamefont{Ohtaka}, \bibfnamefont{K.}}, and
  \bibinfo{author}{\bibfnamefont{S.}~\bibnamefont{Yamaguti}},
  \bibinfo{year}{2001}, \bibinfo{journal}{Optics and Spectroscopy}
  \textbf{\bibinfo{volume}{91}}, \bibinfo{pages}{506}.

\bibitem[{\citenamefont{Oleshko}(2008)}]{O08}
\bibinfo{author}{\bibnamefont{Oleshko}, \bibfnamefont{V.~P.}},
  \bibinfo{year}{2008}, \bibinfo{journal}{Plasmonics}
  \textbf{\bibinfo{volume}{3}}, \bibinfo{pages}{41}.

\bibitem[{\citenamefont{Olsen and Kolbenstvedt}(1980)}]{OK1980}
\bibinfo{author}{\bibnamefont{Olsen}, \bibfnamefont{H.~A.}}, and
  \bibinfo{author}{\bibfnamefont{H.}~\bibnamefont{Kolbenstvedt}},
  \bibinfo{year}{1980}, \bibinfo{journal}{Phys.\ Rev.\ A}
  \textbf{\bibinfo{volume}{21}}, \bibinfo{pages}{1987}.

\bibitem[{\citenamefont{Onishchenko}
  \emph{et~al.}(2002)\citenamefont{Onishchenko, {D. Yu. Sidorenko}, and
  Sotnikov}}]{OSS02}
\bibinfo{author}{\bibnamefont{Onishchenko}, \bibfnamefont{I.~N.}},
  \bibinfo{author}{\bibnamefont{{D. Yu. Sidorenko}}}, and
  \bibinfo{author}{\bibfnamefont{G.~V.} \bibnamefont{Sotnikov}},
  \bibinfo{year}{2002}, \bibinfo{journal}{Phys.\ Rev.\ E}
  \textbf{\bibinfo{volume}{65}}, \bibinfo{pages}{066501}.

\bibitem[{\citenamefont{Ortega} \emph{et~al.}(1998)\citenamefont{Ortega,
  {Garc\'{\i}a de Abajo}, Echenique, Manke, Kalka, {M. D\"{a}hne}, Ochs,
  Molodtsov, and Rubio}}]{paper031}
\bibinfo{author}{\bibnamefont{Ortega}, \bibfnamefont{J.~E.}},
  \bibinfo{author}{\bibfnamefont{F.~J.} \bibnamefont{{Garc\'{\i}a de Abajo}}},
  \bibinfo{author}{\bibfnamefont{P.~M.} \bibnamefont{Echenique}},
  \bibinfo{author}{\bibfnamefont{I.}~\bibnamefont{Manke}},
  \bibinfo{author}{\bibfnamefont{T.}~\bibnamefont{Kalka}},
  \bibinfo{author}{\bibnamefont{{M. D\"{a}hne}}},
  \bibinfo{author}{\bibfnamefont{D.}~\bibnamefont{Ochs}},
  \bibinfo{author}{\bibfnamefont{S.~L.} \bibnamefont{Molodtsov}}, and
  \bibinfo{author}{\bibfnamefont{A.}~\bibnamefont{Rubio}},
  \bibinfo{year}{1998}, \bibinfo{journal}{Phys.\ Rev.\ B}
  \textbf{\bibinfo{volume}{58}}, \bibinfo{pages}{2233}.

\bibitem[{\citenamefont{Osterwalder}
  \emph{et~al.}(1990)\citenamefont{Osterwalder, Greber, H\"ufner, and
  Schlapbach}}]{OGH1990}
\bibinfo{author}{\bibnamefont{Osterwalder}, \bibfnamefont{J.}},
  \bibinfo{author}{\bibfnamefont{T.}~\bibnamefont{Greber}},
  \bibinfo{author}{\bibfnamefont{S.}~\bibnamefont{H\"ufner}}, and
  \bibinfo{author}{\bibfnamefont{L.}~\bibnamefont{Schlapbach}},
  \bibinfo{year}{1990}, \bibinfo{journal}{Phys.\ Rev.\ B}
  \textbf{\bibinfo{volume}{41}}, \bibinfo{pages}{12495}.

\bibitem[{\citenamefont{Otto}(1967)}]{O1967}
\bibinfo{author}{\bibnamefont{Otto}, \bibfnamefont{A.}}, \bibinfo{year}{1967},
  \bibinfo{journal}{Phys.~Status~Solidi} \textbf{\bibinfo{volume}{22}},
  \bibinfo{pages}{401}.

\bibitem[{\citenamefont{Ouyang} \emph{et~al.}(1992)\citenamefont{Ouyang,
  Batson, and Isaacson}}]{OBI92}
\bibinfo{author}{\bibnamefont{Ouyang}, \bibfnamefont{F.}},
  \bibinfo{author}{\bibfnamefont{P.~E.} \bibnamefont{Batson}}, and
  \bibinfo{author}{\bibfnamefont{M.}~\bibnamefont{Isaacson}},
  \bibinfo{year}{1992}, \bibinfo{journal}{Phys.\ Rev.\ B}
  \textbf{\bibinfo{volume}{46}}, \bibinfo{pages}{15421}.

\bibitem[{\citenamefont{Ouyang and Isaacson}(1989{\natexlab{a}})}]{OI1989_2}
\bibinfo{author}{\bibnamefont{Ouyang}, \bibfnamefont{F.}}, and
  \bibinfo{author}{\bibfnamefont{M.}~\bibnamefont{Isaacson}},
  \bibinfo{year}{1989}{\natexlab{a}}, \bibinfo{journal}{Ultramicroscopy}
  \textbf{\bibinfo{volume}{31}}, \bibinfo{pages}{345}.

\bibitem[{\citenamefont{Ouyang and Isaacson}(1989{\natexlab{b}})}]{OI1989}
\bibinfo{author}{\bibnamefont{Ouyang}, \bibfnamefont{F.}}, and
  \bibinfo{author}{\bibfnamefont{M.}~\bibnamefont{Isaacson}},
  \bibinfo{year}{1989}{\natexlab{b}}, \bibinfo{journal}{Philos.\ Mag.\ B}
  \textbf{\bibinfo{volume}{60}}, \bibinfo{pages}{481}.

\bibitem[{\citenamefont{Oxley and Allen}(1998)}]{OA98}
\bibinfo{author}{\bibnamefont{Oxley}, \bibfnamefont{M.~P.}}, and
  \bibinfo{author}{\bibfnamefont{L.~J.} \bibnamefont{Allen}},
  \bibinfo{year}{1998}, \bibinfo{journal}{Phys.\ Rev.\ B}
  \textbf{\bibinfo{volume}{57}}, \bibinfo{pages}{3273}.

\bibitem[{\citenamefont{Oxley and Pennycook}(2008)}]{OP08}
\bibinfo{author}{\bibnamefont{Oxley}, \bibfnamefont{M.~P.}}, and
  \bibinfo{author}{\bibfnamefont{S.~J.} \bibnamefont{Pennycook}},
  \bibinfo{year}{2008}, \bibinfo{journal}{Micron}
  \textbf{\bibinfo{volume}{39}}, \bibinfo{pages}{676}.

\bibitem[{\citenamefont{Ozawa}(1990)}]{O1990}
\bibinfo{author}{\bibnamefont{Ozawa}, \bibfnamefont{L.}}, \bibinfo{year}{1990},
  \emph{\bibinfo{title}{Cathodoluminescence: Theory and Applications}}
  (\bibinfo{publisher}{VCH Publishers}, \bibinfo{address}{New York}).

\bibitem[{\citenamefont{Ozbay}(2006)}]{O06}
\bibinfo{author}{\bibnamefont{Ozbay}, \bibfnamefont{E.}}, \bibinfo{year}{2006},
  \bibinfo{journal}{Science} \textbf{\bibinfo{volume}{311}},
  \bibinfo{pages}{189}.

\bibitem[{\citenamefont{Palik}(1985)}]{P1985}
\bibinfo{author}{\bibnamefont{Palik}, \bibfnamefont{E.~D.}},
  \bibinfo{year}{1985}, \emph{\bibinfo{title}{Handbook of Optical Constants of
  Solids}} (\bibinfo{publisher}{Academic Press}, \bibinfo{address}{New York}).

\bibitem[{\citenamefont{Palik}(1991)}]{P91}
\bibinfo{author}{\bibnamefont{Palik}, \bibfnamefont{E.~D.}},
  \bibinfo{year}{1991}, \emph{\bibinfo{title}{Handbook of Optical Constants of
  Solids II}} (\bibinfo{publisher}{Academic Press}, \bibinfo{address}{San
  Diego}).

\bibitem[{\citenamefont{Palmer and Rous}(1992)}]{PR92}
\bibinfo{author}{\bibnamefont{Palmer}, \bibfnamefont{R.~E.}}, and
  \bibinfo{author}{\bibfnamefont{P.~J.} \bibnamefont{Rous}},
  \bibinfo{year}{1992}, \bibinfo{journal}{Rev.\ Mod.\ Phys.}
  \textbf{\bibinfo{volume}{64}}, \bibinfo{pages}{383}.

\bibitem[{\citenamefont{Patro}(1982)}]{P1982}
\bibinfo{author}{\bibnamefont{Patro}, \bibfnamefont{D.~N.}},
  \bibinfo{year}{1982}, \bibinfo{journal}{Phys.\ Rev.\ Lett.}
  \textbf{\bibinfo{volume}{49}}, \bibinfo{pages}{1083}.

\bibitem[{\citenamefont{Pawlak} \emph{et~al.}(2008)\citenamefont{Pawlak,
  Kolodziejak, Rozniatowski, Diduszko, Kaczkan, Malinowski, Piersa,
  Kisielewski, and Lukasiewicz}}]{PKR08}
\bibinfo{author}{\bibnamefont{Pawlak}, \bibfnamefont{D.~A.}},
  \bibinfo{author}{\bibfnamefont{K.}~\bibnamefont{Kolodziejak}},
  \bibinfo{author}{\bibfnamefont{K.}~\bibnamefont{Rozniatowski}},
  \bibinfo{author}{\bibfnamefont{R.}~\bibnamefont{Diduszko}},
  \bibinfo{author}{\bibfnamefont{M.}~\bibnamefont{Kaczkan}},
  \bibinfo{author}{\bibfnamefont{M.}~\bibnamefont{Malinowski}},
  \bibinfo{author}{\bibfnamefont{M.}~\bibnamefont{Piersa}},
  \bibinfo{author}{\bibfnamefont{J.}~\bibnamefont{Kisielewski}}, and
  \bibinfo{author}{\bibfnamefont{T.}~\bibnamefont{Lukasiewicz}},
  \bibinfo{year}{2008}, \bibinfo{journal}{Crystal Growth \& Design}
  \textbf{\bibinfo{volume}{8}}, \bibinfo{pages}{1243}.

\bibitem[{\citenamefont{Peale} \emph{et~al.}(2008)\citenamefont{Peale,
  Lopatiuk, Cleary, Santos, Henderson, Clark, Chernyak, Winningham, {Del
  Barco}, Heinrich, and Buchwald}}]{PLC08}
\bibinfo{author}{\bibnamefont{Peale}, \bibfnamefont{R.~E.}},
  \bibinfo{author}{\bibfnamefont{O.}~\bibnamefont{Lopatiuk}},
  \bibinfo{author}{\bibfnamefont{J.}~\bibnamefont{Cleary}},
  \bibinfo{author}{\bibfnamefont{S.}~\bibnamefont{Santos}},
  \bibinfo{author}{\bibfnamefont{J.}~\bibnamefont{Henderson}},
  \bibinfo{author}{\bibfnamefont{D.}~\bibnamefont{Clark}},
  \bibinfo{author}{\bibfnamefont{L.}~\bibnamefont{Chernyak}},
  \bibinfo{author}{\bibfnamefont{T.~A.} \bibnamefont{Winningham}},
  \bibinfo{author}{\bibfnamefont{E.}~\bibnamefont{{Del Barco}}},
  \bibinfo{author}{\bibfnamefont{H.}~\bibnamefont{Heinrich}}, and
  \bibinfo{author}{\bibfnamefont{W.~R.} \bibnamefont{Buchwald}},
  \bibinfo{year}{2008}, \bibinfo{journal}{J.\ Opt.\ Soc.\ Am.\ B}
  \textbf{\bibinfo{volume}{25}}, \bibinfo{pages}{1708}.

\bibitem[{\citenamefont{Pendry and {Mart\'{\i}n-Moreno}}(1994)}]{PM94}
\bibinfo{author}{\bibnamefont{Pendry}, \bibfnamefont{J.~B.}}, and
  \bibinfo{author}{\bibfnamefont{L.}~\bibnamefont{{Mart\'{\i}n-Moreno}}},
  \bibinfo{year}{1994}, \bibinfo{journal}{Phys.\ Rev.\ B}
  \textbf{\bibinfo{volume}{50}}, \bibinfo{pages}{5062}.

\bibitem[{\citenamefont{Peng} \emph{et~al.}(2001)\citenamefont{Peng, Fehlhaber,
  Bursill, and McCulloch}}]{PFB01}
\bibinfo{author}{\bibnamefont{Peng}, \bibfnamefont{J.~L.}},
  \bibinfo{author}{\bibfnamefont{R.~P.} \bibnamefont{Fehlhaber}},
  \bibinfo{author}{\bibfnamefont{L.~A.} \bibnamefont{Bursill}}, and
  \bibinfo{author}{\bibfnamefont{D.~G.} \bibnamefont{McCulloch}},
  \bibinfo{year}{2001}, \bibinfo{journal}{J.\ Appl.\ Phys.}
  \textbf{\bibinfo{volume}{89}}, \bibinfo{pages}{6204}.

\bibitem[{\citenamefont{Pennycook}(2008)}]{P08}
\bibinfo{author}{\bibnamefont{Pennycook}, \bibfnamefont{S.~J.}},
  \bibinfo{year}{2008}, \bibinfo{journal}{Scanning}
  \textbf{\bibinfo{volume}{30}}, \bibinfo{pages}{287}.

\bibitem[{\citenamefont{Persson and Baratoff}(1992)}]{PB92}
\bibinfo{author}{\bibnamefont{Persson}, \bibfnamefont{B.~N.~J.}}, and
  \bibinfo{author}{\bibfnamefont{A.}~\bibnamefont{Baratoff}},
  \bibinfo{year}{1992}, \bibinfo{journal}{Phys.\ Rev.\ Lett.}
  \textbf{\bibinfo{volume}{68}}, \bibinfo{pages}{3224}.

\bibitem[{\citenamefont{Pettit} \emph{et~al.}(1975)\citenamefont{Pettit,
  Silcox, and Vincent}}]{PSV1975}
\bibinfo{author}{\bibnamefont{Pettit}, \bibfnamefont{R.~B.}},
  \bibinfo{author}{\bibfnamefont{J.}~\bibnamefont{Silcox}}, and
  \bibinfo{author}{\bibfnamefont{R.}~\bibnamefont{Vincent}},
  \bibinfo{year}{1975}, \bibinfo{journal}{Phys.\ Rev.\ B}
  \textbf{\bibinfo{volume}{11}}, \bibinfo{pages}{3116}.

\bibitem[{\citenamefont{Pfl\"uger and Fink}(1991)}]{PF91}
\bibinfo{author}{\bibnamefont{Pfl\"uger}, \bibfnamefont{J.}}, and
  \bibinfo{author}{\bibfnamefont{J.}~\bibnamefont{Fink}}, \bibinfo{year}{1991},
  \emph{\bibinfo{title}{Handbook of Optical Constants in Solids II}}
  (\bibinfo{publisher}{Academic Press}, \bibinfo{address}{San Diego}), chapter
  \bibinfo{chapter}{Determination of optical constants by high-energy,
  electron-energy-loss spectroscopy (EELS)}, pp. \bibinfo{pages}{293--311}.

\bibitem[{\citenamefont{Pfl\"uger} \emph{et~al.}(1984)\citenamefont{Pfl\"uger,
  Fink, Weber, Bohnen, and Crecelius}}]{PFW1984}
\bibinfo{author}{\bibnamefont{Pfl\"uger}, \bibfnamefont{J.}},
  \bibinfo{author}{\bibfnamefont{J.}~\bibnamefont{Fink}},
  \bibinfo{author}{\bibfnamefont{W.}~\bibnamefont{Weber}},
  \bibinfo{author}{\bibfnamefont{K.~P.} \bibnamefont{Bohnen}}, and
  \bibinfo{author}{\bibfnamefont{G.}~\bibnamefont{Crecelius}},
  \bibinfo{year}{1984}, \bibinfo{journal}{Phys.\ Rev.\ B}
  \textbf{\bibinfo{volume}{30}}, \bibinfo{pages}{1155}.

\bibitem[{\citenamefont{Pfl\"uger} \emph{et~al.}(1985)\citenamefont{Pfl\"uger,
  Fink, Weber, Bohnen, and Crecelius}}]{PFW1985}
\bibinfo{author}{\bibnamefont{Pfl\"uger}, \bibfnamefont{J.}},
  \bibinfo{author}{\bibfnamefont{J.}~\bibnamefont{Fink}},
  \bibinfo{author}{\bibfnamefont{W.}~\bibnamefont{Weber}},
  \bibinfo{author}{\bibfnamefont{K.~P.} \bibnamefont{Bohnen}}, and
  \bibinfo{author}{\bibfnamefont{G.}~\bibnamefont{Crecelius}},
  \bibinfo{year}{1985}, \bibinfo{journal}{Phys.\ Rev.\ B}
  \textbf{\bibinfo{volume}{31}}, \bibinfo{pages}{1244}.

\bibitem[{\citenamefont{Pijper and Kruit}(1991)}]{PK91}
\bibinfo{author}{\bibnamefont{Pijper}, \bibfnamefont{F.~J.}}, and
  \bibinfo{author}{\bibfnamefont{P.}~\bibnamefont{Kruit}},
  \bibinfo{year}{1991}, \bibinfo{journal}{Phys.\ Rev.\ B}
  \textbf{\bibinfo{volume}{44}}, \bibinfo{pages}{9192}.

\bibitem[{\citenamefont{Pines and Bohm}(1952)}]{PB1952}
\bibinfo{author}{\bibnamefont{Pines}, \bibfnamefont{D.}}, and
  \bibinfo{author}{\bibfnamefont{D.}~\bibnamefont{Bohm}}, \bibinfo{year}{1952},
  \bibinfo{journal}{Phys.\ Rev.} \textbf{\bibinfo{volume}{85}},
  \bibinfo{pages}{338}.

\bibitem[{\citenamefont{Pines and Nozi\`{e}res}(1966)}]{PN1966}
\bibinfo{author}{\bibnamefont{Pines}, \bibfnamefont{D.}}, and
  \bibinfo{author}{\bibfnamefont{P.}~\bibnamefont{Nozi\`{e}res}},
  \bibinfo{year}{1966}, \emph{\bibinfo{title}{The Theory of Quantum Liquids}}
  (\bibinfo{publisher}{W. A. Benjamin, Inc.}, \bibinfo{address}{New York}).

\bibitem[{\citenamefont{Pitarke} \emph{et~al.}(2007)\citenamefont{Pitarke,
  Silkin, Chulkov, and Echenique}}]{PSC07}
\bibinfo{author}{\bibnamefont{Pitarke}, \bibfnamefont{J.~M.}},
  \bibinfo{author}{\bibfnamefont{V.~M.} \bibnamefont{Silkin}},
  \bibinfo{author}{\bibfnamefont{E.~V.} \bibnamefont{Chulkov}}, and
  \bibinfo{author}{\bibfnamefont{P.~M.} \bibnamefont{Echenique}},
  \bibinfo{year}{2007}, \bibinfo{journal}{Rep.\ Prog.\ Phys.}
  \textbf{\bibinfo{volume}{70}}, \bibinfo{pages}{1}.

\bibitem[{\citenamefont{Pochon} \emph{et~al.}(2004)\citenamefont{Pochon,
  MacDonald, Knize, and Zheludev}}]{PMK04}
\bibinfo{author}{\bibnamefont{Pochon}, \bibfnamefont{S.}},
  \bibinfo{author}{\bibfnamefont{K.~F.} \bibnamefont{MacDonald}},
  \bibinfo{author}{\bibfnamefont{R.~J.} \bibnamefont{Knize}}, and
  \bibinfo{author}{\bibfnamefont{N.~I.} \bibnamefont{Zheludev}},
  \bibinfo{year}{2004}, \bibinfo{journal}{Phys.\ Rev.\ Lett.}
  \textbf{\bibinfo{volume}{92}}, \bibinfo{pages}{145702}.

\bibitem[{\citenamefont{Pogorzelski and Yeh}(1973)}]{PY1973}
\bibinfo{author}{\bibnamefont{Pogorzelski}, \bibfnamefont{R.}}, and
  \bibinfo{author}{\bibfnamefont{C.}~\bibnamefont{Yeh}}, \bibinfo{year}{1973},
  \bibinfo{journal}{Phys.\ Rev.\ A} \textbf{\bibinfo{volume}{8}},
  \bibinfo{pages}{137}.

\bibitem[{\citenamefont{Pokrant} \emph{et~al.}(2006)\citenamefont{Pokrant,
  Pantel, and Cheynet}}]{PPC06}
\bibinfo{author}{\bibnamefont{Pokrant}, \bibfnamefont{S.}},
  \bibinfo{author}{\bibfnamefont{R.}~\bibnamefont{Pantel}}, and
  \bibinfo{author}{\bibfnamefont{M.}~\bibnamefont{Cheynet}},
  \bibinfo{year}{2006}, \bibinfo{journal}{Microelectron.\ Eng.}
  \textbf{\bibinfo{volume}{83}}, \bibinfo{pages}{2364–}.

\bibitem[{\citenamefont{Potapov} \emph{et~al.}(2007)\citenamefont{Potapov,
  Verbeeck, Schattschneider, Lichte, and {van Dyck}}}]{PVS07}
\bibinfo{author}{\bibnamefont{Potapov}, \bibfnamefont{P.~L.}},
  \bibinfo{author}{\bibfnamefont{J.}~\bibnamefont{Verbeeck}},
  \bibinfo{author}{\bibfnamefont{P.}~\bibnamefont{Schattschneider}},
  \bibinfo{author}{\bibfnamefont{H.}~\bibnamefont{Lichte}}, and
  \bibinfo{author}{\bibfnamefont{D.}~\bibnamefont{{van Dyck}}},
  \bibinfo{year}{2007}, \bibinfo{journal}{Ultramicroscopy}
  \textbf{\bibinfo{volume}{107}}, \bibinfo{pages}{559}.

\bibitem[{\citenamefont{Potylitsyn}(1998)}]{P98}
\bibinfo{author}{\bibnamefont{Potylitsyn}, \bibfnamefont{A.~P.}},
  \bibinfo{year}{1998}, \bibinfo{journal}{Nucl.\ Instrum.\ Methods\ Phys.\
  Res.\ B} \textbf{\bibinfo{volume}{145}}, \bibinfo{pages}{169}.

\bibitem[{\citenamefont{Powell}(1965)}]{P1965}
\bibinfo{author}{\bibnamefont{Powell}, \bibfnamefont{C.~J.}},
  \bibinfo{year}{1965}, \bibinfo{journal}{Phys.\ Rev.\ Lett.}
  \textbf{\bibinfo{volume}{15}}, \bibinfo{pages}{852}.

\bibitem[{\citenamefont{Powell}(1968)}]{P1968}
\bibinfo{author}{\bibnamefont{Powell}, \bibfnamefont{C.~J.}},
  \bibinfo{year}{1968}, \bibinfo{journal}{Phys.\ Rev.}
  \textbf{\bibinfo{volume}{175}}, \bibinfo{pages}{972}.

\bibitem[{\citenamefont{Powell and Swan}(1959)}]{PS1959}
\bibinfo{author}{\bibnamefont{Powell}, \bibfnamefont{C.~J.}}, and
  \bibinfo{author}{\bibfnamefont{J.~B.} \bibnamefont{Swan}},
  \bibinfo{year}{1959}, \bibinfo{journal}{Phys.\ Rev.}
  \textbf{\bibinfo{volume}{115}}, \bibinfo{pages}{869}.

\bibitem[{\citenamefont{Pratesi} \emph{et~al.}(2003)\citenamefont{Pratesi,
  Guidice, Vishnevsky, Manfredotti, and Cipriani}}]{PGV03}
\bibinfo{author}{\bibnamefont{Pratesi}, \bibfnamefont{G.}},
  \bibinfo{author}{\bibfnamefont{A.~L.} \bibnamefont{Guidice}},
  \bibinfo{author}{\bibfnamefont{S.}~\bibnamefont{Vishnevsky}},
  \bibinfo{author}{\bibfnamefont{C.}~\bibnamefont{Manfredotti}}, and
  \bibinfo{author}{\bibfnamefont{C.}~\bibnamefont{Cipriani}},
  \bibinfo{year}{2003}, \bibinfo{journal}{Am.\ Mineral.}
  \textbf{\bibinfo{volume}{88}}, \bibinfo{pages}{1778}.

\bibitem[{\citenamefont{Purcell}(1946)}]{P1946}
\bibinfo{author}{\bibnamefont{Purcell}, \bibfnamefont{E.~M.}},
  \bibinfo{year}{1946}, \bibinfo{journal}{Phys.\ Rev.}
  \textbf{\bibinfo{volume}{69}}, \bibinfo{pages}{681}.

\bibitem[{\citenamefont{Purcell and Pennypacker}(1973)}]{PP1973}
\bibinfo{author}{\bibnamefont{Purcell}, \bibfnamefont{E.~M.}}, and
  \bibinfo{author}{\bibfnamefont{C.~R.} \bibnamefont{Pennypacker}},
  \bibinfo{year}{1973}, \bibinfo{journal}{Astrophys.\ J.}
  \textbf{\bibinfo{volume}{186}}, \bibinfo{pages}{705}.

\bibitem[{\citenamefont{Raether}(1967)}]{R1967_2}
\bibinfo{author}{\bibnamefont{Raether}, \bibfnamefont{H.}},
  \bibinfo{year}{1967}, \bibinfo{journal}{Surf.\ Sci.}
  \textbf{\bibinfo{volume}{8}}, \bibinfo{pages}{233}.

\bibitem[{\citenamefont{Raether}(1980)}]{R1980}
\bibinfo{author}{\bibnamefont{Raether}, \bibfnamefont{H.}},
  \bibinfo{year}{1980}, \emph{\bibinfo{title}{Excitation of Plasmons and
  Interband Transitions by Electrons}}, volume~\bibinfo{volume}{88} of
  \emph{\bibinfo{series}{Springer Tracks in Modern Physics}}
  (\bibinfo{publisher}{Springer-Verlag}, \bibinfo{address}{Berlin}).

\bibitem[{\citenamefont{Raether}(1988)}]{R1988}
\bibinfo{author}{\bibnamefont{Raether}, \bibfnamefont{H.}},
  \bibinfo{year}{1988}, \emph{\bibinfo{title}{Surface Plasmons on Smooth and
  Rough Surfaces and on Gratings}}, volume \bibinfo{volume}{111} of
  \emph{\bibinfo{series}{Springer Tracks in Modern Physics}}
  (\bibinfo{publisher}{Springer-Verlag}, \bibinfo{address}{Berlin}).

\bibitem[{\citenamefont{Rafferty and Brown}(1998)}]{RB98}
\bibinfo{author}{\bibnamefont{Rafferty}, \bibfnamefont{B.}}, and
  \bibinfo{author}{\bibfnamefont{L.~M.} \bibnamefont{Brown}},
  \bibinfo{year}{1998}, \bibinfo{journal}{Phys.\ Rev.\ B}
  \textbf{\bibinfo{volume}{58}}, \bibinfo{pages}{10326}.

\bibitem[{\citenamefont{Rang} \emph{et~al.}(2008)\citenamefont{Rang, Jones,
  Zhou, Li, Wiley, Xia, and Raschke}}]{RJZ08}
\bibinfo{author}{\bibnamefont{Rang}, \bibfnamefont{M.}},
  \bibinfo{author}{\bibfnamefont{A.~C.} \bibnamefont{Jones}},
  \bibinfo{author}{\bibfnamefont{F.}~\bibnamefont{Zhou}},
  \bibinfo{author}{\bibfnamefont{Z.~Y.} \bibnamefont{Li}},
  \bibinfo{author}{\bibfnamefont{B.~J.} \bibnamefont{Wiley}},
  \bibinfo{author}{\bibfnamefont{Y.}~\bibnamefont{Xia}}, and
  \bibinfo{author}{\bibfnamefont{M.~B.} \bibnamefont{Raschke}},
  \bibinfo{year}{2008}, \bibinfo{journal}{Nano\ Lett.}
  \textbf{\bibinfo{volume}{8}}, \bibinfo{pages}{3357}.

\bibitem[{\citenamefont{Rapoport and Khattak}(1988)}]{RK1988}
\bibinfo{author}{\bibnamefont{Rapoport}, \bibfnamefont{W.~R.}}, and
  \bibinfo{author}{\bibfnamefont{C.~P.} \bibnamefont{Khattak}},
  \bibinfo{year}{1988}, \bibinfo{journal}{Appl.\ Opt.}
  \textbf{\bibinfo{volume}{27}}, \bibinfo{pages}{2677}.

\bibitem[{\citenamefont{Reed} \emph{et~al.}(1999)\citenamefont{Reed, Chen,
  MacDonald, Silcox, and Bertsch}}]{RCM99}
\bibinfo{author}{\bibnamefont{Reed}, \bibfnamefont{B.~W.}},
  \bibinfo{author}{\bibfnamefont{J.~M.} \bibnamefont{Chen}},
  \bibinfo{author}{\bibfnamefont{N.~C.} \bibnamefont{MacDonald}},
  \bibinfo{author}{\bibfnamefont{J.}~\bibnamefont{Silcox}}, and
  \bibinfo{author}{\bibfnamefont{G.~F.} \bibnamefont{Bertsch}},
  \bibinfo{year}{1999}, \bibinfo{journal}{Phys.\ Rev.\ B}
  \textbf{\bibinfo{volume}{60}}, \bibinfo{pages}{5641}.

\bibitem[{\citenamefont{Righini} \emph{et~al.}(2008)\citenamefont{Righini,
  Volpe, Girard, Petrov, and Quidant}}]{RVG08}
\bibinfo{author}{\bibnamefont{Righini}, \bibfnamefont{M.}},
  \bibinfo{author}{\bibfnamefont{G.}~\bibnamefont{Volpe}},
  \bibinfo{author}{\bibfnamefont{C.}~\bibnamefont{Girard}},
  \bibinfo{author}{\bibfnamefont{D.}~\bibnamefont{Petrov}}, and
  \bibinfo{author}{\bibfnamefont{R.}~\bibnamefont{Quidant}},
  \bibinfo{year}{2008}, \bibinfo{journal}{Phys.\ Rev.\ Lett.}
  \textbf{\bibinfo{volume}{100}}, \bibinfo{pages}{186804}.

\bibitem[{\citenamefont{Ritchie}(1957)}]{R1957}
\bibinfo{author}{\bibnamefont{Ritchie}, \bibfnamefont{R.~H.}},
  \bibinfo{year}{1957}, \bibinfo{journal}{Phys.\ Rev.}
  \textbf{\bibinfo{volume}{106}}, \bibinfo{pages}{874}.

\bibitem[{\citenamefont{Ritchie} \emph{et~al.}(1968)\citenamefont{Ritchie,
  Arakawa, Cowan, and Hamm}}]{RAC1968}
\bibinfo{author}{\bibnamefont{Ritchie}, \bibfnamefont{R.~H.}},
  \bibinfo{author}{\bibfnamefont{E.~T.} \bibnamefont{Arakawa}},
  \bibinfo{author}{\bibfnamefont{J.~J.} \bibnamefont{Cowan}}, and
  \bibinfo{author}{\bibfnamefont{R.~N.} \bibnamefont{Hamm}},
  \bibinfo{year}{1968}, \bibinfo{journal}{Phys.\ Rev.\ Lett.}
  \textbf{\bibinfo{volume}{21}}, \bibinfo{pages}{1530}.

\bibitem[{\citenamefont{Ritchie} \emph{et~al.}(1964)\citenamefont{Ritchie,
  Ashley, and Emerson}}]{RAE1965}
\bibinfo{author}{\bibnamefont{Ritchie}, \bibfnamefont{R.~H.}},
  \bibinfo{author}{\bibfnamefont{J.~C.} \bibnamefont{Ashley}}, and
  \bibinfo{author}{\bibfnamefont{L.~C.} \bibnamefont{Emerson}},
  \bibinfo{year}{1964}, \bibinfo{journal}{Phys.\ Rev.}
  \textbf{\bibinfo{volume}{135}}, \bibinfo{pages}{A759}.

\bibitem[{\citenamefont{Ritchie and Eldridge}(1962)}]{RE1962}
\bibinfo{author}{\bibnamefont{Ritchie}, \bibfnamefont{R.~H.}}, and
  \bibinfo{author}{\bibfnamefont{H.~B.} \bibnamefont{Eldridge}},
  \bibinfo{year}{1962}, \bibinfo{journal}{Phys.\ Rev.}
  \textbf{\bibinfo{volume}{126}}, \bibinfo{pages}{1935}.

\bibitem[{\citenamefont{Ritchie and Howie}(1988)}]{RH1988}
\bibinfo{author}{\bibnamefont{Ritchie}, \bibfnamefont{R.~H.}}, and
  \bibinfo{author}{\bibfnamefont{A.}~\bibnamefont{Howie}},
  \bibinfo{year}{1988}, \bibinfo{journal}{Philos.\ Mag.\ A}
  \textbf{\bibinfo{volume}{58}}, \bibinfo{pages}{753}.

\bibitem[{\citenamefont{Ritchie and Marusak}(1966)}]{RM1966}
\bibinfo{author}{\bibnamefont{Ritchie}, \bibfnamefont{R.~H.}}, and
  \bibinfo{author}{\bibfnamefont{A.~L.} \bibnamefont{Marusak}},
  \bibinfo{year}{1966}, \bibinfo{journal}{Surf.\ Sci.}
  \textbf{\bibinfo{volume}{4}}, \bibinfo{pages}{234}.

\bibitem[{\citenamefont{Rivacoba} \emph{et~al.}(2000)\citenamefont{Rivacoba,
  Zabala, and Aizpurua}}]{RZA00}
\bibinfo{author}{\bibnamefont{Rivacoba}, \bibfnamefont{A.}},
  \bibinfo{author}{\bibfnamefont{N.}~\bibnamefont{Zabala}}, and
  \bibinfo{author}{\bibfnamefont{J.}~\bibnamefont{Aizpurua}},
  \bibinfo{year}{2000}, \bibinfo{journal}{Prog. Surf. Sci.}
  \textbf{\bibinfo{volume}{65}}, \bibinfo{pages}{1}.

\bibitem[{\citenamefont{Rivacoba} \emph{et~al.}(1992)\citenamefont{Rivacoba,
  Zabala, and Echenique}}]{RZE92}
\bibinfo{author}{\bibnamefont{Rivacoba}, \bibfnamefont{A.}},
  \bibinfo{author}{\bibfnamefont{N.}~\bibnamefont{Zabala}}, and
  \bibinfo{author}{\bibfnamefont{P.~M.} \bibnamefont{Echenique}},
  \bibinfo{year}{1992}, \bibinfo{journal}{Phys.\ Rev.\ Lett.}
  \textbf{\bibinfo{volume}{69}}, \bibinfo{pages}{3362}.

\bibitem[{\citenamefont{Rocca}(1995)}]{R95}
\bibinfo{author}{\bibnamefont{Rocca}, \bibfnamefont{M.}}, \bibinfo{year}{1995},
  \bibinfo{journal}{Surf.\ Sci.\ Rep.} \textbf{\bibinfo{volume}{22}},
  \bibinfo{pages}{1}.

\bibitem[{\citenamefont{Rodt} \emph{et~al.}(2005)\citenamefont{Rodt, Schliwa,
  P\"otschke, Guffarth, and Bimberg}}]{RSP05}
\bibinfo{author}{\bibnamefont{Rodt}, \bibfnamefont{S.}},
  \bibinfo{author}{\bibfnamefont{A.}~\bibnamefont{Schliwa}},
  \bibinfo{author}{\bibfnamefont{K.}~\bibnamefont{P\"otschke}},
  \bibinfo{author}{\bibfnamefont{F.}~\bibnamefont{Guffarth}}, and
  \bibinfo{author}{\bibfnamefont{D.}~\bibnamefont{Bimberg}},
  \bibinfo{year}{2005}, \bibinfo{journal}{Phys.\ Rev.\ B}
  \textbf{\bibinfo{volume}{71}}, \bibinfo{pages}{155325}.

\bibitem[{\citenamefont{Rogacheva} \emph{et~al.}(2006)\citenamefont{Rogacheva,
  Fedotov, Schwanecke, and Zheludev}}]{RFS06}
\bibinfo{author}{\bibnamefont{Rogacheva}, \bibfnamefont{A.~V.}},
  \bibinfo{author}{\bibfnamefont{V.~A.} \bibnamefont{Fedotov}},
  \bibinfo{author}{\bibfnamefont{A.~S.} \bibnamefont{Schwanecke}}, and
  \bibinfo{author}{\bibfnamefont{N.~I.} \bibnamefont{Zheludev}},
  \bibinfo{year}{2006}, \bibinfo{journal}{Phys.\ Rev.\ Lett.}
  \textbf{\bibinfo{volume}{97}}, \bibinfo{pages}{177401}.

\bibitem[{\citenamefont{Rojas} \emph{et~al.}(1988)\citenamefont{Rojas, Claro,
  and Fuchs}}]{RCF1988}
\bibinfo{author}{\bibnamefont{Rojas}, \bibfnamefont{R.}},
  \bibinfo{author}{\bibfnamefont{F.}~\bibnamefont{Claro}}, and
  \bibinfo{author}{\bibfnamefont{R.}~\bibnamefont{Fuchs}},
  \bibinfo{year}{1988}, \bibinfo{journal}{Phys.\ Rev.\ B}
  \textbf{\bibinfo{volume}{37}}, \bibinfo{pages}{6799}.

\bibitem[{\citenamefont{Romero} \emph{et~al.}(2006)\citenamefont{Romero,
  Aizpurua, Bryant, and {Garc\'{\i}a de Abajo}}}]{paper114}
\bibinfo{author}{\bibnamefont{Romero}, \bibfnamefont{I.}},
  \bibinfo{author}{\bibfnamefont{J.}~\bibnamefont{Aizpurua}},
  \bibinfo{author}{\bibfnamefont{G.~W.} \bibnamefont{Bryant}}, and
  \bibinfo{author}{\bibfnamefont{F.~J.} \bibnamefont{{Garc\'{\i}a de Abajo}}},
  \bibinfo{year}{2006}, \bibinfo{journal}{Opt.\ Express}
  \textbf{\bibinfo{volume}{14}}, \bibinfo{pages}{9988}.

\bibitem[{\citenamefont{R\"{o}sler and Brauer}(1991)}]{RB91}
\bibinfo{author}{\bibnamefont{R\"{o}sler}, \bibfnamefont{M.}}, and
  \bibinfo{author}{\bibfnamefont{W.}~\bibnamefont{Brauer}},
  \bibinfo{year}{1991}, \emph{\bibinfo{title}{Particle Induced Electron
  Emission I}} (\bibinfo{publisher}{Springer-Verlag},
  \bibinfo{address}{Berlin}), volume \bibinfo{volume}{122} of
  \emph{\bibinfo{series}{Springer Tracts in Modern Physics}}, chapter
  \bibinfo{chapter}{Theory of electron emission from nearly-free-electron
  metals by proton and electron bombardment}, pp. \bibinfo{pages}{1--65}.

\bibitem[{\citenamefont{Ruppin}(1978)}]{R1978}
\bibinfo{author}{\bibnamefont{Ruppin}, \bibfnamefont{R.}},
  \bibinfo{year}{1978}, \bibinfo{journal}{J.\ Phys.\ Chem.\ Solids}
  \textbf{\bibinfo{volume}{39}}, \bibinfo{pages}{233}.

\bibitem[{\citenamefont{Ruppin}(1982)}]{R1982}
\bibinfo{author}{\bibnamefont{Ruppin}, \bibfnamefont{R.}},
  \bibinfo{year}{1982}, \bibinfo{journal}{Phys.\ Rev.\ B}
  \textbf{\bibinfo{volume}{26}}, \bibinfo{pages}{3440}.

\bibitem[{\citenamefont{Ruska}(1987)}]{R1987_2}
\bibinfo{author}{\bibnamefont{Ruska}, \bibfnamefont{E.}}, \bibinfo{year}{1987},
  \bibinfo{journal}{Rev.\ Mod.\ Phys.} \textbf{\bibinfo{volume}{59}},
  \bibinfo{pages}{627}.

\bibitem[{\citenamefont{Ruthermann}(1948)}]{R1948}
\bibinfo{author}{\bibnamefont{Ruthermann}, \bibfnamefont{G.}},
  \bibinfo{year}{1948}, \bibinfo{journal}{Ann.\ Phys.}
  \textbf{\bibinfo{volume}{2}}, \bibinfo{pages}{113}.

\bibitem[{\citenamefont{{Ru\v{z}i\v{c}ka} and Zrelov}(1993)}]{rmp-cr2}
\bibinfo{author}{\bibnamefont{{Ru\v{z}i\v{c}ka}}, \bibfnamefont{J.}}, and
  \bibinfo{author}{\bibfnamefont{V.~P.} \bibnamefont{Zrelov}},
  \bibinfo{year}{1993}, \bibinfo{journal}{Czech.\ J.\ Phys.}
  \textbf{\bibinfo{volume}{43}}, \bibinfo{pages}{551}.

\bibitem[{\citenamefont{Saito} \emph{et~al.}(1991)\citenamefont{Saito,
  Shinohara, and Ohshita}}]{SSO91}
\bibinfo{author}{\bibnamefont{Saito}, \bibfnamefont{Y.}},
  \bibinfo{author}{\bibfnamefont{H.}~\bibnamefont{Shinohara}}, and
  \bibinfo{author}{\bibfnamefont{A.}~\bibnamefont{Ohshita}},
  \bibinfo{year}{1991}, \bibinfo{journal}{Jpn.\ J.\ Appl.\ Phys.\ Part\
  2-Lett.} \textbf{\bibinfo{volume}{30}}, \bibinfo{pages}{L1068}.

\bibitem[{\citenamefont{Salisbury}(1970)}]{S1970}
\bibinfo{author}{\bibnamefont{Salisbury}, \bibfnamefont{W.~W.}},
  \bibinfo{year}{1970}, \bibinfo{journal}{J.\ Opt.\ Soc.\ Am.}
  \textbf{\bibinfo{volume}{60}}, \bibinfo{pages}{1279}.

\bibitem[{\citenamefont{S\'{a}nchez-Iglesias}
  \emph{et~al.}(2006)\citenamefont{S\'{a}nchez-Iglesias, Pastoriza-Santos,
  P\'{e}rez-Juste, Rodr\'{\i}guez-Gonz\'{a}lez, {Garc\'{\i}a de Abajo}, and
  Liz-Marz\'{a}n}}]{paper115}
\bibinfo{author}{\bibnamefont{S\'{a}nchez-Iglesias}, \bibfnamefont{A.}},
  \bibinfo{author}{\bibfnamefont{I.}~\bibnamefont{Pastoriza-Santos}},
  \bibinfo{author}{\bibfnamefont{J.}~\bibnamefont{P\'{e}rez-Juste}},
  \bibinfo{author}{\bibfnamefont{B.}~\bibnamefont{Rodr\'{\i}guez-Gonz\'{a}lez}%
}, \bibinfo{author}{\bibfnamefont{F.~J.} \bibnamefont{{Garc\'{\i}a de Abajo}}},
  and \bibinfo{author}{\bibfnamefont{L.}~\bibnamefont{Liz-Marz\'{a}n}},
  \bibinfo{year}{2006}, \bibinfo{journal}{Adv.\ Mater.}
  \textbf{\bibinfo{volume}{18}}, \bibinfo{pages}{2529}.

\bibitem[{\citenamefont{Sander} \emph{et~al.}(2001)\citenamefont{Sander,
  Gronsky, Lin, and Dresselhaus}}]{SGL01}
\bibinfo{author}{\bibnamefont{Sander}, \bibfnamefont{M.~S.}},
  \bibinfo{author}{\bibfnamefont{R.}~\bibnamefont{Gronsky}},
  \bibinfo{author}{\bibfnamefont{Y.~M.} \bibnamefont{Lin}}, and
  \bibinfo{author}{\bibfnamefont{M.~S.} \bibnamefont{Dresselhaus}},
  \bibinfo{year}{2001}, \bibinfo{journal}{J.\ Appl.\ Phys.}
  \textbf{\bibinfo{volume}{89}}, \bibinfo{pages}{2733}.

\bibitem[{\citenamefont{Sarid}(1981)}]{S1981}
\bibinfo{author}{\bibnamefont{Sarid}, \bibfnamefont{D.}}, \bibinfo{year}{1981},
  \bibinfo{journal}{Phys.\ Rev.\ Lett.} \textbf{\bibinfo{volume}{47}},
  \bibinfo{pages}{1927}.

\bibitem[{\citenamefont{Sauer} \emph{et~al.}(2000)\citenamefont{Sauer,
  Sternschulte, Wahl, Thonke, and Anthony}}]{SSW00}
\bibinfo{author}{\bibnamefont{Sauer}, \bibfnamefont{R.}},
  \bibinfo{author}{\bibfnamefont{H.}~\bibnamefont{Sternschulte}},
  \bibinfo{author}{\bibfnamefont{S.}~\bibnamefont{Wahl}},
  \bibinfo{author}{\bibfnamefont{K.}~\bibnamefont{Thonke}}, and
  \bibinfo{author}{\bibfnamefont{T.~R.} \bibnamefont{Anthony}},
  \bibinfo{year}{2000}, \bibinfo{journal}{Phys.\ Rev.\ Lett.}
  \textbf{\bibinfo{volume}{84}}, \bibinfo{pages}{4172}.

\bibitem[{\citenamefont{Schaffer} \emph{et~al.}(2009)\citenamefont{Schaffer,
  Hohenester, Tr\"ugler, and Hofer}}]{SHT09}
\bibinfo{author}{\bibnamefont{Schaffer}, \bibfnamefont{B.}},
  \bibinfo{author}{\bibfnamefont{U.}~\bibnamefont{Hohenester}},
  \bibinfo{author}{\bibfnamefont{A.}~\bibnamefont{Tr\"ugler}}, and
  \bibinfo{author}{\bibfnamefont{F.}~\bibnamefont{Hofer}},
  \bibinfo{year}{2009}, \bibinfo{journal}{Phys.\ Rev.\ B}
  \textbf{\bibinfo{volume}{79}}, \bibinfo{pages}{041401(R)}.

\bibitem[{\citenamefont{Schattschneider}
  \emph{et~al.}(1987)\citenamefont{Schattschneider, {F\"odermayr}, and
  Su}}]{SFS87}
\bibinfo{author}{\bibnamefont{Schattschneider}, \bibfnamefont{P.}},
  \bibinfo{author}{\bibfnamefont{F.}~\bibnamefont{{F\"odermayr}}}, and
  \bibinfo{author}{\bibfnamefont{D.~S.} \bibnamefont{Su}},
  \bibinfo{year}{1987}, \bibinfo{journal}{Phys.\ Rev.\ Lett.}
  \textbf{\bibinfo{volume}{59}}, \bibinfo{pages}{724}.

\bibitem[{\citenamefont{Schattschneider}
  \emph{et~al.}(2006)\citenamefont{Schattschneider, Rubino, H\'ebert, Rusz,
  Kune$\check{s}$, Nov\'ak, Carlino, Fabrizioli, Panaccione, and
  Rossi}}]{SRH06}
\bibinfo{author}{\bibnamefont{Schattschneider}, \bibfnamefont{P.}},
  \bibinfo{author}{\bibfnamefont{S.}~\bibnamefont{Rubino}},
  \bibinfo{author}{\bibfnamefont{C.}~\bibnamefont{H\'ebert}},
  \bibinfo{author}{\bibfnamefont{J.}~\bibnamefont{Rusz}},
  \bibinfo{author}{\bibfnamefont{J.}~\bibnamefont{Kune$\check{s}$}},
  \bibinfo{author}{\bibfnamefont{P.}~\bibnamefont{Nov\'ak}},
  \bibinfo{author}{\bibfnamefont{E.}~\bibnamefont{Carlino}},
  \bibinfo{author}{\bibfnamefont{M.}~\bibnamefont{Fabrizioli}},
  \bibinfo{author}{\bibfnamefont{G.}~\bibnamefont{Panaccione}}, and
  \bibinfo{author}{\bibfnamefont{G.}~\bibnamefont{Rossi}},
  \bibinfo{year}{2006}, \bibinfo{journal}{Nature}
  \textbf{\bibinfo{volume}{441}}, \bibinfo{pages}{486}.

\bibitem[{\citenamefont{Schattschneider and Werner}(2005)}]{SW05}
\bibinfo{author}{\bibnamefont{Schattschneider}, \bibfnamefont{P.}}, and
  \bibinfo{author}{\bibfnamefont{W.~S.~M.} \bibnamefont{Werner}},
  \bibinfo{year}{2005}, \bibinfo{journal}{J.\ Electron\ Spectrosc.\ Relat.\
  Phenom.} \textbf{\bibinfo{volume}{143}}, \bibinfo{pages}{81}.

\bibitem[{\citenamefont{Scheinfein}
  \emph{et~al.}(1985)\citenamefont{Scheinfein, Muray, and Isaacson}}]{SMI1985}
\bibinfo{author}{\bibnamefont{Scheinfein}, \bibfnamefont{M.}},
  \bibinfo{author}{\bibfnamefont{A.}~\bibnamefont{Muray}}, and
  \bibinfo{author}{\bibfnamefont{M.}~\bibnamefont{Isaacson}},
  \bibinfo{year}{1985}, \bibinfo{journal}{Ultramicroscopy}
  \textbf{\bibinfo{volume}{16}}, \bibinfo{pages}{233}.

\bibitem[{\citenamefont{Scheinfein}
  \emph{et~al.}(1993)\citenamefont{Scheinfein, Drucker, and Weiss}}]{SDW93}
\bibinfo{author}{\bibnamefont{Scheinfein}, \bibfnamefont{M.~R.}},
  \bibinfo{author}{\bibfnamefont{J.}~\bibnamefont{Drucker}}, and
  \bibinfo{author}{\bibfnamefont{J.~K.} \bibnamefont{Weiss}},
  \bibinfo{year}{1993}, \bibinfo{journal}{Phys.\ Rev.\ B}
  \textbf{\bibinfo{volume}{47}}, \bibinfo{pages}{4068}.

\bibitem[{\citenamefont{Schieber and Sch\"{a}chter}(1998)}]{SS98}
\bibinfo{author}{\bibnamefont{Schieber}, \bibfnamefont{D.}}, and
  \bibinfo{author}{\bibfnamefont{L.}~\bibnamefont{Sch\"{a}chter}},
  \bibinfo{year}{1998}, \bibinfo{journal}{Phys.\ Rev.\ E}
  \textbf{\bibinfo{volume}{57}}, \bibinfo{pages}{6008}.

\bibitem[{\citenamefont{Schieber} \emph{et~al.}(2000)\citenamefont{Schieber,
  Krinsley, and Riciputi}}]{SKR00}
\bibinfo{author}{\bibnamefont{Schieber}, \bibfnamefont{J.}},
  \bibinfo{author}{\bibfnamefont{D.}~\bibnamefont{Krinsley}}, and
  \bibinfo{author}{\bibfnamefont{L.}~\bibnamefont{Riciputi}},
  \bibinfo{year}{2000}, \bibinfo{journal}{Nature}
  \textbf{\bibinfo{volume}{406}}, \bibinfo{pages}{981}.

\bibitem[{\citenamefont{Schmeits}(1989)}]{S1989}
\bibinfo{author}{\bibnamefont{Schmeits}, \bibfnamefont{M.}},
  \bibinfo{year}{1989}, \bibinfo{journal}{Phys.\ Rev.\ B}
  \textbf{\bibinfo{volume}{39}}, \bibinfo{pages}{7567}.

\bibitem[{\citenamefont{Schmeits and Dambly}(1991)}]{SD91}
\bibinfo{author}{\bibnamefont{Schmeits}, \bibfnamefont{M.}}, and
  \bibinfo{author}{\bibfnamefont{L.}~\bibnamefont{Dambly}},
  \bibinfo{year}{1991}, \bibinfo{journal}{Phys.\ Rev.\ B}
  \textbf{\bibinfo{volume}{44}}, \bibinfo{pages}{12706}.

\bibitem[{\citenamefont{Schuster} \emph{et~al.}(1993)\citenamefont{Schuster,
  Swanson, Alex, Bourret, and Simon}}]{SSA93}
\bibinfo{author}{\bibnamefont{Schuster}, \bibfnamefont{S.~C.}},
  \bibinfo{author}{\bibfnamefont{R.~V.} \bibnamefont{Swanson}},
  \bibinfo{author}{\bibfnamefont{L.~A.} \bibnamefont{Alex}},
  \bibinfo{author}{\bibfnamefont{R.~B.} \bibnamefont{Bourret}}, and
  \bibinfo{author}{\bibfnamefont{M.~I.} \bibnamefont{Simon}},
  \bibinfo{year}{1993}, \bibinfo{journal}{Nature}
  \textbf{\bibinfo{volume}{365}}, \bibinfo{pages}{343}.

\bibitem[{\citenamefont{Shibata} \emph{et~al.}(1998)\citenamefont{Shibata,
  Hasebe, Ishi, Ono, Ikezawa, Nakazato, Oyamada, Urasawa, Takahashi, Matsuyama,
  Kobayashi, and Fujita}}]{SHK98}
\bibinfo{author}{\bibnamefont{Shibata}, \bibfnamefont{Y.}},
  \bibinfo{author}{\bibfnamefont{S.}~\bibnamefont{Hasebe}},
  \bibinfo{author}{\bibfnamefont{K.}~\bibnamefont{Ishi}},
  \bibinfo{author}{\bibfnamefont{S.}~\bibnamefont{Ono}},
  \bibinfo{author}{\bibfnamefont{M.}~\bibnamefont{Ikezawa}},
  \bibinfo{author}{\bibfnamefont{T.}~\bibnamefont{Nakazato}},
  \bibinfo{author}{\bibfnamefont{M.}~\bibnamefont{Oyamada}},
  \bibinfo{author}{\bibfnamefont{S.}~\bibnamefont{Urasawa}},
  \bibinfo{author}{\bibfnamefont{T.}~\bibnamefont{Takahashi}},
  \bibinfo{author}{\bibfnamefont{T.}~\bibnamefont{Matsuyama}},
  \bibinfo{author}{\bibfnamefont{K.}~\bibnamefont{Kobayashi}}, and
  \bibinfo{author}{\bibfnamefont{Y.}~\bibnamefont{Fujita}},
  \bibinfo{year}{1998}, \bibinfo{journal}{Phys.\ Rev.\ E}
  \textbf{\bibinfo{volume}{57}}, \bibinfo{pages}{1061}.

\bibitem[{\citenamefont{Shieh and Ritchie}(1970)}]{SR1970}
\bibinfo{author}{\bibnamefont{Shieh}, \bibfnamefont{S.~Y.}}, and
  \bibinfo{author}{\bibfnamefont{R.~H.} \bibnamefont{Ritchie}},
  \bibinfo{year}{1970}, \bibinfo{journal}{Phys.\ Rev.\ B}
  \textbf{\bibinfo{volume}{2}}, \bibinfo{pages}{1646}.

\bibitem[{\citenamefont{Shiles} \emph{et~al.}(1980)\citenamefont{Shiles,
  Sasaki, Inokuti, and Smith}}]{SSI1980}
\bibinfo{author}{\bibnamefont{Shiles}, \bibfnamefont{E.}},
  \bibinfo{author}{\bibfnamefont{T.}~\bibnamefont{Sasaki}},
  \bibinfo{author}{\bibfnamefont{M.}~\bibnamefont{Inokuti}}, and
  \bibinfo{author}{\bibfnamefont{D.~Y.} \bibnamefont{Smith}},
  \bibinfo{year}{1980}, \bibinfo{journal}{Phys.\ Rev.\ B}
  \textbf{\bibinfo{volume}{22}}, \bibinfo{pages}{1612}.

\bibitem[{\citenamefont{Shubina} \emph{et~al.}(2004)\citenamefont{Shubina,
  Ivanov, Jmerik, Solnyshkov, Vekshin, Kop'ev, Vasson, Leymarie, Kavokin,
  Amano, Shimono, Kasic} \emph{et~al.}}]{SIJ04}
\bibinfo{author}{\bibnamefont{Shubina}, \bibfnamefont{T.~V.}},
  \bibinfo{author}{\bibfnamefont{S.~V.} \bibnamefont{Ivanov}},
  \bibinfo{author}{\bibfnamefont{V.~N.} \bibnamefont{Jmerik}},
  \bibinfo{author}{\bibfnamefont{D.~D.} \bibnamefont{Solnyshkov}},
  \bibinfo{author}{\bibfnamefont{V.~A.} \bibnamefont{Vekshin}},
  \bibinfo{author}{\bibfnamefont{P.~S.} \bibnamefont{Kop'ev}},
  \bibinfo{author}{\bibfnamefont{A.}~\bibnamefont{Vasson}},
  \bibinfo{author}{\bibfnamefont{J.}~\bibnamefont{Leymarie}},
  \bibinfo{author}{\bibfnamefont{A.}~\bibnamefont{Kavokin}},
  \bibinfo{author}{\bibfnamefont{H.}~\bibnamefont{Amano}},
  \bibinfo{author}{\bibfnamefont{K.}~\bibnamefont{Shimono}},
  \bibinfo{author}{\bibfnamefont{A.}~\bibnamefont{Kasic}}, \emph{et~al.},
  \bibinfo{year}{2004}, \bibinfo{journal}{Phys.\ Rev.\ Lett.}
  \textbf{\bibinfo{volume}{92}}, \bibinfo{pages}{117407}.

\bibitem[{\citenamefont{Simonsen} \emph{et~al.}(1997)\citenamefont{Simonsen,
  Yubero, and Tougaard}}]{SYT97}
\bibinfo{author}{\bibnamefont{Simonsen}, \bibfnamefont{A.~C.}},
  \bibinfo{author}{\bibfnamefont{F.}~\bibnamefont{Yubero}}, and
  \bibinfo{author}{\bibfnamefont{S.}~\bibnamefont{Tougaard}},
  \bibinfo{year}{1997}, \bibinfo{journal}{Phys.\ Rev.\ B}
  \textbf{\bibinfo{volume}{56}}, \bibinfo{pages}{1612}.

\bibitem[{\citenamefont{Skryabin} \emph{et~al.}(2003)\citenamefont{Skryabin,
  Luan, Knight, and Russell}}]{SLK03}
\bibinfo{author}{\bibnamefont{Skryabin}, \bibfnamefont{D.~V.}},
  \bibinfo{author}{\bibfnamefont{F.}~\bibnamefont{Luan}},
  \bibinfo{author}{\bibfnamefont{J.~C.} \bibnamefont{Knight}}, and
  \bibinfo{author}{\bibfnamefont{P.~S.~J.} \bibnamefont{Russell}},
  \bibinfo{year}{2003}, \bibinfo{journal}{Science}
  \textbf{\bibinfo{volume}{301}}, \bibinfo{pages}{1705}.

\bibitem[{\citenamefont{Smith}(1985)}]{S1985}
\bibinfo{author}{\bibnamefont{Smith}, \bibfnamefont{N.~V.}},
  \bibinfo{year}{1985}, \bibinfo{journal}{Phys.\ Rev.\ B}
  \textbf{\bibinfo{volume}{32}}, \bibinfo{pages}{3549}.

\bibitem[{\citenamefont{Smith and Purcell}(1953)}]{SP1953}
\bibinfo{author}{\bibnamefont{Smith}, \bibfnamefont{S.~J.}}, and
  \bibinfo{author}{\bibfnamefont{E.~M.} \bibnamefont{Purcell}},
  \bibinfo{year}{1953}, \bibinfo{journal}{Phys.\ Rev.}
  \textbf{\bibinfo{volume}{92}}, \bibinfo{pages}{1069}.

\bibitem[{\citenamefont{Sonderegger}
  \emph{et~al.}(2006)\citenamefont{Sonderegger, Feltin, Merano, Crottini,
  Carlin, Sachot, Deveaud, Grandjean, and Gani$\grave{e}$re}}]{SFM06}
\bibinfo{author}{\bibnamefont{Sonderegger}, \bibfnamefont{S.}},
  \bibinfo{author}{\bibfnamefont{E.}~\bibnamefont{Feltin}},
  \bibinfo{author}{\bibfnamefont{M.}~\bibnamefont{Merano}},
  \bibinfo{author}{\bibfnamefont{A.}~\bibnamefont{Crottini}},
  \bibinfo{author}{\bibfnamefont{J.~F.} \bibnamefont{Carlin}},
  \bibinfo{author}{\bibfnamefont{R.}~\bibnamefont{Sachot}},
  \bibinfo{author}{\bibfnamefont{B.}~\bibnamefont{Deveaud}},
  \bibinfo{author}{\bibfnamefont{N.}~\bibnamefont{Grandjean}}, and
  \bibinfo{author}{\bibfnamefont{J.~D.} \bibnamefont{Gani$\grave{e}$re}},
  \bibinfo{year}{2006}, \bibinfo{journal}{Appl.\ Phys.\ Lett.}
  \textbf{\bibinfo{volume}{89}}, \bibinfo{pages}{232109}.

\bibitem[{\citenamefont{Sosa} \emph{et~al.}(2001)\citenamefont{Sosa, Mendoza,
  and Barrera}}]{SMB01}
\bibinfo{author}{\bibnamefont{Sosa}, \bibfnamefont{I.~O.}},
  \bibinfo{author}{\bibfnamefont{C.~I.} \bibnamefont{Mendoza}}, and
  \bibinfo{author}{\bibfnamefont{R.~G.} \bibnamefont{Barrera}},
  \bibinfo{year}{2001}, \bibinfo{journal}{Phys.\ Rev.\ B}
  \textbf{\bibinfo{volume}{63}}, \bibinfo{pages}{144201}.

\bibitem[{\citenamefont{Stefanou} \emph{et~al.}(2000)\citenamefont{Stefanou,
  Yannopapas, and Modinos}}]{SYM00}
\bibinfo{author}{\bibnamefont{Stefanou}, \bibfnamefont{N.}},
  \bibinfo{author}{\bibfnamefont{V.}~\bibnamefont{Yannopapas}}, and
  \bibinfo{author}{\bibfnamefont{A.}~\bibnamefont{Modinos}},
  \bibinfo{year}{2000}, \bibinfo{journal}{Comput.\ Phys.\ Commun.}
  \textbf{\bibinfo{volume}{132}}, \bibinfo{pages}{189}.

\bibitem[{\citenamefont{Steinmann}(1960)}]{S1960}
\bibinfo{author}{\bibnamefont{Steinmann}, \bibfnamefont{W.}},
  \bibinfo{year}{1960}, \bibinfo{journal}{Phys.\ Rev.\ Lett.}
  \textbf{\bibinfo{volume}{5}}, \bibinfo{pages}{470}.

\bibitem[{\citenamefont{{St\'ephan}}
  \emph{et~al.}(2002)\citenamefont{{St\'ephan}, Taverna, Kociak, Suenaga,
  Henrard, and Colliex}}]{STK02}
\bibinfo{author}{\bibnamefont{{St\'ephan}}, \bibfnamefont{O.}},
  \bibinfo{author}{\bibfnamefont{D.}~\bibnamefont{Taverna}},
  \bibinfo{author}{\bibfnamefont{M.}~\bibnamefont{Kociak}},
  \bibinfo{author}{\bibfnamefont{K.}~\bibnamefont{Suenaga}},
  \bibinfo{author}{\bibfnamefont{L.}~\bibnamefont{Henrard}}, and
  \bibinfo{author}{\bibfnamefont{C.}~\bibnamefont{Colliex}},
  \bibinfo{year}{2002}, \bibinfo{journal}{Phys.\ Rev.\ B}
  \textbf{\bibinfo{volume}{66}}, \bibinfo{pages}{155422}.

\bibitem[{\citenamefont{Stern and Ferrell}(1960)}]{SF1960}
\bibinfo{author}{\bibnamefont{Stern}, \bibfnamefont{E.~A.}}, and
  \bibinfo{author}{\bibfnamefont{R.~A.} \bibnamefont{Ferrell}},
  \bibinfo{year}{1960}, \bibinfo{journal}{Phys.\ Rev.}
  \textbf{\bibinfo{volume}{120}}, \bibinfo{pages}{130}.

\bibitem[{\citenamefont{Stevens} \emph{et~al.}(2001)\citenamefont{Stevens,
  Wahlstrand, Kuhl, and Merlin}}]{SWK01}
\bibinfo{author}{\bibnamefont{Stevens}, \bibfnamefont{T.~E.}},
  \bibinfo{author}{\bibfnamefont{J.~K.} \bibnamefont{Wahlstrand}},
  \bibinfo{author}{\bibfnamefont{J.}~\bibnamefont{Kuhl}}, and
  \bibinfo{author}{\bibfnamefont{R.}~\bibnamefont{Merlin}},
  \bibinfo{year}{2001}, \bibinfo{journal}{Science}
  \textbf{\bibinfo{volume}{291}}, \bibinfo{pages}{627}.

\bibitem[{\citenamefont{{St\"ockli}}
  \emph{et~al.}(1997)\citenamefont{{St\"ockli}, Bonard, Stadelmann, and
  Ch\^{a}telain}}]{SBS97}
\bibinfo{author}{\bibnamefont{{St\"ockli}}, \bibfnamefont{T.}},
  \bibinfo{author}{\bibfnamefont{J.~M.} \bibnamefont{Bonard}},
  \bibinfo{author}{\bibfnamefont{P.~A.} \bibnamefont{Stadelmann}}, and
  \bibinfo{author}{\bibfnamefont{A.}~\bibnamefont{Ch\^{a}telain}},
  \bibinfo{year}{1997}, \bibinfo{journal}{Z.\ Phys.\ D}
  \textbf{\bibinfo{volume}{40}}, \bibinfo{pages}{425–}.

\bibitem[{\citenamefont{{St\"oger-Pollach}}(2008)}]{S08}
\bibinfo{author}{\bibnamefont{{St\"oger-Pollach}}, \bibfnamefont{M.}},
  \bibinfo{year}{2008}, \bibinfo{journal}{Micron}
  \textbf{\bibinfo{volume}{39}}, \bibinfo{pages}{1092}.

\bibitem[{\citenamefont{St\"oger-Pollach}
  \emph{et~al.}(2006)\citenamefont{St\"oger-Pollach, Franco, Schattschneider,
  Lazar, Schaffer, Grogger, and Zandbergen}}]{SFS06}
\bibinfo{author}{\bibnamefont{St\"oger-Pollach}, \bibfnamefont{M.}},
  \bibinfo{author}{\bibfnamefont{H.}~\bibnamefont{Franco}},
  \bibinfo{author}{\bibfnamefont{P.}~\bibnamefont{Schattschneider}},
  \bibinfo{author}{\bibfnamefont{S.}~\bibnamefont{Lazar}},
  \bibinfo{author}{\bibfnamefont{B.}~\bibnamefont{Schaffer}},
  \bibinfo{author}{\bibfnamefont{W.}~\bibnamefont{Grogger}}, and
  \bibinfo{author}{\bibfnamefont{H.~W.} \bibnamefont{Zandbergen}},
  \bibinfo{year}{2006}, \bibinfo{journal}{Micron}
  \textbf{\bibinfo{volume}{37}}, \bibinfo{pages}{396–}.

\bibitem[{\citenamefont{{St\"oger-Pollach}}
  \emph{et~al.}(2008)\citenamefont{{St\"oger-Pollach}, Laister, and
  Schattschneider}}]{SLS08}
\bibinfo{author}{\bibnamefont{{St\"oger-Pollach}}, \bibfnamefont{M.}},
  \bibinfo{author}{\bibfnamefont{A.}~\bibnamefont{Laister}}, and
  \bibinfo{author}{\bibfnamefont{P.}~\bibnamefont{Schattschneider}},
  \bibinfo{year}{2008}, \bibinfo{journal}{Ultramicroscopy}
  \textbf{\bibinfo{volume}{108}}, \bibinfo{pages}{439}.

\bibitem[{\citenamefont{St\"oger-Pollach and Schattschneider}(2007)}]{SS07}
\bibinfo{author}{\bibnamefont{St\"oger-Pollach}, \bibfnamefont{M.}}, and
  \bibinfo{author}{\bibfnamefont{P.}~\bibnamefont{Schattschneider}},
  \bibinfo{year}{2007}, \bibinfo{journal}{Ultramicroscopy}
  \textbf{\bibinfo{volume}{107}}, \bibinfo{pages}{1178}.

\bibitem[{\citenamefont{Stolojan} \emph{et~al.}(2006)\citenamefont{Stolojan,
  Moreau, Goringe, and Silva}}]{SMG06}
\bibinfo{author}{\bibnamefont{Stolojan}, \bibfnamefont{V.}},
  \bibinfo{author}{\bibfnamefont{P.}~\bibnamefont{Moreau}},
  \bibinfo{author}{\bibfnamefont{M.~J.} \bibnamefont{Goringe}}, and
  \bibinfo{author}{\bibfnamefont{S.~R.~P.} \bibnamefont{Silva}},
  \bibinfo{year}{2006}, \bibinfo{journal}{Appl.\ Phys.\ Lett.}
  \textbf{\bibinfo{volume}{88}}, \bibinfo{pages}{122109}.

\bibitem[{\citenamefont{Su} \emph{et~al.}(2003)\citenamefont{Su, Zandbergen,
  Tiemeijer, Kothleitner, H\"avecker, H\'ebert, Knop-Gericke, Freitag, Hofer,
  and Schl\"ogl}}]{SZT03}
\bibinfo{author}{\bibnamefont{Su}, \bibfnamefont{D.~S.}},
  \bibinfo{author}{\bibfnamefont{H.~W.} \bibnamefont{Zandbergen}},
  \bibinfo{author}{\bibfnamefont{P.~C.} \bibnamefont{Tiemeijer}},
  \bibinfo{author}{\bibfnamefont{G.}~\bibnamefont{Kothleitner}},
  \bibinfo{author}{\bibfnamefont{M.}~\bibnamefont{H\"avecker}},
  \bibinfo{author}{\bibfnamefont{C.}~\bibnamefont{H\'ebert}},
  \bibinfo{author}{\bibfnamefont{A.}~\bibnamefont{Knop-Gericke}},
  \bibinfo{author}{\bibfnamefont{B.~H.} \bibnamefont{Freitag}},
  \bibinfo{author}{\bibfnamefont{F.}~\bibnamefont{Hofer}}, and
  \bibinfo{author}{\bibfnamefont{R.}~\bibnamefont{Schl\"ogl}},
  \bibinfo{year}{2003}, \bibinfo{journal}{Micron}
  \textbf{\bibinfo{volume}{34}}, \bibinfo{pages}{235}.

\bibitem[{\citenamefont{Sun} \emph{et~al.}(1993)\citenamefont{Sun, Shi, and
  Leapman}}]{SSL93}
\bibinfo{author}{\bibnamefont{Sun}, \bibfnamefont{S.}},
  \bibinfo{author}{\bibfnamefont{S.}~\bibnamefont{Shi}}, and
  \bibinfo{author}{\bibfnamefont{R.}~\bibnamefont{Leapman}},
  \bibinfo{year}{1993}, \bibinfo{journal}{Ultramicroscopy}
  \textbf{\bibinfo{volume}{50}}, \bibinfo{pages}{127}.

\bibitem[{\citenamefont{Talley} \emph{et~al.}(2005)\citenamefont{Talley,
  Jackson, Oubre, Grady, Hollars, Lane, Huser, Nordlander, and Halas}}]{TJO05}
\bibinfo{author}{\bibnamefont{Talley}, \bibfnamefont{C.~E.}},
  \bibinfo{author}{\bibfnamefont{J.~B.} \bibnamefont{Jackson}},
  \bibinfo{author}{\bibfnamefont{C.}~\bibnamefont{Oubre}},
  \bibinfo{author}{\bibfnamefont{N.~K.} \bibnamefont{Grady}},
  \bibinfo{author}{\bibfnamefont{C.~W.} \bibnamefont{Hollars}},
  \bibinfo{author}{\bibfnamefont{S.~M.} \bibnamefont{Lane}},
  \bibinfo{author}{\bibfnamefont{T.~R.} \bibnamefont{Huser}},
  \bibinfo{author}{\bibfnamefont{P.}~\bibnamefont{Nordlander}}, and
  \bibinfo{author}{\bibfnamefont{N.~J.} \bibnamefont{Halas}},
  \bibinfo{year}{2005}, \bibinfo{journal}{Nano\ Lett.}
  \textbf{\bibinfo{volume}{5}}, \bibinfo{pages}{1569}.

\bibitem[{\citenamefont{Tamm}(1939)}]{T1939_2}
\bibinfo{author}{\bibnamefont{Tamm}, \bibfnamefont{I.}}, \bibinfo{year}{1939},
  \bibinfo{journal}{J.\ Phys.\ (Moscow)} \textbf{\bibinfo{volume}{1}},
  \bibinfo{pages}{439}.

\bibitem[{\citenamefont{Tanuma} \emph{et~al.}(1994)\citenamefont{Tanuma,
  Powell, and Penn}}]{TPP93}
\bibinfo{author}{\bibnamefont{Tanuma}, \bibfnamefont{S.}},
  \bibinfo{author}{\bibfnamefont{C.~J.} \bibnamefont{Powell}}, and
  \bibinfo{author}{\bibfnamefont{D.~R.} \bibnamefont{Penn}},
  \bibinfo{year}{1994}, \bibinfo{journal}{Surf.\ Interface\ Anal.}
  \textbf{\bibinfo{volume}{21}}, \bibinfo{pages}{165}.

\bibitem[{\citenamefont{Taverna} \emph{et~al.}(2002)\citenamefont{Taverna,
  Kociak, Charbois, and Henrard}}]{TKC02}
\bibinfo{author}{\bibnamefont{Taverna}, \bibfnamefont{D.}},
  \bibinfo{author}{\bibfnamefont{M.}~\bibnamefont{Kociak}},
  \bibinfo{author}{\bibfnamefont{V.}~\bibnamefont{Charbois}}, and
  \bibinfo{author}{\bibfnamefont{L.}~\bibnamefont{Henrard}},
  \bibinfo{year}{2002}, \bibinfo{journal}{Phys.\ Rev.\ B}
  \textbf{\bibinfo{volume}{66}}, \bibinfo{pages}{235419}.

\bibitem[{\citenamefont{Temkin}(1998)}]{T98}
\bibinfo{author}{\bibnamefont{Temkin}, \bibfnamefont{R.}},
  \bibinfo{year}{1998}, \bibinfo{journal}{Science}
  \textbf{\bibinfo{volume}{280}}, \bibinfo{pages}{854}.

\bibitem[{\citenamefont{Teng and Stern}(1967)}]{TS1967}
\bibinfo{author}{\bibnamefont{Teng}, \bibfnamefont{Y.~Y.}}, and
  \bibinfo{author}{\bibfnamefont{E.~A.} \bibnamefont{Stern}},
  \bibinfo{year}{1967}, \bibinfo{journal}{Phys.\ Rev.\ Lett.}
  \textbf{\bibinfo{volume}{19}}, \bibinfo{pages}{511}.

\bibitem[{\citenamefont{Ter-Mikaelian}(1972)}]{T1972}
\bibinfo{author}{\bibnamefont{Ter-Mikaelian}, \bibfnamefont{M.~L.}},
  \bibinfo{year}{1972}, \emph{\bibinfo{title}{High-Energy Electromagnetic
  Processes in Condensed Media}} (\bibinfo{publisher}{Wiley},
  \bibinfo{address}{New York}).

\bibitem[{\citenamefont{Terauchi} \emph{et~al.}(1999)\citenamefont{Terauchi,
  Tanaka, Tsuno, and Ishida}}]{TTT99}
\bibinfo{author}{\bibnamefont{Terauchi}, \bibfnamefont{M.}},
  \bibinfo{author}{\bibfnamefont{M.}~\bibnamefont{Tanaka}},
  \bibinfo{author}{\bibfnamefont{K.}~\bibnamefont{Tsuno}}, and
  \bibinfo{author}{\bibfnamefont{M.}~\bibnamefont{Ishida}},
  \bibinfo{year}{1999}, \bibinfo{journal}{J.\ Microsc.}
  \textbf{\bibinfo{volume}{194}}, \bibinfo{pages}{203–}.

\bibitem[{\citenamefont{Toma\v{s}} \emph{et~al.}(1972)\citenamefont{Toma\v{s},
  Lucas, and {\v{S}unji\'{c}}}}]{TLS1972}
\bibinfo{author}{\bibnamefont{Toma\v{s}}, \bibfnamefont{M.~S.}},
  \bibinfo{author}{\bibfnamefont{A.~A.} \bibnamefont{Lucas}}, and
  \bibinfo{author}{\bibfnamefont{M.}~\bibnamefont{{\v{S}unji\'{c}}}},
  \bibinfo{year}{1972}, \bibinfo{journal}{Solid\ State\ Commun.}
  \textbf{\bibinfo{volume}{10}}, \bibinfo{pages}{1181}.

\bibitem[{\citenamefont{Toma\v{s}} \emph{et~al.}(1974)\citenamefont{Toma\v{s},
  Lucas, {\v{S}unji\'{c}}, and Jureti\'{c}}}]{TLS1974}
\bibinfo{author}{\bibnamefont{Toma\v{s}}, \bibfnamefont{M.~S.}},
  \bibinfo{author}{\bibfnamefont{A.~A.} \bibnamefont{Lucas}},
  \bibinfo{author}{\bibfnamefont{M.}~\bibnamefont{{\v{S}unji\'{c}}}}, and
  \bibinfo{author}{\bibfnamefont{D.}~\bibnamefont{Jureti\'{c}}},
  \bibinfo{year}{1974}, \bibinfo{journal}{Phys.\ Rev.\ B}
  \textbf{\bibinfo{volume}{9}}, \bibinfo{pages}{1489}.

\bibitem[{\citenamefont{Tomita} \emph{et~al.}(2006)\citenamefont{Tomita, Yoda,
  Uchiyama, Ishii, Sasa, Kaneko, and Kudo}}]{TYU06}
\bibinfo{author}{\bibnamefont{Tomita}, \bibfnamefont{S.}},
  \bibinfo{author}{\bibfnamefont{S.}~\bibnamefont{Yoda}},
  \bibinfo{author}{\bibfnamefont{R.}~\bibnamefont{Uchiyama}},
  \bibinfo{author}{\bibfnamefont{S.}~\bibnamefont{Ishii}},
  \bibinfo{author}{\bibfnamefont{K.}~\bibnamefont{Sasa}},
  \bibinfo{author}{\bibfnamefont{T.}~\bibnamefont{Kaneko}}, and
  \bibinfo{author}{\bibfnamefont{H.}~\bibnamefont{Kudo}}, \bibinfo{year}{2006},
  \bibinfo{journal}{Phys.\ Rev.\ A} \textbf{\bibinfo{volume}{73}},
  \bibinfo{pages}{060901(R)}.

\bibitem[{\citenamefont{{Toraldo di Francia}}(1960)}]{T1960}
\bibinfo{author}{\bibnamefont{{Toraldo di Francia}}, \bibfnamefont{G.}},
  \bibinfo{year}{1960}, \bibinfo{journal}{Nuevo Cimento}
  \textbf{\bibinfo{volume}{16}}, \bibinfo{pages}{61}.

\bibitem[{\citenamefont{Tr\"ugler and Hohenester}(2008)}]{TH08}
\bibinfo{author}{\bibnamefont{Tr\"ugler}, \bibfnamefont{A.}}, and
  \bibinfo{author}{\bibfnamefont{U.}~\bibnamefont{Hohenester}},
  \bibinfo{year}{2008}, \bibinfo{journal}{Phys.\ Rev.\ B}
  \textbf{\bibinfo{volume}{77}}, \bibinfo{pages}{115403}.

\bibitem[{\citenamefont{Tu} \emph{et~al.}(2006)\citenamefont{Tu, Kwei, and
  Tung}}]{TKT06}
\bibinfo{author}{\bibnamefont{Tu}, \bibfnamefont{Y.~H.}},
  \bibinfo{author}{\bibfnamefont{C.~M.} \bibnamefont{Kwei}}, and
  \bibinfo{author}{\bibfnamefont{C.~J.} \bibnamefont{Tung}},
  \bibinfo{year}{2006}, \bibinfo{journal}{Surf.\ Sci.}
  \textbf{\bibinfo{volume}{600}}, \bibinfo{pages}{820}.

\bibitem[{\citenamefont{Ugarte} \emph{et~al.}(1992)\citenamefont{Ugarte,
  Colliex, and Trebbia}}]{UCT92}
\bibinfo{author}{\bibnamefont{Ugarte}, \bibfnamefont{D.}},
  \bibinfo{author}{\bibfnamefont{C.}~\bibnamefont{Colliex}}, and
  \bibinfo{author}{\bibfnamefont{P.}~\bibnamefont{Trebbia}},
  \bibinfo{year}{1992}, \bibinfo{journal}{Phys.\ Rev.\ B}
  \textbf{\bibinfo{volume}{45}}, \bibinfo{pages}{4332}.

\bibitem[{\citenamefont{Uhlemann and Haider}(1998)}]{UH98}
\bibinfo{author}{\bibnamefont{Uhlemann}, \bibfnamefont{S.}}, and
  \bibinfo{author}{\bibfnamefont{M.}~\bibnamefont{Haider}},
  \bibinfo{year}{1998}, \bibinfo{journal}{Ultramicroscopy}
  \textbf{\bibinfo{volume}{72}}, \bibinfo{pages}{109}.

\bibitem[{\citenamefont{Ulrich and Tacke}(1973)}]{UT1973}
\bibinfo{author}{\bibnamefont{Ulrich}, \bibfnamefont{R.}}, and
  \bibinfo{author}{\bibfnamefont{M.}~\bibnamefont{Tacke}},
  \bibinfo{year}{1973}, \bibinfo{journal}{Appl.\ Phys.\ Lett.}
  \textbf{\bibinfo{volume}{22}}, \bibinfo{pages}{251}.

\bibitem[{\citenamefont{Urata} \emph{et~al.}(1998)\citenamefont{Urata,
  Goldstein, Kimmitt, Naumov, Platt, and Walsh}}]{UGK98}
\bibinfo{author}{\bibnamefont{Urata}, \bibfnamefont{J.}},
  \bibinfo{author}{\bibfnamefont{M.}~\bibnamefont{Goldstein}},
  \bibinfo{author}{\bibfnamefont{M.~F.} \bibnamefont{Kimmitt}},
  \bibinfo{author}{\bibfnamefont{A.}~\bibnamefont{Naumov}},
  \bibinfo{author}{\bibfnamefont{C.}~\bibnamefont{Platt}}, and
  \bibinfo{author}{\bibfnamefont{J.~E.} \bibnamefont{Walsh}},
  \bibinfo{year}{1998}, \bibinfo{journal}{Phys.\ Rev.\ Lett.}
  \textbf{\bibinfo{volume}{80}}, \bibinfo{pages}{516}.

\bibitem[{\citenamefont{{van Attekum} and Trooster}(1978)}]{VT1978}
\bibinfo{author}{\bibnamefont{{van Attekum}}, \bibfnamefont{P.~M. T.~M.}}, and
  \bibinfo{author}{\bibfnamefont{J.~M.} \bibnamefont{Trooster}},
  \bibinfo{year}{1978}, \bibinfo{journal}{Phys.\ Rev.\ B}
  \textbf{\bibinfo{volume}{18}}, \bibinfo{pages}{3872}.

\bibitem[{\citenamefont{{van Benthem}} \emph{et~al.}(2001)\citenamefont{{van
  Benthem}, {Els\"asser}, and French}}]{BEF01}
\bibinfo{author}{\bibnamefont{{van Benthem}}, \bibfnamefont{K.}},
  \bibinfo{author}{\bibfnamefont{C.}~\bibnamefont{{Els\"asser}}}, and
  \bibinfo{author}{\bibfnamefont{R.~H.} \bibnamefont{French}},
  \bibinfo{year}{2001}, \bibinfo{journal}{J.\ Appl.\ Phys.}
  \textbf{\bibinfo{volume}{90}}, \bibinfo{pages}{6156}.

\bibitem[{\citenamefont{{van de Hulst}}(1981)}]{V1981}
\bibinfo{author}{\bibnamefont{{van de Hulst}}, \bibfnamefont{H.~C.}},
  \bibinfo{year}{1981}, \emph{\bibinfo{title}{Light Scattering by Small
  Particles}} (\bibinfo{publisher}{Dover}, \bibinfo{address}{New York}).

\bibitem[{\citenamefont{{van den Berg}}(1973)}]{V1973}
\bibinfo{author}{\bibnamefont{{van den Berg}}, \bibfnamefont{P.~M.}},
  \bibinfo{year}{1973}, \bibinfo{journal}{J.\ Opt.\ Soc.\ Am.}
  \textbf{\bibinfo{volume}{63}}, \bibinfo{pages}{1588}.

\bibitem[{\citenamefont{{van den Berg} and Tan}(1974)}]{VT1974}
\bibinfo{author}{\bibnamefont{{van den Berg}}, \bibfnamefont{P.~M.}}, and
  \bibinfo{author}{\bibfnamefont{T.~H.} \bibnamefont{Tan}},
  \bibinfo{year}{1974}, \bibinfo{journal}{J.\ Opt.\ Soc.\ Am.}
  \textbf{\bibinfo{volume}{64}}, \bibinfo{pages}{325}.

\bibitem[{\citenamefont{{Van Hove}}(1953)}]{V1953}
\bibinfo{author}{\bibnamefont{{Van Hove}}, \bibfnamefont{L.}},
  \bibinfo{year}{1953}, \bibinfo{journal}{Phys.\ Rev.}
  \textbf{\bibinfo{volume}{89}}, \bibinfo{pages}{1189}.

\bibitem[{\citenamefont{{van Wijngaarden}}
  \emph{et~al.}(2006)\citenamefont{{van Wijngaarden}, Verhagen, Polman, Ross,
  Lezec, and Atwater}}]{VVP06}
\bibinfo{author}{\bibnamefont{{van Wijngaarden}}, \bibfnamefont{J.~T.}},
  \bibinfo{author}{\bibfnamefont{E.}~\bibnamefont{Verhagen}},
  \bibinfo{author}{\bibfnamefont{A.}~\bibnamefont{Polman}},
  \bibinfo{author}{\bibfnamefont{C.~E.} \bibnamefont{Ross}},
  \bibinfo{author}{\bibfnamefont{H.~J.} \bibnamefont{Lezec}}, and
  \bibinfo{author}{\bibfnamefont{H.~A.} \bibnamefont{Atwater}},
  \bibinfo{year}{2006}, \bibinfo{journal}{Appl.\ Phys.\ Lett.}
  \textbf{\bibinfo{volume}{88}}, \bibinfo{pages}{221111}.

\bibitem[{\citenamefont{Varela} \emph{et~al.}(2004)\citenamefont{Varela,
  Findlay, Lupini, Christen, Borisevich, Dellby, Krivanek, Nellist, Oxley,
  Allen, and Pennycook}}]{VFL04}
\bibinfo{author}{\bibnamefont{Varela}, \bibfnamefont{M.}},
  \bibinfo{author}{\bibfnamefont{S.~D.} \bibnamefont{Findlay}},
  \bibinfo{author}{\bibfnamefont{A.~R.} \bibnamefont{Lupini}},
  \bibinfo{author}{\bibfnamefont{H.~M.} \bibnamefont{Christen}},
  \bibinfo{author}{\bibfnamefont{A.~Y.} \bibnamefont{Borisevich}},
  \bibinfo{author}{\bibfnamefont{N.}~\bibnamefont{Dellby}},
  \bibinfo{author}{\bibfnamefont{O.~L.} \bibnamefont{Krivanek}},
  \bibinfo{author}{\bibfnamefont{P.~D.} \bibnamefont{Nellist}},
  \bibinfo{author}{\bibfnamefont{M.~P.} \bibnamefont{Oxley}},
  \bibinfo{author}{\bibfnamefont{L.~J.} \bibnamefont{Allen}}, and
  \bibinfo{author}{\bibfnamefont{S.~J.} \bibnamefont{Pennycook}},
  \bibinfo{year}{2004}, \bibinfo{journal}{Phys.\ Rev.\ Lett.}
  \textbf{\bibinfo{volume}{92}}, \bibinfo{pages}{095502}.

\bibitem[{\citenamefont{Varela} \emph{et~al.}(2007)\citenamefont{Varela, Oxley,
  Roberts, Garcia-Barriocanal, Lupini, Rashkeev, Leon, Krishnan, Santamaria,
  Pantelides, and Pennycook}}]{VOR07}
\bibinfo{author}{\bibnamefont{Varela}, \bibfnamefont{M.}},
  \bibinfo{author}{\bibfnamefont{M.~P.} \bibnamefont{Oxley}},
  \bibinfo{author}{\bibfnamefont{K.~G.} \bibnamefont{Roberts}},
  \bibinfo{author}{\bibfnamefont{J.}~\bibnamefont{Garcia-Barriocanal}},
  \bibinfo{author}{\bibfnamefont{A.~R.} \bibnamefont{Lupini}},
  \bibinfo{author}{\bibfnamefont{S.~N.} \bibnamefont{Rashkeev}},
  \bibinfo{author}{\bibfnamefont{C.}~\bibnamefont{Leon}},
  \bibinfo{author}{\bibfnamefont{K.~M.} \bibnamefont{Krishnan}},
  \bibinfo{author}{\bibfnamefont{J.}~\bibnamefont{Santamaria}},
  \bibinfo{author}{\bibfnamefont{S.~T.} \bibnamefont{Pantelides}}, and
  \bibinfo{author}{\bibfnamefont{S.~J.} \bibnamefont{Pennycook}},
  \bibinfo{year}{2007}, \bibinfo{journal}{Microsc.\ Microanal.}
  \textbf{\bibinfo{volume}{13}}, \bibinfo{pages}{142}.

\bibitem[{\citenamefont{Vast} \emph{et~al.}(2002)\citenamefont{Vast, Reining,
  Olevano, Schattschneider, and Jouffrey}}]{VRO02}
\bibinfo{author}{\bibnamefont{Vast}, \bibfnamefont{N.}},
  \bibinfo{author}{\bibfnamefont{L.}~\bibnamefont{Reining}},
  \bibinfo{author}{\bibfnamefont{V.}~\bibnamefont{Olevano}},
  \bibinfo{author}{\bibfnamefont{P.}~\bibnamefont{Schattschneider}}, and
  \bibinfo{author}{\bibfnamefont{B.}~\bibnamefont{Jouffrey}},
  \bibinfo{year}{2002}, \bibinfo{journal}{Phys.\ Rev.\ Lett.}
  \textbf{\bibinfo{volume}{88}}, \bibinfo{pages}{037601}.

\bibitem[{\citenamefont{Verbeeck}(2006)}]{V06}
\bibinfo{author}{\bibnamefont{Verbeeck}, \bibfnamefont{J.}},
  \bibinfo{year}{2006}, \bibinfo{journal}{Ultramicroscopy}
  \textbf{\bibinfo{volume}{106}}, \bibinfo{pages}{461–}.

\bibitem[{\citenamefont{Verbeeck} \emph{et~al.}(2008)\citenamefont{Verbeeck,
  Bertoni, and Schattschneider}}]{VBS08}
\bibinfo{author}{\bibnamefont{Verbeeck}, \bibfnamefont{J.}},
  \bibinfo{author}{\bibfnamefont{G.}~\bibnamefont{Bertoni}}, and
  \bibinfo{author}{\bibfnamefont{P.}~\bibnamefont{Schattschneider}},
  \bibinfo{year}{2008}, \bibinfo{journal}{Ultramicroscopy}
  \textbf{\bibinfo{volume}{108}}, \bibinfo{pages}{263}.

\bibitem[{\citenamefont{Veselago}(1968)}]{V1968}
\bibinfo{author}{\bibnamefont{Veselago}, \bibfnamefont{V.~G.}},
  \bibinfo{year}{1968}, \bibinfo{journal}{Sov.\ Phys.\ Usp.}
  \textbf{\bibinfo{volume}{10}}, \bibinfo{pages}{509}.

\bibitem[{\citenamefont{Vesseur} \emph{et~al.}(2007)\citenamefont{Vesseur, {de
  Waele}, Kuttge, and Polman}}]{VDK07}
\bibinfo{author}{\bibnamefont{Vesseur}, \bibfnamefont{E.~J.~R.}},
  \bibinfo{author}{\bibfnamefont{R.}~\bibnamefont{{de Waele}}},
  \bibinfo{author}{\bibfnamefont{M.}~\bibnamefont{Kuttge}}, and
  \bibinfo{author}{\bibfnamefont{A.}~\bibnamefont{Polman}},
  \bibinfo{year}{2007}, \bibinfo{journal}{Nano\ Lett.}
  \textbf{\bibinfo{volume}{7}}, \bibinfo{pages}{2843}.

\bibitem[{\citenamefont{Vesseur} \emph{et~al.}(2008)\citenamefont{Vesseur, {de
  Waele}, Lezec, Atwater, {Garc\'{\i}a de Abajo}, and Polman}}]{paper147}
\bibinfo{author}{\bibnamefont{Vesseur}, \bibfnamefont{E.~J.~R.}},
  \bibinfo{author}{\bibfnamefont{R.}~\bibnamefont{{de Waele}}},
  \bibinfo{author}{\bibfnamefont{H.~J.} \bibnamefont{Lezec}},
  \bibinfo{author}{\bibfnamefont{H.~A.} \bibnamefont{Atwater}},
  \bibinfo{author}{\bibfnamefont{F.~J.} \bibnamefont{{Garc\'{\i}a de Abajo}}},
  and \bibinfo{author}{\bibfnamefont{A.}~\bibnamefont{Polman}},
  \bibinfo{year}{2008}, \bibinfo{journal}{Appl.\ Phys.\ Lett.}
  \textbf{\bibinfo{volume}{92}}, \bibinfo{pages}{083110}.

\bibitem[{\citenamefont{Vincent and Silcox}(1973)}]{VS1973}
\bibinfo{author}{\bibnamefont{Vincent}, \bibfnamefont{R.}}, and
  \bibinfo{author}{\bibfnamefont{J.}~\bibnamefont{Silcox}},
  \bibinfo{year}{1973}, \bibinfo{journal}{Phys.\ Rev.\ Lett.}
  \textbf{\bibinfo{volume}{31}}, \bibinfo{pages}{1487}.

\bibitem[{\citenamefont{{Von Festenberg} and Kr\"{o}ger}(1968)}]{VK1968}
\bibinfo{author}{\bibnamefont{{Von Festenberg}}, \bibfnamefont{C.}}, and
  \bibinfo{author}{\bibfnamefont{E.}~\bibnamefont{Kr\"{o}ger}},
  \bibinfo{year}{1968}, \bibinfo{journal}{Phys.\ Lett.\ A}
  \textbf{\bibinfo{volume}{26}}, \bibinfo{pages}{339}.

\bibitem[{\citenamefont{Wagner}(1966)}]{W1966}
\bibinfo{author}{\bibnamefont{Wagner}, \bibfnamefont{D.}},
  \bibinfo{year}{1966}, \bibinfo{journal}{Z.\ Naturforsch.\ A}
  \textbf{\bibinfo{volume}{21}}, \bibinfo{pages}{634}.

\bibitem[{\citenamefont{Walls and Howie}(1989)}]{WH1989}
\bibinfo{author}{\bibnamefont{Walls}, \bibfnamefont{M.~G.}}, and
  \bibinfo{author}{\bibfnamefont{A.}~\bibnamefont{Howie}},
  \bibinfo{year}{1989}, \bibinfo{journal}{Ultramicroscopy}
  \textbf{\bibinfo{volume}{28}}, \bibinfo{pages}{40}.

\bibitem[{\citenamefont{Walsh}(1991)}]{W91_2}
\bibinfo{author}{\bibnamefont{Walsh}, \bibfnamefont{C.~A.}},
  \bibinfo{year}{1991}, \bibinfo{journal}{Philos.\ Mag.\ B}
  \textbf{\bibinfo{volume}{63}}, \bibinfo{pages}{1063}.

\bibitem[{\citenamefont{Wang}
  \emph{et~al.}(2008{\natexlab{a}})\citenamefont{Wang, {D'Alfonso}, Findlay,
  Allen, and Bleloch}}]{WDF08}
\bibinfo{author}{\bibnamefont{Wang}, \bibfnamefont{P.}},
  \bibinfo{author}{\bibfnamefont{A.~J.} \bibnamefont{{D'Alfonso}}},
  \bibinfo{author}{\bibfnamefont{S.~D.} \bibnamefont{Findlay}},
  \bibinfo{author}{\bibfnamefont{L.~J.} \bibnamefont{Allen}}, and
  \bibinfo{author}{\bibfnamefont{A.~L.} \bibnamefont{Bleloch}},
  \bibinfo{year}{2008}{\natexlab{a}}, \bibinfo{journal}{Phys.\ Rev.\ Lett.}
  \textbf{\bibinfo{volume}{101}}, \bibinfo{pages}{236102}.

\bibitem[{\citenamefont{Wang} \emph{et~al.}(1993)\citenamefont{Wang, Zhang, Yu,
  and Harmon}}]{WZY93}
\bibinfo{author}{\bibnamefont{Wang}, \bibfnamefont{X.}},
  \bibinfo{author}{\bibfnamefont{X.~G.} \bibnamefont{Zhang}},
  \bibinfo{author}{\bibfnamefont{Q.}~\bibnamefont{Yu}}, and
  \bibinfo{author}{\bibfnamefont{B.~N.} \bibnamefont{Harmon}},
  \bibinfo{year}{1993}, \bibinfo{journal}{Phys.\ Rev.\ B}
  \textbf{\bibinfo{volume}{47}}, \bibinfo{pages}{4161}.

\bibitem[{\citenamefont{Wang}(1996)}]{W96_2}
\bibinfo{author}{\bibnamefont{Wang}, \bibfnamefont{Z.~L.}},
  \bibinfo{year}{1996}, \bibinfo{journal}{Micron}
  \textbf{\bibinfo{volume}{27}}, \bibinfo{pages}{265}.

\bibitem[{\citenamefont{Wang}
  \emph{et~al.}(2008{\natexlab{b}})\citenamefont{Wang, Chan, Li, and
  Hao}}]{WCL08}
\bibinfo{author}{\bibnamefont{Wang}, \bibfnamefont{Z.~L.}},
  \bibinfo{author}{\bibfnamefont{H.~L.~W.} \bibnamefont{Chan}},
  \bibinfo{author}{\bibfnamefont{H.~L.} \bibnamefont{Li}}, and
  \bibinfo{author}{\bibfnamefont{J.~H.} \bibnamefont{Hao}},
  \bibinfo{year}{2008}{\natexlab{b}}, \bibinfo{journal}{Appl.\ Phys.\ Lett.}
  \textbf{\bibinfo{volume}{93}}, \bibinfo{pages}{141106}.

\bibitem[{\citenamefont{Wang and Cowley}(1987{\natexlab{a}})}]{WC1987}
\bibinfo{author}{\bibnamefont{Wang}, \bibfnamefont{Z.~L.}}, and
  \bibinfo{author}{\bibfnamefont{J.~M.} \bibnamefont{Cowley}},
  \bibinfo{year}{1987}{\natexlab{a}}, \bibinfo{journal}{Ultramicroscopy}
  \textbf{\bibinfo{volume}{21}}, \bibinfo{pages}{335}.

\bibitem[{\citenamefont{Wang and Cowley}(1987{\natexlab{b}})}]{WC1987_2}
\bibinfo{author}{\bibnamefont{Wang}, \bibfnamefont{Z.~L.}}, and
  \bibinfo{author}{\bibfnamefont{J.~M.} \bibnamefont{Cowley}},
  \bibinfo{year}{1987}{\natexlab{b}}, \bibinfo{journal}{Ultramicroscopy}
  \textbf{\bibinfo{volume}{21}}, \bibinfo{pages}{347}.

\bibitem[{\citenamefont{Wang and Cowley}(1987{\natexlab{c}})}]{WC1987_3}
\bibinfo{author}{\bibnamefont{Wang}, \bibfnamefont{Z.~L.}}, and
  \bibinfo{author}{\bibfnamefont{J.~M.} \bibnamefont{Cowley}},
  \bibinfo{year}{1987}{\natexlab{c}}, \bibinfo{journal}{Ultramicroscopy}
  \textbf{\bibinfo{volume}{23}}, \bibinfo{pages}{97}.

\bibitem[{\citenamefont{Wang and Cowley}(1987{\natexlab{d}})}]{WC1987_4}
\bibinfo{author}{\bibnamefont{Wang}, \bibfnamefont{Z.~L.}}, and
  \bibinfo{author}{\bibfnamefont{J.~M.} \bibnamefont{Cowley}},
  \bibinfo{year}{1987}{\natexlab{d}}, \bibinfo{journal}{Ultramicroscopy}
  \textbf{\bibinfo{volume}{21}}, \bibinfo{pages}{77}.

\bibitem[{\citenamefont{Wang and Cowley}(1988)}]{WC1988}
\bibinfo{author}{\bibnamefont{Wang}, \bibfnamefont{Z.~L.}}, and
  \bibinfo{author}{\bibfnamefont{J.~M.} \bibnamefont{Cowley}},
  \bibinfo{year}{1988}, \bibinfo{journal}{Surf.\ Sci.}
  \textbf{\bibinfo{volume}{193}}, \bibinfo{pages}{501}.

\bibitem[{\citenamefont{Ward} \emph{et~al.}(2006)\citenamefont{Ward, Notte, and
  Economou}}]{WNE06}
\bibinfo{author}{\bibnamefont{Ward}, \bibfnamefont{B.~W.}},
  \bibinfo{author}{\bibfnamefont{J.~A.} \bibnamefont{Notte}}, and
  \bibinfo{author}{\bibfnamefont{N.~P.} \bibnamefont{Economou}},
  \bibinfo{year}{2006}, \bibinfo{journal}{J.\ Vac.\ Sci.\ Technol.\ B}
  \textbf{\bibinfo{volume}{24}}, \bibinfo{pages}{2871}.

\bibitem[{\citenamefont{Warmack} \emph{et~al.}(1984)\citenamefont{Warmack,
  Becker, Anderson, Ritchie, Chu, Little, and Ferrell}}]{WBA1984}
\bibinfo{author}{\bibnamefont{Warmack}, \bibfnamefont{R.~J.}},
  \bibinfo{author}{\bibfnamefont{R.~S.} \bibnamefont{Becker}},
  \bibinfo{author}{\bibfnamefont{V.~E.} \bibnamefont{Anderson}},
  \bibinfo{author}{\bibfnamefont{R.~H.} \bibnamefont{Ritchie}},
  \bibinfo{author}{\bibfnamefont{Y.~T.} \bibnamefont{Chu}},
  \bibinfo{author}{\bibfnamefont{J.}~\bibnamefont{Little}}, and
  \bibinfo{author}{\bibfnamefont{T.~L.} \bibnamefont{Ferrell}},
  \bibinfo{year}{1984}, \bibinfo{journal}{Phys.\ Rev.\ B}
  \textbf{\bibinfo{volume}{29}}, \bibinfo{pages}{4375}.

\bibitem[{\citenamefont{Watanabe}(1956)}]{W1956}
\bibinfo{author}{\bibnamefont{Watanabe}, \bibfnamefont{H.}},
  \bibinfo{year}{1956}, \bibinfo{journal}{J.\ Phys.\ Soc.\ Jpn.}
  \textbf{\bibinfo{volume}{11}}, \bibinfo{pages}{112}.

\bibitem[{\citenamefont{Went} \emph{et~al.}(2008)\citenamefont{Went, Vos, and
  Werner}}]{WVW08}
\bibinfo{author}{\bibnamefont{Went}, \bibfnamefont{M.~R.}},
  \bibinfo{author}{\bibfnamefont{M.}~\bibnamefont{Vos}}, and
  \bibinfo{author}{\bibfnamefont{W.~S.~M.} \bibnamefont{Werner}},
  \bibinfo{year}{2008}, \bibinfo{journal}{Surf.\ Sci.}
  \textbf{\bibinfo{volume}{602}}, \bibinfo{pages}{2069}.

\bibitem[{\citenamefont{Werner}(2006)}]{W06}
\bibinfo{author}{\bibnamefont{Werner}, \bibfnamefont{W.~S.~M.}},
  \bibinfo{year}{2006}, \bibinfo{journal}{Appl.\ Phys.\ Lett.}
  \textbf{\bibinfo{volume}{89}}, \bibinfo{pages}{213106}.

\bibitem[{\citenamefont{Werner} \emph{et~al.}(2007)\citenamefont{Werner, Went,
  and Vos}}]{WWV07}
\bibinfo{author}{\bibnamefont{Werner}, \bibfnamefont{W.~S.~M.}},
  \bibinfo{author}{\bibfnamefont{M.~R.} \bibnamefont{Went}}, and
  \bibinfo{author}{\bibfnamefont{M.}~\bibnamefont{Vos}}, \bibinfo{year}{2007},
  \bibinfo{journal}{Surf.\ Sci.} \textbf{\bibinfo{volume}{601}},
  \bibinfo{pages}{L109}.

\bibitem[{\citenamefont{Williams} \emph{et~al.}(2000)\citenamefont{Williams,
  {L\'evy-Cl\'ement}, Albu-Yaron, Brun, and Colliex}}]{WLA00}
\bibinfo{author}{\bibnamefont{Williams}, \bibfnamefont{P.}},
  \bibinfo{author}{\bibfnamefont{C.}~\bibnamefont{{L\'evy-Cl\'ement}}},
  \bibinfo{author}{\bibfnamefont{A.}~\bibnamefont{Albu-Yaron}},
  \bibinfo{author}{\bibfnamefont{N.}~\bibnamefont{Brun}}, and
  \bibinfo{author}{\bibfnamefont{C.}~\bibnamefont{Colliex}},
  \bibinfo{year}{2000}, \bibinfo{journal}{J.\ Porous Mater.}
  \textbf{\bibinfo{volume}{7}}, \bibinfo{pages}{159}.

\bibitem[{\citenamefont{Winter}(2002)}]{W02}
\bibinfo{author}{\bibnamefont{Winter}, \bibfnamefont{H.}},
  \bibinfo{year}{2002}, \bibinfo{journal}{Phys.\ Rep.}
  \textbf{\bibinfo{volume}{367}}, \bibinfo{pages}{387}.

\bibitem[{\citenamefont{Winther and Alder}(1979)}]{WK1979}
\bibinfo{author}{\bibnamefont{Winther}, \bibfnamefont{A.}}, and
  \bibinfo{author}{\bibfnamefont{K.}~\bibnamefont{Alder}},
  \bibinfo{year}{1979}, \bibinfo{journal}{Nucl.\ Phys.\ A}
  \textbf{\bibinfo{volume}{319}}, \bibinfo{pages}{518}.

\bibitem[{\citenamefont{Woods} \emph{et~al.}(1995)\citenamefont{Woods, Walsh,
  Stoner, Kirk, and Fernow}}]{WWS95}
\bibinfo{author}{\bibnamefont{Woods}, \bibfnamefont{K.~J.}},
  \bibinfo{author}{\bibfnamefont{J.~E.} \bibnamefont{Walsh}},
  \bibinfo{author}{\bibfnamefont{R.~E.} \bibnamefont{Stoner}},
  \bibinfo{author}{\bibfnamefont{H.~G.} \bibnamefont{Kirk}}, and
  \bibinfo{author}{\bibfnamefont{R.~C.} \bibnamefont{Fernow}},
  \bibinfo{year}{1995}, \bibinfo{journal}{Phys.\ Rev.\ Lett.}
  \textbf{\bibinfo{volume}{74}}, \bibinfo{pages}{3808}.

\bibitem[{\citenamefont{Xu} \emph{et~al.}(1999)\citenamefont{Xu, Bjerneld,
  {K\"all}, and {B\"orjesson}}}]{XBK99}
\bibinfo{author}{\bibnamefont{Xu}, \bibfnamefont{H.}},
  \bibinfo{author}{\bibfnamefont{E.~J.} \bibnamefont{Bjerneld}},
  \bibinfo{author}{\bibfnamefont{M.}~\bibnamefont{{K\"all}}}, and
  \bibinfo{author}{\bibfnamefont{L.}~\bibnamefont{{B\"orjesson}}},
  \bibinfo{year}{1999}, \bibinfo{journal}{Phys.\ Rev.\ Lett.}
  \textbf{\bibinfo{volume}{83}}, \bibinfo{pages}{4357}.

\bibitem[{\citenamefont{Xu} \emph{et~al.}(2008)\citenamefont{Xu, Dong, and
  Zhang}}]{XDZ08}
\bibinfo{author}{\bibnamefont{Xu}, \bibfnamefont{J.}},
  \bibinfo{author}{\bibfnamefont{Y.}~\bibnamefont{Dong}}, and
  \bibinfo{author}{\bibfnamefont{X.}~\bibnamefont{Zhang}},
  \bibinfo{year}{2008}, \bibinfo{journal}{Phys.\ Rev.\ E}
  \textbf{\bibinfo{volume}{78}}, \bibinfo{pages}{046601}.

\bibitem[{\citenamefont{Xu and Zhang}(2008)}]{XZ08}
\bibinfo{author}{\bibnamefont{Xu}, \bibfnamefont{J.}}, and
  \bibinfo{author}{\bibfnamefont{X.}~\bibnamefont{Zhang}},
  \bibinfo{year}{2008}, \bibinfo{journal}{Phys.\ Lett.\ A}
  \textbf{\bibinfo{volume}{372}}, \bibinfo{pages}{1129}.

\bibitem[{\citenamefont{Yacobi and Holt}(1986)}]{HY1986}
\bibinfo{author}{\bibnamefont{Yacobi}, \bibfnamefont{B.~G.}}, and
  \bibinfo{author}{\bibfnamefont{D.~B.} \bibnamefont{Holt}},
  \bibinfo{year}{1986}, \bibinfo{journal}{J.\ Appl.\ Phys.}
  \textbf{\bibinfo{volume}{59}}, \bibinfo{pages}{R1}.

\bibitem[{\citenamefont{Yamaguti} \emph{et~al.}(2002)\citenamefont{Yamaguti,
  Inoue, Haeberl\'{e}, and Ohtaka}}]{YIH02}
\bibinfo{author}{\bibnamefont{Yamaguti}, \bibfnamefont{S.}},
  \bibinfo{author}{\bibfnamefont{J.}~\bibnamefont{Inoue}},
  \bibinfo{author}{\bibfnamefont{O.}~\bibnamefont{Haeberl\'{e}}}, and
  \bibinfo{author}{\bibfnamefont{K.}~\bibnamefont{Ohtaka}},
  \bibinfo{year}{2002}, \bibinfo{journal}{Phys.\ Rev.\ B}
  \textbf{\bibinfo{volume}{66}}, \bibinfo{pages}{195202}.

\bibitem[{\citenamefont{Yamamoto} \emph{et~al.}(2004)\citenamefont{Yamamoto,
  Sakakibara, Yano, Segawa, Shibata, Ishi, Ohsaka, Hara, Kondo, Miyazaki,
  Hinode, Matsuyama} \emph{et~al.}}]{YSY04}
\bibinfo{author}{\bibnamefont{Yamamoto}, \bibfnamefont{K.}},
  \bibinfo{author}{\bibfnamefont{R.}~\bibnamefont{Sakakibara}},
  \bibinfo{author}{\bibfnamefont{S.}~\bibnamefont{Yano}},
  \bibinfo{author}{\bibfnamefont{Y.}~\bibnamefont{Segawa}},
  \bibinfo{author}{\bibfnamefont{Y.}~\bibnamefont{Shibata}},
  \bibinfo{author}{\bibfnamefont{K.}~\bibnamefont{Ishi}},
  \bibinfo{author}{\bibfnamefont{T.}~\bibnamefont{Ohsaka}},
  \bibinfo{author}{\bibfnamefont{T.}~\bibnamefont{Hara}},
  \bibinfo{author}{\bibfnamefont{Y.}~\bibnamefont{Kondo}},
  \bibinfo{author}{\bibfnamefont{H.}~\bibnamefont{Miyazaki}},
  \bibinfo{author}{\bibfnamefont{F.}~\bibnamefont{Hinode}},
  \bibinfo{author}{\bibfnamefont{T.}~\bibnamefont{Matsuyama}}, \emph{et~al.},
  \bibinfo{year}{2004}, \bibinfo{journal}{Phys.\ Rev.\ E}
  \textbf{\bibinfo{volume}{69}}, \bibinfo{pages}{045601(R)}.

\bibitem[{\citenamefont{Yamamoto}
  \emph{et~al.}(2001{\natexlab{a}})\citenamefont{Yamamoto, Araya, and
  {Garc\'{\i}a de Abajo}}}]{paper064}
\bibinfo{author}{\bibnamefont{Yamamoto}, \bibfnamefont{N.}},
  \bibinfo{author}{\bibfnamefont{K.}~\bibnamefont{Araya}}, and
  \bibinfo{author}{\bibfnamefont{F.~J.} \bibnamefont{{Garc\'{\i}a de Abajo}}},
  \bibinfo{year}{2001}{\natexlab{a}}, \bibinfo{journal}{Phys.\ Rev.\ B}
  \textbf{\bibinfo{volume}{64}}, \bibinfo{pages}{205419}.

\bibitem[{\citenamefont{Yamamoto}
  \emph{et~al.}(2001{\natexlab{b}})\citenamefont{Yamamoto, Araya, Toda, and
  Sugiyama}}]{YAT01}
\bibinfo{author}{\bibnamefont{Yamamoto}, \bibfnamefont{N.}},
  \bibinfo{author}{\bibfnamefont{K.}~\bibnamefont{Araya}},
  \bibinfo{author}{\bibfnamefont{A.}~\bibnamefont{Toda}}, and
  \bibinfo{author}{\bibfnamefont{H.}~\bibnamefont{Sugiyama}},
  \bibinfo{year}{2001}{\natexlab{b}}, \bibinfo{journal}{Surf.\ Interface\
  Anal.} \textbf{\bibinfo{volume}{31}}, \bibinfo{pages}{79–}.

\bibitem[{\citenamefont{Yamamoto}
  \emph{et~al.}(2006{\natexlab{a}})\citenamefont{Yamamoto, Bhunia, and
  Watanabe}}]{YBW06}
\bibinfo{author}{\bibnamefont{Yamamoto}, \bibfnamefont{N.}},
  \bibinfo{author}{\bibfnamefont{S.}~\bibnamefont{Bhunia}}, and
  \bibinfo{author}{\bibfnamefont{Y.}~\bibnamefont{Watanabe}},
  \bibinfo{year}{2006}{\natexlab{a}}, \bibinfo{journal}{Appl.\ Phys.\ Lett.}
  \textbf{\bibinfo{volume}{88}}, \bibinfo{pages}{153106}.

\bibitem[{\citenamefont{Yamamoto}
  \emph{et~al.}(2006{\natexlab{b}})\citenamefont{Yamamoto, Nakano, and
  Suzuki}}]{YNS06}
\bibinfo{author}{\bibnamefont{Yamamoto}, \bibfnamefont{N.}},
  \bibinfo{author}{\bibfnamefont{M.}~\bibnamefont{Nakano}}, and
  \bibinfo{author}{\bibfnamefont{T.}~\bibnamefont{Suzuki}},
  \bibinfo{year}{2006}{\natexlab{b}}, \bibinfo{journal}{Surf.\ Interface\
  Anal.} \textbf{\bibinfo{volume}{38}}, \bibinfo{pages}{1725–}.

\bibitem[{\citenamefont{Yamamoto}
  \emph{et~al.}(1996{\natexlab{a}})\citenamefont{Yamamoto, Sugiyama, and
  Toda}}]{YST96}
\bibinfo{author}{\bibnamefont{Yamamoto}, \bibfnamefont{N.}},
  \bibinfo{author}{\bibfnamefont{H.}~\bibnamefont{Sugiyama}}, and
  \bibinfo{author}{\bibfnamefont{A.}~\bibnamefont{Toda}},
  \bibinfo{year}{1996}{\natexlab{a}}, \bibinfo{journal}{Proc.\ R.\ Soc.\
  London,\ Ser.\ A} \textbf{\bibinfo{volume}{452}}, \bibinfo{pages}{2279}.

\bibitem[{\citenamefont{Yamamoto and Suzuki}(2008)}]{YS08}
\bibinfo{author}{\bibnamefont{Yamamoto}, \bibfnamefont{N.}}, and
  \bibinfo{author}{\bibfnamefont{T.}~\bibnamefont{Suzuki}},
  \bibinfo{year}{2008}, \bibinfo{journal}{Appl.\ Phys.\ Lett.}
  \textbf{\bibinfo{volume}{93}}, \bibinfo{pages}{093114}.

\bibitem[{\citenamefont{Yamamoto}
  \emph{et~al.}(1996{\natexlab{b}})\citenamefont{Yamamoto, Toda, and
  Araya}}]{YTA96}
\bibinfo{author}{\bibnamefont{Yamamoto}, \bibfnamefont{N.}},
  \bibinfo{author}{\bibfnamefont{A.}~\bibnamefont{Toda}}, and
  \bibinfo{author}{\bibfnamefont{K.}~\bibnamefont{Araya}},
  \bibinfo{year}{1996}{\natexlab{b}}, \bibinfo{journal}{J.\ Electron\ Microsc.}
  \textbf{\bibinfo{volume}{45}}, \bibinfo{pages}{64}.

\bibitem[{\citenamefont{Yurtsever} \emph{et~al.}(2008)\citenamefont{Yurtsever,
  Couillard, and Muller}}]{YCM08}
\bibinfo{author}{\bibnamefont{Yurtsever}, \bibfnamefont{A.}},
  \bibinfo{author}{\bibfnamefont{M.}~\bibnamefont{Couillard}}, and
  \bibinfo{author}{\bibfnamefont{D.~A.} \bibnamefont{Muller}},
  \bibinfo{year}{2008}, \bibinfo{journal}{Phys.\ Rev.\ Lett.}
  \textbf{\bibinfo{volume}{100}}, \bibinfo{pages}{217402}.

\bibitem[{\citenamefont{Zabala and Echenique}(1990)}]{ZE1990}
\bibinfo{author}{\bibnamefont{Zabala}, \bibfnamefont{N.}}, and
  \bibinfo{author}{\bibfnamefont{P.~M.} \bibnamefont{Echenique}},
  \bibinfo{year}{1990}, \bibinfo{journal}{Ultramicroscopy}
  \textbf{\bibinfo{volume}{32}}, \bibinfo{pages}{327}.

\bibitem[{\citenamefont{Zabala} \emph{et~al.}(1989)\citenamefont{Zabala,
  Rivacoba, and Echenique}}]{ZRE1989}
\bibinfo{author}{\bibnamefont{Zabala}, \bibfnamefont{N.}},
  \bibinfo{author}{\bibfnamefont{A.}~\bibnamefont{Rivacoba}}, and
  \bibinfo{author}{\bibfnamefont{P.~M.} \bibnamefont{Echenique}},
  \bibinfo{year}{1989}, \bibinfo{journal}{Surf.\ Sci.}
  \textbf{\bibinfo{volume}{209}}, \bibinfo{pages}{465}.

\bibitem[{\citenamefont{Zabala} \emph{et~al.}(1997)\citenamefont{Zabala,
  Rivacoba, and Echenique}}]{ZRE97}
\bibinfo{author}{\bibnamefont{Zabala}, \bibfnamefont{N.}},
  \bibinfo{author}{\bibfnamefont{A.}~\bibnamefont{Rivacoba}}, and
  \bibinfo{author}{\bibfnamefont{P.~M.} \bibnamefont{Echenique}},
  \bibinfo{year}{1997}, \bibinfo{journal}{Phys.\ Rev.\ B}
  \textbf{\bibinfo{volume}{56}}, \bibinfo{pages}{7623}.

\bibitem[{\citenamefont{Zaremba}(1985)}]{Z1985}
\bibinfo{author}{\bibnamefont{Zaremba}, \bibfnamefont{E.}},
  \bibinfo{year}{1985}, \bibinfo{journal}{Surf.\ Sci.}
  \textbf{\bibinfo{volume}{151}}, \bibinfo{pages}{91}.

\bibitem[{\citenamefont{Zhang} \emph{et~al.}(2008)\citenamefont{Zhang, Erni,
  Verbeeck, and {Van Tendeloo}}}]{ZEV08}
\bibinfo{author}{\bibnamefont{Zhang}, \bibfnamefont{L.}},
  \bibinfo{author}{\bibfnamefont{R.}~\bibnamefont{Erni}},
  \bibinfo{author}{\bibfnamefont{J.}~\bibnamefont{Verbeeck}}, and
  \bibinfo{author}{\bibfnamefont{G.}~\bibnamefont{{Van Tendeloo}}},
  \bibinfo{year}{2008}, \bibinfo{journal}{Phys.\ Rev.\ B}
  \textbf{\bibinfo{volume}{77}}, \bibinfo{pages}{195119}.

\bibitem[{\citenamefont{Zhao and L\"u}(2008)}]{ZL08}
\bibinfo{author}{\bibnamefont{Zhao}, \bibfnamefont{Z.}}, and
  \bibinfo{author}{\bibfnamefont{B.}~\bibnamefont{L\"u}}, \bibinfo{year}{2008},
  \bibinfo{journal}{Opt.\ Quant.\ Electr.} \textbf{\bibinfo{volume}{40}},
  \bibinfo{pages}{615}.

\bibitem[{\citenamefont{Zheludev}(2008)}]{Z08}
\bibinfo{author}{\bibnamefont{Zheludev}, \bibfnamefont{N.~I.}},
  \bibinfo{year}{2008}, \bibinfo{journal}{Nat.\ Mater.}
  \textbf{\bibinfo{volume}{7}}, \bibinfo{pages}{420}.

\bibitem[{\citenamefont{Zia} \emph{et~al.}(2006)\citenamefont{Zia, Schuller,
  Chandran, and Brongersma}}]{ZSC06}
\bibinfo{author}{\bibnamefont{Zia}, \bibfnamefont{R.}},
  \bibinfo{author}{\bibfnamefont{J.~A.} \bibnamefont{Schuller}},
  \bibinfo{author}{\bibfnamefont{A.}~\bibnamefont{Chandran}}, and
  \bibinfo{author}{\bibfnamefont{M.~L.} \bibnamefont{Brongersma}},
  \bibinfo{year}{2006}, \bibinfo{journal}{Mater.\ Today}
  \textbf{\bibinfo{volume}{9}}, \bibinfo{pages}{20}.

\bibitem[{\citenamefont{Ziegler}(1999)}]{Z99}
\bibinfo{author}{\bibnamefont{Ziegler}, \bibfnamefont{J.~F.}},
  \bibinfo{year}{1999}, \bibinfo{journal}{J.\ Appl.\ Phys.}
  \textbf{\bibinfo{volume}{85}}, \bibinfo{pages}{1249}.

\bibitem[{\citenamefont{Zrelov and {Ru\v{z}i\v{c}ka}}(1989)}]{rmp-cr1}
\bibinfo{author}{\bibnamefont{Zrelov}, \bibfnamefont{V.~P.}}, and
  \bibinfo{author}{\bibfnamefont{J.}~\bibnamefont{{Ru\v{z}i\v{c}ka}}},
  \bibinfo{year}{1989}, \bibinfo{journal}{Czech.\ J.\ Phys.\ Sect.\ B}
  \textbf{\bibinfo{volume}{39}}, \bibinfo{pages}{368}.

\bibitem[{\citenamefont{Zuloaga} \emph{et~al.}(2009)\citenamefont{Zuloaga,
  Prodan, and Nordlander}}]{ZPN09}
\bibinfo{author}{\bibnamefont{Zuloaga}, \bibfnamefont{J.}},
  \bibinfo{author}{\bibfnamefont{E.}~\bibnamefont{Prodan}}, and
  \bibinfo{author}{\bibfnamefont{P.}~\bibnamefont{Nordlander}},
  \bibinfo{year}{2009}, \bibinfo{journal}{Nano\ Lett.}
  \textbf{\bibinfo{volume}{9}}, \bibinfo{pages}{887}.

\end{thebibliography}

\end{document}